\documentclass[a4paper,twoside,bindnopdf,ams,booktabs]{myhepthesis}

\usepackage{amsfonts,amssymb,amsmath,amsthm,bm,euscript,array,longtable,ltxtable,cite,lscape,hhline,
indentfirst,framed}

\usepackage{ifpdf}
\ifpdf
\usepackage{graphicx}
\else
\usepackage[dvips]{epsfig,color}
\fi
\usepackage[all]{xy}
\usepackage[symbol]{footmisc}
\usepackage{thesis}

\usepackage[utf8]{inputenc}
\usepackage[english,greek]{babel}

\def\en#1{\textlatin{#1}}
\newcommand{\sen}{\selectlanguage{english}}

\sen
\definethesis%
{{\Large {\sc Dimensional Reduction\\ of\\[0.7ex] Supersymmetric Gauge Theories}}}
{Dr. Theodoros Grammatikopoulos\\
\vskip 0.3cm
{\small
National Technical University of Athens\\
School of Applied Sciences\\
Physics Department}
}
\newcommand{\be}{\begin{equation}}
\newcommand{\ee}{\end{equation}}
\newcommand{\beq}{\begin{equation}}
\newcommand{\eeq}{\end{equation}}
\newcommand{\beqnn}{\begin{equation*}}
\newcommand{\eeqnn}{\end{equation*}}
\newcommand{\beqa}{\begin{eqnarray}}
\newcommand{\eeqa}{\end{eqnarray}}
\newcommand{\eq}[1]{(\ref{#1})}

\def\nn{\nonumber}
\def\bea{\begin{eqnarray}}
\def\eea{\end{eqnarray}}

\def\a{\alpha} 
\def\b{\beta} 
\def\c{\gamma} 
\def\d{\delta}
\def\e{\epsilon} \def\vare{\varepsilon}
\def\f{\phi} \def\F{\Phi}\def\vf{\varphi}
\def\h{\eta}
\def\k{\kappa}
\def\l{\lambda} \def\L{\Lambda} 
\def\m{\mu}
\def\n{\nu}
\def\o{\omega}\def\O{\Omega}
\def\p{\pi} 

\def\s{\sigma}\def\S{\Sigma}
\def\t{\tau}
\def\th{\theta}

\def\z{\zeta}
\def\cA{{\cal A}}

\def\cC{{\cal C}}
\def\cD{{\cal D}}

\def\cH{{\cal H}}
\def\cI{{\cal I}}
\def\cJ{{\cal J}}

\def\cL{{\cal L}}

\def\cN{{\cal N}}
\def\cO{{\cal O}}
\def\cP{{\cal P}}

\def\cR{{\cal R}}
\def\cS{{\cal S}}

\def\cU{{\cal U}}

\def\cX{{\cal X}}

\def\cZ{{\cal Z}}

\def\R{{\mathbb R}}

\def\N{{\mathbb N}}
\def\Z{{\mathbb Z}}

\def\one{\mbox{1 \kern-.59em {\rm \en{l}}}}

\def\mg{\mathfrak{g}}
\def\mmu{\mathfrak{u}}
\def\msu{\mathfrak{su}}

\def\mso{\mathfrak{so}}
\def\mk{\mathfrak{K}}

\def\adj{{\rm adj}}
\def\Der{{\rm Der}}
\def\dim{{\rm dim}}
\def\Curv{{\rm Curv}}
\def\J{{\rm J}}
\def\rank{{\rm rank}}

\def\tld#1{\tilde{#1}}
\newcommand{\Tr}{{\rm Tr}}

\newcommand{\NN}{{\rm N}}

\def\Xwhat{\widehat{X}}

\def\sD{\scriptstyle{D}}
\def\ssM{\scriptscriptstyle{M}}
\def\ssN{\scriptscriptstyle{N}}
\def\ssK{\scriptscriptstyle{K}}
\def\ssL{\scriptscriptstyle{\L}}

\def\ssX{\scriptscriptstyle{X}}

\def\obar#1{\overline {#1}}
\def\mbf#1{\mathbf {#1}}
\def\bb#1{\mathbb {#1}}
\def\lra{{\leftrightarrow}}
\def\ep{i\epsilon}

\newcommand\prt{\partial}
\newcommand\wdg{\wedge}

\def\vev#1{\langle #1 \rangle} 

\def\HH{{\rm H}}                     

\def\AAA{SU(4)}
\def\AA{SU(3)}
\def\A#1{SU^{#1}(2)}
\def\Y{U(1)}

\def\ubr#1{\underbrace{#1}}

\def\kbar{{\mathchar'26\mkern-9muk}}            
\def\bit{\begin{itemize}}
\def\eit{\end{itemize}}
\def\({\left(} \def\){\right)}
\def\dsum{\oplus}\def\tens{\otimes} 
 
\def\rep{rep.}

\begin{document}
\begin{frontmatter}
\sen
\thesistitlepage%
{epeaek-2}
{National Technical University of Athens\\
School of Applied Sciences\\[0.3ex]
Physics Department}
{EMP-logo}
{A dissertation submitted to the National Technical University of Athens\\
for the degree of Doctor of Philosophy in Physics.
}
{~}
\begin{abstract}

Main objective of the present dissertation is the determination of all the possible low energy
models which emerge in four dimensions by the dimensional reduction of a gauge theory over multiple
connected coset spaces. The higher dimensional gauge theory is chosen to be the one that the
Heterotic string theory suggests: (i)~it is defined in ten dimensions, (ii)~it is based on the
$E_{8}\times E_{8}$ symmetry group and (iii)~it is $\cN=1$ globally supersymmetric. The search of
all four-dimensional gauge theories resulting from the aforementioned dimensional reduction, is
restricted only to models which are potentially interesting from a phenomenological point of view.
This requirement constrain these models to come from one of the known Grand Unified Theories (GUTs)
in an intermediate stage of the spontaneous symmetry breaking. Main result of my study is that
extensions of the Standard Model (SM) which are based on the Pati-Salam group structure can
be obtained in four dimensions.

A second direction of research which is discussed in this dissertation is based on the following
conclusions of a previous research work: (i)~It is possible to obtain four-dimensional theories with
a non-abelian gauge symmetry by the dimensional reduction of a higher dimensional $U(1)$
noncommutative theory and (ii)~the particle physics models resulting from this particular
dimensional reduction are renormalizable. The objective of the present dissertation in this
direction is the study of the last remark. Starting with the most general renormalizable gauge
theory with scalar fields (consistent with the dimensional reduction over a fuzzy sphere) which can
be defined in four dimensions, it turns out that fuzzy extra dimensions emerge dynamically. This is
supported by the calculation of the spectrum of vector and scalar bosons. In this way, the
renormalizability of the four-dimensional low energy models resulting from the dimensional reduction
over fuzzy coset spaces is verified.

Finally, it is extremely interesting to assume noncommutative characteristics not only for the
internal space of a higher dimensional theory, but also for the four-dimensional Minkowski space,
$M^{4}$, providing that these appear only in an elementary length scale which is assumed to be the
Planck length. In this framework, the approach of linearized noncommutative gravity was examined and
the connection of the algebra describing the noncommutative space with its geometry was
studied. It turned out that the linear perturbation which describes the noncommutativity of the
algebra contributes to the Ricci curvature tensor in a non-trivial way. This conclusion suggests a
possible fundamental connection between the noncommutative geometry and the theory of gravity, which
have to be investigated further.
\end{abstract}
\cleardoublepage
\begin{center}
{\large Theodoros Grammatikopoulos}\\
\vspace*{1.5cm}
{\Large \sc{Dimensional Reduction\\of\\[1ex]Supersymmetric Gauge Theories}}
\end{center}
\vspace*{0.8cm}
A dissertation submitted to the Department of Physics of the National Technical University of Athens in
partial fulfillment of the requirements for the degree of Doctor of Philosophy in Physics.
\vspace*{1cm}
\begin{center}
\begin{tabular}{ll}
\textbf{Thesis supervisor:}                           &\begin{tabular}[t]{l}
							Prof. George Zoupanos
							\end{tabular}\\[1ex]
\textbf{Thesis committee:}
                                                      &\begin{tabular}[t]{l}
                                                         Prof. George Zoupanos\\
                                                         Prof. Ioannis Bakas\\
                                                         Assoc. Prof. George Koutsoumbas\\
						         Prof. Athanasios Lahanas\\
                                                         Assoc. Prof. Nikos Tracas\\
                                                         Assist. Prof. Alex Kehagias\\
						         Assist. Prof. Konstantinos
                                                                        Anagnostopoulos
                                                        \end{tabular}
\end{tabular}\\
\vspace*{2.5cm}
Athens, December 2007
\end{center}
\cleardoublepage
\vspace*{3cm}
\begin{center}
{\Large{\textbf{Acknowledgments}}}
\end{center}
\vspace*{1cm}%

I would like to thank my supervisor Prof. George Zoupanos for his guidance and support during my Ph.D.
studies. I would also like to thank the members of my advisory committee, Prof. Ioannis Bakas and  Assoc.
Prof. George Koutsoumbas, and the members of my evaluation committee, Prof. Athanasios Lahanas and Assoc.
Profs. Nikos Tracas, Alex Kehagias and Konstantinos Anagnostopoulos. I am also grateful to all my teachers
during my M.Sc. and Ph.D. time.

I also thank Profs. {John Madore} and {Maja Buric} for their collaboration and guidance on the interesting
problems of the noncommutative differential geometry and gravity.

I thank Drs. {Harold Steinacker} and {Paolo Aschieri} for their collaboration and guidance on the study of
Matrix models over fuzzy spaces.

I thank the Ph.D. student George Douzas for his collaboration in the classification of the
CSDR-resulting low-energy models.

I thank Dr. Pantelis Manousselis and the Ph.D. student Athanasios Chatzistavrakidis for useful discussions on
CSDR-related topics.

Finally, I thank my parents Grigorios and Irini and my brother Gerasimos for their constant support
throughout these years.

\vskip 1.3cm
\emph{
The project was supported by the EPEAEK II program ``Irakleitos'' and co-funded by the European
Unions ($75\%$) and 
the Hellenic state ($25\%$).
}


\tableofcontents
\end{frontmatter}
\begin{mainmatter}
%
%
%
%
%
\selectlanguage{english}
\setcounter{chapter}{0}
\chapter{Introduction}

The quest for unification of all observed interactions at low energies has been going on for many
years. The successful unification of electromagnetic and weak interactions was achieved of the
semisimple gauge group $SU(2)\times U(1)$, which was spontaneously broken to $U(1)$ at a scale of
$\cO(100 GeV)$~\cite{Weinberg:1967tq,Glashow:1970gm}. For the spontaneous symmetry
breaking of the theory, a scalar field sector was introduced in an ad hoc manner. An attractive
framework to unify the strong and the electroweak interactions is provided by Grand Unified Theories
(GUTs), which make use of a simple gauge group, e.g. $SU(5)$,
$SO(10)$, $E_{6}$~\cite{Georgi:1974sy,Fritzsch:1974nn,Mohapatra:1985xm}. In GUTs a further scale of
$\cO(10^{15} GeV)$, related to the superstrong symmetry breaking of the theory, had to be introduced
in addition to electroweak symmetry breaking scale. Then the scalar sector had to be enlarged even
further. In addition a new complication appeared, the so-called hierarchy problem, related
to the huge difference between the two scales~\cite{Weinberg:1978ym}.

Although the GUT unification scale is not very far from the Planck scale, GUTs do not incorporate
the gravitational interactions. On the other hand, the earliest unification attempts of Kaluza and
Klein~\cite{Kaluza:1921tu,Klein:1926tv} included gravity and electromagnetism, which were the
established interactions at that time. At the heart of the Kaluza-Klein scheme lies the assumption
that space-time has more than four dimensions, which has been considered too speculative.

The Kaluza-Klein proposal was to reduce a pure gravity theory from five dimensions to four, which
led to a $U(1)$ gauge theory, identified with electromagnetism, coupled to gravity. A revival of
interest in the Kaluza-Klein scheme started after the
realisation~\cite{Bailin:1987jd,Salam:1981xd} that non-abelian gauge groups appear naturally when
one further extend the space-time dimensions. With the assumption that the total space-time manifold
can be written as a direct product $M^{D}=M^{4}\times B$, where $B$ is a compact Riemannian space
with a non-abelian isometry group $S$,  dimensional reduction of the theory leads to gravity coupled
to a Yang-Mills theory with a gauge group containing $S$ and scalars in four dimensions. The
main advantage of this picture is the geometrical unification of gravity with the other interactions
and also the explanation of gauge symmetries. There are, however, some problems in the Kaluza-Klein
framework. One of the most serious obstacle to obtaining a realistic model of the low-energy
interactions seems to be that after adding fermions to the original action it is impossible to
get chiral fermions in four dimensions~\cite{Witten:1983ux}. If, however, one adds suitable matter
fields to the original action - in particular Yang-Mills fields - then one can have massless
fermions and parity violation in the fermion sector~\cite{Horvath:1977st,Chapline:1982wy}. Thus one
is led to introduce Yang-Mills fields in higher dimensions. In fact, in some other popular schemes
such as supergravity~\cite{Duff:1986hr} and superstring theories~\cite{Green:1987sp,Green:1987mn}
the Einstein-Yang-Mills theory appears in the bosonic sector.

It is a common belief that the effects of gravity are negligible for low-energy phenomena.
Therefore, inasmuch as one is interested in describing only the low-energy interactions, one can
take the bold step to neglect gravity altogether, assuming, however, the direct product of
space-time $M^{D}=M^{4}\times B$. Then one starts with a Yang-Mills theory defined on
$M^{D}=M^{4}\times B$, yielding a Yang-Mills-Higgs theory in four dimensions. This provides a
potential unification of low-energy interactions as well as of gauge and Higgs fields. A naive and
crude way to fulfil this requirement is to discard the field dependence on the extra coordinates. A
more elegant one is to allow for a non-trivial dependence on them, but impose the condition that a
symmetry transformation by an element of the isometry group $S$ of the space formed by the extra
dimensions $B$ corresponds to a gauge transformation. Then the Lagrangian will be independent of the
extra coordinates just because it is gauge invariant. This is the basis of the CSDR
scheme~\cite{Forgacs:1979zs,Kapetanakis:1992hf,Kubyshin:1989vd}, which assumes
that $B$ is a compact coset space, $S/R$. The requirement that transformations of the fields under
action of the symmetry group of $S/R$ are compensated by gauge transformations, leads to certain
constraints on the fields.

It is worth recalling that the Coset Space Dimensional Reduction
(CSDR)~\cite{Forgacs:1979zs,Kapetanakis:1992hf,Kubyshin:1989vd} was suggesting from the beginning
that a unification of the gauge and Higgs sectors can be achieved in higher dimensions.
Phenomenologically interesting GUTs which are obtained by the application of the CSDR method have
been reported in~\cite{Kapetanakis:1992hf}. However their surviving scalars transform in the
fundamental of the resulting gauge group and are not suitable for the superstrong symmetry breaking
towards the SM. As a way out to it has been suggested~\cite{Zoupanos:1987wj,Kapetanakis:1992hf} to
take advantage of non-trivial topological properties of the extra compactification coset space,
apply the Hosotani or Wilson flux breaking
mechanism~\cite{Hosotani:1983xw,Hosotani:1983vn,Witten:1985xc} and break the
gauge symmetry of the theory further~\cite{Kapetanakis:1989gd,Kapetanakis:1992hf}. The main
objective of my work is the investigation to which extent applying both methods namely CSDR and
Wilson flux breaking mechanism, one can obtain reasonable low-energy models.

In chapter~\ref{CSDR}  I present the CSDR scheme in sufficient detail to make the dissertation
self-contained and give a simple example of the method. Namely, I discuss the dimensional reduction 
of an $\cN=1$ Yang-Mills-Dirac theory over the six-dimensional sphere $S^{6}$. In
chapter~\ref{sec:WilsonFlux-theory} I recall the Wilson flux breaking mechanism and make some
important remarks in order to apply the method on models resulting from dimensional reduction.
In chapter~\ref{Classification} starting with an $\cN=1$, $E_{8}$ Yang-Mills-Dirac theory defined
in ten dimensions, I classify the low-energy models resulting from CSDR and a subsequent
application of Wilson flux spontaneous symmetry breaking. The space-time on which the theory is
defined can be written in the compactified form $M^{4}\times B$, with $M^{4}$ the ordinary Minkowski
spacetime and $B=S/R$ a six-dimensional homogeneous coset space. I constrain my investigation in
those cases that the dimensional reduction leads to phenomenologically interesting and anomaly-free
four-dimensional GUTs such as $E_{6}$, $SO(10)$. In
chapters~\ref{NC-generalisations-motivation}\,-\,\ref{FuzzyExtraDim} I discuss noncommutative
generalisations of the CSDR scheme with emphasis on the renormalizability of the emergent
four-dimensional theories. Finally, in chapter~\ref{linPhD} I present some aspects of the more
adventurous assumption of promoting the ordinary spacetime to a noncommutative `manifold'.

\chapter{Coset Space Dimensional Reduction}
\label{CSDR}

The celebrated Standard Model (SM) of Elementary Particle Physics had so far outstanding successes
in all its confrontations with experimental results. However the apparent success of the SM is
spoiled by the presence of a plethora of free parameters mostly related to the ad-hoc introduction
of the Higgs and Yukawa sectors in the theory. It is worth recalling that the Coset Space
Dimensional Reduction (CSDR)~\cite{Forgacs:1979zs,Kapetanakis:1992hf,Kubyshin:1989vd} was suggesting
from the beginning that a unification of the gauge and Higgs sectors can be achieved in higher
dimensions. The four-dimensional gauge and Higgs fields are simply the surviving components of the
gauge fields of a pure gauge theory defined in higher dimensions. Here, I describe the
CSDR scheme giving emphasis on the construction of symmetric fields and discuss the
resulting four-dimensional theory. For various details
consult~\cite{Kapetanakis:1992hf,Castellani:1999fz} whereas the consistency of the method have been
proven in~\cite{Chatzistavrakidis:2007by}.

\section{Introduction}

In the CSDR scheme one assume a gauge theory defined over a higher dimensional space which is
compactified in the form of $M^{4}\times (S/R)$, where $S/R$ a compact coset space. Basis of the
theory is to allow the higher dimensional fields to have a non-trivial dependence on the extra
coordinates.  This is realised by imposing the condition that a symmetry transformation by
an element of the isometry group $S$ of the space formed by the extra dimensions $B$ corresponds to
a gauge transformation. Then the Lagrangian will be independent of the extra coordinates just
because it is gauge invariant. The theory provides a gauge-Higgs unification with the
four-dimensional gauge and Higgs fields to be simply the surviving components of the gauge
fields of a pure gauge theory defined in higher dimensions. The introduction of
fermions~\cite{Manton:1981es} was a major development. Then the four-dimensional Yukawa and gauge
interactions of fermions found also a unified description in the gauge interactions of the higher
dimensional theory. Recent improvement in this unified description in high dimensions is to relate
the gauge and fermion fields that have been
introduced~\cite{Manousselis:2000aj,Manousselis:2001xb,Manousselis:2001re}.
A simple way to achieve that is to demand that the higher dimensional gauge theory is\ ${\cal N}= 1$
supersymmetric which requires that the gauge and fermion fields are members of the same
supermultiplet. An additional strong argument towards higher dimensional supersymmetry including
gravity comes from the stability of the corresponding compactifying solutions that lead to the
four-dimensional theory.

In the spirit described above a very welcome additional input is that string theory suggests
furthermore the dimension and the gauge group of the higher dimensional supersymmetric
theory~\cite{Green:1987sp,Green:1987mn,Lust:1989tj}. Further support to this unified description
comes from the fact that the reduction of the theory over coset~\cite{Kapetanakis:1992hf} and CY
spaces~\cite{Green:1987sp,Green:1987mn,Lust:1989tj} provides the four-dimensional theory with
scalars belonging in the fundamental representation (rep.) of the gauge group as are introduced in
the SM. In addition the fact that the SM is a chiral theory lead us to consider $D$-dimensional
supersymmetric gauge theories with $D=4n+2$ \cite{Chapline:1982wy,Kapetanakis:1992hf}, which include
the ten dimensions suggested by the heterotic string
theory~\cite{Green:1987sp,Green:1987mn,Lust:1989tj}.

Concerning supersymmetry, the nature of the four-dimensional theory depends on the corresponding
nature of the compact space used to reduce the higher dimensional theory. Specifically the reduction
over CY spaces leads to supersymmetric
theories~\cite{Green:1987sp,Green:1987mn,Lust:1989tj} in four dimensions, the reduction over
symmetric coset spaces leads to non-supersymmetric theories, while a reduction over non-symmetric
ones leads to softly broken supersymmetric
theories~\cite{Manousselis:2000aj,Manousselis:2001xb,Manousselis:2001re}.

In section~\ref{sec:CSDR} I present the CSDR scheme in sufficient detail to make the
dissertation self-contained. I especially discuss some elements of the coset space geometry, the
reduction of a higher dimensional Yang-Mills-Dirac action and the constraints that the surviving
fields have to fulfil. I finally make some remarks on the four-dimensional Lagrangian. In
section~\ref{UnifiedTheoriesHigherDims-Example} an example of the method is given. I describe
the dimensional reduction of an $\cN=1$, $E_{8}$ Yang-Mills-Dirac theory over the six-dimensional
sphere $SO(7)/SO(6)\sim S^{6}$.

\section{Coset Space Dimensional Reduction\label{sec:CSDR}}

Given a gauge theory defined in higher dimensions the obvious way to dimensionally reduce it is to
demand that the field dependence on the extra coordinates is such that the Lagrangian is independent
of them. A crude way to fulfil this requirement is to discard the field dependence on the extra
coordinates, while an elegant one is to allow for a non-trivial dependence on them, but impose
the condition that a symmetry transformation by an element of the isometry group $S$ of the space
formed by the extra dimensions $B$ corresponds to a gauge transformation. Then the Lagrangian will
be independent of the extra coordinates just because it is gauge invariant. This is the basis of the
CSDR scheme~\cite{Forgacs:1979zs,Kapetanakis:1992hf,Kubyshin:1989vd}, which assumes
that $B$ is a compact coset space, $S/R$.

In the CSDR scheme one starts with a Yang-Mills-Dirac Lagrangian, with gauge group $G$, defined on a
$D$-dimensional space-time $M^{D}$, with metric $g^{MN}$, which is compactified to
$ M^{4}\times S/R$ with $S/R$ a coset space. The metric is assumed to have the form
\begin{equation}
g^{MN}=
\left(\begin{array}{cc}\eta^{\mu\nu}&0\\0&-g^{ab}\end{array}
\right),\label{Compactified-metric}
\end{equation}
where $\eta^{\mu\nu}= diag(1,-1,-1,-1)$ and $g^{ab}$ is the coset space metric. The requirement that
transformations of the fields under the action of the symmetry group of $S/R$ are compensated by
gauge transformations leads to certain constraints on the fields. The solution of these constraints
provides the four-dimensional unconstrained fields as well as the gauge invariance
that remains in the theory after dimensional reduction. Therefore a potential unification of all
low-energy interactions, gauge, Yukawa and Higgs is achieved, which was the first motivation of this
framework.

It is interesting to note that the fields obtained using the CSDR approach are the first terms in 
the expansion of the $D$-dimensional fields in harmonics of the internal space $B$. The effective
field theories resulting from compactification of higher dimensional theories contain also towers of
massive higher harmonics (Kaluza-Klein) excitations, whose contributions at the quantum level alter
the behaviour of the running couplings from logarithmic to power~\cite{Taylor:1988vt}. As a result
the traditional picture of unification of couplings may change drastically~\cite{Dienes:1998vg}.
Higher dimensional theories have also been studied at the quantum level using the continuous Wilson
renormalisation group~\cite{Kobayashi:1998ye,Kubo:1999ua} which can be formulated in any number of
space-time dimensions with results in agreement with the treatment involving massive Kaluza-Klein
excitations.

\subsection{Coset space geometry}
\label{sec:CosetSpaceGeometry}

Here I recall some aspects of the coset space geometry. One can divide the generators of $S$, $
Q_{A}$ in two sets: the generators of $R$, $Q_{i}$ $(i=1, \ldots,\dim (R))$, and the generators of
$S/R$, $ Q_{a}$($a=\dim R+1 \ldots,\dim (S))$, and $\dim(S/R)=\dim (S)-\dim (R) =d$. Then the
commutation relations for the generators of $S$ are the following
\begin{subequations}
\begin{align}
&\left[ Q_{i},Q_{j} \right] = f_{ij}{}^{k} Q_{k}\,, \\
&\left[ Q_{i},Q_{a} \right]= f_{ia}{}^{b}Q_{b}\,,\\
&\left[ Q_{a},Q_{b} \right]=f_{ab}{}^{i}Q_{i}+f_{ab}{}^{c}Q_{c}\,.\label{S/R-comm-rels}
\end{align}
\end{subequations}

So $S/R$ is assumed to be a reductive but in general non-symmetric coset space. When $S/R$ is
symmetric, the $f_{ab}{}^{c}$ in~(\ref{S/R-comm-rels}) vanish.

The above splitting of the $S$ generators can be characterised by determining the decomposition of
the adjoint rep. of $S$ under $R$.
\begin{eqnarray}
S&\supset&R\\
\adj(S)&=&\adj(R)+{\rm {v}}\,,
\end{eqnarray}
where ${\rm {v}}$ corresponds to the coset generators.

The tangent space vectors transform under rotations of $SO(d)$ among themselves. Actually the
generators $Q_{a}$ form a vector on the coset space, so they transform as the vector rep.
of $SO(d)$, ${d}$. On the other hand, as it was stated above, the $Q_{a}$ generators transform
also as ${\rm {v}}$ of $R$, thus there must be an embedding of $R$ into $SO(d)$ such that
${d}={\rm {v}}$. This embedding is determined as soon as the embedding of $R$ into $S$ is
made. Consider the $SO(d)$ commutation relations
\begin{equation}
[\S^{ab},\S^{cd}]=-g^{ac}\S^{bd}+g^{ad}\S^{bc}+g^{bc}\S^{ad}-g^{bd}\S^{ac}\,,\label{SOdCommRel}
\end{equation}
where $\S^{ab}$ are the $SO(d)$ generators and $g^{ab}$ is the metric of the tangent space. Now the
embedding of $R$ into $SO(d)$ is determined by
\begin{equation}
T_{i}=-\frac{1}{2}f_{iab}\S^{ab}\,,
\end{equation}
since one can show that the $T_{i}$ form an $R$-subalgebra of $SO(d)$ using the Jacobi identities
and the commutation relations~(\ref{SOdCommRel}).

Let me now introduce the coordinates on the $M^{4} \times (S/R)$ space
$x^{M}= (x^{\mu},y^{\alpha})$. In order to parametrise the coset I choose a representative
element
\begin{equation}
L(y)=exp(y^{\a}\d_{\a}^{a}Q_{a})\,,\label{CosetRep}
\end{equation}
of each $R$-equivalence class. I use greek indices to denote the coordinates of the coset space
and latin ones for coordinates of the tangent space. The $\d_{\a}^{a}$ is used to connect the
indices of the manifold and those of the tangent space. Consider the action of an
$S$-transformation, $s$, on the representative element $L(y)$\footnote{We assume right cosets
only.}. This will give another element of $S$, which in general will belong to a different
equivalent class, whose representative I denote by $L(y')$. Then an extra transformation $r\in R$ is
needed to bring $L(y')$ to that element of $S$. This can be expressed as
\begin{equation}
L(y)s=r(y,s)L(y')\,.\label{CosetAction}
\end{equation}
The equation determines both $y'$ and $r$ as a function of $y$ and $s$. We use $\omega^{i}$ and
$\omega^{\a}$ to parametrise $S$ and $\f^{i}$ to parametrise $R$, i.e.,
$s=exp(\omega^{i}Q_{i}+\omega^{\a}\d_{\a}^{a}Q_{a})$  and $r=exp(\f^{i}Q_{i})$. If I
consider an infinitesimal $S$-transformation, $s$, in the neighbourhood of the identity I obtain
\begin{equation}
\d y^{\a}=y'^{\a}-y^{\a}=\omega^{\b}\d_{\b}^{a}\xi_{a}^{\a}
+\omega^{i}\xi_{i}^{\alpha}\,,\label{InfinitesimalSTransform}
\end{equation}
where $\xi_{a}^{\a}$ and $\xi_{i}^{\a}$ are vector fields (they are in fact Killing vector fields as
will be shown later), tangential in the direction of the given transformation. The infinitesimal
$R$-transformations to order $\omega$ are
\begin{equation}
\f^{i}=\omega^{\b}\d_{\b}^{a}\Omega_{a}^{i}+\omega^{j}\xi_{j}^{i}\,.\label{InfinitesimalRTransform}
\end{equation}
The coefficients $\Omega_{a}^{i}$ and $\Omega_{i}^{j}$ are sometimes called $R$-compensators, as
they are associated with the compensating $R$-transformation.

Now inserting eqs~(\ref{InfinitesimalSTransform}),~(\ref{InfinitesimalRTransform}) into
eq.~(\ref{CosetAction}) I find that the Killing vectors and the $R$-compensators are given as
\begin{align}
&\xi_{a}^{\a}=\d_{a}^{\a}-\frac{1}{2}y^{\b}\d_{\b}^{b}f_{ab}{}^{c}\d_{c}^{\a}+\ldots,\qquad
\xi_{i}^{\a}=-y^{\b}\d_{\b}^{a}\d_{c}^{\a}f_{ia}{}^{c}+\ldots,\label{KillingVectors}\\
&\Omega_{j}^{i}=\d_{j}^{i}+\ldots,\qquad
\Omega_{\a}^{i}=-\frac{1}{2}y^{\b}\d_{\b}^{b}\d_{\a}^{a}f_{ab}{}^{i}+\ldots\label{Rcompensators}\,.
\end{align}
It is evident then that at $y=0$, $\xi_{a}^{\a}=\d_{a}^{\a}$, $\xi_{i}^{\a}=0$,
$\Omega_{j}^{i}=\d_{j}^{i}$ and $\Omega_{\a}^{i}=0$, a fact I shall use later. Furthermore the
Killing vectors, found in~(\ref{KillingVectors}), obey the algebra
\begin{equation}
\xi_{A}^{\b}\partial_{\b}\xi_{B}^{\a} -\xi_{B}^{\b}\partial_{\b}\xi_{A}^{\a}
=f_{AB}{}^{C}\xi_{C}^{\a}\,.
\end{equation}

The vielbein and the $R$-connection are defined through the Maurer-Cartan form which takes values
in the Lie algebra of $S$
\begin{equation}
L^{-1}(y)dL(y) = e(y)=e^{a}_{\alpha}Q_{a}+e^{i}Q_{i}\,.\label{R-connection}
\end{equation}
where $e(y)$ is a Lie algebra valued one-form. Let me recall here some properties of the exterior
derivative $d$; $d$ is a linear operator satisfying
\begin{itemize}
\item[(i)]
for a function $f$, $df=(\partial_{\a}f)dy^{\a}=(\partial_{\a}f)e^{\a}$,
\item[(ii)]
$d^{2}=0$,
\item[(iii)]
for a $q$-form $u$, $d(u\wdg v)=du\wdg v+(-1)^{q}u\wdg dv$.
\end{itemize}
The one-form $e(y)$ obeys the so called Maurer-Cartan equations
\begin{equation}
de(y)=e(y)\wdg e(y)\,,\label{MaurerCartan}
\end{equation}
which can be easily proved by using eq.~(\ref{R-connection}) and the properties of $d$. Using now 
the commutation relations~(\ref{S/R-comm-rels}) and~(\ref{R-connection}) I can rewrite
eq.~(\ref{MaurerCartan}) in terms of $R$ and $S/R$ components of the one-form $e(y)$ as
\begin{equation}
de^{a}=\frac{1}{2}f_{bc}{}^{a}e^{b}\wdg e^{c}+f_{bi}{}^{a}e^{i},\qquad 
de^{i}=\frac{1}{2}f_{ab}{}^{i}e^{a}\wdg e^{b}+\frac{1}{2}f_{jk}{}^{i}e^{j}\wdg e^{k}\,.
\end{equation}
The Maurer-Cartan equations are very useful to calculate various quantities relevant to characterise
the geometry of the manifold, such as the connection and the curvature. Using the parametrisation
of the coset~(\ref{CosetRep}) and the components of the one-form $e(y)$ can be easily computed, for
infinitesimal $y$, to be
\begin{equation}
e_{\a}^{a}(y)=\d_{\a}^{a}+\frac{1}{2}y^{\b}\d_{\b}^{c}\d_{\a}^{b}f_{cb}{}^{a}+\ldots\,,\qquad
e_{\a}^{i}(y)=\frac{1}{2}y^{\b}\d_{\b}^{c}\d_{\a}^{b}f_{cb}{}^{i}+\ldots\,.
\end{equation}
From the equations above is evident that at $y = 0$, $ e^{a}_{\alpha} = \delta^{a}_{\alpha}$ and
$e^{i}_{\alpha}= 0$, a fact I shall use later.

I proceed by calculating the connection on $S/R$ which is described by a connection-form
$\theta^{a}_{\ b}$. In the general case where torsion may be non-zero, one calculate first the
torsionless part $\omega^{a}_{\ b}$ by setting the torsion form $T^{a}$ equal to zero,
\begin{equation}
T^{a} = de^{a} + \omega^{a}_{\ b} \wedge e^{b} = 0\,,
\end{equation}
while using the Maurer-Cartan equation,
\begin{equation}
de^{a} = \frac{1}{2}f^{a}{}_{bc}e^{b}\wedge e^{c} +f^{a}{}_{bi}e^{b}\wedge e^{i}\,,
\end{equation}
I see that the condition of having vanishing torsion is solved by
\begin{equation}
\omega^{a}{}_{b}= -f^{a}{}_{ib}e^{i}-D^{a}{}_{bc}e^{c}\,,
\end{equation}
where
$$
D^{a}{}_{bc}=\frac{1}{2}g^{ad}[f_{db}{}^{e}g_{ec}+f_{cb}{}^{e} g_{de}- f_{cd}{}^{e}g_{be}]\,.
$$ 
The $D$'s can be related to $f$'s by a rescaling~\cite{Kapetanakis:1992hf} 
$$
D^{a}{}_{bc}=(\lambda^{a}\lambda^{b}/\lambda^{c})f^{a}{}_{bc}\,,
$$ 
where the $\lambda$'s depend on the coset radii. Note that in general the rescalings change the
antisymmetry properties of $f$'s, while in the case of equal radii
$D^{a}{}_{bc}=\frac{1}{2}f^{a}{}_{bc}$. Note also that the connection-form $\omega^{a}{}_{b}$ is
$S$-invariant. This means that parallel transport commutes with  the $S$
action~\cite{Castellani:1999fz}. Then the most general form of an $S$-invariant connection on $S/R$
would be
\begin{equation}
\omega^{a}{}_{b} = f^{a}{}_{ib}e^{i}+J^{a}{}_{cb}e^{c}\,,
\end{equation}
with $J$ an $R$-invariant tensor, i.e.
$$ \delta J_{cb}{}^{a}=-f_{ic}{}^{d}J_{db}{}^{a}+
f_{id}{}^{a}J_{cb}{}^{d}-f_{ib}{}^{d}J_{cd}{}^{a}=0\,. $$ 
This condition is satisfied by the $D$'s as can be proven using the Jacobi identity.

In the case of non-vanishing torsion one has
\begin{equation}
T^{a} = de^{a} + \theta^{a}{}_{b} \wedge e^{b}\,,
\end{equation}
where 
$$
\theta^{a}{}_{b}=\omega^{a}{}_{b}+\tau^{a}{}_{b}\,,
$$
with
\begin{equation}
\tau^{a}{}_{b} = - \frac{1}{2} \Sigma^{a}{}_{bc}e^{c}\,,
\end{equation}
while the contorsion $ \Sigma^{a}{}_{bc} $ is given by
\begin{equation}
\Sigma^{a}{}_{bc} = T^{a}{}_{bc}+T_{bc}{}^{a}-T_{cb}{}^{a}
\end{equation}
in terms of the torsion components $ T^{a}{}_{bc} $. Therefore in general and for the case of non-symmetric cosets the
connection-form $ \theta^{a}{}_{b}$ is
\begin{equation}
\theta^{a}{}_{b} = -f^{a}{}_{ib}e^{i} -(D^{a}{}_{bc}+\frac{1}{2}\Sigma^{a}{}_{bc})e^{c}= -f^{a}{}_{ib}e^{i}-G^{a}{}_{bc}e^{c}\,.
\label{Connection}
\end{equation}
The natural choice of torsion which would generalise the case of equal
radii~\cite{Lust:1986ix,Castellani:1986rg,Gavrilik:1999xr,MuellerHoissen:1987cq,Batakis:1989gb},
$T^{a}{}_{bc}=\eta f^{a}{}_{bc}$ would be $T^{a}{}_{bc}=2\tau D^{a}{}_{bc}$ except that the $D$'s do
not have the required symmetry properties. Therefore one has to define $\Sigma$ as a combination of
$D$'s which makes $\Sigma$ completely antisymmetric and $S$-invariant according to the definition
given above, i.e.
\begin{equation}
\Sigma_{abc} \equiv 2\tau(D_{abc}+D_{bca}-D_{cba})\,.\label{Sigma}
\end{equation}
In this general case the Riemann curvature two-form is given by~\cite{Kapetanakis:1992hf}
\begin{equation}
R^{a}{}_{b}=\left[-\frac{1}{2}f_{ib}{}^{a}f_{de}{}^{i}-
\frac{1}{2}G_{cb}{}^{a}f_{de}{}^{c}+ \frac{1}{2}(G_{dc}{}^{a}G_{eb}{}^{c}-G_{ec}{}^{a}G_{db}{}^{c})\right]e^{d} \wedge e^{e}\,,
\end{equation}
whereas the Ricci tensor $R_{ab}=R^{d}{}_{adb}$ is
\begin{equation}
R_{ab}=G_{ba}{}^{c}G_{dc}{}^{d}-G_{bc}{}^{d}G_{da }{}^{c}-G_{ca}{}^{d}f_{db }{}^{c}
-f_{ia }{}^{d}f_{db }{}^{i}\,.
\end{equation}
By choosing vanishing  parameter $\tau$ in the eqs.~(\ref{Sigma}) and~(\ref{Connection}) above one
obtains the { \it Riemannian connection}, 
$\theta_{R\ \ b}^{\ a}=-f^{a}{}_{ib}e^{i}-D^{a}{}_{bc}e^{c}$. On the other hand, by adjusting the
radii and $\tau$ one can obtain the { \it canonical connection},
$\theta_{C \ \ b}^{\ a} =-f^{a}{}_{bi}e^{i}$ which is an $R$-gauge
field~\cite{Lust:1986ix,Castellani:1986rg}. In general though the $\theta^{a}{}_{b}$ connection is
an $SO(6)$ field, i.e. lives on the tangent space of the six-dimensional cosets I consider and
describes their general holonomy. In subsection \ref{sec:4D-theory} I will show how the
$G^{ab}{}_{c}$ term of eq.~(\ref{Connection}) it is connected with the geometrical and torsion
contributions that the masses of the surviving four-dimensional gaugini acquire. Since I am
interested here in four-dimensional models without light supersymmetric particles I keep
$\theta^{a}{}_{b}$ general. Concerning the Ricci tensor, $R_{ab}$ one can make appropriate
adjustments of the torsion to set it equal to zero~\cite{Lust:1986ix,Castellani:1986rg}, thus
defining a {\it Ricci flattening connection}.

\subsection[Reduction of a $D$-dimensional YM-Dirac Lagrangian]{Reduction of a
$D$-dimensional Yang-Mills-Dirac\\ Lagrangian}\label{sec:CSDR-rules}

The group $S$ acts as a symmetry group on the extra coordinates. The CSDR scheme demands that an
$S$-transformation of the extra $d$ coordinates is a gauge transformation of the fields that are
defined on $M^{4}\times (S/R)$, thus a gauge invariant Lagrangian written on this space is
independent of the extra coordinates.

To see this in detail let me consider a $D$-dimensional Yang-Mills-Dirac theory with gauge group $G$
defined on a manifold $M^{D}$ which as stated will be compactified to $M^{4}\times (S/R)$, $D=4+d$,
$d=\dim(S)-\dim(R)$
\begin{equation}
A=\int d^{4}xd^{d}y\sqrt{-g}\Bigl[-\frac{1}{4}
Tr\left(F_{MN}F_{K\Lambda}\right)g^{MK}g^{N\Lambda}
+\frac{i}{2}\overline{\psi}\Gamma^{M}D_{M}\psi\Bigr]\,,
\end{equation}
where
\begin{equation}
D_{M}= \partial_{M}-\theta_{M}-A_{M}\,,
\end{equation}
with
\begin{equation}
\theta_{M}=\frac{1}{2}\theta_{MN\Lambda}\Sigma^{N\Lambda}
\end{equation}
the spin connection of $M^{D}$, and
\begin{equation}
F_{MN}
=\partial_{M}A_{N}-\partial_{N}A_{M}-\left[A_{M},A_{N}\right]\,,
\end{equation}
where $M$, $N$ run over the $D$-dimensional space. The fields $A_{M}$ and $\psi$ are, as
explained, symmetric in the sense that any transformation under symmetries of $S/R$  is compensated
by gauge transformations. The fermion fields can be in any rep. $F$ of $G$ unless a
further symmetry is required. Here, since I assume dimensional reductions of $\cN=1$ supersymmetric
gauge theory, the higher dimensional fermions have to transform in the adjoint of the
higher dimensional gauge group. To be more specific let $\xi_{A}^{\alpha}$,
$A =1,\ldots,\dim (S)$, be the Killing vectors which generate the symmetries of $S/R$ and $W_{A}$
the compensating gauge transformation associated with $\xi_{A}$. Define next the infinitesimal
coordinate transformation as $\delta_{A} \equiv L_{\xi_{A}}$, the Lie derivative with respect to
$\xi$, then one has for the scalar, vector and spinor fields,
\begin{align}
&\delta_{A}\phi=\xi_{A}^{\alpha}\partial_{\alpha}\phi=D(W_{A})\phi\,,
\nonumber \\
&\delta_{A}A_{\alpha}=\xi_{A}^{\beta}\partial_{\beta}A_{\alpha}+\partial_{\alpha}
\xi_{A}^{\beta}A_{\beta}=\partial_{\alpha}W_{A}-[W_{A},A_{\alpha}]\,,\label{CSDR-constraints}\\
&\delta_{A}\psi=\xi_{A}^{\alpha}\psi-\frac{1}{2}G_{Abc}\Sigma^{bc}\psi=
D(W_{A})\psi\,. \nonumber
\end{align}
$W_{A}$ depend only on internal coordinates $y$ and $D(W_{A})$ represents a gauge transformation in 
the appropriate rep. of the fields. $G_{Abc}$ represents a tangent space rotation of the spinor
fields. The variations $\delta_{A}$ satisfy, $[\delta_{A},\delta_{B}]=f_{AB}{}^{C}\delta_{C}$ and
lead to the following consistency relation for $W_{A}$'s,
\begin{equation}
\xi_{A}^{\alpha}\partial_{\alpha}W_{B}-\xi_{B}^{\alpha}\partial_{\alpha}
W_{A}-\left[W_{A},W_{B}\right]=f_{AB}{}^{C}W_{C}\,.\label{ConsistencyRelation}
\end{equation}
Let me know examine how the $W_{A}$ will change when the fields are transformed under the gauge
group. Consider the gauge transformation of the scalar field $\f$, $\f\to\f^{(g)}=D(g)\f$. Then the
$W_{A}$ needed to compensate an $S$-transformation acting on $\f^{(g)}$ will be
\begin{equation}
{W}^{(g)}_{A} = g\,W_{A}\,g^{-1}+(\delta_{A}g)g^{-1}\,.\label{W-gauge-transf}
\end{equation}
The requirement that $W_{A}$ transforms according to eq.~(\ref{W-gauge-transf}) under
gauge transformation ensure that the constraints~(\ref{CSDR-constraints}) remain invariant under
general coordinate and gauge transformations.

In order to solve the constraints I make use of the transitivity of the action of $S$ on $S/R$.
Then the value of a symmetric field at any point on $S/R$ is determined by its value at the origin
and an $S$-transformation. Therefore a convenient point to do the calculations is the origin $y=0$
while using the gauge freedom~(\ref{W-gauge-transf}) one can make the choice
\begin{equation}
W_{a}(y^{\a}=0)=0\,.
\end{equation}
Under these assumptions eq.~(\ref{ConsistencyRelation}) yields
\begin{align}
&\partial_{a}W_{b}-\partial_{b}W_{a}=f_{ab}{}^{i}W_{i}\,\label{Wsol1},\\
&\partial_{a}W_{i}=0\,\label{Wsol2},\\
&[W_{i},W_{j}]=-f_{ij}{}^{k}W_{k}\,.\label{Wsol3}
\end{align}
From eq.~(\ref{Wsol2}) I see that the $W_{i}$ are constants over the coset. Defining
\begin{equation}
J_{i}\equiv -W_{i}\,,\label{RinGSubalgebra}
\end{equation}
eq.~(\ref{RinGSubalgebra}) implies that the $J_{i}$ form the algebra of $R$. Since the $W$ live by
definition in the Lie algebra of the gauge group $G$, eq.~(\ref{Wsol3}) makes sense only if $R$ is
embedded in $G$. In that case the $J_{i}$ are the generators of an $R$-subgroup, $R_{G}$ of $G$. The
eq.~(\ref{Wsol1}) will be useful in the calculation of the potential.

I proceed by analysing the constraints on the fields in the theory. A gauge field $A_{M}$ on $M^{D}$
splits into $A_{\mu}$ on $M^{4}$ and $A_{a}$ on $S/R$; $A_{\mu}$ behaves as a scalar under
$S$-transformations and lies in the adjoint rep. of $G$. From the first of
eqs~(\ref{CSDR-constraints}) one obtains at $y=0$
\begin{equation}
\partial_{a}A_{\mu}=0\,,\qquad [J_{i},A_{\mu}]=0\,.\label{AmuSol}
\end{equation}
The first of the above equations indicates that the four-dimensional gauge field is completely
independent of the coset space coordinates. Furthermore the gauge group in four dimensions is
dictated by the second of eqs~(\ref{AmuSol}). Since $A_{\mu}$ is commutes with $J_{i}$, which are
the generators of $R_{G}$ in $G$, the surviving gauge symmetry $H$ is that subgroup of $G$ which
commutes with $R$. In other words is the centraliser of $R$ in $G$, i.e., $H=C_{G}(R_{G})$.

The remaining components of the higher dimensional gauge field $A_{\a}$ become vectors under the
coset space transformations. As they will be the scalar fields in the resulting four-dimensional
theory one can write $\F_{a}=e_{a}^{\a}A_{\a}$. The second of the eqs~(\ref{CSDR-constraints}) at
$y=0$ implies
\begin{subequations}
\begin{align}
&\partial_{a}\,\F_{b}-\partial_{b}W_{a}=\frac{1}{2}f_{ab}{}^{c}\,\F_{c}\,,\label{AaSol-1}\\
&[J_{i},\F_{a}]=\f_{ia}{}^{c}\F_{c}\,.\label{AaSol-2}
\end{align}
\end{subequations}
Eq.~(\ref{AaSol-1}) will be useful when I will calculate the potential of the theory. From
eq~(\ref{AaSol-2}) I see that the $\F_{a}$ act as interwining operator connecting the induced
reps of $R$ in $G$ and in $S$. Indeed, I have already shown that the $J_{i}$ form an
$R$-subalgebra of $G$. Denoting by $G_{s}$ the generators of the gauge group $G$ and its structure
constants by $g_{str}$, one can write $\F_{a}=\F_{a}^{s}G_{s}$ and eq.~(\ref{AaSol-2}) takes the
form
\begin{equation}
\F_{a}^{s}g_{ist}=f_{ia}{}^{c}\F_{c}^{t}\,,
\end{equation}
or
\begin{equation}
\F_{a}^{s}(M_{i})_{st}=(M'_{i})_{ac}\F_{c}^{t}\,,\label{FSol-1}
\end{equation}
where $(M_{i})_{st}=-g_{ist}$ and $(M'_{i})_{ac}=-f_{ia}{}^{c}$\,.

In general $M_{i}$ and $M'_{i}$ are reducible representations of $R$. With a suitable choice of
basis in each case one can write $M_{i}$ and $M'_{i}$ in a block diagonal form. Then each submatrix
form an irreducible representation (irrep.) of $R$. Let me consider the submatrices $M_{p}$ and
$M'_{q}$ corresponding to two irreps of $R$. Then restricting eq.~(\ref{FSol-1}) to these particular
submatrices I obtain
\begin{equation}
\F_{a}^{s(qp)}{(M_{p})}_{st}={(M'_{q})}_{ac}\F'^{t(qp)}_{c}\,.\label{FSol-2}
\end{equation}
Since $M_{p}$ and $M'_{q}$ are of irreps, $\F^{(qp)}$ must have linearly
independent rows and columns. Then if $M_{p}$ and $M'_{q}$ are of different dimension, then
obviously the $\F^{(qp)}$ has to vanish. Furthermore, if $M_{p}$ and $M'_{q}$ have the same
dimension but are different irreps, $\F^{(qp)}$ vanishes again because otherwise eq.~(\ref{FSol-2})
implies that $M_{p}$ and $M'_{q}$ are related by a change of basis. Finally, if $M_{p}$ and $M'_{q}$
are the same rep., then eq.~(\ref{FSol-2}) states that $\F^{(qp)}$ commutes with all the matrices in
that rep. and by Schur's lemma it must be multiple of the identity matrix, i.e.,
$\F^{(qp)}=\f^{(qp)}(x)\one$. This shows that in order to find the rep. of the gauge group $H$ under
which the $\f$ transform in four dimensions, one has to decompose $S$ under $R$
\begin{eqnarray}
S&\supset& R \nonumber \\
\adj\,S&=& \adj\,R +{\rm v}\label{RinS-embedding}
\end{eqnarray}
and the gauge group $G$ according to the embedding
\begin{eqnarray}
G&\supset& R_{G} \times H \nonumber \\
\adj\,G&=&(\adj\,R,1)+(1,\adj\,H)+\sum(r_{i},h_{i})\,.\label{RinG-embedding}
\end{eqnarray}
Then if ${\rm v}=\sum s_{i}$, where each $s_{i}$ is an irrep. of $R$, there survives an $h_{i}$
multiplet for every pair $(r_{i},s_{i})$, where $r_{i}$ and $s_{i}$ are identical irreps of $R$.

Turning next to the fermion fields
\cite{Kapetanakis:1992hf,Manton:1981es,Chapline:1982wy,Wetterich:1982ed,Palla:1983re,Pilch:1984xx,
Forgacs:1985vp,Barnes:1986ea} similarly to scalars, they act as intertwining operators between
induced reps acting on $G$ and the tangent space of $S/R$, $SO(d)$. Proceeding along similar lines
as in the case of scalars to obtain the rep. of $H$ under which the four-dimensional
fermions transform, I have to decompose the rep. $F$ of the initial gauge group in which
the fermions are assigned under $R_{G} \times H$, i.e.
\begin{equation}
F= \sum (t_{i},h_{i})\,,\label{Fdecomp}
\end{equation}
and the spinor of $SO(d)$ under $R$
\begin{equation}
\sigma_{d} = \sum \sigma_{j}\,.\label{Frule}
\end{equation}
Then for each pair $t_{i}$ and $\sigma_{i}$, where $t_{i}$ and $\sigma_{i}$ are identical irreps
there is an $h_{i}$ multiplet of spinor fields in the four-dimensional theory. In order however  to
obtain chiral fermions in the effective theory one has to impose further requirements. I first
impose the Weyl condition in $D$ dimensions. In $D = 4n+2$ dimensions which is the case at hand, the
decomposition of the left handed, say spinor under $SU(2) \times SU(2) \times SO(d)$ is
\begin{equation}
\sigma _{D} = (2,1,\sigma_{d}) + (1,2,\overline{\sigma}_{d})\,.
\end{equation}
Furthermore in order to be $\sigma_{d}\neq\obar{\sigma}_{d}$ the coset space $S/R$ must be such that
$\rank(R)=\rank(S)$~\cite{Kapetanakis:1992hf,Bott:1965}. The six-dimensional coset spaces which
satisfy this condition are listed in tables~\ref{SO6VectorSpinorContentSCosets}
and~\ref{SO6VectorSpinorContentNSCosets}. Then under the $SO(d)\supset R$ decomposition one has
\begin{equation}
\sigma_{d} = \sum \sigma_{k},~\overline{\sigma}_{d}= \sum \overline{\sigma}_{k}\,.
\end{equation}
In the following chapters I assume that the higher dimensional theory is $\cN=1$ supersymmetric.
Therefore the higher dimensional fermionic fields have to be considered transforming in the adjoint 
of $E_{8}$ which is vectorlike. In this case each term $(t_{i},h_{i})$ in eq.~(\ref{Fdecomp}) will
be either self-conjugate or it will have a partner $(\overline{t}_{i},\overline{h}_{i})$. According
to the rule described in eqs.~(\ref{Fdecomp}),~(\ref{Frule}) and considering $\sigma_{d}$ I will
have in four dimensions left-handed fermions transforming as $ f_{L} = \sum h^{L}_{k}$. It is
important to notice that since $\sigma_{d}$ is non self-conjugate, $f_{L}$ is non self-conjugate
too. Similarly from $\overline{\sigma}_{d}$ I will obtain the right-handed rep. $ f_{R}= \sum
\overline{h}^{R}_{k}$ but as I have assumed that $F$ is vector-like, $\overline{h}^{R}_{k}\sim
h^{L}_{k}$. Therefore there will appear two sets of Weyl fermions with the same quantum numbers
under $H$. This is already a chiral theory but still one can go further and try to impose the
Majorana condition in order to eliminate the doubling of the fermion spectrum. However this is not
required in the present case of interest where I apply the Hosotani mechanism for the further
breaking of the gauge symmetry, as I will explain in chapter~\ref{sec:WilsonFlux-theory}.

An important requirement is that the resulting four-dimensional particle physics models should be
anomaly free. Starting with an anomaly free theory in higher dimensions,
Witten~\cite{Witten:1984dg} has given the condition to be fulfilled in order to obtain
anomaly free four-dimensional theories. The condition restricts the allowed embeddings of $R$ into 
$G$ by relating them with the embedding of $R$ into  $SO(6)$~\cite{Pilch:1985qf,Kapetanakis:1992hf}.
To be more specific if $\L_{a}$ are the generators of $R$ into $G$ and $T_{a}$ are the generators of
$R$ into $SO(6)$ the condition reads
\begin{equation}
Tr(\L_{a}\L_{b})=30\,Tr(T_{a}T_{b})\,.\label{WittenCond}
\end{equation}
According to ref.~\cite{Pilch:1985qf} the anomaly cancellation condition~(\ref{WittenCond}) is
automatically satisfied for the choice of embedding
\begin{equation}
E_{8}\supset SO(6) \supset R\,,\label{RinSO6inGembedding}
\end{equation}
which I adopt here. Furthermore, concerning the abelian group factors of the
four-dimensional gauge theory, note that the corresponding gauge bosons surviving in four
dimensions become massive at the compactification scale~\cite{Green:1984bx,Witten:1984dg} and
therefore, they do not contribute in the anomalies; they correspond only to global symmetries.

\setlength{\arraycolsep}{1pt}
\setlength{\captionmargin}{0pt}
\begin{table}
\caption[\textbf{Six-dimensional symmetric cosets spaces with $\rank(R)=\rank(S)$.}]
{\textbf{Six-dimensional symmetric cosets spaces with $\rank(R)=\rank(S)$.}
\emph{The  freely acting discrete symmetries ${\rm Z}(S)$ and ${\rm W}$ for each case are listed. The transformation
properties  under $R$ are also noted.}\label{SO6VectorSpinorContentSCosets}}
\begin{scriptsize}
\begin{center}
\begin{math}
\begin{array}{|c!{\vrule width 1pt}c!{\vrule width 1pt} c|c!{\vrule width 1pt} l|l|}\hline
{\bf Case}
&{\mbf 6D}~{\bf Coset~Spaces} 
& {\mbf Z(S)}
& {\mbf W}
& {\mbf V}
& {\mbf F}  \\ \hhline{*{6}{=}}                                             
{\bf a}
&\frac{SO(7)}{SO(6)}
& {\bb Z}_{2}
& {\bb Z}_{2}
&\begin{array}{l}
{\mbf 6} \lra {\mbf 6}
\end{array}
&\begin{array}{l}
{\mbf 4} \lra \obar{{\mbf 4}}
 \end{array}\\\hline
{\bf b}
&\frac{SU(4)}{SU(3)\times U(1)}
& {\bb Z}_{4}
& \one
& \begin{array}{l}
{\mbf 6}={\mbf 3}_{(-2)} + \obar{{\mbf 3}}_{(2)} \\
                                     -
\end{array}
& \begin{array}{l}
{\mbf 4}={\mbf 1}_{(3)}+{\mbf 3}_{(-1)}\\
              -
\end{array}\\
\hline
{\bf c}
&\frac{Sp(4)}{(SU(2)\times U(1))_{max}} &{\bb Z}_{2}
&{\bb Z}_{2}
&\begin{array}{l}
{\mbf 6}={\mbf 3}_{(-2)} +{\mbf 3}_{(2)}\\
{\mbf 3}_{(-2)}\lra{\mbf 3}_{(2)}
\end{array}
&\begin{array}{l}
{\mbf 4}={\mbf 1}_{(3)}+{\mbf 3}_{(-1)}\\
{\mbf 1}_{(3)}\lra{\mbf 1}_{(-3)}~~{\mbf 3}_{(-1)}\lra{\mbf 3}_{(1)}
\end{array}\\
\hline
{\bf d}
&\left(\frac{SU(3)}{SU(2)\times U(1)}\right)\times\left(\frac{SU(2)}{U(1)}\right) & {\bb Z}_{2}\times {\bb Z}_{3} & {\bb
Z}_{2}
&\begin{array}{l}
\begin{array}{l@{}l@{}l@{}}
{\mbf 6}&=&{\mbf 1}_{(0,2a)}+{\mbf 1}_{(0,-2a)}\\
           &+&{\mbf 2}_{(b,0)}+{\mbf 2}_{(-b,0)}
\end{array}\\
{\mbf 1}_{(0,2a)}\lra{\mbf 1}_{(0,-2a)}
\end{array}
&\begin{array}{l}
{\mbf 4}={\mbf 2}_{(0,a)}+{\mbf 1}_{(b,-a)}+{\mbf 1}_{(-b,-a)}\\
{\mbf 2}_{(0,a)}\lra{\mbf 2}_{(0,-a)}\\
{\mbf 1}_{(b,-a)}\lra{\mbf 1}_{(b,a)}\\
{\mbf 1}_{(-b,-a)}\lra{\mbf 1}_{(-b,a)}
\end{array}\\
\hline
{\bf e}
&\left(\frac{Sp(4)}{SU(2)\times SU(2)}\right)\times\left(\frac{SU(2)}{U(1)}\right)
&({\bb Z}_{2})^{2} 
& ({\bb Z}_{2})^{2}
&\begin{array}{l}
{\mbf 6}=({\mbf 2},{\mbf 2})_{(0)}+({\mbf 1},{\mbf 1})_{(2)}+({\mbf 1},{\mbf 1})_{(-2)}\\
\mbox{(${\bb Z}_{2}$ of $SU(2)/U(1)$)}\\
({\mbf 1},{\mbf 1})_{(2)}\lra({\mbf 1},{\mbf 1})_{(-2)}
\end{array}
&\begin{array}{l}
{\mbf 4}=({\mbf 2},{\mbf 1})_{(1)}+({\mbf 1},{\mbf 2})_{(-1)}\\
({\mbf 2},{\mbf 1})_{(1)}\lra({\mbf 2},{\mbf 1})_{(-1)}\\
({\mbf 1}, {\mbf 2})_{(1)}\lra({\mbf 1},{\mbf 2})_{(-1)}
\end{array}\\
\hline
{\bf f}
&\left(\frac{SU(2)}{U(1)}\right)^{3}
& ({\bb Z}_{2})^{3} 
&({\bb Z}_{2})^{3}
&\begin{array}{l}
\begin{array}{l@{}l@{}l@{}}
{\mbf 6}&=&(2a,0,0)+(0,2b,0)+(0,0,2c)\\
            &+&(-2a,0,0)+(0,-2b,0)+(0,0,-2c)\\
\end{array}\\
\mbox{each ${\bb Z}_{2}$ changes the sign of $a$,$b$, $c$}
\end{array}
&\begin{array}{l}
\begin{array}{l@{}l@{}l@{}}
{\mbf 4}&=&(a,b,c)+(-a,-b,c)\\
           &+&(-a,b,-c)+(a,-b,-c)
\end{array}\\
\mbox{each ${\bb Z}_{2}$ changes the sign of}\\
\mbox{$a$,$b$, $c$}
\end{array}\\
\hline
\end{array}
\end{math}
\end{center}
\end{scriptsize}
\end{table}
\setlength{\arraycolsep}{5pt}
\setlength{\captionmargin}{0pt}
\begin{table}
\caption[\textbf{Six-dimensional non-symmetric cosets spaces with $\rank(R)=\rank(S)$.}]
{\textbf{Six-dimensional non-symmetric cosets spaces with $\rank(R)=\rank(S)$.}
\emph{The available freely acting discrete symmetries ${\rm Z}(S)$ and ${\rm W}$ for each case are listed. The transformation
properties under $R$ are also noted.}}\label{SO6VectorSpinorContentNSCosets}
\begin{scriptsize}
\begin{center}
\begin{math}
\begin{array}{|c!{\vrule width 1pt}c!{\vrule width 1pt}c|c!{\vrule width 1pt}l|l|}\hline
{\bf Case}
&{\mbf 6D}~{\bf Coset~Spaces} 
& {\mbf Z(S)}
& {\mbf W}
& {\mbf V}
& {\mbf F}  \\ \hhline{*{6}{=}}                                               
{\bf a'}
&\frac{G_{2}}{SU(3)} 
&{\mbf 1}
&{\bb Z}_{2}
&\begin{array}{l}
{\mbf 6}={\mbf 3}+\obar{{\mbf 3}}\\
{\mbf 3}\lra\obar{{\mbf 3}}
\end{array}
&\begin{array}{l}
{\mbf 4}={\mbf 1}+{\mbf 3}\\
{\mbf 1}\lra{\mbf 1}\\
{\mbf 3}\lra\obar{{\mbf 3}}
\end{array}\\
\hline
{\bf b'}
&\frac{Sp(4)}{(SU(2) \times U(1))_{nonmax}}
&{\bb Z}_{2}
&{\bb Z}_{2}
&\begin{array}{l}
{\mbf 6}={\mbf 1}_{(2)}+{\mbf 1}_{(-2)} + {\mbf 2}_{(1)}+{\mbf 2}_{(-1)}\\
{\mbf 1}_{(2)}\lra{\mbf 1}_{(-2)}\\
{\mbf 2}_{(1)}\lra{\mbf 2}_{(-1)}
\end{array}
&\begin{array}{l}
{\mbf 4}={\mbf 1}_{(0)}+{\mbf 1}_{(2)}+{\mbf 2}_{(-1)}\\
{\mbf 1}_{(2)}\lra{\mbf 1}_{(-2)}~~{\mbf 1}_{(0)}\lra{\mbf 1}_{(0)}\\
{\mbf 2}_{(1)}\lra{\mbf 2}_{(-1)}
\end{array}\\
\hline
{\bf c'}
&\frac{SU(3)}{U(1)\times U(1)}
&{\bb Z}_{3}
&{\mbf S}_{3}
&\begin{array}{l@{}l@{}l@{}}
{\mbf 6}&=&(a,c)+(b,d)+(a+b,c+d)\\
           &+&(-a,-c)+(-b,-d)\\
           &+&(-a-b,-c-d)
\end{array}
&\begin{array}{l@{}l@{}l@{}}
{\mbf 4}&=&(0,0)\\
           &+&(a,c)+(b,d)+(-a-b,-c-d)
\end{array}\\[25pt]
\ \ 
& \  \
&\   \ 
&{\bb Z}_{2}
&\begin{array}{l}
(b,d)\lra(-b,-d)\\
(a+b,c+d)\lra(a,c)\\
(-a,-c)\lra(-a-b,-c-d)
\end{array}
&\begin{array}{l}
(b,d)\lra(-b,-d)\\
(a,c)\lra(a+b,c+d)\\
(-a-b,-c-d)\lra(-a,-c)
\end{array}\\[25pt]
\  \
& \  \
&\  \
&{\bb Z}_{2}
&\begin{array}{l}
(b,d)\lra(a+b,c+d)\\
(a,c)\lra(-a,-c)\\
(-b,-d)\lra(-a-b,-c-d)
\end{array}
&\begin{array}{l}
(b,d)\lra(a+b,c+d)\\
(a,c)\lra(-a,-c)\\
(-a-c,-b-d)\lra(-b,-d)
\end{array}\\ [25pt]
\ \
&\  \
& \ \
&{\bb Z}_{2}
&\begin{array}{l}
(b,d)\lra(-a,-c)\\
(a+b,c+d)\lra(-a-b,-c-d)\\
(a,c)\lra(-b,-d)
\end{array}
&\begin{array}{l}
(b,d)\lra(-a,-c)\\
(a,c)\lra(-b,-d)\\
(-a-b,-c-d)\lra(a+b,c+d)
\end{array}\\
\hline
\end{array}
\end{math}
\end{center}
\end{scriptsize}
\end{table}

\subsection{The four-dimensional~theory}\label{sec:4D-theory}

Next let me obtain the four-dimensional effective action. Assuming that the metric is block
diagonal, taking into account all the constraints and integrating out the extra coordinates I obtain
in four dimensions the following Lagrangian
\begin{equation}
A=C \int d^{4}x \biggl( -\frac{1}{4} F^{t}_{\mu\nu}{F^{t}}^{\mu\nu}+\frac{1}{2}(D_{\mu}\phi_{a})^{t}
(D^{\mu}\phi^{a})^{t}
+V(\phi)+\frac{i}{2}\overline{\psi}\Gamma^{\mu}D_{\mu}\psi+\frac{i}{2}
\overline{\psi}\Gamma^{a}D_{a}\psi\biggr)\,,\label{YM-Dirac-4Daction}
\end{equation}
where $D_{\mu} = \partial_{\mu} - A_{\mu}$ and $D_{a}= \partial_{a}- \theta_{a}-\phi_{a}$ with  
$\theta_{a}=\frac{1}{2}\theta_{abc}\Sigma^{bc}$ the connection of the coset space and $\Sigma^{bc}$
the $SO(6)$ generators. With $C$ I denote the volume of the coset space. The potential
$V(\phi)$ is given by
\begin{equation}
V(\phi) = - \frac{1}{4} g^{ac}g^{bd}Tr( f _{ab}{}^{C}\phi_{C} -
[\phi_{a},\phi_{b}] ) (f_{cd}{}^{D}\phi_{D} - [\phi_{c},\phi_{d}] )\,,\label{potential}
\end{equation}
where, $A=1,\ldots,\dim (S)$ and $f$ ' s are the structure constants appearing in the commutators of
the generators of the Lie algebra of S. The expression (\ref{potential}) for $V(\phi)$ is only
formal because $\phi_{a}$ must satisfy the constraints coming from eq.~(\ref{CSDR-constraints}),
\begin{equation}
f_{ai}{}^{D}\phi_{D} - [\phi_{a},\phi_{i}] = 0\,,
\end{equation}
where the $\phi_{i}$ generate $R_{G}$. These constraints imply that some components $\phi_{a}$'s
are zero, some are constants and the rest can be identified with the genuine Higgs fields according
to the rules presented in eqs~(\ref{RinS-embedding}) and~(\ref{RinG-embedding}).

When $V(\phi)$ is expressed in terms of the unconstrained independent Higgs fields, it remains a
quartic polynomial which is invariant under gauge transformations of the final gauge group $H$, and
its minimum determines the vacuum expectation values of the Higgs
fields~\cite{Chapline:1980mr,Bais:1985yd,Farakos:1986sm,Farakos:1986cj}. The
minimisation of the potential is in general a difficult problem. If however $S$ has an
isomorphic image $S_{G}$ in $G$ which contains $R_{G}$ in a consistent way then it is possible to
allow the $\phi_{a}$ to become generators of $S_{G}$. That is 
$\overline{\phi}_{a} =<\phi^{i}>Q_{ai} = Q_{a}$ with $<\phi^{i}>Q_{ai}$ suitable combinations of $G$
generators, $Q_{a}$ a generator of $S_{G}$ and $a$ is also a coset-space index. Then
\begin{align*}
\overline{F}_{ab}&=f_{ab}{}^{i}Q_{i}+f_{ab}{}^{c}\overline{\phi}_{c}
-[\overline{\phi}_{a},\overline{\phi}_{b}]\\
&= f_{ab}{}^{i}Q_{i}+ f_{ab}{}^{c}Q_{c}- [Q_{a},Q_{b}] = 0
\end{align*}
because of the commutation relations of $S$. Thus I  have proven that $V(\phi=\overline{\phi})=0$
which furthermore is the minimum, because $V$ is positive definite. Furthermore, the
four-dimensional gauge group $H$ breaks further by these non-zero vacuum expectation values of the
Higgs fields to the centraliser $K$ of the image of $S$ in $G$, i.e. $K=C_{G}(S)$
\cite{Kapetanakis:1992hf,Chapline:1980mr,Bais:1985yd,Farakos:1986sm,Farakos:1986cj}. This can been
seen if I examine a gauge transformation of $\phi_{a}$ by an element $h$ of $H$. Then I have 
$$ \phi_{a} \rightarrow h\phi_{a}h^{-1},\qquad h \in H $$ We note that the v.e.v. of the Higgs
fields is gauge invariant for the set of $h$'s that commute with $S$. That is $h$ belongs to
a subgroup $K$ of $H$ which is the centraliser of $S_{G}$ in $G$. It should be stressed that the
four-dimensional fermions of this class of models acquire large masses due to geometrical
contributions at the compactification scale~\cite{Kapetanakis:1992hf,Barnes:1986ea}.
In general it can be proven~\cite{Kapetanakis:1992hf} that dimensional reduction over a
symmetric coset space always gives a potential of spontaneous breaking form which is not the case
of non-symmetric cosets of more than one radii.

In the fermion part of the Lagrangian the first term is just the kinetic term of fermions, while the
second is the Yukawa term~\cite{Barnes:1986ea,Kapetanakis:1990tk}. The last term in
(\ref{YM-Dirac-4Daction}) can be written as
\begin{equation}
L_{D}= \frac{i}{2}\overline{\psi}\Gamma^{a}(\partial_{a}-
\frac{1}{2}f_{ibc}e^{i}_{\gamma}e^{\gamma}_{a}\Sigma^{bc}-
\frac{1}{2}G_{abc}\Sigma^{bc}- \phi_{a}) \psi
=\frac{i}{2}\overline{\psi}\Gamma^{a}\nabla_{a}\psi+
\overline{\psi}V\psi\,,\label{Dirac-4Daction}
\end{equation}
where
\begin{align}
&\nabla_{a} =  \partial_{a} -
\frac{1}{2}f_{ibc}e^{i}_{\gamma}e^{\gamma}_{a}\Sigma^{bc} - \phi_{a},\label{CovDerivative}\\
&V=-\frac{i}{4}\Gamma^{a}G_{abc}\Sigma^{bc}\,,\label{GeometricMass}
\end{align}
and $G_{abc}$ is given in eq.~(\ref{Connection}) as
$G^{a}{}_{bc}=D^{a}{}_{bc}+\frac{1}{2}\Sigma^{bc}$. I have already noticed that according to the
CSDR constraints, $\partial_{a}\psi= 0$. Furthermore one can consider the Lagrangian
at the point $y=0$, due to its invariance under $S$-transformations, and according to the
discussion in subsection~\ref{sec:CosetSpaceGeometry}~$e^{i}_{\gamma}=0$ at that point.
Therefore (\ref{CovDerivative}) becomes just $\nabla_{a}= \phi_{a}$ and the term
$\frac{i}{2}\overline{\psi}\Gamma^{a}\nabla_{a}\psi $ in eq.~(\ref{Dirac-4Daction}) is exactly the
Yukawa term. 

Let me examine now the last term appearing in (\ref{Dirac-4Daction}). One can show easily that the
operator $V$ anticommutes with the six-dimensional helicity operator~\cite{Kapetanakis:1992hf}.
Furthermore one can show that $V$ commutes with the $T_{i}=-\frac{1}{2}f_{ibc}\Sigma^{bc}$ [$T_{i}$
close the $R$-subalgebra of $SO(6)$]. In turn I can draw the conclusion, exploiting Schur's lemma,
that the non-vanishing elements of $V$ are only those which appear in the decomposition of both
$SO(6)$ irreps $4$ and $\overline{4}$, e.g. the singlets. Since this term is of pure geometric
nature, I reach the conclusion that the singlets in $4$ and $\overline{4}$ will acquire large
geometrical masses, a fact that has serious phenomenological implications. First note this is
characteristic of the non-symmetric cosets only. In~\cite{Manousselis:2001xb,Manousselis:2001re} was
found that dimensional reduction of supersymmetric theories defined in higher dimensions over
non-symmetric coset spaces results in particle physics models with a softly broken supersymmetry.
The surviving four-dimensional fermions coming from the identification of singlets of $SO(6)$ irreps
$\mbf{4}$ live necessarily in the adjoint of $H$ as was stated in subsection \ref{sec:CSDR-rules}
thus being the gaugini of the model. Then, according to our previous argument, will receive masses
comparable to the compactification scale. In the case of the symmetric cosets though, the $V$
operator is absent since $f_{ab}{}^{c}$ are vanishing by definition.

\section{Remarks on Grand Unified theories resulting from CSDR}
\label{UnifiedTheoriesHigherDims-Example}

Here I make few remarks on models resulting from the coset space dimensional reduction of an
$\cN=1$, $E_{8}$ gauge theory which is defined on a ten-dimensional compactified space
$M^{D}=M^{4}\times (S/R)$. The coset spaces $S/R$ I consider are listed in
the first column of tables~\ref{SO6VectorSpinorContentSCosets}
and~\ref{SO6VectorSpinorContentNSCosets}. In order to obtain four-dimensional GUTs potentially with
phenomenological interest, namely $\cH=E_{6}$, $SO(10)$ and $SU(5)$, is sufficient to
consider only embeddings of the isotropy group $R$ of the coset space in
\begin{subequations}
\begin{alignat}{2}
\cR&=C_{E_{8}}(\cH)= SU(3)\,,\qquad                       & \mbox{for}\quad
&\cH=E_{6}\,,\label{Rcases-1} \\
\cR&=C_{E_{8}}(\cH)=SO(6)\thicksim SU(4)\,,\qquad & \mbox{for}\quad &\cH=SO(10)\,,\label{Rcases-2}\\
\cR&=C_{E_{8}}(\cH)=SU(5)\,,\qquad                       & \mbox{for}\quad 
&\cH=SU(5)\,.\label{Rcases-3}
\end{alignat}
\end{subequations}

As it was noted in subsection~\ref{sec:CSDR-rules} the anomaly cancellation
condition~(\ref{WittenCond})  is satisfied automatically for the choice of embedding
\begin{equation}
E_{8}\supset SO(6)\supset R\,,\label{RinSO6inE8-embedding}
\end{equation}
which I adopt here. This requirement is trivially fullfiled for the case of $R\hookrightarrow\cR$
embeddings of eq.~(\ref{Rcases-2}) which lead to $SO(10)$ GUTs in four dimensions. It is obviously
also satisfied for the case of $R\hookrightarrow\cR$ embeddings of eq.~(\ref{Rcases-1}) since
$SU(3)\subset SO(6)$. The above case leads to $E_{6}$ GUTs in four dimensions. Finally, 
$R\hookrightarrow\cR$ embeddings of eq.~(\ref{Rcases-3}) are excluded since the
requirement~(\ref{RinSO6inE8-embedding}) cannot be satisfied.

\section[Reduction of $\cN=1$, $E_{8}$ YM-Dirac theory over $SO(7)/SO(6)$]
{Reduction of $\cN=1$, $E_{8}$ Yang-Mills-Dirac theory over $SO(7)/SO(6)$}
\label{CSDR-example}

Let a $\cN=1$, $G=E_{8}$ Yang-Mills-Dirac theory defined in ten dimensions
which are compactified in the form $M^{4}\times S^{6}$, $S^{6}\sim SO(7)/SO(6)$. Furthermore let me
consider Weyl fermions belonging in the adjoint of $G=E_{8}$  and the embedding of $R=SO(6)$ into
$E_{8}$ suggested by the decomposition
\begin{equation}
\begin{array}{l@{}l@{}l@{}}                                                          
E_{8}&\supset& SO(16) \supset SO(6)\times SO(10)\\                           
\mbf{248}&=&(\mbf{1}, \mbf{45})+(\mbf{6},\mbf{10})+(\mbf{15},\mbf{1})
                +(\mbf{4},\mbf{16})+(\obar{\mbf {4}},\obar{\mbf {16}})\,.
\end{array}\label{DecompCh1-SO10}
\end{equation}
If only the CSDR mechanism was applied, the resulting four-dimensional gauge group would be
\begin{equation*}
H=C_{E_{8}}(SO(6))=SO(10)\,.
\end{equation*}
According to table~\ref{SO6VectorSpinorContentSCosets}, the $R=SO(6)$ content of vector and spinor 
of $B_{0}=S/R=SO(7)/SO(6)$ is $\mbf{6}$ and $\mbf{4}$, respectively. Then applying the CSDR
rules~(\ref{RinS-embedding}),~(\ref{RinG-embedding})
and~(\ref{Fdecomp}),~(\ref{Frule}) the four-dimensional theory would contain scalars transforming as
$\mbf{10}$ under the $H=SO(10)$ gauge group and two copies of chiral fermions belonging in the
$\mbf{16}_{L}$ of $H$.

According to subsection~\ref{sec:4D-theory}, dimensional reduction over the $SO(7)/SO(6)$ symmetric
coset space leads to a four-dimensional action with a potential of spontaneously symmetry breaking
form. In addition, note that the isometry group of the coset, $SO(7)$, is embeddable in $E_{8}$ as
\begin{equation*}
\begin{array}{c@{}c@{}c@{}}
E_{8}\supset &SO(7)\times &SO(9)\\
\ \                     &\cup              &\cap\\
\ \                     &SO(6) \times  &SO(10)\,.
\end{array}
\end{equation*}
Then, according to the theorem mentioned in subsection~\ref{sec:4D-theory}, the final gauge group is
\begin{equation*}
\cH=C_{E_{8}}(SO(7))=SO(9)\,,
\end{equation*}
i.e. the $\mbf{10}$ of $SO(10)$ obtains a v.e.v. leading to the spontaneous symmetry breaking
\begin{equation}
\begin{array}{l}
SO(10)\to SO(9)\label{SO10SO9-breaking}\\
\mbf{10}=\langle\mbf{1}\rangle + \mbf{9}\,.
\end{array}
\end{equation}

\chapter{Wilson flux breaking mechanism in CSDR\label{sec:WilsonFlux-theory}}

In the previous section an example of the CSDR mechanism was given. I assumed an $\cN=1$, $E_{8}$ Yang-Mills-Dirac theory defined
on the ten-dimensional space $M^{4}\times S/R$, $S/R=SO(7)/SO(6)$. The resulting four-dimensional gauge theory was an $SO(10)$
GUT with scalars transforming as $\mbf{10}$ of $SO(10)$ and two copies of chiral fermion transforming as $\mbf{16}_{L}$
left-handed multiplets. The surviving scalars of the theory being in the fundamental representation of the gauge group are not
able to provide the appropriate superstrong symmetry breaking towards the SM. This is an intrinsic characteristic of the CSDR
scheme. As described in subsection~\ref{sec:CSDR-rules} the surviving scalars in four dimensions are calculated by comparing the $R$
content of the vector $SO(6)$ under the $SO(6)\supset R$ decomposition with the irreps occurring in the vector of the $G\supset
R_{G}\times H$ embedding. Therefore it is impossible to transform in the adjoint of the surviving gauge group as an appropriate
superstrong symmetry breaking of GUTs requires. As a way out it has been suggested~\cite{Zoupanos:1987wj} to take advantage of
non-trivial topological properties of the compactification coset space, apply the Hosotani or Wilson flux breaking
mechanism~\cite{Hosotani:1983xw,Hosotani:1983vn,Witten:1985xc} and break the gauge symmetry of the theory further. Application of
this mechanism imposes further constraints in the scheme.

Here I first recall the Wilson flux breaking mechanism, I make some remarks on specific cases which potentially
lead to interesting models and I finally calculate the actual symmetry breaking patterns of the GUTs.

\newpage
\section{Wilson flux breaking mechanism}\label{sec:WilsonFluxBreaking-theory}

Let me briefly recall the Wilson flux mechanism for breaking spontaneously a gauge theory. Then instead of considering a gauge
theory on $M^{4}\times B_{0}$, with $B_{0}$ a simply connected manifold, and in my case a coset space $B_{0}=S/R$, I consider a
gauge theory on $M^{4}\times B$, with $B=B_{0}/F^{S/R}$ and $F^{S/R}$ a freely acting discrete symmetry\footnote{By freely acting
I mean that for every element $g\in F$, except the identity, there exists no points of $B_0$ that remain invariant.} of $B_{0}$.
It turns out that $B$ becomes multiply  connected, which means that there will be contours not contractible to a point due to
holes in the manifold. For each element $g\in F^{S/R}$, I  pick up an element $U_{g}$ in $H$, i.e. in the four-dimensional gauge
group of the reduced theory, which can be represented as the Wilson loop
\begin{equation}
U_{g}=\mathcal{P}exp\left(-i~\int_{\gamma_g}T^{a}A_{M}^{a}(x)dx^{M}\right)\,,
\end{equation}
where $A_{M}^{a}(x)$ are vacuum $H$ fields with group generators $T^{a}$, $\gamma_g$ is a contour representing the abstract
element $g$ of $F^{S/R}$, and $\mathcal{P}$ denotes the path ordering.

Now if $\gamma_g$ is chosen not to be contractible to a point, then $U_g\neq 1$ although the vacuum field strength vanishes
everywhere. In this  way an homomorphism of $F^{S/R}$ into $H$ is induced with image $T^{H}$, which is the subgroup of $H$
generated by $\{U_g\}$. A field $f(x)$ on  $B_0$ is obviously equivalent to another field on $B_0$ which obeys $f(g(x))=f(x)$
for every $g\in F^{S/R}$. However in the presence of the gauge group $H$ this statement can be generalised to
\begin{equation}
\label{eq:Wilson-symmetry}
f(g(x))=U_{g}f(x)\,.
\end{equation}
Next, one would like to see which gauge symmetry is preserved by the vacuum. The vacuum has $A_{\mu}^{a}=0$ and I
represent a gauge transformation by a space-dependent matrix $V(x)$ of $H$. In order to keep $A_{\mu}^{a}=0$ and leave the
vacuum invariant, $V(x)$ must be constant. On the other hand, $f\to Vf$ is consistent with equation (\ref{eq:Wilson-symmetry}),
only if $[V,U_g]=0$
for all $g\in F^{S/R}$. Therefore the $H$ breaks towards the centraliser of $T^{H}$ in $H$, $K'=C_{H}(T^{H})$. In addition
the matter fields have to be invariant under the diagonal sum
\begin{equation}
F^{S/R}\oplus T^{H}\,,\label{WFluxSurvivingField}
\end{equation}
in order to satisfy eq.~(\ref{eq:Wilson-symmetry}) and therefore survive in the four-dimensional theory.

\newpage
\section{Further remarks concerning the use of the $F^{S/R}$}
\label{sec:InterestingDiscreteSymmetries}

The discrete symmetries $F^{S/R}$, which act freely on coset spaces $B_0=S/R$ are the center of $S$, $\mathrm{Z}(S)$ and the
$\mathrm{W}=\mathrm{W}_{S}/\mathrm{W}_{R}$, with $\mathrm{W}_{S}$ and $mathrm{W}_{R}$ being the Weyl groups of $S$ and $R$,
respectively~\cite{Kapetanakis:1992hf,Kapetanakis:1989gd,Kozimirov:1989kn,Kozimirov:1989xp}. The freely acting discrete
symmetries, $F^{S/R}$, of the specific six-dimensional coset spaces under discussion are listed in the second and third column of
tables~\ref{SO6VectorSpinorContentSCosets} and~\ref{SO6VectorSpinorContentNSCosets}. The $F^{S/R}$ transformation
properties of the vector and spinor irreps under $R$ are noted in the last two columns of the same tables.

According to the discussion in section~\ref{UnifiedTheoriesHigherDims-Example}, dimensional reduction over the six-dimension\-al
coset spaces listed in tables~\ref{SO6VectorSpinorContentSCosets} and~\ref{SO6VectorSpinorContentSCosets}, leads to $E_{6}$ and
$SO(10)$ GUTs. My approach is to embed the $F^{S/R}$ discrete symmetries into four-dimensional $H=E_{6}$ and $SO(10)$ gauge
groups. I make this choice only for bookeeping reasons since, according to section~\ref{sec:WilsonFluxBreaking-theory},  the
actual topological symmetry breaking takes place in higher dimensions. Few remarks are in order. In both classes of models, namely
$E_{6}$ and $SO(10)$ GUTs, the use of the discrete symmetry of the center of $S$, ${\rm Z}(S)$, cannot lead to phenomenologically
interesting cases since various components of the irreps of the four-dimensional GUTs containing the SM fermions do not survive.
The reason is that the irreps of $H$ remain invariant under the action of the discrete symmetry, ${\rm Z}(S)$, and as
a result the phase factors gained  by the action of $T^{H}$ cannot be compensated. Therefore the complete SM fermion spectrum
cannot be invariant under $F^{S/R}\oplus T^{H}$ and survive. On the other hand, the use of the Weyl discrete symmetry can lead to
better results. Models with potentially interesting fermion spectrum can be obtained employing at least one $\bb{Z}_{2}\subset
{\rm W}$. Then, the fermion content of the four-dimensional theory is found to transform in linear combinations of the two copies
of the CSDR-surviving left-handed fermions. Details  will be given in chapter~\ref{Classification}. As I will discuss there,
employing $\bb{Z}_{2}\times\bb{Z}_{2}\subseteq {\rm W}$ or $\bb{Z}_{2}\times\bb{Z}_{2}\subseteq {\rm W}\times{\rm Z}(S)$ can also
lead to interesting models.

Therefore the interesting cases for further study are
\begin{equation}
F^{S/R}=\begin{cases}
             \bb{Z}_{2}\subseteq{\rm W} \\
             \bb{Z}_{2}\times\bb{Z}_{2}\subseteq {\rm W}\\
             \bb{Z}_{2}\times\bb{Z}_{2}\subseteq {\rm W}\times{\rm Z}(S)\,.
             \end{cases}\label{FCases}
\end{equation}

\section{Symmetry breaking patterns of $E_{6}$-like GUTs}\label{TopologicallyInducedGaugeGroupBreaking-E6}

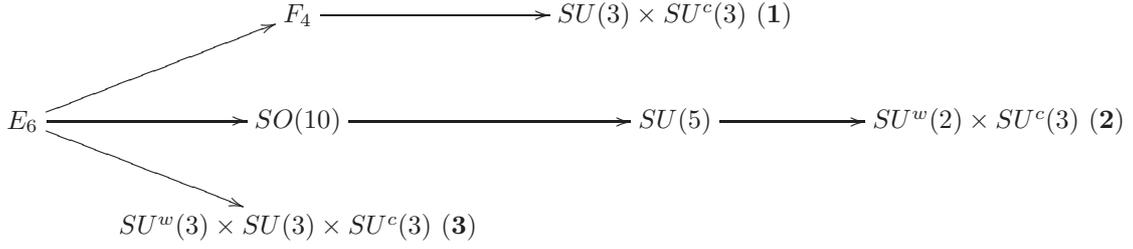
\begin{figure}
\begin{footnotesize}
$$
\xymatrix{
                                 & F_{4}\ar[r]                    & SU(3)\times SU^{c}(3)~(\mbf{1}) &\\
E_{6}\ar[ur]\ar[r]\ar[dr]& SO(10)\ar[r]                 & SU(5)\ar[r]                & SU^{w}(2)\times SU^{c}(3)~(\mbf{2})\\
                                 & SU^{w}(3)\times SU(3)\times SU^{c}(3)~(\mbf{3}) &           & }
$$
\caption{\emph{$E_{6}$ decompositions leading potentially to SM gauge group structure.}
\label{DecompsDiscreteSymm}}
\end{footnotesize}
\end{figure}

Here I determine the image, $T^{H}$, that each of the discrete symmetries of eq.~(\ref{FCases}) induces in the
gauge group $H=E_{6}$. I consider embeddings of the $F^{S/R}$ discrete symmetries  into abelian subgroups of $E_{6}$ 
and examine their topologically induced symmetry breaking patterns~\cite{Witten:1985xc}. These are realised by a diagonal matrix
$U_{g}$ of unit determinant, which as explained in section~\ref{sec:WilsonFluxBreaking-theory}, has to be homomorphic to the
considered discrete symmetry. In fig.~\ref{DecompsDiscreteSymm} I present those $E_{6}$ decompositions which potentially lead to
the SM gauge group structure~\cite{Slansky:1981yr}.

\subsection{The $\bb{Z}_{2}$ case}
\label{TopologicallyInducedGaugeGroupBreaking-E6-Z2}

{\myverysubsection{Embedding $(\mbf{1})$:~
${\bb{Z}_{2}\hookrightarrow SU(3)}$ of ${E_{6}\supset F_{4}\supset SU(3)\times SU^{c}(3)}$\,.}}

Let me consider the maximal subgroups of $E_{6}$ and the corresponding decomposition of fundamental and adjoint irreps
\begin{equation}
\begin{array}{l}
E_{6}\supset F_{4}\supset SU(3)\times SU^{c}(3)\\
\begin{array}{r@{}l@{}l@{}}
\mbf{27}&=&(\mbf{1},\mbf{1})+(\mbf{8},\mbf{1})+(\mbf{3},\mbf{3})+(\obar{\mbf{3}},\obar{\mbf{3}})\,,\\
\mbf{78}&=&(\mbf{8},\mbf{1})+(\mbf{3},\mbf{3})+(\obar{\mbf{3}},\obar{\mbf{3}})
              +(\mbf{8},\mbf{1})+(\mbf{1},\mbf{8})+(\mbf{6},\obar{\mbf{3}})+(\obar{\mbf{6}},\mbf{3})\,.
\end{array}\label{E6toF4toSU3SU3}
\end{array}
\end{equation}
Let me also embed the $F^{S/R}=\bb{Z}_{2}$ discrete symmetry in the $SU(3)$ group factor above. There exist two distinct
possibilities of embedding, either $\bb{Z}_{2}\hookrightarrow U^{I}(1)$ which appears under the
$SU(3)\supset SU(2)\times U^{II}(1)\supset U^{I}(1)\times U^{II}(1)$ decomposition or $\bb{Z}_{2}\hookrightarrow U^{II}(1)$. Since
the former is trivial, namely cannot break the $SU(3)$ appearing in eq.~(\ref{E6toF4toSU3SU3}), only the latter is interesting for
further investigation. This is realised as
\begin{equation}
U_{g}^{(1)}=diag(-1,-1,1)\,.\label{Ug1}
\end{equation}
Indeed $(U_{g}^{(1)})^{2}=\one_{3}$ as required by the $F^{S/R}\mapsto H$ homomorphism and $det(U_{g}^{(1)})=1$ since $U_{g}$ is
an $H$ group element.

Then, the various components of the decomposition of $SU(3)$ irreps under $SU(2)\times U(1)$ acquire
the underbraced phase factors in the following list
\begin{equation}
\begin{array}{l}
SU(3) \supset  SU(2)\times U(1)\\
\begin{array}{r@{}l@{}l@{}}
\mbf{3}&=&\ubr{\mbf{1}_{(-2)}}_{(+1)}+\ubr{\mbf{2}_{(1)}}_{(-1)}\,,\\
\mbf{6}&=&\ubr{\mbf{1}_{(-4)}}_{(+1)}+\ubr{\mbf{2}_{(-1)}}_{(-1)}+\ubr{\mbf{3}_{(2)}}_{(+1)}\,,\\
\mbf{8}&=&\ubr{\mbf{1}_{(0)}}_{(+1)}+\ubr{\mbf{3}_{(0)}}_{(+1)}+\ubr{\mbf{2}_{(-3)}}_{(-1)}+\ubr{\mbf{2}_{(3)}}_{(-1)}\,.
\end{array}\label{SU3toSU2U1HB}
\end{array}
\end{equation}
Consequently the various components of the decomposition of $E_{6}$ irreps~(\ref{E6toF4toSU3SU3}) under
$
F_{4}\supset SU(3)\times SU^{c}(3)\supset (SU(2)\times U(1))\times SU^{c}(3)
$
acquire the underbraced phase factors in the following list
\begin{equation}
\begin{array}{r@{}l@{}l@{}}
E_{6}&\supset& SU(2)\times SU^{c}(3)\times U(1)\\
\mbf{27}&=&\ubr{(\mbf{1},\mbf{1})_{(0)}}_{(+1)}+\ubr{(\mbf{1},\mbf{1})_{(0)}}_{(+1)}
              +\ubr{(\mbf{3},\mbf{1})_{(0)}}_{(+1)}\\
            &+&\ubr{(\mbf{1},\mbf{3})_{(-2)}}_{(+1)}+\ubr{(\mbf{2},\mbf{3})_{(1)}}_{(-1)}
              +\ubr{(\mbf{1},\obar{\mbf{3}})_{(2)}}_{(+1)}+\ubr{(\mbf{2},\obar{\mbf{3}})_{(-1)}}_{(-1)}
              + \ubr{(\mbf{2},\mbf{1})_{(-3)}}_{(-1)}+\ubr{(\mbf{2},\mbf{1})_{(3)}}_{(-1)}\,,\\
\mbf{78}&=&\ubr{(\mbf{1},\mbf{1})_{(0)}}_{(+1)}+\ubr{(\mbf{3},\mbf{1})_{(0)}}_{(+1)}
              +\ubr{(\mbf{1},\mbf{1})_{(0)}}_{(+1)}+\ubr{(\mbf{3},\mbf{1})_{(0)}}_{(+1)}
              +\ubr{(\mbf{1},\mbf{8})_{(0)}}_{(+1)}\\
            &+&\ubr{(\mbf{1},\mbf{3})_{(-2)}}_{(+1)}+\ubr{(\mbf{1},\obar{\mbf{3}})_{(2)}}_{(+1)}
              +\ubr{(\mbf{1},\obar{\mbf{3}})_{(-4)}}_{(+1)}+\ubr{(\mbf{1},\mbf{3})_{(4)}}_{(+1)}\\
           &+&\ubr{(\mbf{2},\mbf{1})_{(-3)}}_{(-1)}+\ubr{(\mbf{2},\mbf{1})_{(3)}}_{(-1)}
             + \ubr{(\mbf{2},\mbf{1})_{(-3)}}_{(-1)}+\ubr{(\mbf{2},\mbf{1})_{(3)}}_{(-1)}\\
           &+&\ubr{(\mbf{2},\mbf{3})_{(1)}}_{(-1)}+\ubr{(\mbf{2},\obar{\mbf{3}})_{(-1)}}_{(-1)}
             +\ubr{(\mbf{2},\mbf{3})_{(1)}}_{(-1)}+\ubr{(\mbf{2},\obar{\mbf{3}})_{(-1)}}_{(-1)}\\
           &+&\ubr{(\obar{\mbf{3}},\mbf{3})_{(-2)}}_{(+1)}+\ubr{(\mbf{3},\obar{\mbf{3}})_{(+2)}}_{(+1)}\,.
\end{array}\label{E6toF4toSU3SU3HB}
\end{equation}
According to the discussion in section~\ref{sec:WilsonFluxBreaking-theory} the four-dimensional gauge group after the topological
breaking is given  by $K'=C_{H}(T^{H})$. Counting the number of singlets under the action of $U_{g}^{(1)}$ in  the $\mbf{78}$
irrep. above suggests that $K'=SO(10)\times U(1)$, a fact which subsequently is determined according to the following
decomposition of the $\mbf{78}$ irrep.
\begin{equation}
\begin{array}{l}
E_{6}\supset SO(10)\times U(1)\\
\begin{array}{r@{}l@{}l@{}}
\mbf{27}&=&\ubr{\mbf{1}_{(-4)}}_{(+1)}+\ubr{\mbf{10}_{(-2)}}_{(+1)}+\ubr{\mbf{16}_{(1)}}_{(-1)}\,,\\
\mbf{78}&=&\ubr{\mbf{1}_{(0)}}_{(+1)}+\ubr{\mbf{45}_{(0)}}_{(+1)}+\ubr{\mbf{16}_{(-3)}}_{(-1)}
+\ubr{\obar{\mbf{16}}_{(3)}}_{(-1)}\,.
\end{array}\label{E6toSO10U1}
\end{array}
\end{equation}

It is interesting to note that although one would naively expect the $E_{6}$ gauge group to break further towards the SM one this
is not the case. The singlets under the action of $U_{g}^{(1)}$ which occur  in the adjoint irrep. of $E_{6}$ in
eq.~(\ref{E6toF4toSU3SU3HB}) add up to provide a larger final unbroken gauge symmetry, namely $SO(10)\times U(1)$.

\newpage
{\myverysubsection{Embedding~$(\mbf{2})$:
$\bb{Z}_{2}\hookrightarrow SU(5)$ of
$$
E_{6}\supset SO(10)\times U(1)\supset SU(5)\times U(1)\times U(1)\,.
$$}}
Similarly, I consider the maximal subgroups of $E_{6}$ and the corresponding decomposition of the fundamental and
adjoint irreps
\begin{equation}
\begin{array}{l}
E_{6}\supset SO(10)\times U(1)\supset SU(5)\times U(1)\times U(1)\\
\begin{array}{r@{}l@{}l@{}}
\mbf{27}&=&\mbf{1}_{(0,-4)}+\mbf{5}_{(2,-2)}+\obar{\mbf{5}}_{(-2,-2)}+\mbf{1}_{(-5,1)}+\obar{\mbf{5}}_{(3,1)}+\mbf{10}_{(-1,1)}\,,
\\
\mbf{78}&=&\mbf{1}_{(0,0)}+\mbf{1}_{(0,0)}+\mbf{24}_{(0,0)}+\mbf{1}_{(-5,-3)}+\mbf{1}_{(5,3)}\\
            &+&\mbf{5}_{(-3,3)}+\obar{\mbf{5}}_{(3,-3)}
              +\mbf{10}_{(4,0)}+\obar{\mbf{10}}_{(-4,0)}+\mbf{10}_{(-1,-3)}+\obar{\mbf{10}}_{(1,3)}\,.
\end{array}\label{E6toSO10U1toSU5U1U1}
\end{array}
\end{equation}
My choice is to embed the $\bb{Z}_{2}$ discrete symmetry in an abelian $SU(5)$ subgroup in a way that is realised by
the diagonal matrix
\begin{equation}
U_{g}^{(2)}=diag(-1,-1,1,1,1)\,.\label{Ug2}
\end{equation}
Then the various components of the $SU(5)$ irreps decomposed under the $SU(2)\times SU(3)\times U(1)$ decomposition acquire the
underbraced phase factors in the following list
\begin{equation}
\begin{array}{l}
SU(5)\supset SU(2)\times SU(3)\times U(1)\\
\begin{array}{r@{}l@{}l@{}}
\mbf{5}&=&\ubr{(\mbf{2},\mbf{1})_{(3)}}_{(-1)}+\ubr{(\mbf{1},\mbf{3})_{(-2)}}_{(+1)}\,,\\
\mbf{10}&=&\ubr{(\mbf{1},\mbf{1})_{(6)}}_{(+1)}+\ubr{(\mbf{1},\obar{\mbf{3}})_{(-4)}}_{(+1)}
              +\ubr{(\mbf{2},\mbf{3})_{(1)}}_{(-1)}\,,\\
\mbf{24}&=&\ubr{(\mbf{1},\mbf{1})_{(0)}}_{(+1)}+\ubr{(\mbf{3},\mbf{1})_{(0)}}_{(+1)}+\ubr{(\mbf{1},\mbf{8})_{(0)}}_{(+1)}
              +\ubr{(\mbf{2},\mbf{3})_{(-5)}}_{(-1)}+\ubr{(\mbf{2},\obar{\mbf {3}})_ {(5)}}_{(-1)}\,.
\end{array}\label{SU5toSU2SU3U1HB}
\end{array}
\end{equation}
 It can be proven, along the lines of the previous case $(\mbf{1})$, that $U_{g}^{(2)}$ leads to the breaking 
$E_{6}\to SU(2)\times SU(6)$
\begin{equation}
\begin{array}{l}
E_{6}\supset SU(2)\times SU(6)\\
\begin{array}{l@{}l@{}l@{}}
\mbf{27}&=&\ubr{(\mbf{2},\obar{\mbf{6}})}_{(-1)}+\ubr{(\mbf{1},\mbf{15})}_{(+1)}\,,\\
\mbf{78}&=&\ubr{(\mbf{3},\mbf{1})}_{(+1)}+\ubr{(\mbf{1},\mbf{35})}_{(+1)}+\ubr{(\mbf{2},\mbf{20})}_{(-1)}\,,
\end{array}\label{E6toSU2SU6HB}
\end{array}
\end{equation}
i.e. I find again an enhancement of the final gauge group as compared to the naively expected one.

Note that other choices of $\bb{Z}_{2}$ into $SU(5)$ embeddings either lead to trivial or to phenomenologically uninteresting
results.

\newpage
{\myverysubsection{Embedding $(\mbf{3})$:~
$\bb{Z}_{2}\hookrightarrow SU(3)$ of $E_{6}\supset SU^{w}(3)\times SU(3)\times SU^{c}(3)$\,.}}

I consider the maximal subgroup of $E_{6}$ and the corresponding decomposition of fundamental and adjoint irreps
\begin{equation}
\begin{array}{l}
E_{6}\supset SU^{w}(3)\times SU(3)\times SU^{c}(3)\\
\begin{array}{l@{}l@{}l@{}}
\mbf{27}&=&(\obar{\mbf{3}},\mbf{3},\mbf{1})+(\mbf{3},\mbf{1},\mbf{3})+(\mbf{1},\obar{\mbf{3}},\obar{\mbf{3}})\,,\\
\mbf{78}&=&(\mbf{8},\mbf{1},\mbf{1})+(\mbf{1},\mbf{8},\mbf{1})+(\mbf{1},\mbf{1},\mbf{8})+
(\mbf{3},\mbf{3},\obar{\mbf{3}})+(\obar{\mbf{3}},\obar{\mbf{3}},\mbf{3})\,.\label{E6-SU3trits}
\end{array}
\end{array}
\end{equation}
Furthermore I assume an $\bb{Z}_{2}\hookrightarrow SU(3)$ embedding, which is realised by
\begin{equation}
U_{g}^{(3)}=(\one_{3})\otimes diag(-1, -1, 1)\otimes(\one_{3})\,.\label{Ug3}
\end{equation}
Although this choice of embedding is not enough to lead to the SM gauge group structure, the results will be usefull for the
discussion of the $\bb{Z}_{2}\times\bb{Z}_{2}'$ case which is presented in subsection~\ref{E6-Z2Z2}. With the choice of embedding
realised by the eq.~(\ref{Ug3}) the second $SU(3)$ decomposes under $SU(2)\times U(1)$ as in eq.~(\ref{SU3toSU2U1HB})
and leads to the breaking~(\ref{E6toSU2SU6HB}), as before.
As was mentioned in case~($\mbf{1}$) the choice of embedding $\bb{Z}_{2}\hookrightarrow U^{I}(1)$, which appears under the
decomposition $SU(3)\supset SU(2)\times U^{II}(1)\supset U^{I}(1)\times U^{II}(1)$ of eq.~(\ref{E6-SU3trits}), cannot break the
$SU(3)$ group factor and it is not an interesting case for further investigation.

In table~\ref{E6-Breaking-Z2} I summarise the above results, concerning the topologically induced symmetry breaking
patterns of the $E_{6}$ gauge group.

\begin{table}
\caption[\textbf{Embeddings of $\bb{Z}_{2}$ discrete symmetry in $E_{6}$ GUT and its symmetry breaking
patterns.}]
{\textbf{Embeddings of $\bb{Z}_{2}$ discrete symmetry in $E_{6}$ GUT and its symmetry breaking
patterns.} \emph{$U_{g}^{(1)}=diag(-1,-1,1)$ and $U_{g}^{(2)}=diag(-1,-1,1,1,1)$ as in text.}
\label{E6-Breaking-Z2}}
\begin{center}
\begin{math}
\begin{array}{|c!{\vrule width 1pt}c|c|c!{\vrule width 1pt}c|c|c|}\hline
{\bf Embedd.}
& U_{g}
&\mbf{K'}
\\ \hhline{*{3}{=}}
\mbf{1}
&
U_{g}^{(1)}
&
SO(10)\times U(1)
\\\hline
\mbf{2}
&
U_{g}^{(2)}
&
SU(2)\times SU(6)
\\\hline
\mbf{3}
&
\one_{3}\otimes U_{g}^{(1)}\otimes\one_{3}
&
SU(2)\times SU(6)
\\\hline
\end{array}
\end{math}
\end{center}
 \end{table}

\subsection{The $\bb{Z}_{2}\times\bb{Z}_{2}'$ case}\label{E6-Z2Z2}
\label{TopologicallyInducedGaugeGroupBreaking-E6-Z2Z2}

{\myverysubsection{Embedding~$(\mbf{2'})$:
$\bb{Z}_{2}\hookrightarrow SO(10)$~and~$\bb{Z}_{2}'\hookrightarrow SU(5)$~of 
$$
E_{6}\supset SO(10)\times U(1)\supset SU(5)\times U(1)\times U(1)\,.
$$}}
Here I embed the $\bb{Z}_{2}$ of the $\bb{Z}_{2}\times\bb{Z}_{2}'$ discrete symmetry in the $SU(5)$ appearing under
the  decomposition $E_{6}\supset SO(10)\times U(1)\supset SU(5)\times U(1)\times U(1)$ as in case ($\mbf{2}$) above.
Furthermore I embed the  $\bb{Z}_{2}'$ discrete symmetry in the $SO(10)$ as
\begin{equation}
U_{g}'=-\one_{10}\,. 
\end{equation}
This leads to the breaking $E_{6}\supset SU(2)\times SU(6)$ as before but with the signs of the phase factors, which appear  in
eq. (\ref{E6toSU2SU6HB}), being reversed under the action of $U_{g}^{(2)}U_{g}'$.

{\myverysubsection{
Embedding~$(\mbf{3'})$:
$\bb{Z}_{2}\hookrightarrow SU(3)$~and~$\bb{Z}_{2}'\hookrightarrow SU^{w}(3)$~of
$$
E_{6}\supset SU^{w}(3)\times SU(3)\times SU^{c}(3)\,.
$$
}}
Here I embed the $\bb{Z}_{2}$ of the $\bb{Z}_{2}\times\bb{Z}_{2}'$ discrete symmetry in the $SU(3)$ group factor appearing under
the $E_{6}\supset SU(3)^{w}\times SU(3)\times SU(3)^{c}$ as in case ($\mbf{3}$) above. Furthermore I embed the $\bb{Z}_{2}'$
discrete symmetry in the $SU(3)^{w}$ group factor in a similar way. Then the embedding ($\mbf{3}'$), which I discuss here, is
realised by considering an element of the $E_{6}$ gauge group
\begin{equation}
U_{g}'U_{g}^{(3)}=diag(-1,-1,1) \otimes diag(-1,-1,1)\otimes (\one_{3})\label{Ug4}\,,
\end{equation}
which leads to the breaking $E_{6}\to SU^{(i)}(2)\times SU^{(ii)}(2)\times SU(4)\times U(1)$ as it is clear from the following
decomposition of $\mbf{78}$ irrep.
\begin{align}
&E_{6}\supset SO(10)\times U(1)\supset  SU^{(i)}(2)\times SU^{(ii)}(2)\times SU(4)\times U(1)\nn\\
&\begin{array}{l@{}l@{}l@{}}
E_{6}&\supset& SU^{(i)}(2)\times SU^{(ii)}(2)\times SU(4)\times U(1)\\
\mbf{78}&=&\ubr{(\mbf{1},\mbf{1},\mbf{1})_{(0)}}_{(+1)}
              +\ubr{(\mbf{1},\mbf{3},\mbf{1})_{(0)}}_{(+1)}+\ubr{(\mbf{3},\mbf{1},\mbf{1})_{(0)}}_{(+1)}
              +\ubr{(\mbf{1},\mbf{1},\mbf{15})_{(0)}}_{(+1)}+\ubr{(\mbf{2},\mbf{2},\mbf{6})_{(0)}}_{(-1)}\\
            &+&\ubr{(\mbf{2},\mbf{1},\mbf{4})_{(-3)}}_{(-1)}+\ubr{(\mbf{2},\mbf{1},\obar{\mbf{4}})_{(3)}}_{(-1)}
              + \ubr{(\mbf{1},\mbf{2},\mbf{4})_{(3)}}_{(-1) }+\ubr{(\mbf{1},\mbf{2},\obar{\mbf{4}})_{(-3)}}_{(-1)}\,.
\end{array}\label{E6toSO10U1toSU4SU2SU2U1HB-78}
\end{align}
Furthermore the irrep. $\mbf{27}$ of $E_{6}$ decomposes under the same breaking as
\begin{equation}
\begin{array}{l@{}l@{}l@{}}
E_{6}&\supset& SU^{(i)}(2)\times SU^{(ii)}(2)\times SU(4)\times U(1)\\
\mbf{27}&=&\ubr{(\mbf{1},\mbf{1},\mbf{1})_{(4)}}_{(+1)}+\ubr{(\mbf{2},\mbf{2},\mbf{1})_{(-2)}}_{(+1)}
              + \ubr{(\mbf{1},\mbf{1},\mbf{6})_{(-2)}}_{(+1)}\\
            &+&\ubr{(\mbf{2},\mbf{1},\mbf{4})_{(1)}}_{(-1)}+\ubr{(\mbf{1},\mbf{2},\obar{\mbf{4}})_{(1)}}_{(-1)}\,.\\
\end{array}
\end{equation}

In table~\ref{E6-Breaking-Z2Z2} I summarise the above results, concerning the topologically induced symmetry breaking
patterns of the $E_{6}$ gauge group.

\begin{table}
\caption[\textbf{Embeddings of $\bb{Z}_{2}\times\bb{Z}_{2}'$ discrete symmetries in $E_{6}$ GUT and its symmetry breaking
patterns.}]
{\textbf{Embeddings of $\bb{Z}_{2}\times\bb{Z}_{2}'$ discrete symmetries in $E_{6}$ GUT and its symmetry breaking
patterns.} \emph{$U_{g}^{(1)}=diag(-1,-1,1)$ and $U_{g}^{(2)}=diag(-1,-1,1,1,1)$ as in text.}
\label{E6-Breaking-Z2Z2}}
\begin{center}
\begin{math}
\begin{array}{|c!{\vrule width 1pt}c|c|c!{\vrule width 1pt}c|c|c|}\hline
{\bf Embedd.}
& U_{g}
&U_{g}'
&\mbf{K'}
\\ \hhline{*{4}{=}}
\mbf{2'}
&
U_{g}^{(2)}
&
-\one_{10}
&
SU(2)\times SU(6)
\\\hline
\mbf{3'}
&
U_{g}^{(1)}\otimes\one_{3}\otimes \one_{3}
&
\one_{3}\otimes U_{g}^{(1)}\otimes \one_{3}
&
SU^{(i)}(2)\times SU^{(ii)}(2)\times SU(4)\times U(1)
\\\hline
\end{array}
\end{math}
\end{center}
\end{table}

\newpage
\section{Symmetry breaking pattern of  $SO(10)$-like GUTs}
\label{TopologicallyInducedGaugeGroupBreaking-SO10}

Here I determine the image, $T^{H}$, that each of the discrete symmetries of eq.~(\ref{FCases}) induces in the
gauge group $H=SO(10)$. I consider embeddings of the $F^{S/R}$ discrete symmetries  into abelian subgroups of $SO(10)$ GUTs and
examine their topologically induced symmetry breaking patterns. The interesting $F^{S/R}\hookrightarrow SO(10)$
embeddings are those which potentially lead to SM gauge group structure, i.e.
$$
SO(10)\supset SU(5)\times U^{II}(1)\supset SU^{w}(2)\times SU^{c}(3)\times U^{I}(1)\times U^{II}(1)\,.
$$

\subsection{The $\bb{Z}_{2}$ case}
\label{TopologicallyInducedGaugeGroupBreaking-SO10-Z2}

{\myverysubsection{ Embedding~$(\mbf{1})$}:~$\bb{Z}_{2}\hookrightarrow SU(5)$ of $SO(10)\supset SU(5)\times U(1)$\,.}

In the present case I assume the maximal subgroup of $SO(10)$
\begin{equation}
\begin{array}{l}
SO(10)\supset SU(5)\times U^{II}(1)\\
\begin{array}{r@{}l@{}l@{}}
\mbf{10}&=&\mbf{5}_{(2)}+\obar{\mbf{5}}_{(-2)}\,,\\
\mbf{16}&=&\mbf{1}_{(-5)}+\obar{\mbf{5}}_{(3)}+\mbf{10}_{(-1)}\,,\\
\mbf{45}&=&\mbf{1}_{(0)}+\mbf{24}_{(0)}+\mbf{10}_{(4)}+\obar{\mbf{10}}_{(-4)}\,,
\end{array}\label{SO10SU5U1}
\end{array}
\end{equation}
and embed a  $\bb{Z}_{2}\hookrightarrow SU(5)$ which is realised as in eq.~(\ref{Ug2}). Then, the $\mbf{5}$, $\mbf{10}$ and
$\mbf{24}$ irreps of $SU(5)$ under the $SU(5)\supset SU(2)\times SU(3)\times U(1)$ decomposition read as in
eq.~(\ref{SU5toSU2SU3U1HB}) and lead to the breaking $SO(10)\to SU^{a}(2)\times SU^{b}(2)\times SU(4)$
\begin{equation}
\begin{array}{l}
SO(10) \supset  SU^{(i)}(2)\times SU^{(ii)}(2)\times SU(4)\\
\begin{array}{r@{}l@{}l@{}}
\mbf{10}&=&\ubr{(\mbf{2},\mbf{2},\mbf{1})}_{(-1)}+\ubr{(\mbf{1},\mbf{1},\mbf{6})}_{(+1)}\,,\\
\mbf{16}&=&\ubr{(\mbf{2},\mbf{1},\mbf{4})}_{(-1)}+\ubr{(\mbf{1},\mbf{2},\obar{\mbf{4}})}_{(+1)}\,,\\
\mbf{45}&=&\ubr{(\mbf{3},\mbf{1},\mbf{1})}_{(+1)}+\ubr{(\mbf{1},\mbf{3},\mbf{1})}_{(+1)}+\ubr{(\mbf{1},\mbf{1},\mbf{15})}_{(+1)}
              +\ubr{(\mbf{2},\mbf{2},\mbf {6})}_{(-1)}\,.
\end{array}\label{SO10toSU2SU2SU4HB}
\end{array}
\end{equation}

Note again that although one would naively expect the $SO(10)$ gauge group to break towards SM, this is not the case.

For completeness in table~\ref{SO10-Breaking-Z2}  I present the above case.

\begin{table}
\caption{\textbf{Embedding of $\bb{Z}_{2}$ discrete symmetry in $SO(10)$ GUT and its symmetry breaking pattern.}
\emph{$U_{g}^{(2)}=diag(-1,-1,1,1,1)$ as in text.}
\label{SO10-Breaking-Z2}}
\begin{center}
\begin{math}
\begin{array}{|c!{\vrule width 1pt}c|c|c!{\vrule width 1pt}c|c|c|}\hline
{\bf Embedd.}
& U_{g}
&\mbf{K'}
\\ \hhline{*{3}{=}}
\mbf{1}
&
U_{g}^{(2)}
&
SU^{(i)}(2)\times SU^{(ii)}(2)\times SU(4)
\\\hline
\end{array}
\end{math}
\end{center}
\end{table}

\subsection{The $\bb{Z}_{2}\times \bb{Z}_{2}'$ case.}
\label{TopologicallyInducedGaugeGroupBreaking-SO10-Z2Z2}

{\myverysubsection{ Embedding~$(\mbf{1'})$}:~$\bb{Z}_{2}\hookrightarrow SU(5)$ and $\bb{Z}_{2}'\hookrightarrow SO(10)$ 
of $SO(10)\supset SU(5)\times U(1)$\,.}

Note that a second $\bb{Z}_{2}$ cannot break the $K'=SU^{a}(2)\times SU^{b}(2)\times SU(4)$ further. However by choosing the
non-trivial embedding $U_{g}'=-\one_{10}$ of $\bb{Z}_{2}$ in the $SO(10)$ the phase factors appearing in
eq.~(\ref{SO10toSU2SU2SU4HB}) have their signs reversed under the action of $U_{g}^{(2)}U_{g}'$.

Again in table~\ref{SO10-Breaking-Z2Z2} I present the above case.

\begin{table}
\caption{\textbf{Embedding of $\bb{Z}_{2}\times\bb{Z}_{2}'$ discrete symmetries in $SO(10)$ GUT and its symmetry breaking
pattern.} \emph{$U_{g}^{(2)}=diag(-1,-1,1,1,1)$ as in text.}
\label{SO10-Breaking-Z2Z2}}
\begin{center}
\begin{math}
\begin{array}{|c!{\vrule width 1pt}c|c|c!{\vrule width 1pt}c|c|c|}\hline
{\bf Embedd.}
& U_{g}
& U_{g}'
&\mbf{K'}
\\ \hhline{*{4}{=}}
\mbf{1'}
&
U_{g}^{(2)}
&
-\one_{10}
&
SU^{(i)}(2)\times SU^{(ii)}(2)\times SU(4)
\\\hline
\end{array}
\end{math}
\end{center}
\end{table}

\section[Reduction of $\cN=1$, $E_{8}$ YM-Dirac over $SO(7)/SO(6)$ revisited]
{Reduction of $\cN=1$, $E_{8}$ Yang-Mills-Dirac over $SO(7)/SO(6)$ revisited}
\label{WilsonFluxBreakingMechanism-example}

In section~\ref{CSDR-example} an illuminating example of application of the CSDR scheme was presented. I discussed there the
dimensional reduction of a $\cN=1$, $E_{8}$ Yang-Mills-Dirac theory over the coset space $S/R=SO(7)/SO(6)\sim S^{6}$. The gauge
group of the resulting four-dimensional theory was found to be $H=SO(10)$. The theory was provided with scalars belonging in the
$\mbf{10}$ of $SO(10)$ and with two copies of chiral fermions transforming as $\mbf{16}_{L}$ under the same gauge group. The
scalars of the theory being in the fundamental rep. of the $H$ gauge group are not able to provide the appropriate superstrong
symmetry breaking towards the SM. However one can use the freely acting discrete symmetries, $F^{S/R}$, of the coset space
$S/R=SO(7)/SO(6)$, assume non-trivial topological properties for it and break the gauge symmetry further according
to section~\ref{sec:WilsonFluxBreaking-theory}.

To be more specific the $F^{S/R}$ discrete symmetries of the coset space $SO(7)/SO(6)$ (case `$\mbf{a}$' in
table~\ref{SO6VectorSpinorContentSCosets}) are that of Weyl, $\rm{W}=\bb{Z}_{2}$, and the center of $S$,
$\rm{Z}(S)=\bb{Z}_{2}$. As it was explained in section~\ref{sec:InterestingDiscreteSymmetries} the use of  ${\rm Z}(S)$ is
excluded. On the other hand, according to the discussion in subsection~\ref{TopologicallyInducedGaugeGroupBreaking-SO10-Z2}
the ${\rm W}$ discrete symmetry leads to a four-dimensional theory with the following gauge group
\begin{equation*}
K'=C_{H}(T^{H})=SU^{(i)}(2)\times SU^{(ii)}(2)\times SU(4)\,.
\end{equation*}
According to section~\ref{sec:WilsonFluxBreaking-theory}, the surviving field content has to be invariant under the combined
action of the considered discrete symmetry itself, $F^{S/R}$, and its induced image in the $H$ gauge group,
$T^{H}$~[eq.~(\ref{WFluxSurvivingField})]. Using the ${\rm W}=\bb{Z}_{2}$ discrete symmetry, the irrep. $\mbf{10}$ of $SO(10)$
decomposes under the $SU^{(i)}(2)\times SU^{(ii)}(2)\times SU(4)$ is given in eq.~(\ref{SO10toSU2SU2SU4HB}),
\begin{equation}
\begin{array}{l}
SO(10) \supset  SU^{(i)}(2)\times SU^{(ii)}(2)\times SU(4)\\
\begin{array}{r@{}l@{}l@{}}
\mbf{10}&=&\ubr{(\mbf{2},\mbf{2},\mbf{1})}_{(-1)}+\ubr{(\mbf{1},\mbf{1},\mbf{6})}_{(+1)}\,.
\end{array}\label{SO10toSU2SU2SU4HB-10}
\end{array}
\end{equation}
Then, recalling that the $R=SO(6)$ content of the vector $B_{0}=SO(7)/SO(6)$ is invariant under the action of ${\rm W}$
(table~\ref{SO6VectorSpinorContentSCosets}), I conclude that the four-dimensional theory contains scalars transforming according
to
$$
(\mbf{1},\mbf{1},\mbf{6})
$$
of $K'$.
Similarly, the irrep. $\mbf{16}$ of $SO(10)$ decomposes under the $K'$ as
\begin{equation}
\begin{array}{l}
SO(10) \supset  SU^{(i)}(2)\times SU^{(ii)}(2)\times SU(4)\\
\begin{array}{r@{}l@{}l@{}}
\mbf{16}&=&\ubr{(\mbf{2},\mbf{1},\mbf{4})}_{(-1)}+\ubr{(\mbf{1},\mbf{2},\obar{\mbf{4}})}_{(+1)}\,.
\end{array}\label{SO10toSU2SU2SU4HB-16}
\end{array}
\end{equation}
In this case the spinor of the tangent space of $SO(7)/SO(6)$ decomposed under $R=SO(6)$ is obviously $\mbf{4}$.
Then, since the ${\rm W}$ transformation properties for the spinor is $\mbf{4}\lra\obar{\mbf{4}}$
(table~\ref{SO6VectorSpinorContentSCosets}), the fermion content of the four-dimensional theory
transforms as
\begin{equation}
(\mbf{2},\mbf{1},\mbf{4})_{L}-(\mbf{2},\mbf{1}, \mbf{4})'_{L} \qquad\mbox{and}\qquad
(\mbf{1},\mbf{2},\mbf{4})_{L}+(\mbf{1},\mbf{2}, \mbf{4})'_{L}\label{4d-fermions-H1-1a}
\end{equation}
under $K'$.

According to the discussion in subsection~\ref{sec:4D-theory}, dimensional reduction over the $\tfrac{SO(7)}{SO(6)}$ symmetric coset
space leads to a four-dimensional potential with  spontaneously symmetry breaking form. However, since the four-dimensional
scalar fields belong in the $(\mbf{1},\mbf{1},\mbf{6})$ under the $K'$ gauge group obtaining a v.e.v. break the $SU(3)$ colour. Therefore, employing the ${\rm W}$ discrete symmetry is not an interesting case to investigate further.

On the other hand, if I use the ${\rm W}\times{\rm Z}(S)=\bb{Z}_{2}\times\bb{Z}_{2}$ discrete symmetry, the Wilson flux
breaking mechanism leads again to the Pati-Salam gauge group, $K'$ (see
subsection~\ref{TopologicallyInducedGaugeGroupBreaking-SO10-Z2Z2}). However in this case, all the underbraced phase factors of
eqs.~(\ref{SO10toSU2SU2SU4HB-10}) and~(\ref{SO10toSU2SU2SU4HB-16}) are multiplied by $-1$. As before, the surviving fields in four
dimensions have to be  invariant under the action $F^{S/R}\oplus T^{H}$. Therefore the four-dimensional theory now contains
scalars transforming according to
$$
(\mbf{2},\mbf{2},\mbf{1})
$$
of $K'$, and two copies of chiral fermions transforming as in eq.~(\ref{4d-fermions-H1-1a}) but with the signs of the linear
combinations reversed.

Note that, if only the CSDR mechanism was applied, the final gauge group would be  (see section~\ref{CSDR-example})
\begin{equation*}
\cH=C_{E_{8}}(SO(7))=SO(9)\,,
\end{equation*}
i.e.  the $\mbf{10}$ of $SO(10)$ would obtain a v.e.v. and lead to the spontaneous symmetry breaking
\begin{equation}
\begin{array}{l}
SO(10)\to SO(9)\\
\mbf{10}=\langle\mbf{1}\rangle + \mbf{9}\,.
\end{array}\label{SO10SO9-breaking-2}
\end{equation}

However now I employ the Wilson flux breaking mechanism which breaks the gauge symmetry further in higher dimensions.
It is instructive to understand the spontaneous symmetry breaking indicated in eq.~(\ref{SO10SO9-breaking-2}) in this context too.
A straightforward examination of the gauge group structure and the representations of the scalars that are involved, suggests the
breaking indicated in eq.~(\ref{SO10toSU2SU2SU4HB-10}) is realised in the present context as
\begin{equation}
\begin{array}{r@{}c@{}l@{}}
SU^{(i)}(2)\times SU^{(ii)}(2)\times SU(4)&~\to~& SU^{diag}(2)\times SU(4)\\
(\mbf{2},\mbf{2},\mbf{1})&~=~&\langle(\mbf{1},\mbf{1})\rangle+(\mbf{3},\mbf{1})\,,
\end{array}\label{SU2SU2SU4toSU2SU4}
\end{equation}
i.e. the final gauge group of the four-dimensional theory is
$$
K=SU^{diag}(2)\times SU(4)\,.
$$
Accordingly, the fermions transform as
\begin{equation*}
(\mbf{2},\mbf{4})_{L}+(\mbf{2},\mbf{4})'_{L}\qquad\mbox{and}\qquad
(\mbf{2},\mbf{4})_{L}-(\mbf{2},\mbf{4})'_{L}
\end{equation*}
under $K$.


\chapter{Classification of semi-realistic particle physics models}
\label{Classification}

Here starting from an $\cN=1$, $E_{8}$ Yang-Mills-Dirac theory defined in ten dimensions, I provide 
a complete classification of semi-realistic particle physics models resulting from CSDR and a 
subsequent application of the Wilson flux breaking mechanism. According to our requirements  in
section~\ref{UnifiedTheoriesHigherDims-Example} the dimensional reduction of this theory over the
coset spaces $S/R$ which are listed in the first column of
tables~\ref{SO6VectorSpinorContentSCosets} and~\ref{SO6VectorSpinorContentNSCosets}, leads to
anomaly free $E_{6}$ and $SO(10)$ GUTs in four dimensions. Recall also that the four-dimensional
surviving scalars transform in the fundamental of the resulting gauge group and are not suitable for
the superstrong symmetry breaking towards the SM. One way out was discussed in
chapter~\ref{sec:WilsonFlux-theory}, namely the Wilson flux breaking mechanism. Here, I investigate
to which extent applying both methods, CSDR and Wilson flux breaking mechanism one can obtain
reasonable low energy models.

\section[Dimensional reduction over symmetric coset spaces]
{Dimensional reduction over symmetric coset\\spaces}
\label{DimReduction-SCosets}

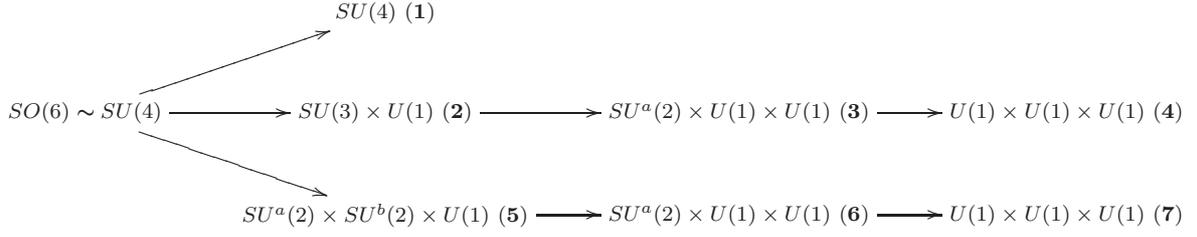
\begin{figure}
\begin{scriptsize}
$$
\xymatrix{
&\AAA~(\mbf{1})            &                                                      &\\
SO(6)\thicksim\AAA\ar[ur]\ar[dr]\ar[r]   &\AA\times\Y~(\mbf{2})\ar[r] &\A{a}\times\Y\times\Y~(\mbf{3})\ar[r]
											    & \Y\times\Y\times\Y~(\mbf{4})\\
&\A{a}\times\A{b}\times\Y~(\mbf{5})\ar[r]  &\A{a}\times\Y\times\Y~(\mbf{6})\ar[r] &\Y\times\Y\times\Y~(\mbf{7}) }
$$
\end{scriptsize}
\caption{\emph{Possible $E_{8}\supset SO(6)\supset R$ embeddings for the symmetric coset spaces, $S/R$, of
table~\ref{SO6VectorSpinorContentSCosets}.} 
\label{A3DecompChannels}}
\end{figure}

Here, I consider all the possible embeddings  $E_{8}\supset SO(6)\supset R$ for the six-dimensional
\textit{symmetric} coset spaces, $S/R$, listed in the first column of
table~\ref{SO6VectorSpinorContentSCosets}\footnote{I have excluded the study of
dimensional reduction over the $Sp(4)/(SU(2)\times U(1))_{max}$ coset space which does not admit
fermions.}. These embeddings are presented in figure~\ref{A3DecompChannels}. It is worth noting that
in all cases the dimensional reduction of the initial gauge theory leads to an $SO(10)$ GUT
according to the concluding remarks in subsection~\ref{sec:CSDR-rules}. The result of my
examination in the present section is that the additional use of Wilson flux breaking mechanism 
summarize leads to four-dimensional theories of Pati-Salam type. In the following
subsections~\ref{Case-1a}\,-\,\ref{Case-5e} I present details of my study and the corresponding
results which I  in tables~\ref{SO10ChannelSCosetsTable-CSDR}
and~\ref{SO10ChannelSCosetsTable-CSDR-HOSOTANI}\footnote{For convenience I label the cases examined
in the following subsections as `\rm{Case No.x}'. `\rm{No}' denotes the embedding $R\hookrightarrow
E_{8}$ and the `\rm{x}' the coset space I use. The same label is also used in
tables~\ref{SO10ChannelSCosetsTable-CSDR} and~\ref{SO10ChannelSCosetsTable-CSDR-HOSOTANI}.}.

\subsection[Reduction of $G=E_{8}$ using $SO(7)/SO(6)$.]
{Reduction of $G=E_{8}$ over $B=B_{0}/F^{B_{0}}$, $B_{0}=S/R=SO(7)/SO(6)$. (Case $\mbf{1a}$)
\label{Case-1a}}

I consider the dimensional reduction of an $\cN=1$, $E_{8}$ Yang-Mills-Dirac theory over the
six-dimensional sphere, $S^{6}\sim SO(7)/SO(6)$. This model was presented in
section~\ref{CSDR-example} as an example of applying the CSDR scheme. The assumed embedding of $R$
into $G=E_{8}$ is the one suggested by the maximal subgroup $E_{8}\supset SO(6)\times SO(10)$ and it
is denoted as $(\mbf{1})$ in figure~\ref{A3DecompChannels}. In
section~\ref{WilsonFluxBreakingMechanism-example} I applied the Hosotani breaking mechanism to
obtain a reasonable low energy model. I summarize the results of my examination in the first row of
tables~\ref{SO10ChannelSCosetsTable-CSDR} and~\ref{SO10ChannelSCosetsTable-CSDR-HOSOTANI}.

\subsection[Reduction of $G=E_{8}$ using $SU(4)/(SU(3)\times U(1))$.]
{Reduction of $G=E_{8}$ over $B=B_{0}/F^{B_{0}}$, $B_{0}=S/R=SU(4)/(SU(3)\times U(1))$. 
(Case $\mbf{2b}$)}
I consider again Weyl fermions belonging in the adjoint of  $G=E_{8}$ and the embedding of
$R=SU(3)\times U(1)$ into $E_{8}$ suggested by the decomposition\footnote{This decomposition is in
accordance with the Slansky tables\cite{Slansky:1981yr} but with
opposite $U(1)$ charge.}.
\begin{align}
&E_{8}\supset SO(16)\supset SO(6)\times SO(10)\backsim SU(4)\times SO(10)\supset SU(3)\times U^{I}(1)\times SO(10)\nn\\
&\begin{array}{l@{}l@{}l@{}}
 E_{8}&\supset& (SU(3)\times U^{I}(1))\times SO(10)\\
 \mbf{248}&=&(\mbf{1},\mbf{1})_{(0)}+(\mbf{1}, \mbf{45})_{(0)}+(\mbf{8},\mbf{1})_{(0)}  
                 +(\mbf{3},\mbf{10})_{(-2)}+(\obar{\mbf{3}},\mbf{10})_{(2)}\\                
               &+&(\mbf{3},\mbf{1})_{(4)}+(\obar{\mbf{3}},\mbf{1})_{(-4)}
                 +(\mbf{1},\mbf{16})_{(-3)}+(\mbf{1},\obar{\mbf{16}})_{(3)}\\
               &+&(\mbf{3},\mbf {16})_{(1)}+(\obar{\mbf {3}},\obar{\mbf {16}})_{(-1)}\,.
 \end{array}\label{DecompCh2-SO10}
\end{align}

If only the the CSDR mechanism was applied, the resulting four-dimensional gauge group would be
\begin{equation*}
H=C_{E_{8}}(SU(3)\times U^{I}(1))=SO(10)\,\Big(\times U^{I}(1)\Big)\,,
\end{equation*}
where the additional $U(1)$ factor in the parenthesis corresponds to a global symmetry, according to
the concluding remarks in subsection~\ref{sec:CSDR-rules}. The $R=SU(3)\times U^{I}(1)$ content of
the vector and spinor of $B_{0}=S/R=SU(4)/(SU(3)\times U(1))$ can be read in the last two columns of
table~\ref{SO6VectorSpinorContentSCosets}. Then according to the CSDR rules, the theory would
contain scalars belonging in the $\mbf{10}_{(-2)}$, $\mbf{10}_{(2)}$ of $H$ and two copies of chiral
fermions transforming as $\mbf{16}_{L(3)}$ and $\mbf{16}_{L(-1)}$ under the same gauge group.
%

The freely acting discrete symmetries of the coset space $SU(4)/(SU(3)\times U(1))$ are not included
in the list~(\ref{FCases}) of those ones that are worth to be examined further.

\subsection[Reduction of $G=E_{8}$ using $SU(3)/(SU(2)\times U(1))\times(SU(2)/U(1))$.]
{Reduction of $G=E_{8}$ over $B=B_{0}/F^{B_{0}}$, $B_{0}=S/R=SU(3)/(SU(2)\times U(1))\times(SU(2)/U(1))$.
(Cases $\mbf{3d}$, $\mbf{6d}$) }
\label{Case-3d}

I consider again Weyl fermions belonging in the adjoint of  $G=E_{8}$ and the following 
decomposition
\begin{equation}
\begin{array}{l@{}l@{}l@{}}
 E_{8} \supset SO(16)~&\supset&~ SO(6)\times SO(10)\backsim SU(4)\times SO(10)
                       \supset (SU'(3)\times U^{II}(1))\times SO(10)\\ 
                     ~&\supset&~ (SU^{a}(2)\times U^{I}(1)\times U^{II}(1))\times SO(10)
\end{array}\label{DecompCh3-SO10}
\end{equation}
or
\begin{equation}
\begin{array}{l@{}l@{}l@{}}
E_{8} \supset SO(16)~&\supset&~ SO(6)\times SO(10)\backsim SU(4)\times SO(10)\\
                    ~&\supset&~(SU^{a}(2)\times SU^{b}(2)\times U^{II}(1))\times SO(10)\\
                    ~&\supset&~(SU^{a}(2)\times U^{I}(1) \times U^{II}(1))\times SO(10)\,.
\end{array}\label{DecompCh6-SO10}
\end{equation}
In both cases I can properly redefine the $U(1)$ charges, and consequently choose an embedding of $R=SU(2)\times
U^{I}(1)\times U^{II}(1)$ into $E_{8}$ as follows
\begin{equation}
\begin{array}{l@{}l@{}l@{}}
 E_{8}&\supset& (SU^{a}(2)\times U^{I'}(1)\times U^{II'}(1))\times SO(10)\\
\mbf{248}&=&(\mbf{1}, \mbf{1})_{(0, 0)}+(\mbf{1}, \mbf{1})_{(0, 0)}             
 	        + (\mbf{3}, \mbf{1})_{(0, 0)}+(\mbf{1}, \mbf{45})_{(0, 0)}\\           
  	      &+&(\mbf{1}, \mbf{1})_{(-2 b, 0)}+(\mbf{1}, \mbf{1})_{(2 b, 0)}
                + (\mbf{2}, \mbf{1})_{(-b, 2 a)}+(\mbf{2}, \mbf{1})_{(b, -2 a)}\\
              &+& (\mbf{2}, \mbf{1})_{(-b, -2 a)}+(\mbf{2}, \mbf{1})_{(b, 2 a)}
                + (\mbf{1}, \mbf{10})_{(0, -2 a)}+(\mbf{1}, \mbf{10})_{(0, 2 a)}\\
              &+& (\mbf{2}, \mbf{10})_{(b, 0)}+(\mbf{2}, \mbf{10})_{(-b, 0)}
                + (\mbf{1}, \mbf{16})_{(b, -a)}+(\mbf{1}, \overline{\mbf{16}})_{(-b, a)}\\
              &+& (\mbf{1}, \mbf{16})_{(-b, -a)}+(\mbf{1}, \overline{\mbf{16}})_{(b, a)}
                + (\mbf{2}, \mbf{16})_{(0, a)}+(\mbf{2}, \overline{\mbf{16}})_{(0, -a)}\,.
\end{array}\label{DecompCh3''-SO10}
\end{equation}

Here, $a$ and $b$ are the $U(1)$ charges of vector and fermion content of the coset space
$B_{0}=S/R=SU(3)/(SU(2)\times U^{I'}(1))\times (SU(2)/U^{II'}(1))$, shown in the last two columns of
table~\ref{SO6VectorSpinorContentSCosets} (case `$\mbf{d}$'). Then, if only the CSDR mechanism was applied, 
the resulting four-dimensional gauge group would be
\begin{equation*}
H=C_{E_{8}}(SU^{a}(2)\times U^{I'}(1) \times U^{II'}(1))=SO(10)\,\Big(\times U^{I'}(1)\times U^{II'}(1)\Big)\,,
\end{equation*}
where the additional $U(1)$ factors in the parenthesis correspond to global symmetries. According to the CSDR 
rules, the four-dimensional model would contain scalars belonging in $\mbf{10}_{(0, -2 a)}$,
$\mbf{10}_{(0,2a)}$, $\mbf{10}_{(b, 0)}$ and $\mbf{10}_{(-b, 0)}$ of $H$ and two copies of chiral fermions
transforming as $\mbf{16}_{L(b, -a)}$, $\mbf{16}_{L(-b, -a)}$ and $\mbf{16}_{L(0, a)}$ under the same gauge
group.

The freely acting discrete symmetries of the coset space under discussion, are the center of $S$, 
$\rm{Z}(S)=\bb{Z}_{3}\times \bb{Z}_{2}$ and the Weyl symmetry, $\rm{W}=\bb{Z}_{2}$. Then according to the
list~(\ref{FCases}) the interesting cases to be examined further are the following two.

In the first case I employ the $\rm{W}=\bb{Z}_{2}$ discrete symmetry which leads to a four-dimensional theory 
with gauge symmetry group
\begin{equation*}
K'= SU^{(i)}(2)\times SU^{(ii)}(2)\times SU(4)\,\Big( \times U^{I'}(1)\times U^{II'}(1)\Big)\,.
\end{equation*}
Similarly to the case discussed in subsection~\ref{Case-1a}, the surviving scalars transform as
\begin{equation}
(\mbf{1},\mbf{1},\mbf{6})_{(b, 0)}\qquad\mbox{and}\qquad (\mbf{1},\mbf{1},\mbf{6})_{(-b, 0)}
\label{Case-3d-scalars}
\end{equation} 
under $K'$ which are the only ones that are invariant under the action of ${\rm W}$ (table~\ref{SO6VectorSpinorContentSCosets}).
Furthermore, taking into account the ${\rm W}$ transformation properties listed in the last column of
table~\ref{SO6VectorSpinorContentSCosets}, as well as the decomposition of $\mbf{16}$ irrep. of $SO(10)$
under $SU^{(i)}(2)\times SU^{(ii)}(2)\times SU(4)$ [see eq.~(\ref{SO10toSU2SU2SU4HB-16})], I conclude that the
four-dimensional fermions transform as
\begin{gather}
\begin{aligned}
&(\mbf{2},\mbf{1},\mbf{4})_{L(b,-a)}-(\mbf{2},\mbf{1},\mbf{4})'_{L(b,-a)}\,,\\
&(\mbf{1},\mbf{2},\mbf{4})_{L(b, -a)}+(\mbf{1},\mbf{2},\mbf{4})'_{L(b, -a)}\,,\\
\end{aligned}\nn\\
\begin{aligned}
&(\mbf{2},\mbf{1},\mbf{4})_{L(-b,-a)}-(\mbf{2},\mbf{1},\mbf{4})'_{L(-b, -a)}\,,\\
&(\mbf{1},\mbf{2},\mbf{4})_{L(-b, -a)}+(\mbf{1},\mbf{2},\mbf{4})'_{L(-b, -a)}\,,\\
\end{aligned}\qquad
\begin{aligned}
&(\mbf{2},\mbf{1},\mbf{4})_{L(0, a)}-(\mbf{2},\mbf{1},\mbf{4})'_{L(0, a)}\,,\\
&(\mbf{1},\mbf{2},\mbf{4})_{L(0, a)}+(\mbf{1},\mbf{2},\mbf{4})'_{L(0, a)}\,,
\end{aligned}\label{4d-fermions-H-3d}
\end{gather}
under $K'$.

Once more I have spontaneous symmetry breaking (since the coset space is symmetric) which breaks the 
$SU(3)$-colour [since the scalars transform as in~(\ref{Case-3d-scalars}) under the $K'$ gauge group].
Therefore, employing the $\rm{W}$ discrete symmetry is not an interesting case for further investigation.

In the second case I use the $\bb{Z}_{2}\times\bb{Z}_{2}$ subgroup of the ${\rm W}\times{\rm Z}(S)$ 
combination of discrete symmetries. The surviving scalars of the four-dimensional theory belong in the
\begin{equation*}
(\mbf{2},\mbf{2},\mbf{1})_{(b, 0)}\,,\qquad\mbox{and}\qquad (\mbf{2},\mbf{2},\mbf{1})_{(-b, 0)}
\end{equation*}
of the $K'$ gauge group which remain the same as before. The fermions, on the other hand, transform as those
in eq.~(\ref{4d-fermions-H-3d}) but with the signs of the linear combinations reversed. The final gauge group
after the spontaneous symmetry breaking of the theory is found to be
$$
K=SU^{diag}(2)\times SU(4)\,\Big(\times U^{I'}(1)\times U^{II'}(1)\Big)\,.
$$
and its fermions transform as
\begin{gather}
\begin{aligned}
&(\mbf{2},\mbf{4})_{(b,-a)}+(\mbf{2},\mbf{4})'_{(b,-a)}\,,\\
&(\mbf{2},\mbf{4})_{(b, -a)}-(\mbf{2},\mbf{4})'_{(b, -a)}\,,\\
\end{aligned}\nn\\
\begin{aligned}
&(\mbf{2},\mbf{4})_{(-b,-a)}+(\mbf{2},\mbf{4})'_{(-b, -a)}\,,\\
&(\mbf{2},\mbf{4})_{(-b, -a)}-(\mbf{2},\mbf{4})'_{(-b, -a)}\,,\\
\end{aligned}\qquad
\begin{aligned}
&(\mbf{2},\mbf{4})_{(0, a)}+(\mbf{2},\mbf{4})'_{(0, a)}\,,\\
&(\mbf{2},\mbf{4})_{(0, a)}-(\mbf{2},\mbf{4})'_{(0, a)}\,,
\end{aligned}
\end{gather}
under $K$.

\subsection[Reduction of $G=E_{8}$ using $(SU(2)/U(1))^{3}$.]
{Reduction of $G=E_{8}$ over $B=B_{0}/F^{B_{0}}$, $B_{0}=S/R=(SU(2)/U(1))^{3}$. 
(Cases $\mbf{4f}$, $\mbf{7f}$)}
\label{Case-4f}

I consider again Weyl fermions belonging in the adjoint of  $G=E_{8}$ and the following decomposition
\begin{equation}
\begin{array}{l@{}l@{}l@{}}
E_{8}\supset SO(16)~&\supset&~ SO(6)\times SO(10)\backsim SU(4)\times SO(10)
         \supset  SU'(3)\times U^{III}(1)\times SO(10)\\
         ~&\supset&~ (SU^{a}(2)\times U^{II}(1)\times U^{III}(1))\times SO(10)\\
         ~&\supset&~SO(10)\times U^{I}(1)\times U^{II}(1)\times U^{III}(1)
\end{array}\label{DecompCh4-SO10}
\end{equation}
or
\begin{equation}
\begin{array}{l@{}l@{}l@{}}
E_{8}\supset SO(16)~&\supset&~ SO(6)\times SO(10)\backsim SU(4)\times SO(10)\\
                   ~&\supset&~ (SU^{a}(2)\times SU^{b}(2)\times U^{III}(1))\times SO(10)\\
                   ~&\supset&~ SO(10)\times U^{I}(1)\times U^{II}(1)\times U^{III}(1)\,.
\end{array}\label{DecompCh7-SO10}
\end{equation}
In both cases I can properly redefine the $U(1)$ charges, and consequently choose an embedding of 
$R=SU(2)\times U^{I}(1)\times U^{II}(1)$ into $E_{8}$ as follows
\begin{equation}
 \begin{array}{l@{}l@{}l@{}}
 E_{8}&\supset& SO(10)\times U^{I'}(1)\times U^{II'}(1)\times U^{III'}(1)\\
\mbf{248}&=&\mbf{1}_{(0, 0, 0)}+\mbf{1}_{(0, 0, 0)}+\mbf{1}_{(0, 0, 0)}   
                + \mbf{45}_{(0, 0, 0)}\\
              &+&\mbf{1}_{(-2 a, 2 b, 0)}+\mbf{1}_{(2 a, -2 b, 0)}
                +  \mbf{1}_{(-2 a, -2 b, 0)}+\mbf{1}_{(2 a, 2 b, 0)}\\
              &+&\mbf{1}_{(-2 a, 0, -2 c)}+\mbf{1}_{(2 a, 0, 2 c)}
                + \mbf{1}_{(0, -2 b, -2 c)}+\mbf{1}_{(0, 2 b, 2 c)}\\
              &+&\mbf{1}_{(-2 a, 0, 2 c)}+\mbf{1}_{(2 a, 0, -2 c)}
                + \mbf{1}_{(0, -2 b, 2 c)}+\mbf{1}_{(0, 2 b, -2 c)}\\
              &+&\mbf{10}_{(0, 0, 2 c)}+\mbf{10}_{(0, 0, -2 c)}
                + \mbf{10}_{(0, 2 b, 0)}+\mbf{10}_{(0, -2 b, 0)}\\
              &+&\mbf{10}_{(2 a, 0, 0)}+\mbf{10}_{(-2 a, 0, 0)}
                + \mbf{16}_{(a, b, c)}+\overline{\mbf{16}}_{(-a, -b, -c)}\\
              &+&\mbf{16}_{(-a, -b, c)}+ \overline{\mbf{16}}_{(a, b, -c)}
                + \mbf{16}_{(-a, b, -c)}+\overline{\mbf{16}}_{(a, -b, c)}\\
              &+&\mbf{16}_{(a, -b, -c)}+\overline{\mbf{16}}_{(-a, b, c)}\,.
\end{array}\label{DecompCh4'-SO10}
\end{equation}

Then, if only the CSDR mechanism was applied, the four-dimensional gauge group would be
\begin{equation*}
H=C_{E_{8}}(U^{I'}(1)\times U^{II'}(1)\times U^{III'}(1))=SO(10)\,\Big(\times U^{I'}(1)\times U^{II'}(1)\times U^{III'}(1)\Big)\,.
\end{equation*}
The same comment as in the previous cases holds for the additional $U(1)$ factors in the parenthesis.
The $R=U^{I'}(1)\times U^{II'}(1)\times U^{III'}(1)$ content of vector and spinor of 
$B_{0}=S/R=(SU(2)/U^{I'}(1))\times (SU(2)/U^{II'}(1))\times (SU(2)/U^{III'}(1))$ can be read in the last two 
columns of table~\ref{SO6VectorSpinorContentSCosets} (case `$\mbf{f}$'). According to the CSDR rules then, the
resulting four-dimensional theory  would contain scalars belonging in $\mbf{10}_{(2 a, 0, 0)}$, 
$\mbf{10}_{(-2a, 0,0)}$, $\mbf{10}_{(0, 2 b,0)}$, $\mbf{10}_{(0, -2 b, 0)}$, $\mbf{10}_{(0, 0, 2 c)}$ and
$\mbf{10}_{(0, 0, -2 c)}$ of $H$ and two copies of fermions transforming as $\mbf{16}_{L(a, b, c)}$,
$\mbf{16}_{L(-a, -b, c)}$, $\mbf{16}_{L(-a, b, -c)}$ and $\mbf{16}_{L(a, -b, -c)}$ under the same gauge group.
%

The freely acting discrete symmetries, $F^{S/R}$, of the coset space $(SU(2)/U(1))^{3}\sim (S^{2})^{3}$ are 
the center of $S$, $\rm{Z}(S)=(\bb{Z}_{2})^{3}$ and the Weyl discrete symmetry, $\rm{W}=(\bb{Z}_{2})^{3}$.
Then according to the list~(\ref{FCases}) the interesting cases to be examined further are the following.
 
First, let me mod out the $(S^{2})^{3}$ coset space by the $\bb{Z}_{2}\subset{\rm W}$ and consider the 
multiply connected manifold $S^{2}/\bb{Z}_{2}\times S^{2}\times S^{2}$. Then, the resulting four-dimensional
gauge group will be
\begin{equation*}
K'=SU^{(i)}(2)\times SU^{(ii)}(2)\times SU(4)\,\Big(\times U^{I'}(1)\times U^{II'}(1)\times U^{III'}(1)\Big)\,.
\end{equation*}
The four-dimensional theory will contain scalars which belong in
\begin{align*}
&(\mbf{1},\mbf{1},\mbf{6})_{(0, 2 b, 0)}\,,\qquad (\mbf{1},\mbf{1},\mbf{6})_{(0, -2 b, 0)}\,,\\
&(\mbf{1},\mbf{1},\mbf{6})_{(0, 0, 2 c)}\,, \qquad (\mbf{1},\mbf{1},\mbf{6})_{(0, 0, -2 c)}
\end{align*}
of $K'$; these are the only ones that are invariant under the action of the considered 
$\bb{Z}_{2}\subset{\rm W}$. However, linear combinations between the two copies of the CSDR-surviving
left-handed fermions have no definite properties under the abelian factors of the $K'$ gauge group and they do
not survive. As a result, the model is not an interesting case for further investigation.

Second, if I employ the $\bb{Z}_{2}\times\bb{Z}_{2}\subset{\rm W}$ discrete symmetry and consider the manifold
$S^{2}/\bb{Z}_{2}\times S^{2}/\bb{Z}_{2}\times S^{2}$, the resulting four-dimensional theory has the same 
gauge symmetry as before, i.e. $K'$. Similarly as before, scalars transform as
\begin{equation*}
(\mbf{1},\mbf{1},\mbf{6})_{(0, 0, 2 c)}\,, \qquad (\mbf{1},\mbf{1},\mbf{6})_{(0, 0, -2 c)}
\end{equation*}
under $K'$. However, no fermions survive in the four-dimensional theory and the model is again not an 
interesting case to examine further.

Finally, if I employ the $\bb{Z}_{2}\times\bb{Z}_{2}\subset{\rm W}\times{\rm Z}(S)$ discrete symmetry, the
four-dimensional theory contains scalars which belong in
\begin{align*}
&(\mbf{2},\mbf{2},\mbf{1})_{(0, 2 b, 0)}\,, \qquad   (\mbf{2},\mbf{2},\mbf{1})_{(0, -2 b, 0)}\,,\\
&(\mbf{2},\mbf{2},\mbf{1})_{(0, 0, 2 c)}\,,  \qquad  (\mbf{2},\mbf{2},\mbf{1})_{(0, 0, -2 c)}
\end{align*}
of $K'$ but no fermions. The model is again not an interesting case for further study.

Therefore although the above studied cases have been obtained using discrete symmetries which are included in 
the list~(\ref{FCases}), no fermion fields survive in the four-dimensional theory. The reason is that I employ
here only a subgroup of the Weyl discrete symmetry ${\rm W}=(\bb{Z}_{2})^{3}$ and I cannot form linear
combinations among the two copies of the CSDR-surviving left-handed fermions which are invariant under
eq.~(\ref{WFluxSurvivingField}). The use of the whole ${\rm W}$ discrete symmetry, on the other hand, would
lead to four-dimensional theories with smaller gauge symmetry than the one of SM.

\subsection[Reduction of $G=E_{8}$ using $Sp(4)/(SU(2)\times SU(2))\times(SU(2)/U(1))$.]
{Reduction of $G=E_{8}$ over $B=B_{0}/F^{B_{0}}$, $B_{0}=S/R=Sp(4)/(SU(2)\times SU(2))\times(SU(2)/U(1))$.
(Case $\mbf{5e}$)
\label{Case-5e}}

Finally, I consider  Weyl fermions in the adjoint of $G=E_{8}$ and the embedding of 
$R=SU(2)\times SU(2)\times U(1)$ into $E_{8}$ suggested by
\begin{align}
&\begin{array}{l@{}l@{}l@{}}
E_{8}    \supset SO(16)~&\supset&~ SO(6)\times SO(10)\backsim SU(4)\times SO(10)\\
                                 ~&\supset&~(SU^{a}(2)\times SU^{b}(2)\times U^{I}(1))\times SO(10)
\end{array}\nn\\
&\begin{array}{l@{}l@{}l@{}}
E_{8}&\supset& (SU^{a}(2)\times SU^{b}(2)\times U^{I}(1))\times SO(10)\\
 \mbf{248}&=&(\mbf{1},\mbf{1},\mbf{1})_{(0)}+(\mbf{1},\mbf{1},\mbf{45})_{(0)}
                      +(\mbf{3},\mbf{1},\mbf{1})_{(0)}+(\mbf{1},\mbf{3},\mbf{1})_{(0)}\\
                   &+&(\mbf{2},\mbf{2},\mbf{1})_{(2)}+(\mbf{2},\mbf{2},\mbf{1})_{(-2)}
                      +(\mbf{1},\mbf{1},\mbf{10})_{(2)}+(\mbf{1},\mbf{1},\mbf{10})_{(-2)}\\
                   &+&(\mbf{2},\mbf{2},\mbf{10})_{(0)}                                                     
                      +(\mbf{2},\mbf{1},\mbf{16})_{(1)}+(\mbf{2},\mbf{1},\obar{\mbf{16}})_{(-1)}\\
                    &+&(\mbf{1},\mbf{2},\mbf{16})_{(-1)}+(\mbf{1},\mbf{2},\obar{\mbf{16}})_{(1)}\,.
 \end{array}\label{DecompCh5-SO10}
\end{align}
If only the CSDR mechanism was applied, the resulting four-dimensional gauge group would be
\begin{equation*}
H=C_{E_{8}}(SU^{a}(2)\times SU^{b}(2)\times U^{I}(1))=SO(10)\,\Big(\times U^{I}(1)\Big)\,.
\end{equation*}
The $R=SU^{a}(2)\times  SU^{b}(2)\times U^{II}(1)$ content of vector and spinor of 
$B_{0}=S/R=Sp(4)/(SU^{a}(2)\times SU^{b}(2))\times(SU(2)/U(1))$
can be read in the last two columns of table \ref{SO6VectorSpinorContentSCosets}. According to the CSDR rules 
the resulting four-dimensional theory would contain scalars belonging in $\mbf{10}_{(0)}$, $\mbf{10}_{(2)}$
and $\mbf{10}_{(-2)}$ of $H$ and two copies of chiral fermions transforming as $\mbf{16}_{L(1)}$
and $\mbf{16}_{L(-1)}$ under the same gauge group.

The freely acting discrete symmetries of the coset space $(Sp(4)/SU(2)\times SU(2))\times(SU(2)/U(1))$ 
(case `$\mbf{e}$' in table~\ref{SO6VectorSpinorContentSCosets}), are the center of $S$,
$\rm{Z}(S)=(\bb{Z}_{2})^{2}$ and the Weyl discrete symmetry, $\rm{W}=(\bb{Z}_{2})^{2}$. According to the
list~(\ref{FCases}) the interesting cases to be examined further are the following.

First, if I employ  the Weyl discrete symmetry, $\rm{W}=(\bb{Z}_{2})^{2}$, leads to a four-dimensional theory 
with a gauge symmetry group
\begin{equation*}
K'=SU^{(i)}(2)\times SU^{(ii)}(2)\times SU(4)\,\Big(\times U^{I}(1)\Big)\,.
\end{equation*}
The surviving scalars of the theory belong in
\begin{equation*}
({\mbf 2},{\mbf 2},{\mbf 1})_{(0)}
\end{equation*}
of $K'$, whereas its fermions transform as
\begin{equation}
\begin{aligned}
&({\mbf 2},{\mbf 1},{\mbf 4})_{L(1)}-({\mbf 2},{\mbf 1},{\mbf 4})'_{L(1)}\,,\\
&({\mbf 1},{\mbf 2},{\mbf 4})_{L(1)}+({\mbf 1},{\mbf 2},{\mbf 4})'_{L(1)}\,,
\end{aligned}\qquad
\begin{aligned}
&({\mbf 2},{\mbf 1},{\mbf 4})_{L(-1)}-({\mbf 2},{\mbf 1},{\mbf 4})'_{L(-1)}\,,\\
&({\mbf 1},{\mbf 2},{\mbf 4})_{L(-1)}+({\mbf 1},{\mbf 2},{\mbf 4})'_{L(-1)}\,
\end{aligned}\label{Case7eFermions}
\end{equation}
under $K'$.

Second, if I employ a  $\bb{Z}_{2}\times\bb{Z}_{2}$ subgroup of the ${\rm W}\times{\rm Z}(S)$ combination of
discrete symmetries, leads to a four-dimensional model with scalars belonging in $({\mbf 2},{\mbf 2},{\mbf
1})_{(0)}$ and fermions transforming as in eq.~(\ref{Case7eFermions}) but with the signs of the linear
combinations reversed. 

Finally, in table~\ref{SO10ChannelSCosetsTable-CSDR-HOSOTANI} I also report the less interesting case 
$\bb{Z}_{2}\subseteq{\rm W}$.

Concerning the spontaneous symmetry breaking of theory, note that for the interesting cases of the
$\rm{W}=(\bb{Z}_{2})^{2}$ and $\bb{Z}_{2}\times\bb{Z}_{2}\subset {\rm W}\times {\rm Z}(S)$ discrete
symmetries, the final unbroken gauge group is found to be
$$
K=SU^{diag}(2)\times SU(4)\,\Big(\times U(1)\Big)\,.
$$
Then, for the case of ${\rm W}$ discrete symmetry, the fermions of the model transform as
\begin{equation}
\begin{aligned}
&({\mbf 2},{\mbf 4})_{(1)}-({\mbf 2},{\mbf 4})'_{(1)}\,,\\
&({\mbf 2},{\mbf 4})_{(1)}+({\mbf 2},{\mbf 4})'_{(1)}\,,
\end{aligned}\qquad
\begin{aligned}
&({\mbf 2},{\mbf 4})_{(-1)}-({\mbf 2},{\mbf 4})'_{(-1)}\,,\\
&({\mbf 2},{\mbf 4})_{(-1)}+({\mbf 2},{\mbf 4})'_{(-1)}\,,
\end{aligned}
\end{equation}
under $K$, whereas for the case of $\bb{Z}_{2}\times\bb{Z}_{2}\subset{\rm W}\times{\rm Z}(S)$ the fermions 
belong in similar linear combinations as above but with their signs reversed.

\section{Dimensional reduction over non-symmetric coset spaces}
\label{DimReduction-NSCosets}

According to the discussion in section~\ref{UnifiedTheoriesHigherDims-Example} I have to consider all the 
possible embeddings $E_{8}\supset SO(6)\supset R$, for the six-dimensional \textit{non-symmetric} cosets,
$S/R$, of table~\ref{SO6VectorSpinorContentNSCosets}. It is worth noting that the embedding of $R$ in all
cases of six-dimensional non-symmetric cosets are obtained by the following chain of maximal subgroups of
$SO(6)$
\begin{equation}
SO(6)\sim SU(4)\supset SU(3)\times U(1)\supset SU^{a}(2)\times U(1)\times U(1)\supset U(1)\times U(1)\times U(1)\,.
\end{equation}
It is also important to recall from the discussion in sections~\ref{sec:CSDR-rules} and~\ref{UnifiedTheoriesHigherDims-Example}
that in all these cases the dimensional reduction of the initial gauge theory leads to an $E_{6}$ GUT. The 
result of my examination in the present section is that the additional use of the Wilson flux breaking
mechanism leads to four-dimensional gauge theories based on three different varieties of groups, namely
$SO(10)\times U(1)$, $SU(2)\times SU(6)$ or $SU^{(i)}(2)\times SU^{(ii)}(2)\times SU(4)\times U(1)$. In the
following subsections~\ref{Case-2a'}\,-\,\ref{Case-4c'} I present details of my examination and summarize the
corresponding results in tables~\ref{E6ChannelTable-CSDR} and~\ref{E6ChannelTable-CSDR-HOSOTANI}\footnote{I
follow the same notation as in the examination of the symmetric cosets.}.

\subsection[Reduction of $G=E_{8}$ using $G_{2}/SU(3)$.]
{Reduction of $G=E_{8}$ over $B=B_{0}/F^{B_{0}}$, $B_{0}=S/R=G_{2}/SU(3)$. (Case $\mbf{2a'}$)}
\label{Case-2a'}

I consider Weyl fermions belonging in the adjoint of  $G=E_{8}$ and identify the $R$ with the $SU(3)$ 
appearing in the decomposition~(\ref{DecompCh2-SO10}). Then, if only the CSDR mechanism was applied, the
resulting four-dimensional gauge group would be
$$
H=C_{E_{8}}(SU(3))=E_{6}\,,
$$
i.e. it appears an enhancement of the gauge group, a fact which was noticed earlier in several examples in
subsections~\ref{TopologicallyInducedGaugeGroupBreaking-E6-Z2} and 
\ref{TopologicallyInducedGaugeGroupBreaking-E6-Z2Z2}. This observation suggests that I could have considered
the following more obvious embedding of $R=SU(3)$ into $E_{8}$,
\begin{equation}
\begin{array}{l@{}l@{}l@{}}
E_{8}&\supset& SU'(3) \times E_{6}\\
\mbf{248}&=&(\mbf{8},\mbf{1})+(\mbf{1},\mbf{78})
                +(\mbf{3},\mbf{27})+(\obar{\mbf{3}},\obar{\mbf{27}})\,.
\end{array}\label{DecompCh2-SO10-E6-2'}
\end{equation}
The $R=SU(3)$ content of vector and spinor of $B_{0}=S/R=G_{2}/SU(3)$ coset is $\mbf{3}+\obar{\mbf{3}}$ and
$\mbf{1}+\mbf{3}$, respectively. According to the CSDR rules, the four-dimensional theory would contain
scalars belonging in $\mbf{27}$ and $\obar{\mbf{27}}$ of $H$, two copies of chiral fermions transforming as
$\mbf{27}_{L}$ under the same gauge group and a set of fermions in the $\mbf{78}$ irrep., since the
dimensional reduction over non-symmetric coset preserves the supersymmetric
spectrum~\cite{Manousselis:2000aj,Manousselis:2001xb,Manousselis:2001re}.
%

The freely acting discrete symmetry, $F^{S/R}$, of the coset space $G_{2}/SU(3)$ is the Weyl,
$\rm{W}=\bb{Z}_{2}$ (case `$\mbf{a'}$' in table~\ref{SO6VectorSpinorContentNSCosets}).
Then, following the discussion in subsection~\ref{TopologicallyInducedGaugeGroupBreaking-E6-Z2}, the Wilson 
flux breaking mechanism leads to a four-dimensional theory either with gauge group
\begin{equation}
(i)\qquad K'^{(\mbf{1})}=C_{H}(T^{H})=SO(10)\times U(1)\,,\label{Case2a'-WFB-1}
\end{equation}
in case I embed the $\bb{Z}_{2}$ into the $E_{6}$ gauge group as in the embedding ($\mbf{1}$) of
subsection~\ref{TopologicallyInducedGaugeGroupBreaking-E6-Z2}, or 
\begin{equation}
(ii)\qquad K'^{(\mbf{2},\mbf{3})}=C_{H}(T^{H})=SU(2)\times SU(6)\,,\label{Case2a'-WFB-2}
\end{equation}
in case I choose to embed the discrete symmetry as in the embeddings ($\mbf{2}$) or ($\mbf{3}$) of the same 
subsection [the superscript in the $K'$'s above refer to the embeddings ($\mbf{1}$), ($\mbf{2}$) or
($\mbf{3}$)].

Making an analysis along the lines presented earlier in the case of symmetric cosets, I determine the particle
content of the two models, which is presented in table~\ref{E6ChannelTable-CSDR-HOSOTANI}. In both cases the
gauge symmetry of the four-dimensional theory cannot be broken further due to the absence of scalars.

\subsection[Reduction of $G=E_{8}$ using $Sp(4)/(SU(2)\times U(1))_{nonmax}$.]
{Reduction of $G=E_{8}$ over $B=B_{0}/F^{B_{0}}$, $B_{0}=S/R=Sp(4)/(SU(2)\times U(1))_{nonmax}$. (Case $\mbf{3b'}$)}

I consider Weyl fermions belonging in the adjoint of $G=E_{8}$ and the decomposition~(\ref{DecompCh3-SO10}). 
In order the $R$ to be embedded in $E_{8}$ as in eq.~(\ref{RinSO6inE8-embedding}), I identify it with the
$SU(2)\times U^{I}(1)$ appearing in the decomposition~(\ref{DecompCh3-SO10}). Then, if only the CSDR mechanism
was applied, the resulting gauge group would be 
\begin{equation}
H=C_{E_{8}}(SU^{a}(2)\times U^{I}(1))=E_{6}\,\Big(\times U^{I}(1)\Big)\,.
\label{Case-3b'-CSDR-gauge-group}
\end{equation}
Note that again appears an enhancement of the gauge group. Similarly with previously discussed cases, the 
additional $U(1)$ factor in the parenthesis corresponds only to a global symmetry.
The observation~(\ref{Case-3b'-CSDR-gauge-group}) suggests that we could have considered the following
embedding of $R=SU(2)\times U(1)$ into $E_{8}$\footnote{This decomposition is in accordance with the Slansky
tables but with opposite $U(1)$ charge.},
\begin{equation}
 \begin{array}{l@{}l@{}l@{}}
 E_{8}&\supset& SU'(3)\times E_{6}\supset SU^{a}(2)\times U^{I'}(1) \times E_{6}\\
\mbf{248}&=&(\mbf{1},\mbf{1})_{(0)}+(\mbf{1},\mbf{78})_{(0)}        
                + (\mbf{3},\mbf{1})_{(0)}                                            
                + (\mbf{2},\mbf{1})_{(-3)}+(\mbf{2},\mbf{1})_{(3)}\\
              &+&(\mbf{1},\mbf{27})_{(2)}+(\mbf{1},\obar{\mbf{27}})_{(-2)}
                +(\mbf{2},\mbf{27})_{(-1)}+(\mbf{2},\obar{\mbf{27}})_{(1)}\,.
\end{array}\label{DecompCh3'-SO10}
\end{equation}
The $R=SU(2)\times U^{I}(1)$ content of vector and spinor of 
$B_{0}=S/R=Sp(4)/(SU(2)\times U^{I}(1))_{non-max}$ can be read in the last two columns of table
\ref{SO6VectorSpinorContentNSCosets} (case `$\mbf{b'}$'). According to the CSDR rules then, the surviving
scalars in four dimensions would transform as $\mbf{27}_{(-2)}$, $\mbf{27}_{(1)}$, $\obar{\mbf{27}}_{(2)}$ and
$\obar{\mbf{27}}_{(-1)}$ under $H=E_{6}(\times U^{I}(1))$. The four-dimensional theory would also contain
fermions belonging  in $\mbf{78}_{(0)}$ of $H$ (gaugini of the model), two copies of left-handed fermions
belonging in $\mbf{27}_{L(2)}$ and $\mbf{27}_{L(-1)}$ and one fermion singlet transforming as $\mbf{1}_{(0)}$
under the same gauge group.
%

The freely acting discrete symmetries, $F^{S/R}$, of the coset space $Sp(4)/(SU(2)\times U(1))_{non-max}$,
are the center of $S$, $\rm{Z}(S)=\bb{Z}_{2}$ and the Weyl, $\rm{W}=\bb{Z}_{2}$. Then, employing the $\rm{W}$
discrete symmetry, I find that the resulting four-dimensional gauge group is either
\begin{eqnarray}
(i)&\qquad& K'^{(\mbf{1})}=C_{H}(T^{H})=SO(10)\times U(1)\,\Big(\times U^{I}(1)\Big)\,,\qquad\mbox{or}\\[3ex]
(ii)&\qquad& K'^{(\mbf{2},\mbf{3})}=C_{H}(T^{H})=SU(2)\times SU(6)\,\Big(\times U^{I}(1)\Big)\,,
\end{eqnarray}
depending on the embedding of $\bb{Z}_{2}\hookrightarrow E_{6}$ I choose to consider
(see subsection~\ref{TopologicallyInducedGaugeGroupBreaking-E6-Z2}).

On the other hand, if I employ  the $\rm{W}\times\rm{Z}(S)=\bb{Z}_{2}\times\bb{Z}_{2}$ combination of discrete
symmetries, the resulting four-dimensional gauge group is either
\begin{equation}
(iii)\qquad K'^{(\mbf{2'})}=C_{H}(T^{H})=SU(2)\times SU(6)\,\Big(\times U^{I}(1)\Big)\,,\label{Case3b'-WFB-3}
\end{equation}
in case I embed the $(\bb{Z}_{2}\times\bb{Z}_{2})$ into the $E_{6}$ gauge group as in the embedding 
($\mbf{2'}$) of subsection~\ref{TopologicallyInducedGaugeGroupBreaking-E6-Z2Z2}, or
\begin{equation}
(iv)\qquad K'^{(\mbf{3'})}=C_{H}(T^{H})=SU^{(i)}(2)\times SU^{(ii)}(2)\times SU(4)\times U(1)\,\Big(\times
U^{I}(1)\Big)\,,\label{Case3b'-WFB-4}
\end{equation}
in case I choose to embed the discrete symmetry as in the embedding ($\mbf{3'}$) of the same subsection.

Making a similar analysis as before, I determine the particle content of the four different models, which is 
presented in table~\ref{E6ChannelTable-CSDR-HOSOTANI}. In all cases the gauge symmetry of the resulting
four-dimensional theory cannot be broken further by a Higgs mechanism due to the absence of scalars.

\subsection[Reduction of $G=E_{8}$ using $SU(3)/(U(1)\times U(1))$.]
{Reduction of $G=E_{8}$ over $B=B_{0}/F^{B_{0}}$, $B_{0}=S/R=SU(3)/(U^(1)\times U(1))$. (Case $\mbf{4c'}$)}
\label{Case-4c'}

I consider Weyl fermions in the adjoint of $G=E_{8}$ and the decomposition~(\ref{DecompCh4-SO10}). In order 
the $R=U(1)\times U(1)$, to be embedded in $E_{8}$ as in eq.~(\ref{RinSO6inE8-embedding}) one has to identify
it with the $U^{I}(1)\times U^{II}(1)$ appearing in the decomposition~(\ref{DecompCh4-SO10}). Then, if only
the CSDR mechanism was applied, the resulting four-dimensional gauge group would be
\begin{equation}
H=C_{E_{8}}(U^{I}(1)\times U^{II}(1))=E_{6}\,\Big(\times U^{I}(1)\times U^{II}(1)\Big)\,.
\label{Case-4c'-CSDR-gauge-group}
\end{equation}
Note again that an enhancement of the gauge group appears, whereas the additional $U(1)$ factors correspond 
to global symmetries. The observation~(\ref{Case-4c'-CSDR-gauge-group}) suggests that I could have considered
the following embedding of $R=U(1)\times U(1)$ ito $E_{8}$,
\begin{align}
&E_{8}\supset SU(3)\times E_{6}\supset (SU^{a}(2)\times U^{II}(1))\times E_{6}\supset E_{6}\times U^{I}(1)\times U^{II}(1)\\
&\begin{array}{l@{}l@{}l@{}}
 E_{8}&\supset&  E_{6}\times U^{I}(1) \times U^{II}(1)\\
\mbf{248}&=&\mbf{1}_{(0,0)}+\mbf{1}_{(0,0)}
                      +\mbf{78}_{(0,0)}
                      +\mbf{1}_{(-2,0)}+\mbf{1}_{(2,0)}
                      +\mbf{1}_{(-1,3)}+\mbf{1}_{(1,-3)}\\
                   &+&\mbf{1}_{(1,3)}+\mbf{1}_{(-1,-3)}
                      +\mbf{27}_{(0,-2)}+\obar{\mbf{27}}_{(0,2)}
                      +\mbf{27}_{(-1,1)}+\obar{\mbf{27}}_{(1,-1)}\\
                   &+&\mbf{27}_{(1,1)}+\obar{\mbf{27}}_{(-1,-1)}\,.
\end{array}\label{DecompCh3-E6}
\end{align}
The $R=U^{I}(1)\times U^{II}(1)$ content of vector and spinor of $B_{0}=S/R=SU(3)/(U^{I}(1)\times U^{II}(1)$ 
can be read in the last two columns of table~\ref{SO6VectorSpinorContentNSCosets} (case `$\mbf{c'}$'). The
closure of the tensor products of the reps appeared in the~(\ref{DecompCh3-E6}) above and especially the
\begin{gather*} 
\mbf{27}_{(0,-2)}\times\mbf{27}_{(-1,1)}=\obar{\mbf{27}}_{(-1,-1)}+\mbf{351}_{a(-1,-1)}+\mbf{351}'_{s(-1,-1)}
\end{gather*}
suggests the identification
\begin{equation*}
\mbf{27}_{(0,-2)}\lra (a,c)\,,\qquad  \mbf{27}_{(-1,1)}\lra (b,d)\,,
\end{equation*}
for the $U(1)$ charges of the $R$ vector and spinor content, i.e. $a=0$, $c=-2$, $b=-1$ and $d=1$. Then, 
according to the CSDR rules, the four-dimensional theory would contain scalars which belong in
$\mbf{27}_{(0,-2)}$, $\mbf{27}_{(-1,1)}$, $\mbf{27}_{(1,1)}$, $\obar{\mbf{27}}_{(0,2)}$,
$\obar{\mbf{27}}_{(1,-1)}$,  and $\obar{\mbf{27}}_{(-1,-1)}$ of $H=E_{6}(\times U^{I}(1)\times U^{II}(1))$.
The resulting four-dimensional theory would also contain gaugini transforming as $\mbf{78}_{(0,0)}$ under $H$,
two copies of left-handed fermions belonging in $\mbf{27}_{L(0,-2)}$, $\mbf{27}_{L(-1,1)}$,
$\mbf{27}_{L(1,1)}$ and two fermion singlets belonging in $\mbf{1}_{(0,0)}$ and $\mbf{1}_{(0,0)}$ of the same
gauge group.
%

The freely acting discrete symmetries, $F^{S/R}$, of the coset space $SU(3)/(U(1)\times U(1))$ (case  
`$\mbf{c'}$' in table~\ref{SO6VectorSpinorContentNSCosets}), are the center of $S$, $\rm{Z}(S)=\bb{Z}_{3}$ and
the Weyl, $\rm{W}=\mbf{S}_{3}$. Then according to the list~(\ref{FCases}) only the $\bb{Z}_{2}\subset{\rm W}$
discrete symmetry is an interesting case to be examined further.

Then, employing the $\bb{Z}_{2}$ subgroup of the ${\rm W}=\mbf{S}_{3}$ discrete symmetry leads to a
four-dimensional theory either with gauge group
\begin{eqnarray}
(i)&\qquad& K'^{(\mbf{1})}=C_{H}(T^{H})=SO(10)\times U(1)\,\Big(\times U^{I}(1)\times U^{II}(1)\Big)\,,\qquad\mbox{or}\qquad\\[3ex]
(ii)&\qquad&K'=C_{H}(T^{H})=SU(2)\times SU(6)\,\Big(\times U^{I}(1)\times U^{II}(1)\Big)
\end{eqnarray}
depending on the embedding of $\bb{Z}_{2}\hookrightarrow E_{6}$ I choose to consider
(see sec.~\ref{TopologicallyInducedGaugeGroupBreaking-E6-Z2}).

Making a similar analysis as before, I determine the particle content of the two models as follows.

\paragraph{Case (i).}
The resulting four-dimensional theory contains gaugini which transform as
\begin{equation*}
\mbf{1}_{(0,0,0)}\,,\qquad\mbf{45}_{(0,0,0)}
\end{equation*}
under $K'^{(\mbf{1})}$, a set of fermion singlets which belong in
\begin{equation*}
\mbf{1}_{(0,0,0)}\,,\qquad\mbf{1}_{(0,0,0)}\,,
\end{equation*}
of $K'^{(\mbf{1})}$ and a set of chiral fermions which belong in one of the linear combinations
\begin{equation*}
\left\{
\begin{array}{l}
\mbf{1}_{L(-4,0,-2)}+\mbf{1}'_{L(-4,0,-2)}\,,\\
\mbf{10}_{L(-2,0,-2)}+\mbf{10}'_{L(-2,0,-2)}\,,\\
\mbf{16}_{L(1,0,-2)}-\mbf{16}'_{L(1,0,-2)}\,,
\end{array}\right\}\,,\qquad
\left\{
\begin{array}{l}
\mbf{1}_{L(-4,-1,1)}+\mbf{1}'_{L(-4,-1,1)}\,,\\
\mbf{10}_{L(-2,-1,1)}+\mbf{10}'_{L(-2,-1,1)}\,,\\
\mbf{16}_{L(1,-1,1)}-\mbf{16}'_{L(1,-1,1)}\,,\\
\end{array}\right\}\,,
\end{equation*}
or
\begin{equation*}
\left\{
\begin{array}{l}
\mbf{1}_{L(-4,1,1)}+\mbf{1}'_{L(-4,1,1)}\,,\\
\mbf{10}_{L(-2,1,1)}+\mbf{10}'_{L(-2,1,1)}\,,\\
\mbf{16}_{L(1,1,1)}-\mbf{16}'_{L(1,1,1)}
\end{array}\right\}
\end{equation*}
of the same gauge group, depending on the $\bb{Z}_{2}$ subgroup of $\mbf{S}_{3}$ that I choose to consider
(see table \ref{SO6VectorSpinorContentNSCosets}).

\paragraph{Case (ii).}
The resulting four-dimensional theory contains gaugini which transform as
\begin{equation*}
\mbf{1}_{(0,0)}\,,\qquad\mbf{1}_{(0,0)}\,,\qquad(\mbf{3},\mbf{1})_{(0,0)}\,,\qquad (\mbf{1},\mbf{35})_{(0,0)}
\end{equation*}
under $K'^{(\mbf{2},\mbf{3})}$, a set of fermion singlets which belong in
\begin{equation*}
\mbf{1}_{(0,0)}\,,\qquad\mbf{1}_{(0,0)}\,,
\end{equation*}
of $K'^{(\mbf{2},\mbf{3})}$ and a set of chiral fermions which belong in one of the linear combinations
\begin{equation*}
\left\{
\begin{array}{l}
(\mbf{1},\mbf{15})_{L(0,-2)}+(\mbf{1},\mbf{15})'_{L(0,-2)}\,,\\
(\mbf{2},\obar{\mbf{6}})_{L(0,-2)}-(\mbf{2},\obar{\mbf{6}})'_{L(0,-2)}\,,
\end{array}\right\}\,,\qquad
\left\{
\begin{array}{l}
(\mbf{1},\mbf{15})_{L(-1,1)}+(\mbf{1},\mbf{15})'_{L(-1,1)}\,,\\
(\mbf{2},\obar{\mbf{6}})_{L(-1,1)}-(\mbf{2},\obar{\mbf{6}})'_{L(-1,1)}\,,
\end{array}\right\}\,,
\end{equation*}
or
\begin{equation*}
\left\{
\begin{array}{l}
(\mbf{1},\mbf{15})_{L(1,1)}+(\mbf{1},\mbf{15})'_{L(1,1)}\,,\\
(\mbf{2},\obar{\mbf{6}})_{L(1,1)}-(\mbf{2},\obar{\mbf{6}})'_{L(1,1)}\,,
\end{array}\right\}
\end{equation*}
of the same gauge group, depending on the $\bb{Z}_{2}$ subgroup of $\mbf{S}_{3}$ that I choose to consider
(see table \ref{SO6VectorSpinorContentNSCosets}).

Note that in both cases the gauge symmetry of the four-dimensional theory cannot be broken further by a Higgs 
mechanism due to the absence of scalars.

Finally, if I have used either the symmetric group of $3$ permutations, $\mbf{S}_{3}$, or its subgroup
$\bb{Z}_{3}\subset\mbf{S}_{3}$, I could not form linear combinations among the two copies of the
CSDR-surviving left-handed fermions and no fermions would survive in four dimensions.

\section{Discussion}

The CSDR is a consistent dimensional reduction scheme
\cite{Chatzistavrakidis:2007by,Chatzistavrakidis:2007pp,Coquereaux:1984ca,Chaichian:1986mf,
Dvali:2001qr}, as well as an elegant framework to incorporate in a unified manner the gauge and the
ad-hoc Higgs sector of spontaneously broken four-dimensional gauge theories using the extra
dimensions. The kinetic terms of fermions were easily included in the same unified description. A
striking feature of the scheme concerning fermions was the discovery that chiral ones can be
introduced~\cite{Manton:1981es} and moreover they could result even from vector-like reps of the
higher dimensional gauge theory~\cite{Chapline:1982wy,Kapetanakis:1992hf}. This possibility is due
to the presence of non-trivial background gauge configurations required by the CSDR principle, in
accordance with the index theorem. Another striking feature of the theory is the possibility that
the softly broken sector of the four-dimensional supersymmetric theories can result from a
higher-dimensional $\cN=1$ supersymmetric gauge theory with only a vector supermultiplet, when is
dimensionally reduced over non-symmetric coset
spaces~\cite{Manousselis:2000aj,Manousselis:2001xb,Manousselis:2001re,Manousselis:2004xd}.
Another interesting feature useful in realistic model searches is the possibility to deform the
metric in certain non-symmetric coset spaces and introduce more than one
scales~\cite{Kapetanakis:1992hf,Farakos:1986sm,Farakos:1986cj}.

Recently there exist a revival of interest in the study of compactifications with internal manifolds
six-dimensional non-symmetric coset spaces possessing an $SU(3)$-structure within the framework of
flux compactifications~\cite{Behrndt:2004km,Koerber:2008rx,House:2005yc,
Lust:1986ix,Castellani:1986rg,Govindarajan:1986kb,Govindarajan:1986iz,Micu:2004tz,Frey:2005zz,
Manousselis:2005xa}. In this framework the CSDR of the heterotic ten-dimensional gauge theory is an
extremely interesting problem. Here, starting with a supersymmetric $\cN=1$, $E_{8}$ gauge theory in
ten dimensions I made a complete classification of the models obtained in four dimensions after
reducing the theory over all multiply connected six-dimensional coset spaces, resulting by moding
out all the freely acting discrete symmetries on these manifolds, and using the Wilson flux breaking
mechanism in an exhaustive way. The results of the extended investigation have been presented
in~\cite{Douzas:2008va,Douzas:2007zz}. Despite some partial success, my result is that the two
mechanisms used to break the gauge symmetry, i.e. the geometric breaking of the CSDR and the
topological of the Hosotani mechanism are not enough to lead the four-dimensional theory to the SM
or some interesting extension as the MSSM. However, in the old CSDR framework one can
think of some new sources of gauge symmetry breaking, such as new scalars coming from a gauge theory
defined even in higher dimensions~\cite{Koca:1984dr,Jittoh:2008jc}. Much more interesting
is to extend my examination in a future study of the full ten-dimensional $E_{8}\times E_{8}$ gauge
theory of the heterotic string. Moreover in that case one does not have to be restricted in the
study of freely acting discrete symmetries of the coset spaces and can extent the analysis including
orbifolds~\cite{Kim:2006hv,Kim:2006hw,Kim:2007mt,Kim:2007dx,Forste:2005gc, Nilles:2006np,
Lebedev:2006kn,Lebedev:2006tr,Lebedev:2007hv}. More possibilities are offered in
refs~\cite{Blumenhagen:2005pm,Blumenhagen:2005zg,Blumenhagen:2006ux,Blumenhagen:2006wj}.

\chapter{Noncommutative generalisations and Motivation}
\label{NC-generalisations-motivation}

Main disadvantage of the CSDR scheme when applied over continuum homogeneous coset spaces
is the introduction of divergences in the quantum theory when higher modes in the Kaluza-Klein
expansion are kept~\cite{Bailin:1987jd}. However, assuming the extra dimensions to form continuum
and homogeneous spaces is not the only possibility. Noncommutative modifications of the extra
dimensions have been proposed some years ago~\cite{Madore:1989ma}. Their
phenomenological consequences~\cite{Carlson:2001bk} and their connection with string
theory~\cite{Jurco:2001my,Szabo:2006wx} have been also studied. In this more modest version of
noncommutative generalisation one provides the extra dimensions with noncommutative characteristics
while keeping the continuum nature of the ordinary four-dimensional Minkowski space. Special case
of noncommutative spaces are these which are approximated by finite complex matrix algebras which
are defined over them and are known as `\textit{fuzzy spaces}' in the
literature~\cite{Madore:2000aq}. One possibility of dimensional reduction over noncommutative
internal spaces of this kind is to generalise appropriately the CSDR scheme
(Fuzzy-CSDR)~\cite{Aschieri:2003vy}.

In chapter~\ref{NcKK}, I present the necessary mathematical framework and ideas to describe
noncommutative generalisations of the ordinary Kaluza-Klein theories and their dimensional
reduction.

In chapter~\ref{FuzzyCSDR}, I review and further explain the Fuzzy-CSDR scheme. Important
conclusions of the study are (i) the enhancement of the gauge symmetry due to the noncommutative
characteristics of the internal manifold and (ii) the fact that dimensional reduction of gauge
theories over such spaces lead to renormalizable theories in four dimensions.

In chapter~\ref{FuzzyExtraDim}, motivated by the interesting features of the Fuzzy-CSDR scheme, I
examine the inverse problem, i.e. whether obtaining fuzzy extra dimensions as a vacuum solution of a
four-dimensional but renormalizable potential is possible. Indeed, starting from the most general
renormalizable potential in four dimensions, fuzzy extra dimensions are dynamically
generated~\cite{Aschieri:2006uw}. Furthermore the initial gauge symmetry found to break
spontaneously towards phenomenologically interesting patterns. 

In chapter~\ref{linPhD}, I  assume that in the the energy regime where the noncommutativity is
expected to be valid, the ordinary spacetime cannot be described as a continuum and investigate
possible generalisations of Einstein gravity.  Elementary cells of length scale $\mu^{-1}_{P}$
(with $\mu_{P}$ denoting the Planck mass) are expected to form out. Then the ordinary
spacetime is just a limit of this noncommutative `phase'. The purpose of the study of noncommutative
spacetime is, firstly to explore new mathematical ideas and generalisations of the Einstein gravity.
Secondly, having assumed the extra dimensions in the Planck mass regime to appear noncommutative
characteristics would  be no profound reason for this behaviour not to be the case in the ordinary
dimensions too.

Finally, in chapter~\ref{Conclusions} I summarise the conclusions of the present research work.

\chapter{Noncommutative modifications of Kaluza-Klein theory}\label{NcKK}

Here, I present the necessary mathematical framework and ideas to describe noncommutative
modifications of Kaluza-Klein theories and their subsequent dimensional reduction. Furthermore, I
choose to assume a modest version of noncommutative generalisation, providing the extra dimensions
with noncommutative characteristics while keeping the continuum nature of the ordinary
four-dimensional Minkowski space. Following~\cite{Madore:2000aq}, I describe the noncommutative
nature of the extra space by a finite complex matrix algebra, $M_{N}$, defined over it. This
promote, the Kaluza-Klein theory, based on a manifold $M^{4}\times B$ and described by the
$\cC\otimes\cC (B)$ tensor product of associated algebras of continuum  functions to a theory with a
geometry described by the algebra $\cA=\cC\otimes M_{N}$.  The motivation for such generalisation is
the lack of a well-defined notion of the point, embedded in its very nature. It is the same
characteristic of the quantum mechanical phase space, expressed for the first time by the Heisenberg
uncertainty relations. The extra space coordinates, then, become noncommutative operators and by
analogy with quantum mechanics, points are expected to be replaced by elementary cells. {It is this
cellular structure which serve as an ultraviolet cut-off} similar to a lattice structure. As a
result, divergences coming from the extra compactification space are no longer appear in quantum
theory, making this noncommutative modification of Kaluza-Klein theories an appealing configuration
for further investigation.

\section{Introductory remarks and motivation}

The simplest definition of the noncommutative geometry is that is a geometry in which the
coordinates does not commute. The most familiar example of such a space is the quantised version of
a two-dimensional phase space. Basic characteristic of this example is the built-in uncertainty in
the simultaneous measurements of its coordinates. The notion of a point is no longer well-defined;
this is realised by the well-known Heisenberg uncertainty relations. The lack of a definition of a
point  constitutes an essential built-in characteristic of noncommutative geometries, too.
Noncommutative geometries are in this sense `pointless geometries'. The mathematical formulation of
these kind of geometries has been studied in a series of works and a variety of different approaches
have been developed by now. For reviews and references therein consult~\cite{Madore:2000aq,
Connes:1994yd, Connes:2000ti, Castellani:2000xt, Ydri:2001pv}. Among them and following
the example of quantum mechanics, one promotes the ordinary coordinates of space and time to
noncommutative operators. Then the notion of a pure state replaces that of the point and
derivations of the algebra replace vector fields~\cite{DuboisViolette:1988ir,Madore:2000aq}. Then
the notion of points is expected to be replaced by the notion of the elementary cells; each of them
correspond to the minimal measurable `area' in the noncommutative `coordinate' space. Noncommutative
geometries, in this sense realised, are close to the example of lattice gauge theories. These were
developed for the study of the nonperturbative regime of various physical systems, a goal which is
achieved by an \textit{ad-hoc} discretisation of the configuration (and internal) space of the
original theory~\cite{Rothe:2005nw}. However, in lattice approach some of the continuous
transformations of the original theory are not preserved by this very discretisation procedure. This
is not always the case with the noncommutative geometries. The transformation properties of the
original theory could be preserved by the `quantisation' of the commutative space, and in this
sense, noncommutative geometry could be a better description of nonperturbative effects at least at
Planck length-scale. In this regime there is theoretical circumstantial evidence that an elementary
length may exist. Therefore, a quantum-mechanical description of the space may be appropriate.

 On the other hand, the question of existence of hidden extra dimensions and the actual geometry
of  their compactification space (if any), is still an open problem. One of the first negative
answers was given by Kaluza~\cite{Kaluza:1921tu} and Klein~\cite{Klein:1926tv} in their attempt to
introduce extra dimensions in order to unify the gravitational field with electromagnetism.
Later, with the advent of more elaborate gauge fields, it was proposed that this internal space
could be taken as a compact Lie group or something more general as a coset space. Dimensional
reduction over such internal space resulted to non-abelian gauge groups in four
dimensions~\cite{Salam:1981xd}. In the previous chapters, I reviewed such ideas and I
studied in detail the CSDR case. However, the great disadvantage of using homogeneous coset
spaces as the extra dimensions is that they introduce even more divergences in the quantum theory 
and lead to an infinite spectrum of new particles. In fact the structure is strongly redundant and
most of it has to be discarded by a truncation procedure of a possible dimensional
reduction. If on the other hand one use matrix approximations of coset spaces (fuzzy coset  spaces)
for the extra dimensions, which are spaces  of a special noncommutative structure\footnote{Each
fuzzy coset space is described by an appropriate Lie algebra. I discuss the case of the
two-dimensional fuzzy sphere as an example in section~\ref{FuzzySphere}.}, the
theory turns out to be power counting renormalizable; the fuzzy spaces are approximated by matrices
of finite dimensions and only a finite number of counterterms are required to make the Lagrangian
renormalizable.

In section~\ref{Kaluza-Klein}, I give a short formulation of the Kaluza-Klein theory in a suitable
language for its noncommutative generalisation. In section~\ref{NoncommutativeGeometry}, I promote
the extra compactification space of Kaluza-Klein theory to its noncommutative approximation. I
assume that the internal structure of the theory is described by a noncommutative geometry in which
the notion of a point does not exist. As a particular example of this noncommutative generalisation,
I shall choose only internal structures which give rise to a finite spectrum of particles, namely
noncommutative spaces described by finite complex matrix algebras~\cite{Madore:2000aq}. In
section~\ref{Kaluza-KleinTheoryRevisited}, I explain, following~\cite{Madore:1989ma, Madore:1996cb},
how this idea can lead to a noncommutative modification of Kaluza-Klein theory. Finally, in
section~\ref{FuzzySphere}, I give an extended description of the fuzzy sphere `manifold',
$S_{N}^{2}$, and the differential geometry which is defined over it. I stress - among others - the
appearance of elementary cells (elementary measurable `area') and the preservation of the continuum
symmetries of the ordinary sphere. These are built-in characteristics of the matrix approximated
manifolds.

\section{Kaluza-Klein theory\label{Kaluza-Klein}}

In its local aspects Kaluza-Klein theory is described by an extended space-time
$V\equiv M^{D}=M^{4}\times B$, of dimension $D= 4 + d$ and with coordinates $x^{\ssM} = (x^\m,x^a)$.
The $x^\m$ are the coordinates of space-time which I consider here to be Minkowski space; the $x^a$
are the coordinates of the internal space, which in this section will be implicitly supposed to be
space-like and `small'. In section~\ref{Kaluza-KleinTheoryRevisited} it will be of purely algebraic
nature and not necessarily `small'.

Let $\cC(V)$ be the commutative and associative algebra of smooth complex-valued functions on $V$.
I define the sum and product of two functions by the sum and product of the value of the function
at each point. The commutative, associative and distributive rules follow then from those of
$\bb{R}$. It is evident that every such $V$ space can be embedded in an Euclidean space of
sufficiently high dimension $D'> D$. It is defined then by a set of $D'-D$ relations in the
Euclidean coordinates $x^{\ssM}$;  the algebra $\cC(V)$ can be considered as a quotient of the
algebra of smooth function over $\bb{R}^{D'}$ by the ideal generated by these relations. I recall
that an \textit{ideal} of an algebra is a subalgebra which is stable under multiplication by a
general element of the algebra. The algebra $\cC(V)$ has many ideals, for example the subalgebra of
functions which vanish on any close set in $V$. The algebra $M_{N}$ of $N\times N$ complex matrices,
on the other hand has no proper ideals.

Let now $X$ be a smooth vector field on the $V$ manifold considered above. I denote the vector
space of all such $X$ as $\cX(V)$ which is a left $\cC(V)$ module; if $f\in\cC(V)$ and
$X\in\cX(V)$ then $fX\in\cX(V)$.  We recall that a \textit{left (right) module} is a vector space on
which there is a left (right) action of the algebra. Let $\prt_{\ssM}$ be the natural basis of the
vectors on the embedding space $\bb{R}^{D'}$. It can be proved that any $X$ can be written as a
linear combination $X=X^{\ssM}\prt_{\ssM}$ with $X^{\ssM}\in\cC(V)$.  As a $\cC(V)$-module, $\cX(V)$
is finitely generated although not uniquely determined. In the case that $V$ space is a
\textit{parallelizable manifold}, the derivations  $\{e_{\ssM}\}$ defined over $\cX(V)$ are linearly
independent and the linear combination $X=X^{\ssM}e_{\ssM}$, $M=1,\ldots,D$, unique. The set
$\{e_{\ssM}\}$ forms the basis for the globally defined moving frame over the $V$. An embedding
space larger than $\bb{R}^{D}$ is no longer required and the set of derivations $\{e_{\ssM}\}$
happens to coincide with the $\{\prt_{\ssM}\}$ basis noted above.  We note however, that in general
a moving frame can be defined only locally on $V$.

I recall that the definition of the \textit{Lie bracket} $[X,Y]$ of two vector fields is
\begin{equation*}
[X,Y]f=(XY-YX)f=(X^{\ssM}\prt_{\ssM}Y^{\ssN}-Y^{\ssM}\prt_{\ssM}X^{\ssN})\prt_{\ssN}f\,,
\end{equation*}
where $f$ an arbitrary element of $\cC(V)$. Note that the Lie bracket of two vector fields is also a
vector field. For the particular case of the parallelizable manifolds I can write the Lie bracket
as
\begin{equation}
[e_{\ssM},e_{\ssN}]=C^{\ssK}{}_{\ssM\ssN}e_{\ssK}\,,
\end{equation}
where the structure constants $C^{\ssK}{}_{\ssM\ssN}$ are elements of $\cC(V)$.

Any vector field can be defined as \textit{derivation} of the algebra $\cC(V)$, if there is a linear
map of $\cC(V)$ onto itself which satisfies the \textit{Leibniz rule}
\begin{equation}
X(f~g)=(X\,f)g+f(X\,g)\,. \label{LeibnizRule}
\end{equation}
Then, I can identify the vector space $\cX(V)$ with the space of derivations
\begin{equation}
\cX(V)\equiv\Der(\cC(V))\,.
\end{equation}

I choose the $\cC(V)$ to be an algebra of complex-valued  functions. Using complex conjugation a
$*$-operation: $f\mapsto f^{*}$ can be defined. The algebra is assumed to close under this operation
being a  so-called $*$-\textit{algebra}. I also assume  that the $\cX(V)$ elements satisfy the
reality condition
\begin{equation}
(Xf)^{*}=X(f^{*}), \quad  \forall~X\in\cX(V)~\mbox{and}~\forall~f\in\cC(V)\,.
\end{equation}
It is straightforward to check that if the $f$ elements  of $*$-algebra were hermitian, the
$*$-algebra  would fail to close under noncommutative multiplication; $(fg)^{*}=g^{*}f^{*}$ which is
not equal with $f^{*}g^{*}$. Therefore, although I am interested in real manifolds and their
noncommutative counterparts,  algebras of complex functions are necessary to be considered.

A \textit{differential form} of order $p$ or $p$-\textit{form}  $\a$ is a $p$-linear completely
antisymmetric map of the vector space $\cX(V)$ into $\cC(V)$. In particular if $f\in\cC(V)$ and
$X_{1},\ldots,X_{p}$ are $p$-vector fields then
\begin{equation*}
(\a f)(X_{1},\cdots,X_{p})=(f\a)(X_{1},\cdots,X_{p})=f(\a(X_{1},\cdots,X_{p}))\,,
\end{equation*}
that is the value of $\a(X_{1},\ldots,X_{p})$  at a point of $V$ depends only on the values of the
vector fields at that point. The set $\O^{p}(V)$ of $p$-forms is a $\cC(V)$-module.

The \textit{exterior product} $\a\wdg\b$ of $\a\in\O^{p}(V)$ and $\b\in\O^{q}(V)$,  on the other
hand, is an $\O^{p+q}(V)$ element defined by
\begin{equation}
\a\wdg\b(X_{1},\cdots,X_{p+q})=\frac{1}{(p+q)!}
\sum\e(i,j)\a(X_{i_{1}},\cdots,X_{i_{p}})\b(X_{j_{1}},\cdots,X_{j_{q}})\,,
\end{equation}
where the summation is taken over all the possible  partitions of $(1,\ldots,p+q)$ into
$(i_{1},\ldots,i_{p})$ and $(j_{1},\ldots,j_{q})$ and $\e(i,j)$ is the signature of the
corresponding permutation. It is graded commutative, $\a\wdg\b=(-1)^{pq}\b\wdg\a$, because the
algebra $\cC(V)$ is commutative. This is not generally the case in the noncommutative algebras and I
shall usually write the exterior product of two forms $\a$ and $\b$ simply as $\a\b\equiv\a\wdg\b$.
I also define $\O^{0}(V)=\cC(V)$ and the $\O^{*}(V)$ as the set of all $\O^{p}(V)$, $p=1,\ldots,D$.
In the particular case of the parallelizable manifolds
\begin{equation}
\O^{*}(V)=\cC(V)\otimes \L^{*}\,,
\end{equation}
where $\L^{*}$ is the exterior algebra over the complex numbers generated by the frame.

The \textit{exterior derivative} $d\a$ of $\a\in\O^{p}(V)$ is defined by the formula
\begin{align}\label{exterior-derivative}
d\a(X_{0},\cdots,X_{p})&=\frac{1}{p+1}
\sum_{i=0}^{p}(-1)^{i}X_{i}(\a(X_{0},\cdots,\hat{X}_{i},\cdots,X_{p}))\nn\\
&+\frac{1}{p+1}\sum_{0\leq i<j\leq p}(-1)^{i+j}
\a([X_{i},X_{j}],X_{0},\cdots,\hat{X}_{i},\cdots,\hat{X}_{j},\cdots,X_{p})\,,
\end{align}
where the hat symbol means that the symbol underneath  is omitted. The case $p=0$ is especially
interesting
\begin{equation}\label{p0-exterior-derivative}
df(X)=X\,f
\end{equation}
leading to the $df(\prt_{\ssM})=\prt_{\ssM}f$ if the linear expansion $X=X^{\ssM}\prt_{\ssM}$ of a
vector field is taken into account. This relation has the same content with the 
$df=(\prt_{\ssM}f) dx^{\ssM}$ definition of ordinary calculus differential. One passes from one to
the other by using the particular case $dx^{\ssM}(\prt_{\ssN})=\d^{\ssM}{}_{\ssN}$. Namely, by the
choice of $\{dx^{\ssM}\}$ basis to be dual to the $\{\prt_{\ssM}\}$ one. The derivations form a
vector space (the tangent space) and (\ref{p0-exterior-derivative}) defines the $df$ as an element
of the dual space (contangent space).

Let now $\a\in\O^{p}(V)$ and $\b\in\O^{*}(V)$. Then $d$ satisfies the condition
\begin{equation}
d(\a\wdg\b)=d\a\wdg\b+(-1)^{p}\a\wdg d\b\,,
\end{equation}
which obviously does not satisfy the Leibniz rule. It is a graded derivation of $\O^{*}(V)$. From
(\ref{exterior-derivative}) it follows that
\begin{equation}
d^{2}=0
\end{equation}
and decomposes the set $\O^{*}(V)$ of all forms into a direct sum
\begin{equation}
\O^{*}(V)=\O^{*+}(V)\oplus\O^{*-}(V)
\end{equation}
of even and odd forms respectively. The differential $d$ takes one into another.

The couple $(\O^{*}(\cC(V), d)$ is called a graded differential algebra or a differential calculus
over $\cC(V)$. I shall show later that $\cC(V)$ need not be commutative and $\O^{*}(\cC(V))$ need
not be graded commutative. Over each algebra $\cC(V)$, be it commutative or not, there can exist a
multitude of differential calculi.

However there are two additional elements of the differential  calculus which are important for its
noncommutative generalisation and have not been discussed so far. Firstly, I note that the
\textit{Lie derivative} of vector field $Y$ with respect to the vector field $Y$, to be
$L_{\ssX}Y=[X,Y]$. Indeed, defining a smooth map $\f$ from $V$ to $V'$, it induces a map
\begin{gather*}
\cC(V')\buildrel\f^{*}\over\longrightarrow\cC(V)\,,\qquad \f^{*}f=f\circ\f^{*}
\end{gather*}
which has a natural extension to the set of all forms
\begin{gather*}
\O^{*}(V')\buildrel\f^{*}\over\longrightarrow\O^{*}(V)\,,\qquad \f^{*}(df)=d(\f^{*}f)\,.
\end{gather*}
If $\f$ is a diffeomorphism one can identify $V'$ and $V$  and consider $\f^{*}$ as an automorphism
of the $\cC(V)$ and $\O^{*}(V)$. In this case $\f$ also induce a map $\f_{*}$ of $\cX(V)$ onto
itself by the formula
\begin{equation*}
(\f_{*}X)f=(\f^{*-1}X\f^{*})f,\qquad f\in\cC(V)\,.
\end{equation*}
Letting $\f^{*}$ be a local one-parameter group of diffeomorphisms of $V$ generated by a vector
field $X$, $\f_{t}^{*}f=f+tXf+\cO(t^{2})$, then acting on a vector field $Y$
\begin{equation*}
\f_{t*}Y=Y-t~L_{\ssX}Y+\cO(t^{2})
\end{equation*}
with the Lie derivative of $Y$ be calculated as $L_{\ssX}Y=[X,Y]$.  By requiring that it be a
derivation, the Lie derivative can be extended to a general element of the tensor algebra over
$\cX(V)$. On the other hand, the Lie derivative of an arbitrary function $f\in\cC(V)$ is given by
$L_{\ssX}f=X f$ and the Lie derivative of a $\a\in\O^{p}(V)$ is given by the formula
\begin{equation}
(L_{\ssX}\a)(Y_{1},\ldots,Y_{p})=X\a(Y_{1},\ldots,Y_{p})
-\sum_{1}^{p}\a(Y_{1},\ldots,[X,Y_{i}],\ldots,Y_{p})\ldots)\,.
\end{equation}
Again, by requiring that it be a derivation, the Lie derivative  can be extended to a general
element of the tensor algebra over $\O^{1}(V)$.

Secondly, the \textit{interior product} $i_{\ssX}$ is defined  to be the map of $\O^{p+1}(V)$ into
$\O^{p}(V)$ and is given for $\a\in\O^{p+1}(V)$ by the equation
\begin{equation}
(i_{\ssX}\a)(X_{1},\cdots,X_{p})=(p+1)\a(X,X_{1},\cdots,X_{p})\,.
\end{equation}
 It is actually a graded derivation of $\O^{*}(V)$.  By setting $i_{\ssX}f=0$, one easily derives
the following formula
\begin{equation}
L_{\ssX}=i_{\ssX}d+d\,i_{\ssX}\,,
\end{equation}
which relates the Lie derivative with the differential and the interior product.

To form tensors one must be able to define tensor products,  for example the tensor product
$\O^{1}(V) \otimes_{\cC(V)} \O^{1}(V)$ of $\O^{1}(V)$ with itself. The $\cC(V)$ subscript is to
denote the fact that I identify $\xi f\otimes\eta$ with $\xi \otimes f \eta$ for every element $f$
of the $\cC(V)$ algebra. This is important in the applications of
section~\ref{Kaluza-KleinTheoryRevisited}. It means also that one must be able to multiply the
elements of $\O^{1}(V)$ on the left and on the right by the elements of the algebra $\cC(V)$. If
$\cC(V)$ is commutative of course these two operations are equivalent. When $\cC(V)$ is an algebra
of functions this left/right linearity is equivalent to the property of locality. It means that the
product of a function with a one-form at a point is again a one-form at the same point. This
property
distinguishes the ordinary product from other, non-local, products such as the convolution. In the
noncommutative case there are no points and locality can not be defined; it is replaced by the
property of left and/or right linearity with respect to the algebra.

To define a metric and covariant derivatives on the extended space-time I set 
$\th^{\ssM} =dx^{\ssM}$ in the absence of a gravitational field, that are dual to the basis of
derivations, according to our previous discussion. Then, I have  $d\th^{\ssM}=0$. The extended
Minkowski metric can be defined as the map
\begin{equation}
g(\th^{\ssM} \tens \th^{\ssN}) = g^{\ssM\ssN}
\label{metric-extendedM}
\end{equation}
which associates to each element $\th^{\ssM} \tens \th^{\ssN}$  of the tensor product $\O^{1}(V)
\tens_{\cC(V)} \O^{1}(V)$ the contravariant components $g^{\ssM\ssN}$ of the (extended) Minkowski
metric. There are of course several other definitions of a metric which are equivalent in
the case of ordinary geometry but the one I have given has the advantage of an easy extension to
the noncommutative case. The map $g$ must be bilinear so that I can define for arbitrary one-forms
$\xi=\xi_{\ssM}\theta^{\ssM}$ and $\eta = \eta_{\ssM}\theta^{\ssM}$
\begin{equation}
g(\xi \tens \eta) = \xi_{\ssM} \eta_{\ssN} g(\theta^{\ssM} \tens \theta^{\ssN})
= \xi_{\ssM}\eta_{\ssN} g^{\ssM\ssN}\,.
\end{equation}
In that case the elementary line element defined over the manifold can be rescaled to
\begin{gather}
ds^{2}=\eta_{\m\n}\th^{\m}\th^{\n}+g_{ab}\th^{a}\th^{b}\,,\label{KK-compactification}\\
\eta_{\m\n}=diag(-1,1,1,1)\,,
\qquad\mbox{and}\qquad g_{ab}=diag(1,\ldots,1)\nn
\end{gather}
which define a local comoving frame. This is also possible  in the noncommutative case but for a
frame $\{\th^{a}\}$, the basis of which, commutes with the noncommutative coordinates of the
`manifold'.

I introduce a gauge potential by first defining  a covariant derivative. Let $\psi$ be a
complex-valued function which I shall consider as a `spinor field' with no Dirac structure and let
$\cH$ be the space of such `spinor fields'. A covariant derivative is a rule which associates to
each such $\psi$ in $\cH$ a spinor-one-form $D\psi$. It is a map
\begin{equation}
\cH \buildrel D \over \rightarrow \O^{1}(V) \otimes_{\cC(V)} \cH                   \label{covariant-Der-map}
\end{equation}
from $\cH$ into the tensor product $\O^{1}(V) \otimes_{\cC(V)} \cH$.  In the absence of any
topological complications the function $\psi = 1$ is a spinor field and I can define a covariant
derivative by the rule
\begin{equation}\label{covariant-Der}
D(1) = A \otimes 1.
\end{equation}
The (local) gauge transformations are the complex-valued functions  with unit norm and so $A$ must
be a one-form with values in the Lie algebra of the unitary group $U(1)$, that is, the imaginary
numbers. An arbitrary spinor field $\psi$ can always be written in the form 
$\psi = f \cdot 1 = 1 \cdot f$ where $f$ is an element of the algebra $\cC(V)$. The extension to
$\psi$ of the covariant derivative is given by the Leibniz rule
\begin{equation}
D \psi = df \otimes 1 + A \otimes f = d\psi \otimes 1 + A \otimes \psi\,,
\end{equation}
an equation which I  simply write in the familiar form $D\psi = d\psi + A\,\psi$.  Using the graded
Leibniz rule one have
\begin{equation}
D (\alpha\psi) = d\alpha \otimes \psi + (-1)^p\, \alpha\, D \psi\,, \label{graded}
\end{equation}
the covariant derivative can be extended to higher-order forms and the field strength $F$ defined by
the equation
\begin{equation}\label{FieldStrength}
D^2 \psi = F \psi.
\end{equation}

To introduce the gravitational field it is always possible to  maintain (\ref{metric-extendedM}) but
at the cost of abandoning $d\th^{\ssM}=0$. This is known as the moving-frame
formalism~\cite{Madore:2000aq}. In the presence of gravity the $dx^{\ssM}$ become arbitrary
one-forms $\th^{\ssM}$. The differential $df$ can be written
\begin{equation}
df = (e_{\ssM} f) \theta^{\ssM}                                   \label{Pfaff}
\end{equation}
which is still of  the form $df=(\prt_{\ssM} f)dx^{\ssM}$ provided  one introduces modified
derivations $e_{\ssM}$. An equation
\begin{equation}
df(e_{\ssM}) = e_{\ssM} f                                      \label{Pfaffder}
\end{equation}
equivalent to $df(\prt_{\ssM})=\prt_{\ssM}f$ can be written if one imposes the relations
\begin{equation}
\theta^{\ssM}(e_{\ssN}) = \delta^{\ssM}{}_{\ssN}\,.
\end{equation}
The $\th^{\ssM}$ are a (local) basis of the one-forms dual to the  derivations $e_{\ssM}$ exactly as
the $dx^{\ssM}$ are dual to the $\prt_{\ssM}$. Equation $d\th^{\ssM}=0$ must be replaced by the
structure equations
\begin{equation}
d\theta^{\ssM} = - \frac{1}{2} C^{\ssM}{}_{\ssN\ssK} \theta^{\ssN} \theta^{\ssK}
\end{equation}
which express simply the fact that the differential of a one-form  is a 2-form and can be thus
written out in terms of the (local) basis $\th^{\ssM}\th^{\ssN}$. The structure equations can
normally not be written globally. In the noncommutative case such equations do not in general make
sense because the differential forms need not have a basis.

The covariant derivative can now be defined as follows.  For the considered (local orthonormal)
moving frame, $\{e_{\ssM}\}$, with $M=1,\ldots,D$, $D=d+4$ and a given vector
field $X=X^{\ssM}e_{\ssM}$, the \textit{covariant derivative} $DX$ of $X$ can be defined by the
local expression $DX=(DX^{\ssM})\otimes e_{\ssM}$
\begin{equation}
DX^{\ssM}=dX^{\ssM}+\o^{\ssM}{}_{\ssN}X^{\ssN}\,;
\end{equation}
$\o^{\ssM}{}_{\ssN}$ is the so called \textit{linear connection} a one-form in $V$ with values in
the
Lie $\mso(3,1)\oplus\mso(d)$ algebra for the given compactification. {The transformation
properties of $\o^{\ssM}{}_{\ssN}$ assure that $D$ is a well defined map from $\cX(V)$ to
$\O^{1}(V)\otimes_{C(V)}\cX(V)$}. $\o^{\ssM}{}_{\ssN}$ also has to satisfy the \textit{structure
equations}
\begin{align}
&T^{\ssM}=d\th^{\ssM}+\omega^{\ssM}{}_{\ssN}\wdg\th^{\ssN}\,, \label{Torsion-def}\\
&\O^{\ssM}{}_{\ssN}=d\o^{\ssM}{}_{\ssN}+\o^{\ssM}{}_{\ssK}\wdg\o^{\ssK}{}_{\ssN}\,,
\label{Curv-def}
\end{align}
whereas the \textit{torsion} $T^{\ssM}$ and the  \textit{curvature form}
\begin{equation}
\O^{\ssM}{}_{\ssN}=\frac{1}{2} R^{\ssM}{}_{\ssN\ssK\ssL}\th^{\ssK}\th^{\ssL}
\end{equation}
satisfy the \textit{Bianchi identities}
\begin{align}
&dT^{\ssM}+\o^{\ssM}{}_{\ssN}T^{\ssN}=\O^{\ssM}{}_{\ssN}\wdg\th^{\ssN}\,,\\
&d\O^{\ssM}{}_{\ssN}+\o^{\ssM}{}_{\ssK}\wdg\O^{\ssK}{}_{\ssN}
-\O^{\ssM}{}_{\ssK}\wdg\o^{\ssK}{}_{\ssN}=0\,.
\end{align}
However, under the considered compactification~\eq{KK-compactification},  these relations
are decomposable into four and extra dimensional part. Their decomposition and the emergent
restrictions will be studied in the context of the noncommutative geometry in sections
\ref{NoncommutativeGeometry} and \ref{Kaluza-KleinTheoryRevisited}.


For a general introduction to Kaluza-Klein theory and references therein consult the review articles
by Bailin \& Love~\cite{Bailin:1987jd}, Coquereaux\& Jadczyk~\cite{Coquereaux:1988ne} or the more
recent one~\cite{Duff:1986hr}.

\section{A noncommutative geometry\label{NoncommutativeGeometry}}

According to the discussion in  the previous section, the basic structure of the differential
geometry of a continuum manifold can be also expressed in terms of an algebra of functions defined
on the manifold. Local coordinates can be replaced by generators of the algebra whereas the vector
fields by derivations. This remain valid for the case of the noncommutative spaces, which can be
described by an abstract associative algebra $\cA$ which is not necessarily
commutative. Historically the most important noncommutative algebra in physics was the quantised
space of non-relativistic quantum mechanics. This algebraic approach to quantum mechanics was 
extended and developed by Neumann~\cite{Jordan:1933vh}. Important for the developing intuition of
the commutative limit of noncommutative geometry was the classical limit of quantum-mechanical
systems. The symplectic geometry of quantised phase space and its relation to noncommutative
geometry have been discussed for example, by~\cite{DuboisViolette:1990xq,Dimakis:1992fj}.
Spaces with noncommutative characteristics cannot called `manifolds'; the notion of the
localisation of a point is absent on them. This is the essential feature which makes noncommutative
geometry particularly well suited for the description of the internal structure of a Kaluza-Klein
theory.

On the other hand, an obvious way of a noncommutative generalisation of a classical field theory
would be to replace the objects which describe the theory in the commutative case by their
corresponding ones with noncommutative characteristics. The simplest suggestion is to consider the
$M_{N}$ algebras of finite $N\times N$ complex matrices. For establishing the various ideas of this
`Matrix geometry' and references to original works consult~\cite{Madore:2000aq}.

The motivation for introducing noncommutative geometry in Kaluza-Klein theory lies in the suggestion
that space-time structure cannot be adequately described by ordinary geometry at all length scales,
including those which are presumably relevant when considering hidden dimensions. There is of course
no reason to believe that the extra structure can be described by the simple matrix geometries I
shall consider, although this seems suggestive by the finite particle multiplets observed in nature.

To be more specific, let the $M_{N}$ algebra of $N\times N$ complex matrices and  
$\{\l_{a}\}\in M_{N}$, $a=1,\ldots,N^{2}-1$ be an antihermitean basis of the Lie algebra of  the
special unitary group $SU(N)$. The Killing metric is given by $g_{ab}=-\Tr(\l_{a}\l_{b})$ which is
going to be used for rising and lowering indices. The set $\{\l_{a}\}$ is a  set of generators of
$M_{N}$ algebra which although not minimal is a convenient one; the derivations
\begin{equation}\label{NC-derivation}
e_{a}=\k^{-1}\adj(\l_{a})
\end{equation}
form a basis over the complex numbers for the derivations of $M_{N}$. They satisfy the commutation
relations
\begin{equation}
[e_{a},e_{b}]=m C^{c}{}_{ab} e_{c}\,,
\end{equation}
where the mass scale $m$ is defined as the inverse of the length scale $\k$.

Let $x^{\m}$ be the coordinates of ordinary spacetime, $M^{4}$. Then the set $(x^{\m},\l_{a})$ is
a set of generators of the algebra $\cA=\cC\times M_{N}$. Here, I concentrate on the internal
algebraic structure of the manifold. The exterior derivative of an element $f$ of $M_{N}$, $df$ is
defined as usual by the relation (\ref{Pfaffder}). Since any element $f$ can be written as
linear combination of the derivations, $f=f^{a}e_{a}$ , the relation above would mean
\begin{equation}
d\l^{a}(e_{b})=m[\l_{a},\l_{b}]=mC^{a}{}_{cb}\l^{c}\,.
\end{equation}

The set of $d\l_{a}$ forms a system of generators of $\O^{1}(M_{N})$ as a left or right module but
is not a convenient one. Since the algebra is now noncommutative happens for example to be
$\l_{a}d\l_{b}\neq d\l_{b}\l_{a}$. However because of the particular structure of $M_{N}$ there is
another system of generators being orthogonal to the derivations $e_{a}$
\begin{equation}\label{theta-def}
\th^{a}(e_{b})=\d^{a}{}_{b}\,.
\end{equation}
and form a dual of the derivation space basis. This set of generators is related to the $d\l_{a}$ by
the equations
\begin{equation}\label{th-dl-relation}
d\l^{a}=m~C^{a}{}_{bc}\l^{b}\th^{c},\qquad \th^{a}=\k\l_{b}\l^{a}d\l^{b}\,,
\end{equation}
and it satisfies the same structure equations as the components of Maurer-Cartan form on the
special unitary group $SU(N)$
\begin{equation}\label{Maurer-Cartan}
d\th^{a}=-\frac{1}{2}m~C^{a}{}_{bc}\th^{b}\th^{c}\,.
\end{equation}
The product in the right-hand side of this relation is the product in $\O(M_{N})$. Although
this product is not in general antisymmetric, because of the (\ref{theta-def}) above I have
\begin{equation}
\th^{b}\th^{a}=-\th^{a}\th^{b}\,.
\end{equation}
The $\th^{a}$'s also commute with elements of $M_{N}$ and $\O^{1}(M_{N})$ and can be identified
by the tensor product of $M_{N}$ and the dual of the vector space of derivations. The subalgebra
$\O^{*}(M_{N})$ generated by the $\th^{a}$ is an exterior algebra. Relation (\ref{Maurer-Cartan})
reveals that it is also a differential subalgebra, as I expected. Since I shall only use
$\O^{*}(M_{N})$ in what follows I shall write the product as a wedge product
\begin{equation}
\th^{a}\th^{b}=\th^{a}\wdg\th^{b}\,.
\end{equation}

Let me choose a basis $\th^{\mu}_{\nu}dx^{\nu}$ of $\O^{1}(\cC)$ over $\cC$ and let $e_{\mu}$ be
the Pfaffian derivations dual to $\th^{\mu}$. I set, similarly with section~\ref{Kaluza-Klein},
$M=(\mu,a)$, $1\leq M\leq 4+(N^{2}-1)$, and introduce
$\th^{\ssM}=(\th^{\mu},\th^{a})$ as generators of $\O^{1}(\cA)$ as left or right module and
$e_{\ssM}=(e_{\mu},e_{a})$ as a basis of derivations, $\Der(\cA)$, over $\cC$. I decompose the
$\O^{1}(\cA)$ in a direct sum
\begin{equation}
\O^{1}(\cA)=\O^{1}_{H}\oplus\O^{1}_{V}\,,
\end{equation}
of an horizontal and vertical part which are defined respectively as
\begin{equation}
\O^{1}_{H}=M_{N}\otimes\O^{1}(\cC)\,,\qquad
\O^{1}_{V}=\cC\otimes\O^{1}(M_{N})\,.
\end{equation}
The $\O^{1}_{H}$ part has basis $\th^{\mu}$ whereas the $\O^{1}_{V}$, $\th^{a}$. One can decompose
similarly the exterior derivative also
\begin{equation}
d=d_{H}+d_{V}\,.
\end{equation}

As I have discussed in the previous section the generators $\th^{a}$ of $\O^{1}(M_{N})$ can be
considered as a sort of moving frame. By comparing the relation (\ref{Maurer-Cartan}) with the first
structure equations for this frame,
\begin{equation}
d_{H}\th^{a}=0,\qquad d_{V}\th^{a}+\o^{a}{}_{b}\wdg\th^{b}=T^{a}\,,
\end{equation}
one concludes that for vanishing torsion $T^{a}$ the internal structure is provided with a linear connection
\begin{equation}\label{non-zero-connection}
\o^{a}{}_{b}=-\frac{1}{2}mC^{a}{}_{bc}\th^{c}\,,
\end{equation}
being actually a curved space. Then, I have  a covariant derivative $D_{a}$ which differs from
$e_{a}$. Alternatively, having required the linear connection $\o^{a}{}_{b}$ to vanish, the torsion
would be
\begin{equation}\label{non-zero-torsion}
T^{a}=-\frac{1}{2}mC^{a}{}_{bc}\th^{b}\wdg\th^{c}\,.
\end{equation}
In this case the covariant derivative $D_{a}$ and $e_{a}$ would coincide in the absence of gauge
couplings but the $D_{a}$ would satisfy
\begin{equation}
[D_{a},D_{b}]=m C^{a}{}_{bc}D_{c}
\end{equation}
forming the Lie algebra of the $SU(N)$ group. In the following lines, I choose the
(\ref{non-zero-connection}) solution in order
to avoid an extra torsion term. If $\a$ a one-form $\a=\a_{a}\th^{a}$ then I have by definition, in
the absence of torsion
\begin{equation}
d\a=(D_{a}\a_{b})\th^{a}\th^{b}\,,
\end{equation}
with
\begin{equation}\label{NC-covariant-derivative}
D_{a}\a_{b}=e_{a}\a_{b}-\frac{1}{2}m C^{c}{}_{ab}a_{c}\,.
\end{equation}
Note that this covariant derivative commutes with itself when acting on elements of the algebra; the
ordinary derivative does not. That is for any $f\in\cA$,
\begin{equation}
D_{[a}D_{b]}f=0\,.
\end{equation}

Note that the equations given above are with respect to an arbitrary basis $\l^{a}$ but they are
all tensorial in character with respect to a change of basis
\begin{equation}\label{NC-coordinate-transformation}
\l^{a}\to\l'^{a}=A^{a}{}_{b}\l^{b}, \qquad (A^{a}{}_{b})\in \rm{GL}(N^{2}-1)\,.
\end{equation}
The covariant derivative (\ref{NC-covariant-derivative}), on the other hand, transform as it
should. What is unusual is that the connection transform also as a tensor and each term of
(\ref{NC-covariant-derivative}) transforms as a tensor separately. This is related to the fact that
on the factor $M_{N}$ of the considered algebra the notion of a point is absent; therefore, there is
no analog of local variation. Each $\th^{a}$ corresponds to a globally defined moving frame and its
transformations correspond to the set of global transformations in internal space. The change of
basis (\ref{NC-coordinate-transformation}) is the equivalent in $M_{N}$ of a coordinate
transformation in $\cC$. Of course, I can always choose an appropriate change of basis and set the
Killing metric $g_{ab}$ equal to the Euclidean one $\d_{ab}$.

Finally, one can also consider the automorphisms of $M_{N}$, given by
\begin{equation}
\l^{a}\to\l'^{a}=g^{-1}\l^{a}g,\qquad g\in\rm{GL}(N)\,.
\end{equation}
Restricting the discussion, in favour of brevity, only in the case of infinitesimal transformations
\begin{equation}
\l^{a}\to\l'^{a}=\l^{a}-[f,\l^{a}],\qquad g\backsimeq 1+f\,,
\end{equation}
I see that
\begin{equation}
\th'^{a}=\th^{a}-L_{\ssX}\th^{a}\,,\qquad X=\adj(f)
\end{equation}
and in general for any $N$-form $\a$
\begin{equation}
\a'=\a-L_{\ssX}\a\,.
\end{equation}

\section{Noncommutative Kaluza-Klein theory\label{Kaluza-KleinTheoryRevisited}}

Here, I describe a noncommutative modification of the Kaluza-Klein theory  following the
original work of~\cite{Madore:1989ma}.

Firstly, I introduce the quadratic form of signature $N^{2}+1$ 
\begin{equation}
ds^{2}=g_{\ssM\ssN}\th^{\ssM}\otimes\th^{\ssN}=\h_{\mu\nu}\th^{\mu}\otimes\th^{\nu}
+g_{ab}\th^{a}\otimes\th^{b}\,,
\end{equation}
where $\h_{\mu\nu}$ is the Minkowski metric. I refer to this form as a metric although it contains
two terms of a slightly different nature. There is however a unique metric $g_{V}$ for $M_{N}$ with
respect to which all the derivations $e_{a}$ are Killing derivations and it is defined by 
$L_{\ssX}g_{V}=0$ if $X=X^{a}e_{a}$ be an arbitrary derivation. To within a rescaling the $g_{ab}$
noted above are the components of $g_{V}$.

A general one-form $\th\in\O^{1}_{V}$, on the other hand, can be easily constructed from the
generators $\th^{a}$ as
\begin{equation}
\th=-m\l_{a}\th^{a}\,,
\end{equation}
which from relations (\ref{th-dl-relation}) and (\ref{Maurer-Cartan}) satisfies
the zero-curvature condition
\begin{equation}\label{th-0-curvature}
d\th+\th^{2}=0\,.
\end{equation}
This $\th$ turns out to be gauge invariant as I shall explain  below. It satisfies with respect to
the algebraic exterior derivative $d_{V}$ similar conditions to the ones satisfied by the Maurer
Cartan form with respect to ordinary exterior derivation of the $SU(N)$ gauge group. As I have
described in section \ref{NoncommutativeGeometry}, I have a map of the trace-free elements of $f\in
M_{N}$ onto the derivations of $M_{N}$ given by $X\to X=\adj(f)$. The one-form $\th$ can be defined
without any reference to the $\th^{a}$ as the inverse map: $\th(X)=-f$. The one-form $\th$ turns out
to be invariant under  all derivations of $M_{N}$. To within rescaling by some complex number it is
the only one-form having this property.

In the commutative case a connection $\o$ on the trivial  principal $U(1)$ bundle is an
antihermitean one-form which can be splitted in a horizontal part, a one-form on the base manifold
and a vertical part, the Maurer-Cartan form $d\a$ on $U(1)$. This is
\begin{equation}
\o=A+d\a\,.
\end{equation}
The gauge potential $A\in\O^{1}(\cC)$ can be used to construct  a covariant derivative on the
associated vector bundle. The notion of a vector bundle can be generalised to the noncommutative
case as an $\cA$ module which is a free module of rank $1$ for the case of no topological
complications; it can be identified with $\cA$ itself. This is in fact the most natural
generalisation, since in the noncommutative modification of the theory which consider, the
$M_{N}$ has replaced the $\bb{C}$. Therefore, it is important to note that the $U(N)$ gauge symmetry
I shall use below comes not from the rank of vector bundle but from the finite dimension of $M_{N}$
matrices. The noncommutative generalisation of a gauge potential $A$, according to the section
\ref{NoncommutativeGeometry}, is an antihermitean $\O^{1}(\cA)$ element, which can be splitted again
in horizontal and vertical parts
\begin{equation}\label{NcPotential}
\o=A+\th+\f\,.
\end{equation}
The $A$ is the gauge potential belonging in the  horizontal part $\O^{1}_{H}$ and $\f$ is an element
of $\O^{1}_{V}$. The $\th$ is similar to Maurer-Cartan form.

Let $\cU$ be the unitary elements of $\cA$. In the case  I am considering here, namely
$\cA=\cC\times M_{N}$ this is the group $\cU(N)$ of smooth functions on $M^{4}$ with values in the
unitary group $U(N)$ and I choose it to be the group of local gauge transformations.

A gauge transformation defines a mapping of $\O^{1}(\cA)$ onto itself of the form
\begin{equation}
\o'=g^{-1}\o g+g^{-1}dg\,.
\end{equation}
We also define
\begin{equation}
\th'=g^{-1}\th g+g^{-1}d_{V}g\,,\qquad
A'=g^{-1}A g+g^{-1}d_{H}g
\end{equation}
and so $\f$ transforms under the adjoint action of $\cU(N)$ as
\begin{equation}
\f'=g^{-1}\f~g\,.
\end{equation}
Therefore the $\th$ form remains invariant under the action  of these local gauge transformations
and the transformed potential $\o'$ is again of the form of (\ref{NcPotential}).

The fact that $\th$ is invariant under a gauge transformation  means that it cannot be made zero by
a choice of gauge. I have then a potential with vanishing curvature but which is not gauge
equivalent to zero. If $M_{N}$ were an algebra of functions over a compact manifold, the existence
of a such one-form would be due to the nontrivial topology of the manifold.

I define the curvature two-form $\O$ and the field strength $F$ as usual 
\begin{equation}\label{Curvature-FieldStrength-defs}
\O=d\o+\o^{2},\qquad F=d_{H}A+A^{2}\,.
\end{equation}
Having defined the covariant exterior derivative as
\begin{equation}
D\f=d\f+\o\f+\f\o
\end{equation}
and decomposed it into horizontal and vertical parts, the substitution of (\ref{NcPotential}) to the first of
(\ref{Curvature-FieldStrength-defs}) gives
\begin{equation}
\O=F+D_{H}\phi+(D_{V}\f-\f^{2})\,.
\end{equation}
In terms of  components, with $\f=\f_{a}\th^{a}$, $A=A_{a}\th^{a}$ and with definitions
\begin{equation}
\O=\frac{1}{2}\O_{\ssM\ssN}\th^{\ssM}\wdg\th^{\ssN}\,,\qquad
F=\frac{1}{2}F_{\mu\nu}\th^{\mu}\th^{\nu}\,,
\end{equation}
I find
\begin{align}\label{connection-components-general}
&\O_{\mu\nu}=F_{\mu\nu}\,,\qquad \O_{\mu a}=D_{\mu}\f_{a}\nn\\
&\O_{ab}=[\f_{a},\f_{b}]-mC^{c}{}_{ab}\f^{c}\,.
\end{align}

I describe the Kaluza-Klein construction in three steps. Firstly  one has to identify the internal
structure. Using the one-forms $\th^{\ssM}$ one can consider the algebra $\cA$ as  the algebra of
functions over  a manifold of dimension $4+(N^{2}-1)$: a product of an ordinary manifold of $4$
dimensions and an algebraic structure of dimension $N^{2}-1$. In general, the invariance group of
the complete structure is  $SO(3+N^{2}-1,1)$, but if I restrict to the local rotations which donnot
mix the ordinary $\th^{\mu}$'s with the algebraic $\th^{a}$'s, this group reduces to $SO(3,1)\times
SO(N^{2}-1)$. The group $U(N)$ acts on the algebraic structure through the adjoint representation
\begin{equation*} 
U(N)\to SO(N^{2}-1)\,.
\end{equation*}
Only the group $SU(N)/\bb{Z}_{2}$ acts non trivially and I have an embedding
\begin{equation*}
SU(N)/\bb{Z}_{2}\hookrightarrow SO(N^{2}-1)\,.
\end{equation*}
Therefore, I am forced to suppose that
\begin{equation}\label{embedding-condition}
\o^{0}{}_{a}=0,\qquad A^{0}{}_{\mu}=0\,,
\end{equation}
and that $g\in\cS\cU(N)$, the local gauge transformations. 

With the condition (\ref{embedding-condition}) the connection can be written
explicitly as
\begin{equation}
\o=(A^{a}{}_{\mu}\th^{\mu}-m\,\th^{a}+\f^{a}{}_{b}\th^{b})\l_{a}
\end{equation}
and the derivations of the algebra $\cA$ will be
\begin{equation}
\tld{e}_{\mu}=e_{\mu}+\k A^{a}{}_{\mu}e_{a}\,,\qquad 
\tld{e}_{a}=e_{a}+\k\o^{b}{}_{a}e_{b}=\k\f^{b}{}_{a}e_{b}\,.
\end{equation}
Dual to them one defines the one-forms
\begin{equation}\label{th-forms-InternalCurvature0}
\tld{\th}^{\mu}=\th^{\mu}\,,\qquad 
\tld{\th}^{a}=m\chi^{a}{}_{b}(\th^{b}-\k A^{b}{}_{a}\th^{a})\,,
\end{equation}
where the inverse $\chi^{a}{}_{b}$ to the matrix $\f^{a}{}_{b}$ have been used.

I firstly describe the special case of connections for  which the internal curvature vanishes, i.e.
$\O_{ab}=0$. This is the case which is mostly resembles ordinary Kaluza-Klein theory. From
(\ref{connection-components-general}) either the $\f^{a}{}_{b}$ vanishes or $\f^{a}{}_{b}$ belongs
to the gauge orbit of $m\d^{a}{}_{b}$. These two cases belong to the two stables vacua of the
theory. The set gives rise to a singular set of one-forms $\tld{\th}^{a}$. The second value, which
corresponds to the physical vacuum, yields a frame $\th^{\ssM}$ which is formally similar to the
usual moving frame constructed on a principal $SU(N)$ bundle.

I denote the linear connection in $\O^{1}(\cC)$ as  $\o^{\mu}{}_{\nu}$, an $\mso(3,1)$-valued
one-form. This has to satisfy the structure equations
\begin{equation}
d\th^{\mu}+\o^{\mu}{}_{\nu}\wdg\th^{\nu}=0\,,\qquad
d\o^{\mu}{}_{\nu}+\o^{\mu}{}_{\rho}\wdg\o^{\rho}{}_{\nu}=\O^{\mu}{}_{\nu}\,.
\end{equation}
Generally I have to construct an $\mso(3+(N^{2}-1),1)$-valued  one-form $\tld{\o}^{\ssM}{}_{\ssN}$
on $\O^{1}(\cA)$ satisfying the first structure equation
\begin{equation}\label{1st-StructureEquation}
d\tld{\th}^{\ssM}+\o^{\ssM}{}_{\ssN}\wdg\tld{\th}^{\ssN}=0\,.
\end{equation}
Under the condition of vanishing internal curvature, $\O_{ab}=0$,  the solution to these equation is
given by
\begin{subequations}
\begin{align}
&\tld{\o}^{\mu}{}_{\nu}=
\o^{\mu}{}_{\nu}+\frac{1}{2}\k F_{a}{}^{\mu}{}_{\nu}\tld{\th}^{a}\,,\\
&\tld{\o}^{\mu}{}_{a}=\frac{1}{2}\k F_{a}{}^{\mu}{}_{\nu}\th^{\nu}\,,\\
&\tld{\o}^{a}{}_{b}=
-\frac{1}{2}m C^{a}{}_{bc}\tld{\th}^{c}+\k C^{a}{}_{cb}A^{c}{}_{\mu}\th^{\mu}\,,
\end{align}
\end{subequations}
which except of an additional term in the $\o^{a}{}_{b}$ equation,  this connection is formally the
same with the one constructed on an $SU(N)$ bundle. The extra term is what remains of the covariant
derivative of the Higgs-boson fields.

I consider now a general $SU(N)$ connection with  a general Higgs-boson field ($\O_{ab}\neq 0$
case). Then the matrix $m\chi^{a}{}_{b}$ in (\ref{th-forms-InternalCurvature0}) can be considered as
a transformation of the frame $\tld{\th}^{a}$ away from its physical vacuum value. Let me expand the
curvature components of the extended space over the algebraic basis of the extra dimensional
structure, i.e $\O_{\ssM\ssN}=\O^{a}{}_{\ssM\ssN}\l_{a}$ and define for brevity
$\O'^{a}{}_{\ssM\ssN}=\chi^{a}{}_{b}\O^{b}{}_{\ssM\ssN}$. Then the solutions to
(\ref{1st-StructureEquation}) is given by
\begin{subequations}
\begin{align}
&\tld{\o}^{\mu}{}_{\nu}=
\o^{\mu}{}_{\nu}+\frac{1}{2}\O'_{a}{}^{\mu}{}_{\nu}\tld{\th}^{a}\,,\\
&\tld{\o}^{\mu}{}_{a}=
\frac{1}{2}\O'_{b}{}^{\mu}{}_{a}\th^{b}
+\frac{1}{2}\O'_{a}{}^{\mu}{}_{\ssM}\tld{\th}^{\ssM}\,,\\
&\tld{\o}^{a}{}_{b}=-\frac{1}{2}mC^{a}{}_{bc}\tld{\th}^{c}
+\frac{1}{2}\O'_{c}{}^{a}{}_{b}\tld{\th}^{c}
+\frac{1}{2}(\O'^{a}{}_{\ssM b}-\O'_{b\ssM}{}^{a})\tld{\th}^{\ssM}\,.
\end{align}
\end{subequations}
As in normal Kaluza-Klein theory the connection contains structure constants and terms which vanish
with the curvature.

To complete however the Kaluza-Klein construction, the second structure equation has to be
fullfilled by the connection, i.e.
\begin{equation}\label{2nd-StructureEquation}
d\tld{\o}^{\ssM}{}_{\ssN}+\tld{\o}^{\ssM}{}_{\ssK}\wdg\tld{\o}^{\ssK}{}_{\ssN}
=\tld {\O}^{\ssM }{ }_{\ssN}\,.
\end{equation}
Moreover the equations of motion following from a suitable action have to be also considered.
Invariance under local $SO(3,1)\times SU(N)$ transformations permits an infinite sum of terms
involving arbitrary powers of $\O_{\mu\nu}$, $\O_{\mu a}$, $\O_{ab}$ as well as the Higgs-boson
field $\f_{a}$. It has been shown in~\cite{Deruelle:1986ia} that in the usual Kaluza-Klein with an
internal space that consistent a classical theory with reasonable stable vacua, only a finite part
of the above mention expansion is required. Here I restrict myself to the case of
Einstein-Hilbert term leaving aside the discussion of higher order terms in the modified version of
Einstein gravity I describe. In that case, one finds for the Lagrangian
\begin{align}
\tld{R}=R
&+\frac{1}{4}\O'_{a\ssM\ssN}\O'^{a\ssM\ssN}-\frac{1}{4}m^{2}C_{abc}C^{abc}
+\frac{1}{2}\O'^{a}{}_{\mu b}\O'^{b\mu}{}_{a}+\O'^{a}{}_{\mu a}\O'^{b\mu}{}_{b}\nn\\
             &+\frac{1}{2}(\O'^{a}{}_{bc}+m C^{a}{}_{bc})\O'^{cb}{}_{a}\,.
\end{align}
All of the terms on the right-hand side of the equation are gauge invariant.

The second term in the equation above is a modified version  of the gauge boson Lagrangian. In the
next chapter I shall use it to describe a Yang-Mills-Dirac theory over an extended space with
noncommutative characteristics; furthermore I shall discuss the dimensional reduction of the theory
using generalisation of the CSDR ideas. The third term is an effective cosmological constant. The
last three terms do not appear in usual Kaluza-Klein theories. They modify in an essential way the
Higgs Lagrangian.

\section{The fuzzy sphere example as an extra dimensional
space\label{FuzzySphere}}

Here, I describe as a concrete example of a matrix geometry  the case of the fuzzy
sphere. This geometrical construction was proposed in~\cite{Madore:1991bw}. Thereafter it was
considered in a series of different studies as an extra dimensional
space~\cite{Ramgoolam:2001zx,Ramgoolam:2002wb,Aschieri:2003vy,Hasebe:2003gx,Aschieri:2004vh,
Aschieri:2005wm,Aschieri:2006uw, Karabali:2006eg,Aschieri:2007fb} or as emergent geometry of a
matrix model~\cite{Iso:2001mg,Valtancoli:2002rx,Aschieri:2006uw,Karabali:2006eg}. It is interesting
to note that the fuzzy sphere has been also described as the classical stable vacuum of D-branes
configurations (Myers effect)~\cite{Myers:1999ps}.  For reviews
consult~\cite{Madore:2000aq,Balachandran:2006gf,Ydri:2001pv}.

For the definition of the fuzzy sphere and the gauge theory over it I follow
ref.~\cite{Madore:2000aq} (see also ref.~\cite{Madore:1991bw}).
A fuzzy `manifold' both in general and in the particular  case of the fuzzy sphere is a discrete
matrix approximation to the corresponding continuous manifold. These fuzzy spaces are described by
associative noncommutative algebras which are constructed by singling-out a finite subspace of the
space of functions defined over their corresponding commutative manifolds. This subspace
is need to be invariant under multiplication and therefore the discrete matrix approximation must be
appropriately chosen. An essential feature of this approximation is that the discretise space
preserves its continuum symmetries~\cite{Balachandran:2001dd}, in contrast with the lattice gauge
theories case. In chapter~\ref{FuzzyCSDR}, I will show that this characteristic is of crucial
importance for the formulation of the Fuzzy-CSDR scheme.

Turning now to the description of the fuzzy sphere construction, let me concider the $\bb{R}^{3}$
and the commutative coordinates $\tld{x}^{a}$, $a=1,2,3$. Let $g_{ab}=\d_{ab}$ the standard
Euclidean metric and the two-sphere manifold, $S^{2}$, defined by the constraint
\begin{equation}\label{S2-constraint}
 g_{ab}\tld{x}^{a}\tld{x}^{b}=r^{2}\,,
\end{equation}
with $r$ being the radius of the sphere. Consider $\cP$ the algebra of polynomials in $\tld{x}^{a}$
and let $\cI$ be the ideal generated by the (\ref{S2-constraint}). That is the $\cI$ consists of
elements of $\cP$ with a $(g_{ab}\tld{x}^{a}\tld{x}^{b}-r^{2})$ factor. Then the quotient algebra
$\cA=\cP/\cI$ is dense in the $\cC(S^{2})$. Any element of $\cA$ can be represented as a finite
multipole expansion of the form
\begin{equation}\label{C(S2)-multipole-expansion}
\tld{f}(\tld{x}^{a})=f_{0}+f_{a}\tld{x}^{a}
+\frac{1}{2}f_{ab}\tld{x}^{a}\tld{x}^{b}+\ldots\,,
\end{equation}
where the $f_{a_{1},\ldots,a_{l}}$ are completely symmetric and traceless.

A sequence of noncommutative approximation to the $\cC(S^{2})$ can now be constructed by means of
truncating the multipole expansion (\ref{C(S2)-multipole-expansion}) up to a certain order. Indeed,
if I truncate all functions to the constant term I reduce the algebra $\cC(S^{2})$ to the algebra
of $\cA_{1}=\bb{C}$ of complex numbers and the geometry of $S^{2}$ is reduced to that of a point.
Keeping the expansions up to the linear term in the $\tld{x}^{a}$ forms a four-dimensional vector
space\,\footnote{$f_{a}$ contributes three independent components plus one more from the constant
term.}. However the subspace of monomial functions over $\cC(S^{2})$ is not invariant under ordinary
multiplication. This can be resolved by an appropriate redefinition of the product. Indeed, if I
require the {radical} of $\cA_{2}$ to be equal zero then there are two possible
ways-out. First, I can define the product so that $\cA_{2}$ becomes a {product of four copies of
$\bb{C}$}. This algebra is commutative and the sphere looks like as a set of four points, being
actually a lattice approximation of the original manifold. Alternatively, I can define the product
so that $\cA_{2}$ becomes the $M_{2}$ algebra of complex $2\times 2$ matrices. That is I replace
\begin{equation*}
\tld{x}^{a}\mapsto x^{a}=\kbar r^{-1}(\s^{a}/2)\,.
\end{equation*}
The $\s^{a}$ are the Pauli matrices and the parameter  $\kbar$ must be related to $r$ by the
equation $\kbar^{2}=\frac{1}{3/4}r^{2}$ in order the defining constraint of the sphere
$g_{ab}x^{a}x^{b}=r^{2}$ to be fullfilled. As I explain in more detail below, the algebra
$M_{2}$ describes the sphere poorly and it is \textit{fuzzy}. In fact it is a quantum mechanical
system with the commutative coordinates replaced by the two-dimensional matrix operators forming an
$\msu(2)$ Lie algebra. As such the corresponding geometrical construction is characterised by two
eigenvalues distinguishing only the north and south poles of the sphere. However the three operators
defining the new geometrical construction cannot be simultaneously measured. Therefore, it is
impossible to define the notion of a point over this {fuzzy} sphere.

Suppose next that I keep the term quartic in $\tld{x}^{a}$.  That is I consider the set $\cA_{3}$
of expansions of the form (\ref{C(S2)-multipole-expansion}) with only the coefficients $f_{0}$,
$f_{a}$, $f_{ab}$ nonvanishing. This is a nine-dimensional vector space because of the constraint
(\ref{S2-constraint}). This subspace of functions closes under multiplication by various ways of
product redefinition. Among them, I can choose the product in a way that $\cA_{3}$ becomes equal to
the algebra  $M_{3}$ of complex $3\times 3$ matrices and make the replacement
\begin{equation}
\tld{x}^{a}\mapsto x^{a}=\kbar r^{-1}J^{a}\,,
\end{equation}
with the $J^{a}$ being the three-dimensional irrep. of $\msu(2)$ Lie algebra. The sphere is now less
fuzzy and the equator as well as the north and south pole can be distinguished.

In the general case one can suppress the $N$-th order  in the $\tld{x}^{a}$. The resulting vector
space $\cA_{N}$ is of dimensions $N^{2}$. It closes under multiplication by a new product in
$\tld{x}^{a}$ which make it into the algebra of $M_{N}$ of complex $N\times N$ matrices. Indeed, the
number of components of the completely symmetric tensor $f_{a_{1}\ldots a_{l}}$ is given by
$\NN_{l}=\binom{N+l-1}{l}$. Because of the constraint (\ref{S2-constraint}) $\NN_{l-2}$ of these
will not contribute to the expansion (\ref{C(S2)-multipole-expansion}). Therefore there will be
$\NN_{l}-\NN_{l-2}=2l+1$ independent monomials of order $l$ and $\sum_{l=1}^{N-1}(2l+1)=N^{2}$
components in all. Moreover, the $\cA_{N}$ vector space closes if I make the replacement
\begin{equation}\label{S2N-map}
\tld{x}^{a}\mapsto x^{a}=i\kbar r^{-1} X^{a}
\end{equation}
where the $X^{a}$ are considered to be antihermitean and fulfilling the commutation relations
\begin{equation}
[X^{a},X^{b}]=\vare^{ab}{}_{c}X^{c}\,,
\end{equation}
that is the $N$-dimensional irrep. of $\msu(2)$ Lie algebra. The noncommutative coordinates of the
geometrical construction will now close under
\begin{equation}\label{NC-coords-closure}
[x^{a},x^{b}]=i\kbar C^{ab}{}_{c}x^{c}\,, \qquad
C^{ab}{}_{c}=r^{-1}\vare^{ab}{}_{c}\,.
\end{equation}
This product redefinition turns the $\cA_{N}$ vector  space to the $M_{N}$ algebra of complex
$N\times N$ matrices. The defining constraint of the sphere is satisfied for the $x^{a}$, i.e
$g_{ab}x^{a}x^{b}=r^{2}$, if
\begin{equation}
\kbar=\frac{r^{2}}{\sqrt{C_{2}(N)}}\,,
\end{equation}
where $C_{2}(N)=\frac{1}{4}(N^{2}-1)$ is the eigenvalue  of the second Casimir operator for the case
of $N$-dimensional $\msu(2)$ algebra irrep. Therefore, the (\ref{S2N-map}) map takes the form
\begin{equation}
\tld{x}^{a}\mapsto x^{a}=  \frac{ir}{\sqrt{C_{2}(N)}}~X^{a}\,.
\end{equation}

Note that for large $N$ the $\kbar\approx\frac{2r^{2}}{N}$;  for $N\to\infty$ it tends to zero.
In this limit, $\kbar\to 0$, the $x^{a}$ commute and all of the points of the sphere can be
distinguished. In fact, the $\kbar$ constant is related with the area of elementary cells. Indeed,
the state of a particle on the fuzzy sphere is described as in quantum mechanics by a state
vector $\psi$. An observable associated to the particle is an hermitean element of $M_{N}$ and the
value of the observable $f$ is given by the real number $\psi^{*}f\psi$. Similarly by what
corresponds to the position of the particle is given by the $3$ numbers $\psi^{*}x^{a}\psi$. As
acting on the eigenstate of $x^{3}$ with the largest eigenvalue the commutation relations
(\ref{NC-coords-closure}) become, for large $N$,
\begin{equation*}
[x^{1},x^{2}]=i\kbar\,.
\end{equation*}
Therefore, there are elementary cells of area $2\p\kbar$; the $\kbar$ itself has dimensions of
$(length)^{2}$ whereas the fuzzy sphere can be covered by $\frac{4\pi r^{2}}{2\pi\kbar}=N$ of them.
Since I am considering a noncommutative model of Kaluza-Klein theory  I am tempted to identify
$\kbar$ with the inverse square of the Planck mass, $\kbar=\m_{P}^{-2}$ and consider the $S_{N}^{2}$
as an extra dimensional space which is fundamentally noncommutative in this length-scale.

The fuzzy sphere construction is more conveniently described in the basis provided by the constant
identity matrix $\one$ and the noncommutative spherical harmonics. Suppressing the $N$-th order
in $x^{a}$ of the multipole expansion (\ref{C(S2)-multipole-expansion}) above, (\textit{fuzzyness
level} $N-1$) , these noncommutative spherical harmonics are defined as
\begin{align}
Y^{lm}=r^{-l}\sum_{a} c_{a_{1}\ldots a_{l}}^{lm} X^{a_{1}}\cdots X^{a_{l}},
\qquad  & 0\leq l\leq N-1\nn\\
             & m=-l,-l+1,\ldots ,l-1,l
\end{align}
with $c_{a_{1}\cdots a_{l}}^{lm}$ the traceless and symmetric tensor of the ordinary spherical
harmonics. We choose to normalise these $Y^{lm}$ as
\begin{equation}
\Tr_{N}\left( (Y^{lm})^{\dag}Y^{l'm'}\right)=\d^{ll'}\d^{mm'}\,.
\end{equation}
A generic function on the fuzzy sphere is expanded under this basis as
\begin{equation}
F=\sum_{l=0}^{N-1}\sum_{m=-l}^{l}f_{lm}Y^{lm}\,,
\end{equation}
or in a more compact form
\begin{align}
S_{N}^{2}\cong (\mbf{N})\tens (\mbf{N})
&=(\mbf{1})\dsum (\mbf{3})\dsum\ldots\dsum(\mbf{N-1})\nn\\
&=\{Y^{0,0}\}\dsum\ldots\dsum\{Y^{(N-1),m}\}\,,
\end{align}
i.e. corresponds to an ordinary function on the commutative sphere with a cut-off on the angular
momentum. Obviously this space of truncated functions is closed under the noncommutative $N\times N$
matrix product. Moreover, in the $N\to\infty$ limit one recovers the usual commutative sphere.

A diffeomorphism of $S^{2}$ defines and it is defined  by an automorphism of the algebra of the
smooth functions on $S^{2}$, $\cC(S^{2})$. Then the noncommutative analogue of a diffeomorphism of
$S^{2}$ is therefore an automorphism of $M_{N}$. Since $M_{N}$ is a simple algebra all of its
automorphisms are of the form $f\mapsto f'=g^{-1}fg$ where $g$ is a fixed arbitrary element
of $M_{N}$ which has an inverse. Since I have considered complex-valued functions on $S^{2}$ the
algebra $\cC(S^{2})$ has a $*$-operation, $\tld{f}\mapsto\tld{f}^{*}$ obtained by taking the complex
conjugate of $\tld{f}$. The diffeomorphism of $\cC(S^{2})$ should respect this $*$-operation:
$\tld{f}^{*'}=\tld{f}^{'*}$. It is expected that its corresponding automorphism of $M_{N}$ will
respect this operation too. Therefore for the diffeomorphism is required
that $(g^{-1}fg)^{\dag}=g^{-1}f^{*}g$. As a result $g^{\dag}=g^{-1}$ and the analogue of
diffeomorphisms of $S^{2}$ in the noncommutative case are described by
\begin{equation}\label{AlgebraAutomorphism-x}
x^{'a}=g^{-1}x^{a}g,\qquad g\in SU(N)\,.
\end{equation}
A different choice of $x^{a}$ not connecting the change of coordinates  with the above relation
would be equivalent to a different choice of a differential or topological structure.

A smooth global vector field on $S^{2}$ defines and it is defined by a derivation of the algebra
$\cC(S^{2})$; the noncommutative analogue of a global vector field on $S^{2}$ is a derivation of the
algebra $M_{N}$, namely a linear map $X$ of $M_{N}$ onto itself satisfying the Leibniz rule
(\ref{LeibnizRule}) and the reality condition $(X(f))^{*}=X(f^{*})$. Therefore, as the $M_{N}$
is a simple algebra the derivations are of the form $X=\adj(h)$ where $h$ is a fixed but arbitrary
antihermitean element of $M_{N}$. The change of generators (\ref{AlgebraAutomorphism-x}) takes
\begin{equation}
X^{a}\mapsto X^{'a}=\adj(g^{-1}hg)
\end{equation}
and so all automorphisms of $M_{N}$ are analogue of diffeomorphisms of $\cC(S^{2})$. If $g$ is near
identity I can write 
\begin{equation}
x'^{a}\simeq x^{a}+e^{a}h,\qquad g\simeq 1+\frac{h}{i\kbar}\,.
\end{equation}
An important special case is given by $h=h_{a}x^{a}$. Then I have
\begin{equation}
x'^{a}\simeq x^{a}+C^{a}{}_{bc}h^{b}x^{c}\label{SO3-rot}
\end{equation}
and therefore in the limit corresponds to a $\R^{3}$ rotation around the axis $h_{a}$. The formula
(\ref{SO3-rot}) yields the adjoint action of the Lie algebra of $SO(3)$ on $M_{N}$, which contains
exactly once the irreducible representation of dimension $2j+1$ with $0\leq j\leq N-1$. The fuzzy
sphere construction respects the symmetries of its continuum counterpart.

On the two-sphere there is a natural action of $SU(2)$ which defines three smooth vector fields
$\tld{e}_{a}$, $a=1,2,3$ which are Killing fields with respect to the induced metric.
Correspondingly I single out three derivations of $e_{a}$ of $M_{N}$ for every $N$. The two-sphere
is not a parallelizable manifold and the module of derivations $\Der(S^{2})$ is not a free module on
the three generators $\tld{e}_{a}$, namely not all of them are linear independent. They satisfy the
relation $\tld{x}^{a}\tld{e}_{a}=0$. On the other hand, each of the truncations of the multipole
expansion (\ref{C(S2)-multipole-expansion}) makes the geometry of $S^{2}$ to look like the geometry
of $SU(2)$. In the $N\to\infty$ limit the two pictures coincide as I have already described. The
$SU(2)$ covariant differential calculus of the fuzzy sphere is three dimensional. The three
derivations $e_\a$ along $X_\a$ of a function $ f$ are given\footnote{The considered metric here is
the standard Euclidean one, $g_{ab}=\d_{ab}$ and I can freely lower the `coordinate' indices.} by
\begin{equation}
e_{a}(f)=[\adj(X_{a})](f)=[X_{a},f]=
\frac{1}{i\k}\left[x_{a},f\right]\,\label{derivations}
\end{equation}
with $\k=\kbar r^{-1}$ and therefore expressed  in $(length)^{-1}$ units as in
(\ref{NC-derivation}) of section \ref{NoncommutativeGeometry}. Accordingly the action of the Lie
derivatives on functions is given by
\begin{equation}\label{LDA}
\cL_{a} f = [X_{a},f]=\frac{1}{i\k}\left[x_{a},f\right]\,,
\end{equation}
they satisfy the Leibniz rule and the $SU(2)$ Lie algebra\footnote{This is essentially the Lie
algebra under which the angular momentum operators $\J_{a}=iX_{a}$ close.} relation
\begin{equation}\label{LDCR}
[\cL_{a}, \cL_{b} ] =\vare^{c}{}_{ab}\cL_{c}\,.
\end{equation}
In the $N \to \infty$ limit the derivations $e_\a$ become
\begin{equation}
\tld{e}_{a} = \vare^{c}{}_{ab} x^{b}\prt_{c}
\end{equation}
and only in this commutative limit the tangent space becomes two dimensional. The exterior
derivative is given by
\begin{equation}
d{f} = e_{a}(f)\th^{a}=[X_{a},{f}]\th^{a}=\frac{1}{i\k}\left[x_{a},f\right]\th^{a}
\end{equation}
with $\th^{a}$ the one-forms dual to the vector fields $e_{a}$, $<e_{a},\th^{b}>=\d_{a}{}^{b}$. The
space of one-forms is generated by the $\th^{a}$'s in the sense that for any one-form $\o=\sum_i f_i
(d h_i)\,t_i$ I can always write $\o=\o_{a}\th^{a}$ with given functions $\o_{a}$ depending on the
functions ${f}_i$,  ${h}_i$ and ${t}_i$. From
 $0=\cL_{a} (<e_{b},\th^{c}>)
=<\cL_{a}(e_{b}),\th^{c}>+<e_{b},\cL_{a}(\th^{c})>$ and $\cL_{a}(e_{b})=C_{ab}{}^{c}e_{c}$
[cf. (\ref{LDCR})] I obtain the action of the Lie derivatives on one-forms,
\begin{equation}\label{2.16}
{\cL}_{a}(\th^{b}) = \vare_{a}{}^{b}{}_{c}\th^{c}\,.
\end{equation}
It is then easy to check that the Lie derivative commutes with the exterior differential $d$, i.e.
$SU(2)$ invariance of the exterior differential. On a general one-form $\o=\o_{a}\th^{a}$ I have
\begin{align}
\cL_{b}\o=\cL_{b}(\o_{a}\th^{a})
&=(\cL_{b}\o_{a})\th^{a}-\o_{a}\vare^{a}{}_{bc}\th^{c}\nn\\
&=\left[X_{b},\o_{a}\right]\th^{a}-\o_{a}\vare^{a}{}_{bc}\th^{c}
\end{align}
and therefore
\begin{equation}
(\cL_{b}\o)_{a}=\left[X_{b},\o_{a}\right]-\o_{c}\vare^{c}{}_{ba}\,.\label{fund}
\end{equation}
Similarly, from
$\cL_{b}(v)=\cL_{b}(v^{a}e_{a})=[X_{b},v^{a}]e_{a}+v^{a}\cL_{b}(e_\a)$ I have 
\begin{equation}
 (\cL_{b} v)^\a=\left[X_{b},v^{a}\right]-v^{c}\vare_{cb}{}^{a}\,.\label{lievect}
\end{equation}

There are different approaches to the study of spinor fields  on the fuzzy sphere
\cite{Grosse:1995jt,Carow-Watamura:1996wg}. Here I follow ref. \cite{Madore:2000aq} (section
8.2)\footnote{For a discussion of chiral fermions and index theorems on matrix approximations of
manifolds see ref. \cite{Dolan:2002af}. }. In the case of the product of Minkowski space and the
fuzzy sphere, $M^{4} \times S^2_{N}$, I have seen that the geometry resembles in some aspects
ordinary commutative geometry in seven dimensions. As $N \to \infty$ it returns to the ordinary
six-dimensional geometry. Let $g_{AB}$ be the Minkowski metric in seven dimensions and $\Gamma^{A}$
the associated Dirac matrices which can be in the form 
\begin{equation}
 \Gamma^{A} = (\Gamma^\mu,\Gamma^\a)=(1 \otimes \gamma^{\mu}, \sigma^{\a} \otimes \gamma_{5})\,.
\label{Gamma7}
\end{equation}
The space of spinors must be a left module with respect to the Clifford algebra. It is therefore a
space of functions with values in a vector space ${\cal H}'$ of the form $$ {\cal H}' = {\cal H}
\otimes C^{2} \otimes C^{4}, $$ where ${\cal H} $ is an $M_{N+1}$ module. The geometry resembles but
is not really seven-dimensional, e.g. chirality can be defined and the fuzzy sphere admits chiral
spinors. Therefore the space $H'$ can be decomposed into two subspaces 
${\cal H}'_{\pm} = \frac{1 \pm\Gamma}{2}{\cal H}'$, where $\Gamma$ is the chirality operator of the
fuzzy sphere \cite{Madore:2000aq,Ydri:2001pv}. The same holds for other fuzzy cosets such as
$(SU(3)/U(1) \times U(1))_{F}$ \cite{Trivedi:2000mq}.

In order to define the action of the Lie derivative ${\cal{L}}_{a}$ on a spinor field $\Psi$, I
write
\begin{equation}
\Psi=\z_{\a}\psi_{\a}\,,
\end{equation}
where $\psi_{\a}$ are the components of $\Psi$  in the $\z_{\a}$ basis. Under a spinor rotation
$\psi_{\a}\to S_{\a\b}\psi_{\b}$ the bilinear $\bar\psi \Gamma^{a}\psi$ transforms as a vector
$v^{a}\to \Lambda_{ab}v^{b}$. The Lie derivative on the basis $\z_{\a}$ is given by
\begin{equation}\label{spinlie1}
{\cal L}_{a} \z_{\a} =\z_{\b}\t^\a_{\b\a}\,,
\end{equation}
where 
\begin{equation}
 \tau^{a}=\frac{1}{2}C_{abc}\Gamma^{bc}\,,\qquad
 \Gamma^{bc}= -\frac{1}{4}(\Gamma^{b} \Gamma^{c} - \Gamma^{c}\Gamma^{b})\,. \label{definitions}
\end{equation}
Using that  $\Gamma^{bc}$ are a rep. of the orthogonal algebra and then using the Jacobi identities
for $C_{abc}$ one has
$
[\tau^a,\tau^b]=C_{abc}\tau^c
$
from which it follows that the Lie derivative on spinors gives a representation of the Lie algebra,
\begin{equation}
[ {\cal L}_{a}, {\cal L}_{b} ] \z_\a= C_{abc}\,{\cal L}_{c}\,\z_\a\,.
\end{equation}
On a generic spinor $\Psi$, applying the Leibniz rule  I have
\begin{equation}
{\cL}_{a}\Psi=
\z_{a}[X_{a},\psi_{\a}]+\z_\b \t^\a_{\b\c}\psi_\c
\end{equation}
and of course $ [ {\cal L}_{a}, {\cal L}_{b} ] \Psi=C_{abc}\, {\cal L}_{c}\,\Psi$; I also write
\begin{equation}
\delta_{a}\psi_\a=(\cL_{a}\Psi)_\a= [X_a,\psi_\a]+\t^a_{\a\c}\psi_\c\,.\label{spinlie11}
\end{equation}
The action of the Lie derivative ${\cal{L}}_{a}$ on the adjoint spinor is obtained considering the
adjoint of the above expression, since $(X_{a})^\dagger=-X_{a}$, $(\tau^{a})^\dagger=-\tau^{a}$, 
$[\tau^{a},\Gamma_0]=0$ one has
\begin{equation}\label{spinlie2}
\delta_{a}\bar\psi_\a=  [X_{a},{\bar\psi_\a}] -\bar\psi_\c\t^{a}_{\c\a}\,.
\end{equation}
One can then check that the variations (\ref{spinlie11}) and (\ref{spinlie2}) are consistent with
$\psi^\dagger\Gamma^0\psi$ being a scalar. Finally I have compatibility among the Lie derivatives
(\ref{spinlie11}), (\ref{spinlie2}) and (\ref{fund}):
\begin{align}
\delta_{a}(\bar\psi\Gamma^\mu\psi)&=[X_{a},\bar\psi\Gamma^\mu\psi]\,,\\
\delta_{a}(\bar\psi\Gamma^{d}\psi)
&=(\delta_{a}\bar\psi)\Gamma^{d}\psi+\bar\psi\Gamma^{d}\delta_{a} \psi
=[X_{a},\bar\psi\Gamma^{d}\psi]+\bar\psi[\Gamma^{d},\tau^{a}]\psi\nn\\
&= [X_{a},\bar\psi\Gamma^{d}\psi] -C_{adc}\bar\psi\Gamma^{c}\psi\,.
\end{align}
This immediately generalises to higher tensors ${\bar\psi}\,\Gamma_{\!d_1}\ldots\Gamma_{\!d_i}
\psi$.
These relations and the one derived earlier for the forms and the vector fields are fundamental
for formulating the CSDR principle on fuzzy cosets. In the next chapter details on these ideas are
discussed.

We finally note that the differential geometry on  the product space Minkowski times fuzzy
sphere, $M^{4} \times S^2_{N}$, is easily obtained from that on $M^4$ and on $S^2_{N}$. For example
a one-form $A$ defined on $M^{4} \times S^2_{N}$ is written as
\begin{equation}\label{oneform}
A= A_{\mu} dx^{\mu} + A_{a} \theta^{a}
\end{equation}
with $A_{\mu} =A_{\mu}(x^{\mu}, X_{a} )$ and $A_{a}=A_{a}(x^{\mu}, X_{a})$.

\bigskip
\noindent
{\large\textit{Matrix approximations of other coset spaces}}

The $S^{2}$ manifold is not the only one that can be approximated by a matrix geometry. This is
possible for some other coset spaces too leading to their `fuzzy-fied' analogues. Such cases has
been studied extensively in the recent literature. To be more specific the sphere $S^{2}$ is the
complex projective space $CP^{1}$ . The generalisation of the fuzzy sphere construction to
$CP^{2}$ and its $spin^{c}$ structure was given in ref.~\cite{Grosse:1999ci}, whereas the
generalisation to $CP^{M-1}=SU(M)/U(M-1)$ and to Grassmannian cosets was given in
ref.~\cite{Balachandran:2001dd}.

While a set of coordinates on the sphere is given by the $\bb{R}^3$ coordinates $\tld{x}^{a}$ modulo
the relation $\sum_{a}\tld{x}^{a}\tld{x}^{a}=r^2$, a set of coordinates on $CP^{M-1}$ is given by
$\tld{x}^{a}$, $a=1,\ldots M^2-1$ modulo the relations
\begin{equation}
 \d_{ab}\,\tld{x}^{a}\tld{x}^{b} = \frac{2(M-1)}{M}r^2 \ ,
 \qquad d^{\,c}{}_{ab}\, \tld{x}^{a}\tld{x}^{b} = \frac{{2}(M-2)}{M}r \tld{x}^{c}\,,
\end{equation}
where $d^{\,c}{}_{ab}$ are the components of the symmetric invariant tensor of $SU(M)$. Then
$CP^{M-1}$ is approximated, at fuzziness level $N$, by $n\times n$ dimensional matrices $x_{a}$, $a=
1, \ldots, M^{2}-1$. These are proportional to the generators $X_{a}$ of $SU(M)$ considered in the
$n=\frac{(M-1+N)!}{(M-1)!N!}$ dimensional irrep., obtained from the $N$-fold symmetric
tensor product of the fundamental $M$-dimensional representation of $SU(M)$. As before I set
$x_{a}=\frac{1}{i r}X_{a}$ so that
\begin{equation} 
\sum_{a=1}^{3} {x}_{a} {x}_{a} = -\frac{C_{2}(n)}{r^2}\label{cas2} {}~,
\qquad[ X_{a}, X_{b} ] = C^{c}{}_{ab} X_{c}
\end{equation}
where $C_{2}(n)$ is the quadratic Casimir of the given $n$-dimensional irrep.,  and $r C^{c}{}_{ab}$
are now the $SU(M)$ structure constants. More generally \cite{Trivedi:2000mq} one can consider fuzzy
coset spaces $(S/R)_F$ described by non-commuting coordinates $X_{a}$ that are proportional to the
generators of a given $n$-dimensional irrep. of the compact Lie group $S$ and thus in particular
satisfy the conditions (\ref{cas2}) where now $r C^{c}{}_{ab}$ are the $S$ structure constants (the
extra constraints associated with the given $n$-dimensional irrep. determine the subgroup $R$ of $S$
in $S/R$). The differential calculus on these fuzzy spaces can be constructed as in the case for the
fuzzy sphere. For example there are $\dim(S)$ Lie derivatives, they are given by eq. (\ref{LDA})
and satisfy the relation (\ref{LDCR}). On these fuzzy spaces, the space of spinors is considered to
be a left module with respect to the Clifford algebra given by (\ref{Gamma7}), where now the
$\sigma^{a}$'s are replaced by the $\gamma^{a}$'s, the gamma matrices on $R^{{\dim S}}$; in
particular all the formulae concerning Lie derivatives on spinors remain unchanged.

\bigskip
\noindent
{\large\textit{Noncommutative gauge fields and transformations}}

Gauge fields arise in non-commutative geometry and in particular on fuzzy spaces very naturally;
they are linked to the notion of covariant coordinate \cite{Madore:2000en}. Consider a field
$\phi(X^{a})$ on a fuzzy space described by the non-commuting coordinates $X^{a}$. An infinitesimal
gauge transformation $\d\phi$ of the field $\phi$ with gauge transformation parameter
$\lambda(X^{a})$ is defined by
\begin{equation}
\d\phi(X) = \lambda(X)\phi(X)\,. 
\end{equation}
This is an infinitesimal abelian $U(1)$ gauge transformation if $\lambda(X)$ is just an
antihermitian function of the coordinates $X^{a}$, it is an infinitesimal nonabelian $U(P)$ gauge
transformation if $\lambda(X)$ is valued in $\mmu(P)$, the Lie algebra of hermitian
$P\times P$ matrices; in the following I will always assume $\mmu(P)$ elements to commute
with the coordinates $X^{a}$. The coordinates $X$ are invariant under a gauge transformation
\begin{equation}
\d X_{a} = 0\,;
\end{equation}
multiplication of a field on the left by a coordinate is then not a covariant operation in the
non-commutative case. That is
\begin{equation}
\d(X_{a}\phi) = X_{a}\lambda(X)\phi\,,
\end{equation}
and in general the right hand side is not equal to $\lambda(X)X_{a}\phi$. Following the ideas of
ordinary gauge theory one then introduces covariant coordinates $\vf_{a}$ such that
\begin{equation}\label{covcoord}
\d(\vf_{a}{}\f) = \lambda\vf_{a}{}\f\,,
\end{equation}
this happens if
\begin{equation}
\label{covtr} \d(\vf_{a})=[\lambda,\vf_{a}]\,.
\end{equation}
Setting
\begin{equation}
\vf_{a} \equiv X_{a} + A_{a}
\end{equation}
$A_{a}$ can be interpreted as the gauge potential of the non-commutative theory; then $\vf_{a}$ is
the non-commutative analogue of a covariant derivative. The transformation properties of $A_{a}$
support the interpretation of $A_{a}$ as gauge field; they arise from requirement (\ref{covtr}),
\begin{equation}
\d A_{a} = -[ X_{a}, \lambda] + [\lambda,A_{a}]\,.
\end{equation}
Correspondingly one can define a tensor $F_{ab}$, the analogue of the field strength, as
\begin{eqnarray}
\label{2.33} F_{ab} &=& [ X_{a}, A_{b}] - [ X_{b}, A_{a} ] + [A_{a} , A_{b} ] - C^{c}{}_{ab}A_{c}\\
&=& [ \vf_{a}, \vf_{b}] - C^{c}{}_{ab}\vf_{c}\,.
\end{eqnarray}
This tensor transforms covariantly
\begin{equation}
\d F_{ab} = [\lambda, F_{ab}]\,.
\end{equation}
Similarly, for a spinor $\psi$ in the adjoint representation, the infinitesimal gauge transformation is given by
\begin{equation}
\d \psi = [\lambda, \psi]\,,
\end{equation}
while for a spinor in the fundamental the infinitesimal gauge transformation is given by
\begin{equation}
\d\psi=\lambda\psi\,.
\end{equation}

\chapter{Dimensional Reduction over Fuzzy Coset Spaces\label{FuzzyCSDR}}

I use the  ideas presented in the previous chapter, concerning a specific noncommutative
modification of the Kaluza-Klein theory, to describe an interesting generalisation of the
Coset Space Dimensional Reduction (CSDR) scheme; this was introduced in~\cite{Aschieri:2003vy}~and
further explored in~\cite{Aschieri:2004vh,Aschieri:2005wm}. One start with a Yang-Mills-Dirac theory
defined in a higher dimensional space, $M^{\sD}=M^{4}\times B$, with the internal space being a
coset space which is approximated by a finite matrix algebra, $M_{N}$, i.e. a fuzzy coset space,
$(S/R)_{F}$. Having assumed fuzzy coset spaces as the hidden extra dimensions, the theory turns out
to be power counting renormalizable; the fuzzy spaces are approximated by matrices of finite
dimensions and only a finite number of counterterms are required to make the Lagrangian
renormalizable.

In the spirit of noncommutative geometry other particle models with noncommutative gauge
theory were 
explored~\cite{Connes:1990qp,Martin:1996wh,DuboisViolette:1989at,DuboisViolette:1988ps,
DuboisViolette:1989vq} . After the work of Seiberg and Witten~\cite{Seiberg:1999vs}, where a map (SW
map) between noncommutative and commutative gauge theories has been described, there has been a lot
of activity also in the construction of noncommutative phenomenological Lagrangians, for example
various noncommutative standard model like Lagrangians have been
proposed~\cite{Chaichian:2001py,Calmet:2001na,Aschieri:2002mc}\footnote{These
SM actions are mainly considered as effective actions because they are not renormalizable. The
effective action interpretation is consistent with the SM in \cite{Calmet:2001na,Aschieri:2002mc}
being anomaly free~\cite{Brandt:2003fx}. Noncommutative phenomenology has been discussed
in~\cite{Carlson:2001sw,Behr:2002wx,Hinchliffe:2002km,Schupp:2002up}.}. More recently
noncommutative modifications of the SM model based on the spectral triple formalism have been
proposed~\cite{Barrett:2006zh,Barrett:2006qq,Connes:2006qv,Martins:2006rp} and their connection with
string theory has been examined~\cite{Lizzi:2006te}. These noncommutative models represent
interesting generalisations of the SM and hint at possible new physics. However {they do not
address the usual problem of the SM}, the presence of a plethora of free parameters mostly related
to the ad hoc introduction of the Higgs and Yukawa sectors in the theory. In the dimensional
reduction scheme which I describe here these two sectors emerge automatically in four
dimensions; Higgs particles are the extra dimensional components of the gauge fields defined over
the full higher dimensional theory. The Yukawa terms are obtained by the fermions - gauge fields
coupling terms. Application of CSDR over fuzzy cosets is tempting for further
investigation but the construction of a realistic particle physics model is yet to be explored.

\section{CSDR over fuzzy coset spaces}

The basic idea of the CSDR scheme was described in
chapter~\ref{CSDR}. The solution of the constraints, imposed on the four-dimensional surviving
fields was also given.

Here, I consider a higher dimensional gauge theory defined on a compactified space
$M^{4} \times (S/R)_{F}$, where $(S/R)_{F}$ is the approximation of  $S/R$ by finite $N\times N$
matrices and have noncommutative characteristics. I denote the local coordinates of the considered
extra dimensional space by $x^{\ssM}=(x^{\mu},y^{a})$ to resemble the coordinate parametrisation of
the compactification over ordinary cosets; here the $y^{a}$ are proportional with some $X_{a}$,
$N\times N$ antihermitean matrices. In section~\ref{FuzzySphere}, I recalled the case of fuzzy
sphere as an example of a such noncommutative `manifold'. In a more general set-up one consider
a noncommutative gauge theory with gauge group $G=U(P)$ over $M^{4} \times (S/R)_{F}$. The
implementation of the CSDR scheme in the fuzzy case (Fuzzy-CSDR)- following~\cite{Aschieri:2003vy} -
can be described in three steps:
\begin{enumerate}
\item
State the CSDR principle on fuzzy cosets and reduce it to a set of constraints - the CSDR
constraints (\ref{3.17}), (\ref{3.19}), (\ref{eq7}), (\ref{eq16}), (\ref{eq16bis}) - that the gauge
and matter fields must satisfy.
\item
Reinterpret a Yang-Mills-Dirac action on $M^4\times (S/R)_F$ with $G=U(P)$ gauge group as
actions on $M^4$ with $U(N P)$ gauge group. This is possible by expanding the fields on
 $M^4\times (S/R)_F$ in Kaluza-Klein modes on  $(S/R)_F$. The algebra of functions on $(S/R)_F$
is finite dimensional and one obtain a finite tower of modes; the $(S/R)_F$ is described by
$N\times N$ matrices and a basis for this mode expansion is given by the generators of Lie
algebra $\mmu(N)$. It has been proven that the different modes can be conveniently grouped
together so that an initial $\mg$-valued field on $M^4\times (S/R)_F$ (with $G=U(P)$) is
reinterpreted as a $\mmu(N P)$ valued field on  $M^4$.
\item
Solve the CSDR constraints and obtain the  gauge group and the particle content
of the reduced four-dimensional actions. I present the example of dimensional reduction over
the two-dimensional fuzzy sphere and describe its generalisation over fuzzy cosets of more than $2$
dimensions.
\end{enumerate}

\section{The CSDR principle}\label{CSDRprinciple}

Since the Lie algebra of $S$ acts on the fuzzy space $(S/R)_{F}$, one can state the CSDR principle
in the same way as in the continuum case, i.e. the fields in the theory must be invariant under the
infinitesimal $S$ action up to an infinitesimal gauge transformation
\begin{equation}
{\cal L}_{b} \phi =\delta^{W_b}\phi= W_{b} \phi\qquad
{\cal L}_{b}A = \delta^{W_b}A=-DW_{b}\,,
\label{csdr}
\end{equation}
where $A$ is the one-form gauge potential
$A = A_{\mu}(x^{\mu},y^{a})dx^{\mu} + A_{a}(x^{\mu},y^{a}) \theta^{a}$, and $W_b$ depends only on
the coset coordinates $y^{a}\sim X_{a}$ and (like $A_\mu, A_a$) is antihermitean. I thus write
$W_b=W_b^{(\alpha)}{\cal T}^{\alpha},\,\alpha=1,2\ldots P^2,$ where ${\cal  T}^\alpha$ are hermitean
generators of $U(P)$ and $(W_b^{(\alpha)})^\dagger=-W_b^{(\alpha)}$, here ${}^\dagger$ is hermitean
conjugation on the $X_{a}$'s. The principle gives for the space-time part $A_{\mu}$ [c.f.~\eq{LDA}]
\begin{equation}
{\cal L}_{b}A_{\mu} = [X_{b} , A_{\mu}] = -[A_{\mu}, W_{b} ]\,,
\label{spacetime}
\end{equation}
while for the internal part $A_{a}$ [c.f.~\eq{fund}]
\begin{equation}
[X_{b}, A_{a}] - A_{c}C^{c}{}_{ba} = - [A_{a}, W_{b}] - {\cal L}_{a}W_{b}\,.
\label{internal}
\end{equation}
Taking in account the cyclicity condition of the Lie derivatives
$[{\cL}_a,{\cL}_b]=C^{c}{}_{ab}{\cL}_{c}$, and that from the first of eqs.~(\ref{csdr}) I have 
${\cal L}_{a} {\cal L}_{b} \phi = ({\cal L}_{a}W_{b}) \phi +W_{b}W_{a}\phi$ which lead to
the consistency condition
\begin{equation}
[X_{a}, W_{b}]-[X_{b}, W_{a}] - [W_{a}, W_{b}] = C^{c}{}_{ab}W_{c}\,.
\label{2.59}
\end{equation}
Under the gauge transformation $\phi\to \phi^{(g)}=g\,\phi$ with $g \in G= U(P)$, I have
${\cL}_a\phi^{(g)}=W^{(g)}_a\phi^{(g)}$ and also
${\cL}_a\phi^{(g)}= ({\cL}_{a}g)\,\phi + g\,({\cL}_{a}\phi)$, and therefore
\begin{equation}
W_\a\to W^{(g)}_{a} = g\,W_{a}\,g^{-1} + [X_{a}, g]\,g^{-1}\,.
\label{2.60}
\end{equation}
Now in order to solve the constraints (\ref{spacetime}), (\ref{internal}), (\ref{2.59}) I cannot
follow the strategy adopted in the commutative case where the constraints were studied just at one
point of the coset (say $y^a=0$). This is due to the intrinsic nonlocality of the constraints. On
the other hand the specific properties of the fuzzy case (e.g. the fact that partial derivatives are
realised via commutators, the concept of covariant derivative) allow to simplify and eventually
solve the constraints. Defining
\begin{equation}
\omega_{a} \equiv X_{a} - W_{a}\,,
\end{equation}
one obtain the following form of the consistency condition
(\ref{2.59})
\begin{equation}\label{3.17}
[ \omega_{a} , \omega_{b}] = C^{c}{}_{ab}\,\omega_{c}\,,
\end{equation}
where $\omega_{a}$ transforms as
\begin{equation}
\omega_\a\to \omega^{(g)}_{a} = g\,\omega_{a}\,g^{-1}\,.
\end{equation}
Now eq.~(\ref{spacetime})  reads
\begin{equation}\label{3.19}
[\omega_{b}, A_{\mu}] =0\,.
\end{equation}
Furthermore by considering the covariant coordinate,
\begin{equation}\label{eq17}
\vf_{d} \equiv X_{d} + A_{d}
\end{equation}
one has
\begin{equation}
\vf\to\vf^{(g)}=g\,\vf\,g^{-1} \label{eq18}
\end{equation}
and eq.~(\ref{internal}) simplifies to
\begin{equation}\label{eq7}
[\omega_{b},\vf_{a}]=\vf_{c}C^{c}{}_{ba}\,.
\end{equation}
Therefore eqs.~(\ref{3.17}),~(\ref{3.19}) and~(\ref{eq7}) are the constraints to be solved. Note
that eqs.~(\ref{eq18}) and (\ref{eq7}) have the symmetry
\begin{equation}\label{groundstate} 
\vf_{a} \to \vf_{a} + \omega_{a} \,,
\end{equation}
suggesting that $\omega_{a}$ is a ground state and $\vf_{a}$ the fluctuations around it.
Indeed, the semi-positive definite potential~\eq{pot}, vanishes for the
value $\vf_{a}^{(vac)}=\omega_{a}$.

One proceeds in a similar way for the spinor fields. The CSDR principle relates the Lie derivative
on a spinor $\psi$, which is considered here to transform in the adjoint representation of $G$,
to a gauge transformation; recalling eqs. (\ref{definitions}) and (\ref{spinlie11}) one has
\begin{equation}\label{eq15}
[X_{a}, \psi] + \frac{1}{2}C_{abc}\Gamma^{bc}\psi = [W_{a},\psi]\,,
\end{equation}
where $\psi$ denotes the column vector with entries $\psi_\alpha$. Setting again $\omega_{a} = X_{a}
- W_{a}$ lead to the constraint
\begin{equation}\label{eq16}
-\frac{1}{2}C_{abc} \Gamma^{bc}\psi = [\omega_{a},\psi]\,.
\end{equation}
Having considered spinors which transform in the fundamental rep. of  the gauge group $G$, one
has $[X_{a}, \psi] + \frac{1}{2}C_{abc}\Gamma^{bc}\psi = W_{a}\psi,$. Setting again 
$\omega_{a} = X_{a} - W_{a}$, lead to the constraint
\begin{equation}\label{eq16bis}
-\frac{1}{2}C_{abc} \Gamma^{bc}\psi = \omega_{a}\psi-\psi X_{a}\,.
\end{equation}

\section{Actions and Kaluza-Klein modes }

Here, I consider a pure Yang-Mills action on $M^{4} \times (S/R)_{F}$
{and recall how it is reinterpreted in four dimensions}. The action is
\begin{equation}
{\cal A}_{YM}=\frac{1}{4} \int d^{4}x\, \Tr\, tr_{G}\, F_{MN}F^{MN}\,,
\end{equation}
where $\Tr$ is the usual trace over $N\times N$ matrices and is actually the integral over the
fuzzy coset $(S/R)_F\, $\footnote{{$\,\Tr$ is a good integral because it has the cyclic
property $\Tr(f_1\ldots f_{p-1}f_p)=\Tr(f_pf_1\ldots f_{p-1})$. It is also invariant under the
action of the group $S$, that I recall to be infinitesimally given by 
${\cal L}_{a} f = [{X}_{a}, f].$}}, while $tr_G$ is the gauge group $G$ trace. The
higher dimensional field strength $F_{MN}$ decomposed in four-dimensional space-time and
extra dimensional components reads as follows $(F_{\mu \nu}, F_{\mu b}, F_{ab} )$; explicitly the
various components of the field strength are given by
\begin{equation}
\begin{array}{l@{}l@{}l@{}}
F_{\mu \nu} &=& \partial_{\mu}A_{\nu} - \partial_{\nu}A_{\mu} + [A_{\mu}, A_{\nu}]\,,\\[.3 em]
F_{\mu a}   &=& \partial_{\mu}A_{a} - [X_{a}, A_{\mu}] + [A_{\mu}, A_{a}]
                   =\partial_{\mu}\vf_{a} + [A_{\mu}, \vf_{a}] = D_{\mu}\vf_{a}\,,\\[.3 em]
F_{ab}        &=& [\vf_{a}, \vf_{b}] - C^{c}{}_{ab} \vf_{c}\,;
\end{array}
\end{equation}
they are covariant under local $G$ transformations: $F_{MN}\to g\,F_{MN}\,g^{-1}$, with
$g=g(x^\mu,X^{a})$.

In terms of the suggested decomposition the action reads
\begin{equation}
{\cal A}_{YM}= \int d^{4}x\, \Tr\, tr_{G}\,\left( \frac{1}{4}\,F_{\mu \nu}^2
+ \frac{1}{2}(D_{\mu}\vf_{a})^2\right) - V(\vf)\,,
\label{theYMaction}
\end{equation}
where the potential term $V(\vf)$ is the $F_{ab}$ kinetic term (recall $F_{ab}$ is antihermitean so
that $V(\vf)$ is hermitean and non-negative)
\begin{align}\label{pot}
V(\phi)&=-{1\over 4} \Tr\,tr_G \sum_{ab} F_{ab} F_{ab} \nonumber \\
          &= -{1\over 4} \Tr\,tr_G \sum_{ab} \left( [\vf_{a}, \vf_{b}]
               - C^{c}{}_{ab} \vf_{c}\right)\left([\vf_{a}, \vf_{b}]
              - C^{c}{}_{ab} \vf_{c}\right)\,.
\end{align}

The action (\ref{theYMaction}) {is naturally interpreted as an action in four dimensions.}
The infinitesimal $G$ gauge transformation with gauge parameter $\lambda(x^\mu,X^\a)$ can indeed be
{interpreted just as an $M^4$ gauge transformation}. I write
\begin{equation}
\lambda(x^\mu,X^{a})=\lambda^\alpha(x^\mu,X^{a}){\cal T}^\alpha
=\lambda^{\alpha, h}(x^\mu)T^h{\cal T}^\alpha\,,\label{3.33}
\end{equation}
where ${\cal T}^\alpha$ are hermitean generators of $U(P)$, $\lambda^\alpha(x^\mu,X^{a})$ are
$N\times N$ antihermitean matrices;
these can be expanded in finite symmetric multipole expansion over $X$'s [c.f. section
\ref{FuzzySphere}] and are expressible as $\lambda(x^\mu)^{\alpha ,h}T^h$, where $T^h$ are
antihermitean generators of $U(N)$. The fields $\lambda(x^\mu)^{\alpha ,h}$, with $h=1,\ldots N^2$,
are the Kaluza-Klein modes (KK-modes) of $\lambda(x^\mu, X^{a})^{\alpha}$. Then one considers on
equal footing the indices $h$ and $\alpha$ and interprets the fields on the r.h.s. of (\ref{3.33})
as one field valued in the tensor product Lie algebra $\mmu(N) \otimes \mmu(P)$. This Lie algebra is
indeed $\mmu(NP)$\,\footnote{Proof: The $(NP)^2$ generators $T^h{\cal T}^\alpha$ are $NP\times NP$
antihermitean matrices. Then one just have to show that they are linearly independent. This is easy
since it is equivalent to prove the linear independence of the $(NP)^2$ matrices
$e_{ij}\varepsilon_{\rho\sigma}$ where $i=1,\ldots n$, $\rho=1,\ldots P$ and $e_{ij}$ is the
$N\times N$ matrix having $1$ in the position $(i,j)$ and zero elsewhere, and similarly for the
$P\times P$ matrix $\varepsilon_{\rho\sigma}$.}. Similarly, I can rewrite the gauge field $A_\nu$ as
\begin{equation}
A_\nu(x^\mu,X^{a})=A_\nu^\alpha(x^\mu,X^{a}){\cal T}^\alpha
 =A_\nu^{\alpha,h}(x^\mu)T^h{\cal T}^\alpha\,,
\end{equation}
and interpret it as a $\mmu(NP)$ valued gauge field on $M^4$, and similarly for $\vf_{a}$. Finally
$\Tr\, tr_G$ is the trace over $U(NP)$ matrices in the fundamental representation.

The above analysis applies also to more general actions, and to the field $\omega_{a}$ and therefore
to the CSDR constraints~(\ref{3.17}),~(\ref{3.19}),~(\ref{eq7}),~(\ref{eq16}),~(\ref{eq16bis})
that can now be reinterpreted as constraints on $M^4$ instead of on 
$M^4\times (S/R)_F$. The action (\ref{theYMaction}) and the minima of the potential (\ref{pot}), in
the case $P=1$, have been studied, without CSDR constraints, in refs.
\cite{Dubois-Violette:1989at,Dubois-Violette:1988ps,Dubois-Violette:1989vq,Madore:1992dk,
Madore:1992ej}.

\section{CSDR constraints for the fuzzy sphere}\label{CSDRonFuzzySphere}

Here, I present the solution of the aforementioned CSDR constraints for the case of two-dimensional
fuzzy sphere and extend the results to more general fuzzy cosets. I consider  $(S/R)_{F}=S^2_{N}$,
i.e. the two-dimensional fuzzy sphere approximated by $N \times N$ matrices (fuzziness level $N-1$).
I first examine the simpler case where the gauge group $G$ is just $U(1)$ and I make some comments
on the $G=U(P)$ generalisation afterwards.

\bigskip
\noindent
{\large{\it The $G=U(1)$ case}}

In this case the $\omega_{a}(X_{b})$ that appear in the consistency condition~(\ref{3.17}) are
$N\times N$ antihermitean matrices, i.e. I can interpret them as elements of the $\mmu(N)$ Lie
algebra. On the other hand eqs.~(\ref{3.17}) are the commutation relations of the $\msu(2)$ Lie
algebra. Indeed, according to section~\ref{FuzzySphere}, if $r$ is the radius of the fuzzy
sphere, the structure constants entering in the CSDR constraints of section~\ref{CSDRprinciple} are
given by $C^{a}{}_{bc}=r^{-1}\vare^{a}{}_{bc}$. Then $r\omega_{a}$ are the $\msu(2)$ Lie generators
in a reducible or in the $N$-\dim~irreducible rep. and define an $\msu(2)$ image into the
$\mmu(N)$ Lie algebra.

The four-dimensional gauge symmetry is determined by solving the constraint~\eq{3.19}.
We consider the expansion of the $A_{\mu}(x,X)$ into the Kaluza-Klein modes of the $S_{N}^{2}$.
Recalling from the previous section that in the simplest case of $G=U(1)$ gauge group,  $A_{\mu}$ is
reinterpreted as a four-dimensional $\mmu(N)$-valued field, an embedding
of $\msu(2)\hookrightarrow\mmu(N)$ is required. One possible embedding is the
following. Let $T^h$ with $h = 1, \ldots ,N^{2}$ be the generators of $\mmu(N)$ in the fundamental
representation and with normalisation $Tr(T^hT^k)=-{1\over 2}\delta^{hk}$. These appear in the
expansion $A_{\mu}(x,X)=A_{\mu}^{h}(x)T^{h}$. We can always use the
convention $h= (a,u)$ with $a = 1,2,3$ and $u= 4,5,\ldots, N^{2}$ where the $T^{a}$ satisfy the
$\msu(2)$ Lie algebra
\begin{equation}
[T^{a}, T^{b}] = r  C^{ab}{}_{c}T^{c}\,.
\end{equation}
Then I define an embedding by identifying
\begin{equation}
 r\omega_{a}= T_{a}\,,
\label{embedding}
\end{equation}
i.e. a regular $\msu(2)$ subalgebra of $\mmu(N)$.
Constraint (\ref{3.19}),~$[\omega_{b} , A_{\mu}] = 0$, then implies that the four-dimensional gauge
group $K$ is the centraliser of the image of $SU(2)$ in $U(N)$, i.e.
\begin{equation}
 K=C_{U(N)}(SU(2)) = SU(N-2) \times U^{I}(1)\times U^{II}(1)\,,
\label{4Dim-GaugeGroup-SpecEmbed}
\end{equation}
where $U(N)\simeq SU(N)\times U^{II}(1)$. The functions $A_{\mu}(x)$ are arbitrary functions of $x$
and take values in Lie $\mk$ subalgebra of $\mmu(N)$. 

Concerning constraint (\ref{eq7}), $[\omega_{b} , \vf_{a}] = C^{c}{}_{ba} \vf_{c}$, it is satisfied by choosing
\begin{equation} \label{soleasy}
 \vf_{a}=\vf(x) r \omega_{a}\,
\end{equation}
i.e. the unconstrained degrees of freedom correspond to the scalar field $\vf(x)$ that is a singlet
under the four-dimensional gauge group $K$.

The physical spinor fields (transforming in the adjoint \rep) are obtained by solving the constraint
(\ref{eq16}), $-\frac{1}{2}C_{abc} \Gamma^{bc}\psi = [\omega_{a},\psi]$. In the l.h.s. of this
formula one can say that she (he) has an embedding of $\msu(2)$ in the spin representation of
$\mso(3)$. This embedding is given by the matrices $\tau^{a}= \frac{1}{2}C^{a}{}_{bc} \Gamma^{bc}$;
since $\msu(2)\sim \mso(3)$ this embedding is rather trivial and indeed $\tau^{a}={-i\over 2
r}\sigma^{a}$. Thus the constraint (\ref{eq16}) states
that the spinor $\psi=\psi^{h} T^{h}=\left(\begin{smallmatrix}
							\psi_{1}\\
							\psi_{2}
						     \end{smallmatrix}\right)$
where $T^h\in \mmu(N)$ and $\psi_{1(2)}=\psi_{1(2)}^hT^h$ are four-dimensional spinors, relate
(intertwine) the fundamental rep. of $SU(2)$ to the representations of $SU(2)$  induced by the
embedding (\ref{embedding}) of $SU(2)$ into $U(N)$, i.e. of $SU(2)$ into $SU(N)$.
In formulae
\begin{equation}
\begin{array}{r@{}c@{}l@{}}
SU(N) &\supset& SU(2) \times SU(N-2) \times U(1) \nonumber \\
\mbf{N^{2}-1} &=&(\mbf{1},\mbf{1})_{(0)} \oplus (\mbf{3},\mbf{1})_{(0)}
                                                         \oplus(\mbf{1},\mbf{(N-2)^{2}})_{(0)}\\
          &\oplus& (\mbf{2},\mbf{(N-2)})_{(-N)} \oplus (\mbf{2},\overline{\mbf{(N-2)}})_{(N)}\,.
\end{array}
\end{equation}
Therefore, the fermions that satisfy constraint (\ref{eq16}) transform as
$\mbf{(N-2)}_{(-N,0)}$ and $\overline{\mbf{(N-2)}}_{(N,0)}$ under 
$K= SU(N-2) \times U^{I}(1)\times U^{II}(1)$. In the case of the fuzzy sphere the embedding 
$\msu(2) \hookrightarrow \mso(3)$ is somehow trivial. If I had chosen instead the fuzzy 
$(SU(3)/U(1)\times U(1))_F$, then $\msu(3)$ should be embedded in $\mso(8)$.

In order to write the action for fermions I have to consider the Dirac operator ${\cal D}$ on
$M^4\times S^2_{N}$. This operator can be constructed following the derivation presented in ref.
\cite{Grosse:1995jt} for the Dirac operator on the fuzzy sphere, see also ref. \cite{Ydri:2001pv}.
For fermions in the adjoint I obtain
\begin{equation}
 {\cal D}\psi=i\Gamma^\mu(\partial_\mu+A_\mu)\psi 
                 +i\sigma^{a} [X_{a}+A_{a},\psi] - {1\over r}\psi\,,
\end{equation}
where $\Gamma^\mu$ is defined in (\ref{Gamma7}), and with slight abuse of notation I have written
$\sigma^{a}$ instead of $\sigma^{a}\otimes 1$. Using eq.~(\ref{eq17}) the fermion action,
\begin{equation}\label{fermionact}
{\cal A}_{F} =\int d^{4}x\, \Tr \, \bar{\psi}{\cal D}\psi
\end{equation}
becomes
\begin{equation}
{\cal A}_{F} = \int\! d^{4}x\,\,\, \Tr \, \bar{\psi}\left(i\Gamma^\mu(\partial_\mu+A_\mu) -{1\over r}\right)\psi\,
                  +i\,\Tr\,\bar{\psi}\sigma^{a} [\phi_{a}  , \psi] \,,\label{341}
\end{equation}
where I recognise the fermion masses $1/r$ and the Yukawa interactions.

Using  eqs.~(\ref{soleasy}), (\ref{eq16}) the Yang-Mills action (\ref{theYMaction}) plus the fermion
action
reads
\begin{eqnarray}
{\cal A}_{YM}+{\cal A}_{F}
&=&\int\! d^{4}x\,\,\, {1\over 4}\Tr(F_{\mu \nu}F^{\mu\nu})-{3\over 4}D_{\mu}\vf D^\mu\vf
  - {3\over 8}(\vf^2-r^{-1}\vf)^2\,\nonumber\\
& &+\int\! d^{4}x\,\,\, \Tr \, \bar{\psi}\left(i\Gamma^\mu(\partial_\mu+A_\mu) -{1\over r}\right)\psi\,
-\, \frac{3}{2}\,\Tr \, \bar{\psi}\vf\psi\,.\label{DimReducedYMaction}
\end{eqnarray}

The choice (\ref{embedding}) defines one of the possible embeddings 
of $\msu(2)\hookrightarrow\mmu(N)$ [$\msu(2)$ is embedded in $\mmu(N)$ as a regular subalgebra],
while on the other extreme one can embed $\msu(2)$ in $\mmu(N)$ using the irreducible $N$
dimensional rep. of $SU(2)$
\begin{equation}
T_{a}=r\omega_{a}=X_{a}^{(N)}\,.\label{NDim-irrep}
\end{equation}
Constraint (\ref{3.19}) in this case implies that the four-dimensional gauge group is $U(1)$ so
that $A_\mu(x)$ is $U(1)$ valued. Constraint  (\ref{eq7}) leads again to the scalar singlet
$\vf(x)$. A different $\msu(2)\hookrightarrow\mmu(N)$ embedding can be described by the choice
\begin{equation}
T_{a}=r\omega_{a}=X_{a}^{(N')}\otimes\one_{n}\,\qquad N=N'n\,.\label{NDim-rrep}
\end{equation}
and the surviving four-dimensional gauge group is found to be $U(n)\times U(1)$ as it can be proven
by the explicit calculation of  the massless modes of the $A_{\mu}$ Kaluza-Klein expansion. In the
next section, I give details of the calculation. Note that the extra $U(1)$ comes from the
$U(N)\simeq SU(N)\times U(1)$ as the example I studied before. Obviously, the
$\msu(2)\hookrightarrow\mmu(N)$ embedding of eq.~(\ref{NDim-irrep}) is a special case of the one
described by eq.~(\ref{NDim-rrep}).

Summarising, the surviving four dimensional spinors are given by the~(\ref{eq16}) constraint.
Constraint (\ref{eq7}) gives the surviving four-dimensional scalars $\vf_{a}=\vf(x)\,r\omega_{a}$,
which is in fact the $N$-\dim~irreducible or a reducible rep. of $\msu(2)$. Note that the
semi-positive definite potential is always minimised by this value, a result which is independent of
the chosen $SU(2)\hookrightarrow U(N)$ embedding.

One can also consider the case of a Yang-Mills-Dirac actions with fermions transforming in the
fundamental of the gauge group $G$. Details of the relevant calculation can be found in
\cite{Aschieri:2003vy}.

\bigskip
\noindent
{\large {\it The $G=U(P)$ case}}

In this case $A_{\mu}(x,X)=A_{\mu}^{\a,h}(x)T^{h}{\cal T}^{\a}$ is an $NP\times NP$ antihermitean
matrix and in order to solve the constraint~\eq{3.17} one has to embed 
$\msu(2)\hookrightarrow\mmu(NP)$. All the results of the $G=U(1)$ case holds also here, I just have
to replace $N$ with $NP$. This is true for the fermion sector too, provided that in the higher
dimensional theory the fermions are considered in the adjoint of $U(P)$ (then, in the action
(\ref{fermionact}) I  need to replace $\Tr$ with $\Tr_{\,} tr_{U(P)}$ i.e. $tr_{U(NP)}$). The
case of Yang-Mills-Dirac action with fermion transforming in the fundamental of the gauge group
$G=U(P)$ has been also examined in~\cite{Aschieri:2003vy}.

\section{Kaluza-Klein modes on $S^2_{N}$ and symmetry breaking}\label{sec:KK-FuzzyCSDR}

To see the above arguments in more detail let me determine the spectrum and the representation
content of the gauge field $A_\mu$ in the simplest case for a gauge group $G=U(1)$. The obtained
conclusions are easily generalised also for the case of the gauge group $G=U(P)$. Assuming
the embedding $\msu(2)\hookrightarrow\mmu(N)$,~\eq{NDim-irrep}, I
show that the $U(N)$ gauge symmetry, which the full action functional has after its KK expansion,
breaks spontaneously to $K=U(1)$ in four dimensions [cf \eq{3.19}]. I expand the 
$M^{4}\times S^{2}_{N}$ fields over the KK modes of the extra dimensional fuzzy sphere. It turns out
that the CSDR surviving fields are no other than the massless KK modes.

Since the $X_a$ are considered to be the generators of the fuzzy sphere $S^2_{N}$, I can decompose
the full extra-dimensional $\mmu(N)$-valued gauge fields $A_\mu$ into spherical harmonics
$Y^{lm}(X)$ on the fuzzy sphere $S^2_{N}$ with coordinates $X_a$:
\begin{equation}
A_\mu = A_\mu(x,X) = \sum_{\genfrac{}{}{0pt}{}{0 \leq l\leq N}{|m|\leq l}}
\, A_{\mu,lm}(x)\otimes Y^{lm(N)}(X)\,;
\label{KK-modes-A}
\end{equation}
$Y^{lm(N)}$ are by definition irreps under the $SU(2)$ rotations on $S^2_{N}$, and form a basis of
hermitean $N \times N$ matrices. The $A_{\mu,lm}(x)$ are taken here to be $\mmu(1)$-valued gauge and
vector fields on $M^4$. Using this expansion, I can interpret $A_\mu(x,X)$ as $\mmu(N)$-valued
functions on $M^4\times S^2_{N}$, expanded into the Kaluza-Klein modes (i.e. harmonics) of
$S^2_{N}$.

The scalar fields $\vf_a$,
\begin{equation}
\vf_a (x,X) =\frac{1}{r}\,X_a^{(N)} + A_a(x,X)\,,
\label{covariant-coord-full-embedding}
\end{equation}
with potential (\ref{pot}) are considered as  `covariant coordinates' on $S^2_{N}$
(section \ref{CSDRprinciple}) and take the value $\omega_{a}=(1/r)X_{a}^{(N)}$ in vacuum.
On the other hand, the fluctuations $A_a$ of these covariant coordinates should be interpreted as
$\mmu(N)$ gauge fields on the fuzzy sphere, (see appendix~\ref{sec:fuzzygaugetheory}). Therefore
they can be expanded similarly as
\begin{equation}
A_a = A_a(x,X) = \sum_{\genfrac{}{}{0pt}{}{0 \leq l\leq N}{|m|\leq l}}
\, A_{a,lm}(x)\otimes Y^{lm(N)}(X)
\label{KK-modes-phi}
\end{equation}
and interpreted as functions (or one-form) on $M^4\times S^2_{N}$ taking values in $\mmu(N)$. One
can then interpret $A_M(x,X) = (A_\mu(x,X), A_a(x,X))$ as $\mmu(N)$-valued gauge or vector fields on
$M^4\times S^2_{N}$.

Larger gauge groups are possible to be retrieved in four dimensions by considering different
$\msu(2)$ into $\mmu(N)$  embeddings. Indeed, having assumed the
$\msu(2)\hookrightarrow\mmu(N)$ embedding,~\eq{NDim-rrep}, the `covariant
coordinate' of the two-dimensional fuzzy sphere would be
\begin{equation}
\vf_{a}(x,X)=\frac{1}{r}\,(X_{a}^{(N')}\otimes\one_{n})+A_{a}(x,X)\,.
\label{covariant-coord-general}
\end{equation}
The consistency condition~\eq{3.17} and the CSDR constraints~\eq{3.19},~\eq{eq7}
and~\eq{eq16} remain unmodified. The four and extra dimensional components of the $A_{M}(x,X)$
gauge field are expanded as
\begin{align}
&A_\mu = A_\mu(x,X) = \sum_{\genfrac{}{}{0pt}{}{0 \leq l\leq N}{|m|\leq l}}
                                 A_{\mu,lm}^{(n)}(x)\otimes Y^{lm(N')}(X)\,
						\label{KK-modes-A-NDimIrrep}\\
&A_a = A_a(x,X) =\sum_{\genfrac{}{}{0pt}{}{0 \leq l\leq N}{|m|\leq l}}
				A_{a,lm}^{(n)}(x)\otimes Y^{lm(N')}(X)\,,
						\label{KK-modes-phi-NDimIrrep}
\end{align}
where $Y^{lm(N')}$ are $N'$-\dim~irreps, under the $SU(2)$ rotations on $S^2_{N}$. The
$A_{\mu,lm}^{(n)}(x)$ and $A_{a,lm}^{(n)}(x)$ are taken here to be $\mmu(n)$-valued gauge and vector
fields on $M^4$. As before then, the $A_M(x,X) = (A_\mu(x,X), A_a(x,X))$ is interpreted as
$\mmu(N)$-valued ($N=N'n$) gauge or vector fields on $M^4\times S^2_{N}$.

Given this expansion into KK modes, I will show that only $A_{\mu,00}(x)$ (i.e. the dimensionally
reduced gauge field) becomes a massless $\mmu(1)$ or $\mmu(n)$-valued gauge field in four
dimensions, depending on the `covariant coordinate' configuration I consider
[\eq{covariant-coord-full-embedding} or \eq{covariant-coord-general}]. All other modes
$A_{\mu,lm}(X)$ with $l \geq 1$ constitute a tower of Kaluza-Klein modes with large mass gap, and
decouple for low energies.

The scalar fields $A_a(x,X)$ will be analysed in a similar way below, and provide no additional
massless degrees of freedom in four dimensions. The surviving fields in four dimensions, found by
the previously considered dimensional reduction of the ten-dimensional~Yang-Mills-Dirac theory, are
exactly the massless modes of their KK expansion. Remarkably, the model I describe is fully
renormalizable in spite of its higher-dimensional character, in contrast to the commutative case;
see also~\cite{Aschieri:2005wm}.

\bigskip
\noindent
{\large{ \it Computation of the KK masses}}

To justify these claims, let me compute the masses of the KK modes \eq{KK-modes-A}. They are induced
by the covariant derivatives 
$\int \Tr\, (D_{\mu}\vf_{{a}})^{2} $ in \eq{theYMaction},
\begin{equation}
\int \Tr\, (D_{\mu}\vf_{{a}})^\dagger D_{\mu}\vf_{{a}} =
\int \Tr\, (\partial_{\mu}\vf_{{a}}^\dagger \partial_{\mu}\vf_{{a}} + 2
(\partial_{\mu}\vf_{{a}}^\dagger) [A_\mu,\vf_{a}] +
[A_\mu,\vf_{a}]^\dagger[A_\mu,\vf_{a}])\,.
\label{covar-mass-term}
\end{equation}
The most general scalar field configuration is given by~\eq{covariant-coord-general}.
As usual, the last term in \eq{covar-mass-term} leads to the mass terms for the gauge fields $A_\mu$
in the vacuum
\begin{equation}
\vf_a^{(vac)} =\omega_{a}=\frac{1}{r}\left(X_a^{(N')} \otimes \one_n\right)\,,
\label{phi-vac}
\end{equation}
provided the mixed term which is linear in $A_\mu$ vanishes in a suitable gauge. This is usually
achieved by going to the unitary gauge. In the present case this is complicated by the fact that I
have three scalars in the adjoint, and there is no obvious definition of the unitary gauge; in fact,
there are are too many scalar degrees of freedom as to gauge away that term completely. However, I
can choose a gauge where all quadratic contributions of that term vanish, leaving only cubic
interaction terms. To see this, I insert \eq{covariant-coord-general} into the term
$(\partial_{\mu}\vf_{{a}}^\dagger) [A_\mu,\vf_{a}]$ in \eq{covar-mass-term}, which gives
$$
\int \Tr\, A_\mu [\vf_a,\partial_{\mu}\vf_{{a}}^\dagger] = -
\int \Tr\, A_\mu \Big\{[\Xwhat_a,\partial_{\mu} A_a(x,X)] 
              + [A_a(x,X),\partial_{\mu}A_a(x,X)]\Big\}\,,
$$
where $\Xwhat_{a}=X_{a}^{(N')}\otimes\one_{n}$.
Now I partially fix the gauge by imposing the `internal' Lorentz gauge $[\Xwhat_a, A_a] =0$ at each
point $X$. This is always possible\footnote{Even though this gauge is commonly used in the
literature on the fuzzy sphere, a proof of existence has apparently not been given.  It can be
proved by extremizing the real function $\Tr\,(X_a \vf_a)$ on a given gauge orbit, which is compact;
the e.o.m. then implies $[X_a, \vf_a] =0$.}, and the above simplifies as
\begin{equation}
\int \Tr\, A_\mu [\vf_a,\partial_{\mu}\vf_{{a}}^\dagger] = 
\int \Tr\, A_\mu [A_a(x,y),\partial_{\mu}A_a(x,y)] =: S_{int}\,.
\label{A-lin-term-2}
\end{equation}
This contains only cubic interaction terms, which are irrelevant for the computation of the masses.
I can therefore proceed by setting the vacuum of the model $\vf_{a}^{(vac)}$
[relation~\eq{phi-vac}] and inserting the expansion \eq{KK-modes-A-NDimIrrep} of $A_\mu$ into the
last term of \eq{covar-mass-term}. Recall that $\J_{a}=iX_{a}$ are the angular momentum generators,
satisfying $[\J_{a},\J_{b}]=i\vare^{c}{}_{ab}\J_{c}$. Then since
$$
i[\Xwhat_{a},A_\mu] = \J_a A_\mu=\sum_{l,m}\left(A_{\mu,lm}^{(n)}(x)\otimes\J_{a} Y^{lm(N')}\right)
$$
is
simply the action of $SU(2)$ on the fuzzy sphere, it follows that $\Tr\, [X_{a},A_\mu]
[X_{a},A_\mu]$ is the quadratic Casimir on the modes of $A_\mu$ which are orthogonal, and I obtain
\begin{equation}
 \int \Tr\, (D_{\mu}\vf_{a})^\dagger D_{\mu}\vf_{a}
=\int\Tr\Big(\partial_{\mu}\vf_{a}^\dagger \partial_{\mu}\vf_{a}
+ \frac{1}{r^{2}}\sum_{l,m}l(l+1)\,A_{\mu,lm}^{(n)}(x)^\dagger A_{\mu,lm}^{(n)}(x) \Big)
+ S_{int}\,.
\label{covar-mass-term-2}
\end{equation}
Therefore the four-dimensional $\mmu(n)$ gauge fields $A_{\mu,lm}^{(n)}(x)$ acquire a mass 
\begin{equation}
 m^2_{l} = \frac{1}{r^2}\,\, l(l+1)\,,
\label{KK-masses-Adiag}
\end{equation}
with $r$ being the radius of the fuzzy sphere and therefore of compactification energy scale,
as it is expected for higher KK modes. In particular, only 
$A_{\mu}^{(n)}(x)\equiv A_{\mu,00}^{(n)}(x)$ survives as a massless four-dimensional $\mmu(n)$
gauge field. The low-energy effective action (LEA) for the gauge sector is then given by
\begin{equation}
 S_{LEA}^{(gauge)} = 
\int\,d^4 x\,\frac 1{4g^{2}}\, tr_{K}\,F_{\mu\nu}^\dagger F_{\mu\nu}\,,
\label{LEA-action-FF}
\end{equation}
where $F_{\mu \nu}$ is the field strength of the low-energy $A_{\mu}(x)$gauge fields. These take
values in the Lie algebra generating the $K=SU(n)\times U^{I}(1)\times U^{II}(1)$ gauge group which
describe the symmetry of the theory in four dimensions. The abelian group factors comes from
$U(N)\simeq SU(N)\times U^{II}(1)$ and $U(n)\simeq SU(n)\times U^{I}(1)$. Obviously by setting
$n=1$ in the partition of $N=N'n$, I obtain the same result but with the surviving four-dimensional
gauge group to be $K=U(1)$.

\bigskip
\noindent
{\large{ \it Scalar sector}}

I now expand the most general scalar fields $\vf_a$ into modes, singling out the coefficient of the
`radial mode' as
\begin{equation}
\vf_a(x,X) = X_a^{(N')} \otimes\Big(\,\frac{1}{r}\,\one_n + \vf(x)\Big)
+ \sum_{k} A_{a,k}(X) \otimes\vf_{k}(x)\,.
\label{phi-a-expansion-modes}
\end{equation}
Here $A_{a,k}(x)$ stands for a suitable basis labeled by $k$ of fluctuation modes of gauge fields on
$S^2_{N}$, and $\vf(x)$ resp. $\vf_k(x)$ are $\mmu(n)$-valued.  One expects that all fluctuation
modes in the expansion \eq{phi-a-expansion-modes} have a large mass gap of the order of the KK
scale, which is indeed the case as shown in detail in the~\ref{sec:stability} appendix.
Therefore she (he) can drop all these modes for the low-energy sector. However, the field $\vf(x)$
plays a somewhat special role.  It corresponds to fluctuations of the radius of the internal fuzzy
sphere, which is the order parameter responsible for the SSB $SU(N) \to SU(n)$, and assumes the
value $\one_n$ in \eq{phi-a-expansion-modes}. $\vf(x)$ is therefore the Higgs which acquires a
positive mass term in the broken phase, which can be obtained by inserting
$$
\vf_a(x,X) 
=X_a^{(N')} \otimes\Big(\,\frac{1}{r}\,\one_n + \vf(x)\Big)
$$
into $V(\vf)$,~\eq{pot}. Then the Higgs potential is found to be
\begin{equation}
V(\vf)=\frac{1}{2}NC_{2}(N')(1-r^{-1})^{2}\Big(\vf^{2}(x)+2r^{-1}\vf(x)+r^{-2}\Big)\,.
\label{Higgs-potential}
\end{equation}

In conclusion the model presented here behaves like a $U(n)$ gauge theory on
$M^4\times S^2_{N}$, with the expected tower of KK modes on the fuzzy sphere $S^2_{N}$ of radius
$r$. The Yang-Mills part of the low-energy effective action is given by the lowest KK mode, which is
\begin{multline}
 S_{LEA} = \int d^4 x\, tr_K \Big[ \frac{1}{4g^{2}}\, F_{\mu \nu}^\dagger F_{\mu \nu}
+ NC_2(N')\,D_{\mu}\varphi(x) D_{\mu}\varphi(x) \, \\
-\frac{1}{2} NC_{2}(N')(1-r^{-1})^{2}\Big(\vf^{2}(x)+2r^{-1}\vf(x)+r^{-2}\Big)\Big] + S_{int}
\label{LEA-action}
\end{multline}
for the $U(n)$ gauge field $A_{\mu}(y)\equiv A_{\mu,00}(y)$.

\section{CSDR constraints for fuzzy cosets}

Consider a fuzzy coset $(S/R)_F$ (e.g. fuzzy $CP^M$) described by $n\times n$ matrices, and let the
higher dimensional theory have gauge group $U(P)$. Then constraint~(\ref{3.17}) forms the
Lie algebra of the isometry group $S$ of the coset realized in its $n$-\dim~irrep. To solve
the~(\ref{3.19}) constraint an embedding of $S$ in $U(nP)$ has to be defined. As a
result, the four-dimensional gauge group $K$ is the centraliser of the image $S_{U(nP)}$  of $S$ in
$U(nP)$, $K=C_{U(nP)}(S_{U(nP)})$.

Concerning fermions in the adjoint, in order to solve constraint (\ref{eq16}) one considers the
embedding $$S\hookrightarrow SO(\dim(S))~, $$ which is given by $\tau_{a} = \frac{1}{2}C_{abc}
\Gamma^{bc}$ that satisfies $[\tau_{a},\tau_{b}]=C_{abc}\tau^{c}$. Therefore $\psi$ is an
intertwining operator between induced representations of $S$ in $U(nP)$ and  in $SO(\dim(S))$. To
find the surviving fermions, as in the commutative case~\cite{Kapetanakis:1992hf}, one has to
decompose the adjoint rep. of $U(nP)$ under $S_{U(nP)}\times K$,
\begin{align}
&U(nP) \supset S_{U(nP)} \times K \nonumber \\
&\adj[U(nP)] = \sum_{i} (s_i, k_i)
\end{align}
and the spinor rep. $\sigma$ of $SO(\dim(S))$ under $S$
\begin{align}
& SO(\dim(S)) \supset S\nn \\
& \sigma = \sum_{e} \sigma_{e}~.
\end{align}
Then, for two identical irreps. $s_i = \sigma_e$, there is a $k_i$  multiplet of fermions
surviving in four dimensions, i.e. four-dimensional spinors $\psi(x)$ belonging to the $k_i$
representation of $K$.

\section{Discussion}

I discussed here a generalisation of the CSDR scheme over spaces approximated by the algebra of
finite matrices (fuzzy cosets). I followed a rather modest approach keeping the commutative nature
of ordinary spacetime and assume only the internal space to be a noncommutative one, i.e. a fuzzy
space. The main advantage of this assumption is that even the higher dimensional theory is
renormalizable. Due to the finite dimension of the matrices approximating the fuzzy spaces only a
finite number of counterterms is required for the renormalizability of the theory. Furthermore the
Higgs and Yukawa sectors of a simple particle physics model resulted from the dimensional reduction
itself which is an intrinsic characteristic of the CSDR scheme. To be more specific, as in ordinary
CSDR case, Higgs particles are the extra dimensional components of the gauge fields of the initially
defined ten-dimensional Yang-Mills-Dirac theory. The Yukawa terms are obtained by the coupling terms
between fermions  and gauge fields.

However, the Fuzzy-CSDR has some different features from the ordinary CSDR and leads to new
possibilities to build reasonable low-energy theories which however remain to be investigated. A
major difference between fuzzy and ordinary CSDR is that in the fuzzy case one always embeds $S$ in
the gauge group $G$ instead of embedding just $R$ in $G$. This is due to the fact that the
differential calculus used in the Fuzzy-CSDR is based on $\dim(S)$ derivations instead of the
restricted $\dim(S)-\dim(R)$ used in the ordinary one.  As a result the four-dimensional gauge group
$H = C_G(R)$ appearing in the ordinary CSDR after the geometrical breaking and before the
spontaneous symmetry breaking due to the four-dimensional Higgs fields does not appear in the
Fuzzy-CSDR. In Fuzzy-CSDR the spontaneous symmetry breaking mechanism takes already place by solving
the Fuzzy-CSDR constraints. The four-dimensional potential has the typical `maxican hat' shape, but
it appears already spontaneously broken. Therefore in four dimensions appears only the physical
Higgs field that survives after a spontaneous symmetry breaking. Correspondingly in the Yukawa
sector of the theory one has the results of the spontaneous symmetry breaking, i.e. massive fermions
and Yukawa interactions among fermions and the physical Higgs field. Having massive fermions in the
final theory is a generic feature of CSDR when $S$ is embedded in $G$~\cite{Kapetanakis:1992hf}.
Therefore, if one would like to describe the spontaneous symmetry breaking of the SM in the present
framework, then one would be naturally led to large extra dimensions.

A fundamental difference between the ordinary CSDR and its fuzzy version is the fact that a
non-abelian gauge group $G$ is not really required in high dimensions. Indeed  the presence of a
$U(1)$ in the higher-dimensional theory is enough to obtain non-abelian gauge theories in four
dimensions.

\chapter[Dynamical generation of Fuzzy Extra Dimensions]{Dynamical generation of Fuzzy Extra
Dimensions and Symmetry Breaking}
\label{FuzzyExtraDim}

According to the discussion so far, CSDR over fuzzy coset spaces, $(S/R)_{F}$, leads
to four-dimensional theories  with phenomenologically interesting characteristics. Theories defined
on the assumed compactified space, $M^{4}\times (S/R)_{F}$, were found to be renormalizable.
Motivated by the interesting features of this approach, I examine here the inverse problem, i.e.
whether obtaining fuzzy extra dimensions as a vacuum solution of a four-dimensional but
renormalizable potential is possible. Indeed, starting from the most general renormalizable
potential in four dimensions, fuzzy extra dimensions is dynamically generated~\cite{Aschieri:2006uw}
as the energetically preferred solution . Furthermore the initial gauge symmetry was found to break
spontaneously towards phenomenologically interesting patterns.

In the following sections after a small introduction, I consider a renormalizable
four-dimensional $SU(\cN)$ gauge theory with a suitable multiplet of scalar fields which dynamically
develop extra dimensions in the form of  a two-dimensional fuzzy sphere, $S^2_N$. The potential of
the theory is chosen to be the most general renormalizable in four dimensions and its minimum is
calculated for some part of the parameter space (section~\ref{4-dimAction}). The
tower of massive Kaluza-Klein modes is consistent with an interpretation as gauge theory on
$M^4 \times S_{N}^2$, the scalars being interpreted as gauge fields on $S^2_{N}$. The gauge group is
broken dynamically, and the low-energy content of the model is determined. Depending on the
parameters of the model the low-energy gauge group can be $SU(n)$, or broken further to $SU(n_1)
\times SU(n_2) \times U(1)$, with mass scale determined by the size of the extra
dimension~(section~\ref{sec:KK}). Finally, I make some remarks on the results of the work and their
connection with the Fuzzy-CSDR discussed in the previous chapter.

\section{Introduction}

In the following sections, I consider a renormalizable $SU(\cN)$ gauge theory on four-dimensional
Minkowski space $M^4$, containing three scalars in the adjoint of $SU(\cN)$ that transform as
vectors under an additional global $SO(3)$ symmetry with the most general renormalizable potential. 
Then it can be proven  that the model dynamically develops fuzzy extra dimensions, more precisely a
two-dimensional  fuzzy sphere $S^2_{N}$. The appropriate interpretation is therefore as gauge theory
on $M^4 \times S^2_{N}$. The low-energy effective action is that of a four-dimensional gauge theory
on $M^4$, whose
gauge group and field content is dynamically determined by compactification and dimensional
reduction on the internal sphere  $S^2_{N}$.  An interesting and quite rich pattern of spontaneous
symmetry breaking (SSB) appears, breaking the original $SU(\cN)$ gauge symmetry down to much smaller
and potentially quite interesting low-energy gauge groups.  In particular, I find explicitly the
tower of massive Kaluza-Klein states, which justifies the interpretation as a compactified
higher-dimensional gauge theory. Nevertheless, the model is renormalizable.

The effective geometry, the symmetry breaking pattern and the low-energy gauge group are determined
dynamically in terms of a few free parameters of the potential. Here, I discuss in detail the two
simplest possible vacua with gauge groups $SU(n)$ and $SU(n_1)\times SU(n_2) \times U(1)$.
I find explicitly the tower of massive Kaluza-Klein modes corresponding to the effective geometry.
The mass scale of these massive gauge bosons is determined by the size of the extra dimensions,
which in turn depends on some logarithmically running coupling constants.  In the case of the
$SU(n_1)\times SU(n_2) \times U(1)$ vacuum, I identify in particular massive gauge fields in the
bifundamental, similar as in GUT models with an adjoint Higgs. Moreover, I also identify a
candidate for a further symmetry breaking mechanism, which may lead to a low-energy content of the
theory close to the standard model.

Perhaps the most remarkable aspect of our model is that the geometric interpretation and the
corresponding low-energy degrees of freedom depend in a nontrivial way on the parameters of the
model, which are running under the RG group. Therefore the massless degrees of freedom and their
geometrical interpretation depend on the energy scale. In particular, the low-energy gauge group
generically turns out to be $SU(n_1) \times SU(n_2)\times U(1)$ or $SU(n)$, while
 gauge groups which are products of more than two simple components [apart from $U(1)$] do not
seem to occur in this model. Moreover, the values of $n_1$ and $n_2$ are determined dynamically, and
may well be small such as 3 and 2. A full analysis of the hierarchy of all possible vacua and their
symmetry breaking pattern is not trivial however, and remain to be investigated.
Here, I restrict myself to establish the basic mechanisms and features of the model, and
discuss in the two following sections the two simplest cases (named as `type 1' and `type 2' vacuum) in some
detail.

The construction under discussion  was further developed by the addition of fermions in the
action~\cite{Steinacker:2007ay}. In particular, in the vacua with low-energy gauge group
$SU(n_1) \times SU(n_2)\times U(1)$, the extra-dimensional sphere always carries a magnetic flux
with nonzero monopole number leading to chiral massless fermions. Unfortunately this not possible
for a minimal set of fermions and their spectrum have to be doubled.

The idea to use fuzzy spaces for the extra dimensions is certainly not new. The work I describe
here was motivated by the Fuzzy-CSDR approach discussed in the previous chapter and combined with
lessons from the matrix-model approach to gauge theory on the fuzzy
sphere~\cite{Steinacker:2003sd,Steinacker:2004yu}. This leads in particular to a dynamical
mechanism of determining the vacuum, SSB patterns and background fluxes.  A somewhat similar model
has been studied recently in~\cite{Andrews:2005cv,Andrews:2006aw}, which realises deconstruction and
a `twisted' compactification of an extra fuzzy sphere based on a supersymmetric gauge theory. The
model under discussion is different and does not require supersymmetry, leading to a much richer
pattern of symmetry breaking and effective geometry.  For other relevant work see
e.g.~\cite{Madore:1992ej}.

The dynamical formation of fuzzy spaces found here is also related to recent work studying the
emergence of stable submanifolds in modified IIB matrix models. In particular, previous studies
based on  actions for fuzzy gauge theory different from ours generically only gave results
corresponding to $U(1)$ or $U(\infty)$ gauge groups, see e.g.
\cite{Azuma:2004ie,Azuma:2005bj,Azuma:2004zq} and references therein. 
The dynamical generation of a nontrivial index on noncommutative spaces has also been
observed in~\cite{Aoki:2004sd,Aoki:2006zi} for  different models.

The mechanism under discussion may also be very interesting in the context of the recent observation
\cite{Abel:2005rh} that extra dimensions are very desirable for the application of noncommutative
field theory to particle physics. Other related recent work discussing the implications of the
higher-dimensional point of view on symmetry breaking and Higgs masses can be found in
\cite{Lim:2006bx,Dvali:2001qr,Antoniadis:2002ns,Scrucca:2003ra}. These issues could now be discussed
within a renormalizable framework.

Finally the dynamical or spontaneous generation of extra dimensions occurring in the construction I
present here is strongly suggestive of gravity. Indeed the results of~\cite{Steinacker:2007dq} allow
to understand this mechanism in terms of gravity: the scalar potential defines a matrix-model action
which - using a slight generalisation of~\cite{Steinacker:2007dq} - can be interpreted as
non-abelian
Yang-Mills coupled to dynamical Euclidean gravity in the extra dimensions.

\section{The four-dimensional action}\label{4-dimAction}

Let a $SU(\cN)$ gauge theory on four-dimensional Minkowski space $M^4$ with coordinates
$x^\mu$, $\mu = 0,1,2,3$.  The action under consideration is
\begin{equation}
{\cal S}_{YM}=
\int d^{4}x\, Tr\,\left( \frac{1}{4g^{2}}\, F_{\mu \nu}^\dagger F_{\mu \nu} +
(D_{\mu}\phi_{{a}})^\dagger D_{\mu}\phi_{{a}}\right) - V(\phi)
\label{the4daction}
\end{equation}
where $A_\mu$ are $\msu(\cN)$-valued gauge fields, $D_\mu = \partial_\mu + [A_\mu,.]$, and
\begin{equation}
 \phi_{{a}} = - \phi_{{a}}^\dagger\,, \qquad a=1,2,3
\end{equation}
are three antihermitian scalars transforming under the adjoint action of $SU(\cN)$,
\begin{equation}
 \phi_{{a}} \to U^\dagger \phi_{{a}} U
\end{equation}
where $U = U(x) \in SU(\cN)$. Furthermore, the $\phi_a$ transform as vectors of an additional global
$SO(3)$ symmetry. The potential $V(\phi)$ is taken to be the most general renormalizable
action invariant under the above symmetries, which is
\begin{eqnarray}
V(\phi) &=& 
Tr\, \left( g_1 \phi_a\phi_a \phi_b\phi_b + g_2\phi_a\phi_b\phi_a \phi_b
- g_3 \varepsilon_{a b c} \phi_a \phi_b\phi_c
+ g_4\phi_a \phi_a \right) \nn\\
&& +\frac{g_5}{\cN}\,Tr(\phi_a \phi_a)Tr(\phi_b \phi_b) 
+\frac{g_6}{\cN}\,Tr(\phi_a\phi_b)Tr(\phi_a \phi_b) +g_7\,.
\label{pot-fuzzyextradim}
\end{eqnarray}
This may not look very transparent at first sight, however it can be written in a very intuitive
way. First, I make the scalars dimensionless by rescaling $\phi_a\to\phi_{a}/R$
where $R$ has dimension of length; I will usually suppress $R$ since it can immediately be
reinserted. Then for suitable choice of $R$
\begin{equation}
R= \frac{2 g_2}{g_3}\,,
\label{Radius}
\end{equation}
the potential can be rewritten as
\begin{eqnarray}
 V(\phi)&=&
Tr \( a^2 (\phi_a\phi_a + \tilde b\, \one)^2 + c +\frac 1{\tilde g^2}\, F_{ab}^\dagger F_{ab}\,\)
+\frac{h}{\cN}\, g_{ab} g_{ab}
\label{V-general-2}
\end{eqnarray}
for suitable constants $a,b,c,\tilde g,h$, where
\begin{align}
&F_{{a}{b}}
=[\phi_{{a}}, \phi_{{b}}] - \varepsilon_{abc} \phi_{{c}}\, =\varepsilon_{abc} F_c\,,\nn\\
&\tilde b = b + \frac{d}{\cN} \, Tr(\phi_a \phi_a)\,, \nn\\
&g_{ab} = Tr(\phi_a \phi_b)\,.
\label{const-def}
\end{align}
I shall omit $c$ from now.  The potential is clearly semi-positive definite provided
\begin{equation}
 a^2 = g_1+g_2 >0\,, \qquad \frac 2{\tilde g^2} = - g_2 >0\,, \qquad h \geq 0\,,
\end{equation}
which I assume from now on.  Here $\tilde b = \tilde b(x)$ is a scalar, $g_{ab} = g_{ab}(x)$ is
a symmetric tensor under the global $SO(3)$, and $F_{ab}=F_{ab}(x)$ is a $\msu(\cN)$-valued
antisymmetric tensor field which will be interpreted as field strength in some dynamically generated
extra dimensions below.  In this form, $V(\phi)$ looks like the action of Yang-Mills gauge theory on
a fuzzy sphere in the matrix
formulation~\cite{Steinacker:2003sd,Steinacker:2004yu,Carow-Watamura:1998jn,Presnajder:2003ak}.  The
presence of the first term $a^2 (\phi_a\phi_a + \tilde b)^2$ might seem
strange at first, however I should not simply omit it since it would be reintroduced by
renormalisation. In fact it is necessary for the interpretation as Yang-Mills
action,~\cite{Steinacker:2003sd,Presnajder:2003ak}, and I shall show that it is very welcome on
physical grounds since it dynamically determines and stabilises a vacuum, which can be
interpreted as extra-dimensional fuzzy sphere. In particular, it removes unwanted flat directions.

{
Let me briefly comment on the RG flow of the various constants. Without attempting any precise
computations here, one can see by looking at the potential~\eq{pot-fuzzyextradim} that $g_4$ will be
quadratically divergent at one loop, while $g_1$ and $g_2$ are logarithmically divergent. Moreover,
the only diagrams contributing to the coefficients $g_5, g_6$ of the `nonlocal' terms are
nonplanar, and thus logarithmically divergent but suppressed by $\frac 1\cN$ compared to the other
(planar) diagrams. This justifies the explicit factors $\frac 1\cN$ in~\eq{pot-fuzzyextradim}
and~\eq{const-def}. 
Finally, the only one-loop diagram contributing to $g_3$ is also logarithmically divergent. In terms
of the constants in the potential~\eq{V-general-2}, this implies that $R, a$, $\tilde g$, $d$ and
$h$ are running logarithmically under the RG flux, while $b$ and therefore $\tilde b$ is running
quadratically. The gauge coupling $g$ is of course logarithmically divergent and 
asymptotically free.}

A full analysis of the RG flow of these parameters is complicated by the fact that the vacuum and
the number of massive resp. massless degrees of freedom depends sensitively on the values of these
parameters, as will be discussed below. This indicates that the RG flow of this model will have a
rich and nontrivial structure, with different effective description at different energy scales.

\subsection{The minimum of the potential}
\label{sec:vacua}

Let me try to determine the minimum of the potential~\eq{V-general-2}. This turns out to be a rather
nontrivial task, and the answer depends crucially on the parameters in the potential.

For suitable values of the parameters in the potential, one can immediately write down the vacuum.
Assume  for simplicity $h=0$ in~\eq{V-general-2} . Since $V(\phi) \geq 0$, the global minimum of the
potential is certainly achieved if
\begin{equation}
 F_{{a}{b}} =
 [\phi_{{a}}, \phi_{{b}}] - \varepsilon_{abc} \phi_{{c}}~= 0, \qquad -\phi_a\phi_a = \tilde b\,,
\label{vacuum-trivial-cond}
\end{equation}
because then $V(\phi) =0$. This implies that $\phi_a$ is a rep. of $SU(2)$, with
prescribed Casimir\footnote{Note that $-\phi\cdot \phi = \phi^\dagger\cdot \phi >0$ since the fields
are antihermitian.} $\tilde b$. These equations may or may not have a solution, depending on the
value of $\tilde b$.  Assume first that $\tilde b$ coincides with the quadratic Casimir of
finite-dimensional irrep. of $SU(2)$,
\begin{equation}
 \tilde b = C_2(N) = \frac{1}{4}(N^2-1)
\end{equation}
for some $N \in \N$. If furthermore the dimension $\cN$ of the matrices $\phi_a$ can be written as
\begin{equation}
\cN = N n\,,
\end{equation}
then clearly the solution of~\eq{vacuum-trivial-cond} is given by
\begin{equation}
\phi_a = X_a^{(N)} \otimes \one_{n}
\label{vacuum-trivial}
\end{equation}
up to a gauge transformation, where $X_a^{(N)}$ denote the generator of the $N$-dimensional irrep.
of $SU(2)$. This can be viewed as a special case of~\eq{solution-general} below, consisting of
$n$ copies of the irrep. $\mbf{N}$ of $SU(2)$.

For generic $\tilde b$, the equations~\eq{vacuum-trivial-cond} cannot be satisfied for
finite-dimensional matrices $\phi_a$. The exact vacuum (which certainly exists since the potential
is positive definite) can in principle be found by solving the `vacuum equation'
$\frac{\d V}{\d \phi_a} =0$,
\begin{equation}
a^2 \left\{\phi_a,(\phi\cdot\phi + \tilde b)
+ \frac{d}{\cN}\, Tr(\phi\cdot \phi+\tilde b)\right\} + \frac{2h}{\cN}\,
g_{ab}\phi_b + \frac 1{\tilde g^2}\, (2[F_{ab},\phi_b] + F_{bc}
\varepsilon_{abc}) =0
\label{eom-int}
\end{equation}
where $\phi\cdot\phi = \phi_a \phi_a$. We note that all solutions under consideration will imply
$g_{ab} = \frac{1}{3}\delta_{ab} Tr(\phi\cdot \phi)$, simplifying this expression.

The general solution of \eq{eom-int} is not known. However, it is easy to write down a large class
of solutions: any decomposition of $\cN =n_1 N_1 + ... + n_k N_k$ into irreps of $SU(2)$ with
multiplicities $n_i$ leads to a block-diagonal solution 
\begin{equation}
\phi_a = diag\Big(\a_1\,X_a^{(N_1)}\otimes\one_{n_{1}},\ldots,
\a_k\,X_a^{(N_k)}\otimes\one_{n_{k}}\Big)
\label{solution-general}
\end{equation}
of the vacuum equations~\eq{eom-int}, where $\a_i$ are suitable constants which will be determined
below. There are hence several possibilities for the true vacuum, i.e. the global minimum of the
potential. Since the general solution is not known, I proceed by first determining the solution of
the form~\eq{solution-general} with minimal potential, and then discuss a possible solution
of a different type (`type 3 vacuum').

\paragraph{Type 1 vacuum.}

It is clear that the solution with minimal potential should satisfy~\eq{vacuum-trivial-cond} at
least approximately. It is therefore plausible that the solution~\eq{solution-general} with minimal
potential contains only representations (reps) whose Casimirs are close to $\tilde b$. In
particular, let $N$ be the dimension of the irrep. whose Casimir $C_2(N)\approx \tilde b$ is closest
to $\tilde b$. If furthermore the dimensions match as $\cN = N n$, I expect that the vacuum is given
by $n$ copies of the irrep. $\mbf{N}$, which can be written as 
\begin{equation}
 \phi_a = \a\, X_a^{(N)} \otimes\one_{n}\,.
\label{vacuum-mod1}
\end{equation}
This is a slight generalisation of \eq{vacuum-trivial}, with $\a$ being determined through the
vacuum equations~\eq{eom-int},
 \begin{equation}
 a^2 (\a^2 C_2(N) - \tilde b)(1+d) + \frac{h}{3}\, \a^2 C_2(N) 
-\frac{1}{\tilde{g}^2}\, (\a-1)(1-2\a) =0
\label{eom-alpha}
\end{equation}
A vacuum of the form~\eq{vacuum-mod1} will be denoted as `type 1 vacuum'.  As I will explain in
detail, it has a natural interpretation in terms of a dynamically generated extra-dimensional fuzzy
sphere $S^2_{N}$, by interpreting $X_a^{(N)}$ as generator of a fuzzy sphere
(c.f. section~\ref{FuzzySphere}). Furthermore, I will show in section~\ref{sec:KK1} that this type
1 vacuum~\eq{vacuum-mod1} leads to spontaneous symmetry breaking, with low-energy (unbroken) gauge
group $SU(n)$. The low-energy sector of the model can then be understood as compactification and
dimensional reduction on this internal fuzzy sphere.

Let me discuss equation~\eq{eom-alpha} in more detail. It can of course be solved exactly, but an
expansion around $\a=1$ is more illuminating.  To simplify the analysis I assume $d=h=0$ from now
on, and assume furthermore that $a^2 \approx (1/\,{\tilde{g}^2})$ have the same order of
magnitude\footnote{Otherwise the vacuum of the theory cannot be stabilised among other flat
directions of the potential and the good characteristics of the matrix model I consider are
spoiled.}. Defining the {\em real} number $\tilde N$ by $\tilde b = \tfrac{1}{4}(\tilde{N}^2-1)\,,$
one finds
\begin{equation}
\a = 1 -\frac {m}{N} + \frac{m(m+1)}{N^2} + O(\frac 1{N^3})
\qquad \mbox{where} \,\, m = N - \tilde N
\label{alpha-solution-N}
\end{equation}
assuming $N$ to be large and $m$ small. Notice that $a$ does not enter to leading order. This can be
understood by noting that the first term in~\eq{eom-alpha} is dominating under these assumptions,
which determines $\a$ to be~\eq{alpha-solution-N} to leading order. The potential
$V(\phi)$ is then dominated by the term
\begin{equation}
\frac {1}{\tilde g^2}
\,F_{ab}^\dagger F_{ab} = \frac {1}{2\tilde g^2}\, m^2 \,\one\,\, + \,O(\frac 1{N})\,,
\label{action-m}
\end{equation}
while $(\phi_a\phi_a + \tilde b)^2 = O(\frac 1{N^2})$.  There is a deeper reason for this simple
result: If $\,\tilde N \in \N$, then the solution~\eq{vacuum-mod1} can be interpreted as a fuzzy
sphere $S^2_{\tilde N}$ carrying a magnetic monopole of strength $m$, as shown explicitly
in~\cite{Steinacker:2003sd}; see also~\cite{Karabali:2001te,Balachandran:1999hx}. Then~\eq{action-m}
is indeed the action of the monopole field strength.

\paragraph{Type 2 vacuum.}

It is now easy to see that for suitable parameters, the vacuum will indeed consist of several
distinct blocks. This will typically be the case if $\cN$ is not divisible by the dimension of the
irrep. whose Casimir is closest to $\tilde b$.

Consider again a solution~\eq{solution-general} with $n_i$ blocks of size $N_i = \tilde N +m_i$,
assuming that $\tilde N$ is large and $\tfrac{m_i}{\tilde{N}} \ll 1$.  Generalising~\eq{action-m},
the action is then given by 
\begin{equation}
V(\phi) = Tr \Big( \frac {1}{2\tilde g^2}\,
\sum_i n_i\, m_i^2 \, \one_{N_i} \, + O(\frac 1{N_i}) \Big) \approx
\frac{1}{2\tilde{g}^{2}}\, \frac{\cN}{k}\, \sum_{i} n_i \, m_{i}^{2}\,
\label{action-mi}
\end{equation}
where $k=\sum n_i$ is the total number of irreps, and the solution can be interpreted in terms of
{`instantons' (nonabelian monopoles) on the internal fuzzy sphere}~\cite{Steinacker:2003sd}.
Hence in order to determine the solution of type~\eq{solution-general} with minimal action, one
simply has to minimise $\sum_i n_i \, m_i^2$, where the $m_i \in \Z -\tilde N$ satisfy the
constraint $\sum n_i \,m_i = \cN - k \tilde N$.

It is now easy to see that as long as the approximations used in~\eq{action-mi} are valid, the
vacuum is given by a partition consisting of  blocks with no more than two distinct sizes $N_1, N_2$
which satisfy $N_2 = N_1+1$. This follows from the convexity of~\eq{action-mi}: assume that the
vacuum is given by a configuration with 3 or more different blocks of size 
$N_1 < N_2 < \ldots < N_k$. Then the action~\eq{action-mi} could be lowered by modifying the
configuration as follows: reduce $n_1$ and $n_k$ by one, and add 2 blocks of size
$N_1+1$ and $N_k-1$. This preserves the overall dimension, and it is easy to check (using
convexity) that the action~\eq{action-mi} becomes smaller. This argument can be applied as long as 
there are 3 or more different blocks, or 2 blocks with $|N_2 - N_1| \geq 2$. Therefore if
$\cN$ is large, the solution with minimal potential among all possible
partitions~\eq{solution-general} is given either by a type 1 vacuum, or
takes the form
\begin{equation}
\phi_a= \left(\begin{array}{cc}
                   \a_1\, X_a^{(N_1)}\otimes\one_{n_1} & 0 \\
                   0 & \a_2\,X_a^{(N_2)}\otimes\one_{n_2}
             \end{array}\right)\,,
\label{vacuum-mod2}
\end{equation}
where the integers $N_1, N_2$ satisfy
\begin{equation}
\cN = N_1 n_1 + N_2 n_2\,, \qquad N_2 = N_1+1\,.
\end{equation}
A vacuum of the form~\eq{vacuum-mod2} will be denoted as `type 2 vacuum', and is the generic case.
In particular, the integers $n_1$ and $n_2$ are determined dynamically. This conclusion might be
altered for nonzero $d,h$ or by a violation of the approximations used in~\eq{action-mi}. I shall
show in subsection~\ref{sec:KK2} that this type of vacuum leads to a low-energy (unbroken) gauge group
$SU(n_1) \times SU(n_2) \times U(1)$, and the low-energy sector can be interpreted as dimensional
reduction of a higher-dimensional gauge theory on an internal fuzzy sphere, with features similar to
a GUT model with SSB $SU(n_1+n_2) \to SU(n_1) \times SU(n_2) \times U(1)$ via an adjoint Higgs.
Furthermore, since the vacuum \eq{vacuum-mod2} can be interpreted as a fuzzy sphere with
nontrivial magnetic flux~\cite{Steinacker:2003sd}, one can expect to obtain massless chiral fermions
in the low-energy action~\cite{Steinacker:2007ay}.

In particular, it is interesting to see that gauge groups which are products of more than two simple
components [apart from $U(1)$] do not occur in this model. Furthermore one can easily verify that
in cases in which both partitions of $\cN$ are possible, namely $\cN=Nn$ and
$\cN=n_{1}N_{1}+n_{2}N_{2}$, the latter is energetically preferable.
A numerical study concerning the stability of vacua type 1 and type 2 is presented
in appendix~\ref{stability-VAC1-VAC2}.

\paragraph{Type 3 vacuum.}
Finally, it could be that the vacuum is of a type different from~\eq{solution-general}, e.g. with
off-diagonal corrections such as 
\begin{equation}
\phi_a = \left(\begin{array}{cc}
                    \a_1\, X_a^{(N_1)}\otimes\one_{n_1} & \varphi_a \\
                    - \varphi_a^\dagger & \a_2\,X_a^{(N_2)}\otimes\one_{n_2}
             \end{array}\right)
\label{vacuum-mod3}
\end{equation}
for some small $\varphi_a$.  I shall provide evidence for the existence of such a vacuum
below, and argue that it leads to a further SSB.  This might play a role similar to low-energy
(`electroweak') symmetry breaking, which will be discussed in more detail below. In particular, it
is interesting to note that the $\varphi_a$ will no longer be in the adjoint of the low-energy gauge
group.  A possible way to obtain a SSB scenario close to the standard model is discussed in
subsection~\ref{sec:standardmodel}.

\subsection{Emergence of extra dimensions and the fuzzy sphere}
\label{sec:emergence}

Before discussing these vacua and the corresponding symmetry breaking in more detail, I want to
explain the geometrical interpretation, assuming first that the vacuum has the
form~\eq{vacuum-mod1}. The $X_a^{(N)}$ are then interpreted as coordinate functions (generators)
of a fuzzy sphere $S^2_{N}$, and the `scalar' action
\begin{equation}
 S_{\phi} =
Tr V(\phi) =
 Tr\Big(a^2 (\phi_a\phi_a + \tilde b)^2 + \frac 1{\tilde g^2}\, F_{ab}^\dagger F_{ab}\Big)
\label{S-YM2}
\end{equation}
for $\cN \times \cN$ matrices $\phi_a$ is precisely the action for a $U(n)$ Yang-Mills theory on
$S^2_{N}$ with coupling $\tilde g$, as shown in~\cite{Steinacker:2003sd} and reviewed
in appendix~\ref{sec:fuzzygaugetheory}. In fact, the `unusual' term $(\phi_a\phi_a + \tilde b)^2$ is
essential for this interpretation, since it stabilises the vacuum $\phi_a = X_a^{(N)}$ and gives a
large mass to the extra `radial' scalar field which otherwise arises.  The fluctuations of 
$\phi_a = X_a^{(N)} + A_a$ then provide the components $A_a$ of a higher-dimensional gauge field
$A_M = (A_\mu, A_a)$, and the action~\eq{the4daction} can be interpreted as Yang-Mills theory on the
six-dimensional space $M^4 \times S^2_{N}$, with gauge group depending on the particular vacuum.
Note that e.g. for the `type 1 vacuum', the local gauge transformations $U(\cN)$ can indeed be
interpreted as local $U(n)$ gauge transformations on $M^4 \times S^2_{N}$.

In other words, the scalar degrees of freedom $\phi_a$ conspire to form a fuzzy space in extra
dimensions.  Therefore one interprets the vacuum~\eq{vacuum-mod1} as describing dynamically
generated extra dimensions in the form of a fuzzy sphere $S^2_{N}$, with an induced Yang-Mills
action on $S^2_{N}$. This geometrical interpretation will be fully justified in section~\ref{sec:KK}
by working out the spectrum of Kaluza-Klein modes.  The effective low-energy theory is then given by
the zero modes on $S^2_{N}$, which is analogous to the models considered in~\cite{Aschieri:2003vy}.
However, in the present approach one has a clear dynamical selection of the geometry due to the
first term in~\eq{S-YM2}.

It is interesting to recall here the running of the coupling constants under the RG as discussed
above. The logarithmic running of $R$ implies that the scale of the internal spheres is only mildly
affected by the RG flow. However, $\tilde b$ is running essentially quadratically, hence is
generically large. This is quite welcome here: starting with some large $\cN$, $\tilde b \approx
C_2(\tilde N)$ must indeed be large in order to lead to the geometric interpretation discussed
above. Hence the problems of naturalness or fine-tuning appear to be rather mild here.

\section{Kaluza-Klein modes, dimensional reduction, and symmetry breaking}
\label{sec:KK}

I now study the model~\eq{the4daction} in more detail.  Let me emphasise again that this is a
four-dimensional renormalizable gauge theory, and there is no fuzzy sphere or any other
extra-dimensional structure to start with. I have already discussed possible vacua of the
potential~\eq{S-YM2}, depending on the parameters $a,\tilde b,\tilde g$ and $\cN$. This is a
nontrivial problem, the full solution of which is beyond the discussion of this dissertation. I
restrict myself here to the simplest types of vacua discussed in subsection~\ref{sec:vacua},
and derive some of the properties of the resulting low-energy models, such as the corresponding
low-energy gauge groups and the excitation spectrum.  In particular, I exhibit the tower of
Kaluza-Klein modes in the different cases. This turns out to be consistent with an interpretation in
terms of compactification on an internal sphere, demonstrating without a doubt the emergence of
fuzzy internal dimensions.  In particular, the scalar fields $\phi_a$ become gauge fields on the
fuzzy sphere.

\subsection{Type 1 vacuum and $SU(n)$ gauge group}
\label{sec:KK1}

Let me start with the simplest case, assuming that the vacuum has the form~\eq{vacuum-mod1}. I want
to determine the spectrum and the rep. content of the gauge field $A_\mu$.  The structure
of $\phi_a = \a\, X_a^{(N)} \otimes\one_{n}$ suggests to consider the subgroups $SU(N) \times SU(n)
$ of $SU(\cN)$, where $K:=SU(n)$ is the commutant of $\phi_a$ i.e. the maximal subgroup of
$SU(\cN)$ which commutes with all $\phi_a$, $a=1,2,3$; this follows from Schur's Lemma. $K$ will
turn out to be the effective (low-energy) unbroken four-dimensional gauge group.

One could now proceed in a standard way arguing that $SU(\cN)$ is spontaneously broken to $K$ since
$\phi_a$ takes a v.e.v. as in~\eq{vacuum-mod1}, and elaborate the Higgs mechanism. This is
essentially what will be done below, however in a language which is very close to the picture of
compactification and KK modes on a sphere in extra dimensions. This is appropriate here, and leads
to a description of the low-energy physics of this model as a dimensionally reduced $SU(n)$ gauge
theory. Similar calculations have been also presented in section~\ref{sec:KK-FuzzyCSDR}.

\paragraph{Kaluza-Klein expansion on $S^2_{N}$.}

As in section~\ref{sec:KK-FuzzyCSDR}, I interpret the $X_a^{(N)}$ as generators of the
fuzzy sphere $S^2_{N}$ and decompose the full four-dimensional $\msu(\cN)$-valued gauge fields
$A_\mu$ into spherical harmonics $Y^{lm}(x)$ on the fuzzy sphere $S^2_{N}$ with coordinates
$y^{a}\sim X_{a}$:
\begin{equation}
A_\mu =A_\mu(x,X)= \sum_{\genfrac{}{}{0pt}{}{0 \leq l\leq N}{|m|\leq l}}
\,A_{\mu,lm}^{(n)}(x)\otimes\,Y^{lm(N)}(X)\,.
\label{KK-modes-A-fuzzyextradim}
\end{equation}
The $Y^{lm(N)}$ are by definition irreps under the $SU(2)$ rotations on $S^2_{N}$, and form a basis
of Hermitian $N \times N$ matrices; for more details see section~\ref{FuzzySphere}. The
$A_{\mu,lm}^{(n)}(x)$ turn out to be $\mmu(n)$-valued gauge and vector fields on $M^4$. Using this
expansion, I can interpret $A_\mu(x,X)$ as $\mmu(n)$-valued functions on
$M^4\times S^2_{N}$, expanded into the Kaluza-Klein modes (i.e. harmonics) of $S^2_{N}$.

The scalar fields $\phi_a$ with potential~\eq{S-YM2} and vacuum \eq{vacuum-mod1} should be
interpreted as `covariant coordinates' on $S^2_{N}$ which describe $U(n)$ Yang-Mills theory on
$S^2_{N}$. This means that the fluctuations $A_a$ of these covariant coordinates 
\begin{equation}
\phi_a = \a\, X_a^{(N)} \otimes\one_{n} + A_a
\end{equation}
should be interpreted as gauge fields on the fuzzy sphere, see appendix~\ref{gaugefield-S2N}. They
can be expanded similarly as
\begin{equation}
 A_a =A_a(x,X)=
\sum_{\genfrac{}{}{0pt}{}{0 \leq l\leq N}{|m|\leq l}}
\,A_{a,lm}^{(n)}(x)\otimes Y^{lm(N)}(X) \,,
\label{KK-modes-phi-fuzzyextradim}
\end{equation}
interpreted as functions (or one-form) on $M^4\times S^2_{N}$ taking values in $\mmu(n)$. One can
then interpret $A_M(x,y) = (A_\mu(x,y), A_a(x,y))$ as $\mmu(n)$-valued gauge or vector fields on
$M^4\times S^2_{N}$.

Given this expansion into KK modes, I shall show that only $A_{\mu,00}(y)$ (i.e. the dimensionally
reduced gauge field) becomes a massless $\msu(n)$-valued\footnote{Note that $A_{\mu,00}(y)$ is
traceless, while $A_{\mu,lm}(y)$ is not in general.} gauge field in 4D, while all other modes
$A_{\mu,lm}(y)$ with $l \geq 1$ constitute a tower of Kaluza-Klein modes with large mass gap, and
decouple for low energies. The existence of these KK modes firmly establishes our claim that the
model develops dynamically extra dimensions in the form of $S^2_{N}$. This geometric interpretation
is hence forced upon us, provided the vacuum has the form~\eq{vacuum-mod1}.  The scalar fields
$A_a(x,y)$ will be analysed in a similar way below, and provide no additional massless degrees of
freedom in four dimensions. More complicated vacua will have a similar interpretation.  Remarkably,
our model is fully renormalizable in spite of its higher-dimensional character, in contrast to the
commutative case; see also \cite{Aschieri:2005wm}.

\paragraph{Computation of the KK masses.}

To justify these claims, let me compute the masses of the KK modes~\eq{KK-modes-A-fuzzyextradim}.
They are induced by the covariant derivatives $\int Tr (D_{\mu}\phi_{{a}})^{2} $
in~\eq{the4daction},
\begin{equation}
\int Tr
(D_{\mu}\phi_{{a}})^\dagger D_{\mu}\phi_{{a}} = \int Tr
(\partial_{\mu}\phi_{{a}}^\dagger \partial_{\mu}\phi_{{a}} + 2
(\partial_{\mu}\phi_{{a}}^\dagger) [A_\mu,\phi_{a}] +
[A_\mu,\phi_{a}]^\dagger[A_\mu,\phi_{a}])\,.
\label{covar-mass-term-fuzzyextradim}
\end{equation}
The most general scalar field configuration can be written as 
\begin{equation}
\phi_a(x,X) = \a(x) X_a^{(N)} \otimes \one_n + A_a(x,X)
\label{phi-a-expansion}
\end{equation}
where $A_a(x,X)$ is interpreted as gauge field on the fuzzy sphere $S^2_{N}$ for each $x \in M^4$.
I allow here for a $x$--dependent $\a(x)$ (which could have been absorbed in $A_a(x,X)$), because
it is naturally interpreted as the Higgs field responsible for the symmetry breaking 
$SU(\cN) \to SU(n)$. As usual, the last term in~\eq{covar-mass-term-fuzzyextradim} leads to the mass
terms for the gauge fields $A_\mu$ in the vacuum $\phi_a(x,X) = \a(x) X_a^{(N)} \otimes \one_n$,
provided the mixed term which is linear in $A_\mu$ does not contribute in the mass term. As I noted
in section~\ref{sec:KK-FuzzyCSDR}, this is achieved by imposing an `internal' Lorentz gauge
$[X_a, A_a]=0$ at each point $x$. Then all quadratic contributions of that term vanish, leaving only
cubic interaction terms. Indeed, by inserting ~\eq{phi-a-expansion} into the term
$(\partial_{\mu}\phi_{{a}}^\dagger) [A_\mu,\phi_{a}]$ in~\eq{covar-mass-term-fuzzyextradim}, I
obtain
\begin{multline}
\int Tr A_\mu [\phi_a,\partial_{\mu}\phi_{{a}}^\dagger] =
\int Tr A_\mu \Big\{\a(x) [X_a,\partial_{\mu} A_a(x,X)]
+[A_a(x,X),\partial_{\mu}\a(x)\, X_a]\nn\\
+[A_a(x,X),\partial_{\mu}A_a(x,X)]\Big\}\,,
\end{multline}
which simplifies as
\begin{equation}
\int Tr A_\mu [\phi_a,\partial_{\mu}\phi_{{a}}^\dagger]
= \int Tr A_\mu [A_a(x,X),\partial_{\mu}A_a(x,X)] =: S_{int}\,.
\label{A-lin-term-2-fuzzyextradim}
\end{equation}
This contains only cubic interaction terms, which are irrelevant for the computation of the masses.
Therefore I can proceed by setting $\phi_a = \a X_a^{(N)} \otimes \one_n$ and inserting the
expansion~\eq{KK-modes-A-fuzzyextradim} of $A_\mu$ into the last term
of~\eq{covar-mass-term-fuzzyextradim}. Since the
$Tr [X_{a},A_\mu] [X_{a},A_\mu]$ is the quadratic Casimir on the modes of $A_\mu$ which are
orthogonal, one obtains
\begin{equation}
 \int Tr (D_{\mu}\phi_{{a}})^\dagger D_{\mu}\phi_{{a}}=
 \int Tr \Big(\partial_{\mu}\phi_{{a}}^\dagger \partial_{\mu}\phi_{{a}} + \sum_{l,m} \a^2\, l(l+1)\,
A_{\mu,lm}(y)^\dagger A_{\mu,lm}(y) \Big) + S_{int}\,.
\label{covar-mass-term-2-fuzzyextradim}
\end{equation}
Therefore the four-dimensional $\mmu(n)$ gauge fields $A_{\mu,lm}(y)$ acquire a mass
\begin{equation}
 m^2_{l} = \frac{\a^2 g^2}{R^2}\, l(l+1)
\label{KK-masses-Adiag-fuzzyextradim}
\end{equation}
reinserting the parameter $R$~\eq{Radius} which has dimension length. This is as expected for higher
KK modes, and determines the radius of the internal $S^2$ to be 
\begin{equation}
r_{S^2} = \frac{\a}g R
\label{S-radius}
\end{equation}
where $\a\approx 1$ according to~\eq{alpha-solution-N}. In particular, only
$A_{\mu}(y)\equiv A_{\mu,00}(y)$ survives as a massless four-dimensional $\msu(n)$ gauge
field. The low-energy effective action for the gauge sector is then given by 
\begin{equation}
 S_{LEA} = \int d^4 x\,\frac 1{4g^{2}}\, \Tr_n \ F_{\mu\nu}^\dagger F_{\mu \nu}\,,
\label{LEA-action-FF-fuzzyextradim}
\end{equation}
where $F_{\mu \nu}$ is the field strength of the low-energy $\msu(n)$ gauge fields, dropping all
other KK modes whose mass scale is set by $\frac{1}{R}$.  For $n=1$, there is no massless gauge
field. However I would find a massless $U(1)$ gauge field if I start with a $U(\cN)$ gauge theory
rather than  $SU(\cN)$.

\paragraph{Scalar sector.}

As in section~\ref{sec:KK-FuzzyCSDR}, I expand the most general scalar fields $\phi_a$ into
modes, singling out the coefficient of the `radial mode' as
\begin{equation}
\phi_a(x) =
X_a^{(N)} \otimes (\a \one_n + \varphi(x)) + \sum_{k} A_{a,k}(x) \otimes\varphi_{k}(x)\,.
\label{phi-a-expansion-modes-fuzzyextradim}
\end{equation}
Here $A_{a,k}(x)$ stands for a suitable basis labeled by $k$ of fluctuation modes of gauge fields on
$S^2_{N}$, and $\varphi(x)$ resp. $\varphi_k(x)$ are $\mmu(n)$-valued.  The
fluctuation modes in the expansion~\eq{phi-a-expansion-modes-fuzzyextradim} have a large mass gap of
the order of the KK scale as before. Therefore I can drop these modes for the low-energy sector.
However, the field $\varphi(x)$ plays a somewhat special role.  It corresponds to fluctuations of
the radius of the internal fuzzy sphere, which is the order parameter responsible for the SSB 
$SU(\cN) \to SU(n)$, and assumes the value $\a \one_n$ in \eq{phi-a-expansion-modes-fuzzyextradim}.
$\varphi(x)$ is therefore the Higgs which acquires a positive mass term in the broken phase, which
can be obtained by inserting
$$
\phi_a(x) = X_a^{(N)} \otimes (\a \one_n + \varphi(x))
$$
into $V(\phi)$. This mass is dominated by the first term in~\eq{V-general-2}
(assuming $a^2 \approx
\frac{1}{\tilde{g^2}}$), of order
\begin{equation}
V(\varphi(x)) \approx N\,
\Big(a^2 C_{2}(N)^2 \varphi(x)^{2} + O(\varphi^3)\Big)
\label{Higgs-potential-fuzzyextradim}
\end{equation}
for large $\cN$ and $N$. The full potential for $\varphi$ is of course quartic.

In conclusion, the model under discussion indeed behaves like a $U(n)$ gauge theory on 
$M^4\times S^2_{N}$, with the expected tower of KK modes on the fuzzy sphere $S^2_{N}$ of
radius~\eq{S-radius}. The low-energy effective action is given by  the lowest KK mode, which is
\begin{equation}
S_{LEA} = \int d^4 x\, \Tr_n \Big[ \frac{1}{4g^{2}}\, F_{\mu \nu}^\dagger F_{\mu \nu}
+ N C_2(N)\,D_{\mu}\varphi(x) D_{\mu}\varphi(x)
+N a^2 C_{2}(N)^2 \varphi(x)^{2} \Big] + S_{int}
\label{LEA-action-fuzzyextradim}
\end{equation}
for the $SU(n)$ gauge field $A_{\mu}(x)\equiv A_{\mu,00}(x)$. In~\eq{LEA-action-fuzzyextradim} I
also
keep the Higgs field $\varphi(x)$, even though it acquires a large mass
\begin{equation}
m_\varphi^2 = \frac{a^2}{R^2} \, C_2(N)
\end{equation}
reinserting $R$.

\subsection{Type 2 vacuum and $SU(n_1)\times SU(n_2)\times U(1)$ gauge group}
\label{sec:KK2}

For different parameters in the potential, one can obtain a different vacuum, with different
low-energy gauge group.  Assume now that the vacuum has the form~\eq{vacuum-mod2}.  The structure of
$\phi_a$ suggests to consider the subgroups 
$(SU(N_1) \times SU(n_1)) \times (SU(N_2) \times SU(n_2)) \times U(1)$ of $SU(\cN)$, where
\begin{equation}
K:=SU(n_1)\times SU(n_2)\times U(1)
\end{equation}
is the maximal subgroup of $SU(\cN)$ which commutes with all $\phi_a$, $a=1,2,3$ (this follows
from Schur's Lemma).  Here the $U(1)$ factor is embedded as
\begin{equation}
\mmu(1) \sim\left(\begin{array}{cc}
                          \frac{1}{N_1 n_1}\,\one_{N_1\times n_1} & \\
                                       & -\frac{1}{N_2 n_2}\,\one_{N_2\times n_2} 
                         \end{array}\right)
\label{U1-embed}
\end{equation}
which is traceless.  $K$ will again be the effective (low-energy) four-dimensional gauge group.

I now repeat the above analysis of the KK modes and their effective four-dimensional mass.  First,
I write 
\begin{equation}
 A_{\mu} =
\left(\begin{array}{cc} A_{\mu}^1 & A_{\mu}^+\\ A_{\mu}^- & A_{\mu}^2 \end{array}\right)
\end{equation}
according to~\eq{vacuum-mod2}, where $(A_{\mu}^+)^\dagger = -A_{\mu}^-$. The masses of the gauge
bosons are again induced by the last term in~\eq{covar-mass-term-fuzzyextradim}.  Consider the term
$[\phi_{a},A_\mu] = [\a_1 X^{(N_1)}_a + \a_2 X^{(N_2)}_a,A_\mu]$. For the diagonal fluctuations
$A_{\mu}^{1,2}$, this is simply the adjoint action of $X^{(N_1)}_a$. For the off-diagonal modes
$A_{\mu}^\pm$, I can get some insight by assuming first $\a_1 = \a_2$. Then the above commutator is
$X^{(N_1)} A_{\mu}^+ - A_{\mu}^+ X^{(N_2)}$, reflecting the rep. content 
$A_{\mu}^+ \in (N_1) \otimes (N_2)$ and $A_{\mu}^- \in (N_2) \otimes (N_1)$. Assuming
$N_1-N_2 =k>0$, this implies in particular that there are {\em no zero modes for the off-diagonal
blocks}, rather the lowest angular momentum is $k$. They can be interpreted as being sections on a
monopole bundle with charge $k$ on $S^2_{N_1}$, cf.~\cite{Steinacker:2003sd}.  The
case $\a_{1}\neq \a_2$ requires a more careful analysis as indicated below.  In any case, I can
again expand $A_{\mu}$ into harmonics,
\begin{equation}
A_\mu = A_\mu(x,X)=
\sum_{l,m} \left(\begin{array}{cc}
                                     A_{\mu,lm}^1(x)\,Y^{lm (N_1)}& A_{\mu,lm}^+(x)\,Y^{lm (+)}\\
                                     A_{\mu,lm}^-(x)\,Y^{lm(-)}    & A_{\mu,lm}^2(x)\,Y^{lm (N_2)}
                              \end{array}\right)\,,
\label{KK-modes-A2}
\end{equation}
setting $Y^{lm (N)}= 0$ if $l > 2N$. Then the $A_{\mu,lm}^{1,2}(x)$ are $\mmu(n_1)$ resp.
$\mmu(n_2)$-valued gauge resp. vector fields on $M^4$, while $A_{\mu,lm}^{\pm}(x)$ are vector
fields on $M^4$ which transform in the bifundamental $(n_1,\obar{n}_2)$ resp. $(n_2,\obar{n}_1)$ of
$\mmu(n_1)\times \mmu(n_2)$.

Now I can compute the masses of these fields.  For the diagonal blocks this is the same as in
subsection~\ref{sec:KK1}, while the off-diagonal components can be handled by writing 
\begin{equation}
Tr([\phi_{a},A_\mu][\phi_{a},A_\mu]) = 2 Tr(\phi_{a} A_\mu\phi_{a}
A_\mu - \phi_{a}\phi_{a} A_\mu A_\mu)\,.
\end{equation}
This gives
\begin{eqnarray}
 \int\!\!
Tr (D_{\mu}\phi_{{a}})^\dagger D_{\mu}\phi_{{a}} \!\! &=& \!\!  \int
\!\!Tr \Big(\partial_{\mu}\phi_{{a}}^\dagger \partial_{\mu}\phi_{{a}}
+ \sum_{l\geq 0} ( m^2_{l,1}\, A_{\mu,lm}^{1\dagger}(x)
A_{\mu,lm}^1(x) + m^2_{l,2}\, A_{\mu,lm}^{2\dagger}(x)
A_{\mu,lm}^2(x)) \nn\\
 && \quad + \sum_{l\geq k} 2 m^2_{l;\pm} (A_{\mu,lm}^+(x))^\dagger A_{\mu,lm}^+(x)\Big)
\label{covar-mass-term-break}
\end{eqnarray}
similar as in~\eq{covar-mass-term-2-fuzzyextradim}, with the same gauge choice and omitting cubic
interaction terms. In particular, the diagonal modes acquire a KK mass 
\begin{equation}
 m^2_{l,i} = \frac{\a_i^2 g^2}{R^2}\,l(l+1)
\label{KK-masses-Adiag-12}
\end{equation}
completely analogous to~\eq{KK-masses-Adiag-fuzzyextradim}, while the off-diagonal modes acquire a
mass 
\begin{eqnarray}
m^2_{l;\pm} &=& \frac{g^2}{R^2}\,\(\a_1 \a_2\, l(l+1) + (\a_1-\a_2)(X_2^2\a_2-X_1^2\a_1)\) \nn\\
&\approx & \frac{g^2}{R^2}\, \( l(l+1) + \frac 14 (m_2-m_1)^2 \,\, + O(\frac 1\cN) \)
\label{mass-offdiag}
\end{eqnarray}
using~\eq{alpha-solution-N} for $\a_i \approx 1$. In particular, all masses are positive.

In conclusion, the gauge fields $A_{\mu,lm}^{1,2}(x)$ have massless components
$A_{\mu,00}^{1,2}(x)$ which take values in $\msu(n_i)$ due to the KK-mode $l=0$ (as long as
$n_i>1$), while the bifundamental fields $A_{\mu,lm}^{\pm}(x)$ have no massless components. Note
that the mass scales of the diagonal modes~\eq{KK-masses-Adiag-12} and the off-diagonal
modes~\eq{mass-offdiag} are essentially the same.  This result is similar to the breaking
$SU(n_1+n_2) \to SU(n_1) \times SU(n_2)\times U(1)$ through an adjoint Higgs, such as in the $SU(5)
\to SU(3)\times SU(2)\times U(1)$ GUT model.  In that case, one also obtains massive
(`ultraheavy') gauge fields in the bifundamental, whose mass should therefore be identified in our
scenario with the mass~\eq{mass-offdiag} of the off-diagonal massive KK modes
$A_{\mu,lm}^{\pm}(x)$.  The $U(1)$ factor ~\eq{U1-embed} corresponds to the massless components
$A_{\mu,00}^{1,2}(x)$ above, which is now present even if $n_i=1$. 

The appropriate interpretation of this vacuum is as a gauge theory on $M^4 \times S^2$, compactified
on $S^2$ which carries a magnetic flux with monopole number $|N_1-N_2|$. This leads to a low-energy
action with gauge group $SU(n_1) \times SU(n_2)\times U(1)$. The existence of a magnetic flux is
particularly interesting in the context of fermions, since internal fluxes naturally lead to chiral
massless fermions~\cite{Steinacker:2007ay}. However this not possible for a minimal set of fermions
and their spectrum have to be doubled.

Repeating the analysis of fluctuations for the scalar fields is somewhat messy, and will not be
given here. However since the vacuum~\eq{vacuum-mod2} is assumed to be stable, all fluctuations in
the $\phi_a$ will again be massive with mass presumably given by the KK scale, and can therefore be
omitted for the low-energy theory.  Again, one could interpret the fluctuations $\varphi_{1,2}(y)$
of the radial modes $X_a^{(N_{1,2})} \otimes (\a_{1,2} +\varphi_{1,2}(y))$ as low-energy Higgs in
analogy to~\eq{phi-a-expansion-modes-fuzzyextradim}, responsible for the symmetry breaking 
$SU(n_1+n_2) \to SU(n_1) \times SU(n_2) \times U(1)$.

\subsection{Type 3 vacuum and further symmetry breaking}
\label{sec:KK3}

Finally consider a vacuum of the form~\eq{vacuum-mod3}. The  additional fields $\varphi_{a}$
transform in the bifundamental of $SU(n_1) \times SU(n_2)$ and lead to further SSB.  Of particular
interest is the simplest case 
\begin{equation}
 \phi_a = \left(\begin{array}{cc}
                     \a_1\, X_a^{(N_1)}\otimes\one_{n} & \varphi_a\\
                      - \varphi_a^\dagger & \a_2\, X_a^{(N_2)}
                      \end{array}\right)
\label{vacuum-mod4}
\end{equation}
corresponding to a would-be gauge group $SU(n) \times U(1)$ according to subsection~\ref{sec:KK2},
which will be broken further. Then 
$\varphi_a = \left(\begin{smallmatrix}
                       \varphi_{a,1} \\
                       \vdots \\
                       \varphi_{a,n}
                     \end{smallmatrix}\right) $
lives in the fundamental of $SU(n)$ charged under $U(1)$, and transforms as $(N_1) \otimes (N_2)$
under the $SO(3)$ corresponding to the fuzzy sphere(s). As discussed below, by adding a further
block, one can get somewhat close to the standard model, with $\varphi_a$ being a candidate for a
low-energy Higgs.

 I shall argue that there is indeed such a solution of the equation of motion~\eq{eom-int} for
$|N_1-N_2|=2$. Note that since
$\varphi_a \in (N_1) \otimes (N_2) = (|N_1-N_2|+1) \oplus ... \oplus (N_1+N_2-1)$,
it can transform as a vector under $SO(3)$ only in that case. Hence assume $N_1=N_2+2$, and define
$\varphi_a \in (N_1) \otimes (N_2)$ to  be the unique component which transform as a vector in the
adjoint. One can then show that 
\begin{equation}
 \phi_a \phi_a = -
 \left(\begin{array}{cc}
       \a_1^2\, C_2(N_1)\otimes\one_{n_1} -\frac{h}{N_1}\, & 0 \\
        0 & \a_2^2\, C_2(N_2) -\frac{h}{N_2}\,
 \end{array}
\right)
\label{phiphi-2}
\end{equation}
where $h$ is a normalisation constant, and
\begin{eqnarray}
\varepsilon_{abc}
\phi_b \phi_c &=& \left(\begin{array}{cc}
(\a_1^2- \frac{g_1}{N_1 }\,\frac{h}{C_2(N_1)})\,X_a^{(N_1)} & (\a_1 g_1 + \a_2 g_2) \varphi_a \\
-(\a_1 g_1 + \a_2 g_2) \varphi_a^\dagger &
(\a_2^2-\frac{g_2}{N_2}\,\frac{h}{C_2(N_2)})\,X_a^{(N_2)}
\end{array}\right)
\label{phiphi-comm-2}
\end{eqnarray}
where $g_1=\frac{N_1+1}2, \, g_2=-\frac{N_2-1}2$.  This has the same form as~\eq{vacuum-mod4} but
with different parameters. We now have three parameters $\a_1,\a_2,h$ at our disposal, hence
generically this ansatz will provide solutions of the e.o.m.~\eq{eom-int} which amounts to three
equations for the independent blocks. It remains to be investigated whether these are energetically
favourable.

\paragraph{The commutant  $K$ and further symmetry breaking.}

To determine the low-energy gauge group i.e. the maximal subgroup $K$ commuting with the solution
$\phi_a$ of type~\eq{vacuum-mod4}, consider 
\begin{eqnarray}
\varepsilon_{abc} \phi_b \phi_c &-&
(\a_1 g_1 + \a_2 g_2) \phi_a = \nn\\ && \!\!\!\!\!\!  \!\!\!\!\!\!\!\!\!\!\!\!
\left(\begin{array}{cc} (\a_1^2 - \a_1(\a_1 g_1 + \a_2 g_2)
-\frac{g_1}{N_1 }\, \frac{h}{C_2(N_1)})\,X_a^{(N_1)} & 0 \\ 0 &
\!\!\!\!\!\!\!\!\!\!\!\!  (\a_2^2 - \a_2(\a_1 g_1 + \a_2 g_2) -
\frac{g_2}{N_2 }\,\frac{h}{C_2(N_2)})\,X_a^{(N_2)}
    \end{array}\right)  \nn\\
\label{phiphi-comm-4}
\end{eqnarray}
Unless one of the two coefficients vanishes, this implies that $K$ must commute
with~\eq{phiphi-comm-4}, hence 
$K = \left(\begin{array}{cc} 
         K_1 & 0\\
         0 & K_2
         \end{array}\right)$
is a subgroup of $SU(n_1) \times SU(n_2)\times U(1)$; here I focus on $SU(n_2) = SU(1)$ being
trivial. Then~\eq{vacuum-mod4} implies that $k_1 \varphi_a = \varphi_a k_2$ for $k_i \in K_i$, which
means that $\varphi_a$ is an eigenvector of $k_1$ with eigenvalue $k_2$. Using a $SU(n_1)$ rotation,
I can assume that $\varphi_a^T = (\varphi_{a,1},0, \dots, 0)$. Taking into account the requirement
that $K$ is traceless, it follows that $K \cong K_1 \cong SU(n_1-1) \subset SU(n_1)$.  Therefore the
gauge symmetry is broken to $SU(n_1-1)$. This can be modified by adding a further block as discussed
below.

\subsection{Towards the standard model}
\label{sec:standardmodel}

Generalising the above considerations, I can construct a vacuum which is quite close to the
standard model. Consider 
\begin{equation}
\cN = N_1 n_1 + N_2 n_2 + N_3\,,
\end{equation}
for $n_1=3$ and $n_2=2$. As discussed above, I expect a vacuum of the form 
\begin{equation}
\phi_a =
\left(\begin{array}{ccc}
                 \a_1\, X_a^{(N_1)}\otimes\one_{3}& 0 &0\\
                 0 & \a_2\,X_a^{(N_2)}\otimes\one_{2} & \varphi_a \\
                 0 & -\varphi_a^\dagger & \a_3\,X_a^{(N_3)}
             \end{array}\right)
\label{vacuum-mod5}
\end{equation}
if $\tilde b \approx C_2(N_1)$ and $N_1 \approx N_2 = N_3\pm 2$.
Then the unbroken low-energy gauge group would be 
\begin{equation}
 K = SU(3) \times U(1)_Q \times U(1)_F\,,
\end{equation}
with $U(1)_F$ generated by the traceless generator 
\begin{equation}
 u(1)_F \sim\left(\begin{array}{cc} 
                       \frac{1}{3 N_1}\,\one_{3 N_1} & \\
                       & -\frac 1{D}\,\one_{D} \end{array}\right)
\end{equation}
where $D = 2 N_2 + N_3$, and $U(1)_Q$ generated by the traceless generator
\begin{equation}
u(1)_Q \sim\left(\begin{array}{ccc}
                        \frac{1}{3 N_1}\,\one_{3 N_1} & & \\ 
                        & -\frac 1{N_2} \left(\begin{array}{cc}
                                                     0 & 0\\
                                                      0 & 1 \end{array}\right) 
                         \one_{N_2} &  \\
                        & & 0 \end{array}\right)\,.
\end{equation}
assuming that $\varphi_a^T = (\varphi_{a,1}, 0)$.  This is starting to be reminiscent of the
standard model, and will be studied in greater detail elsewhere. However, I should recall that the
existence of a {\em vacuum} of this form has not been established at this point.

\subsection*{Relation with CSDR}

Let me compare the results of the current approach with the CSDR construction presented
in chapter~\ref{FuzzyCSDR}. I described there the construction of four-dimensional
models starting from gauge theory on $M^4 \times S^2_{\cN}$ by imposing CSDR constraints, which were
appropriately generalised for the case of fuzzy cosets. The solution of the
constraints was boiled down to choosing embeddings $\,\omega_a$, $a=1,2,3$ of $SU(2)$
into $SU(\cN)$. which determine the unbroken gauge field as the commutant of $\,\omega_a$,
and the low-energy (unbroken) Higgs by $\varphi_a \sim \omega_a$.
Obviously, the vacua solutions~\eq{vacuum-mod1} or~\eq{vacuum-mod2} could be also interpreted as solutions of the Fuzzy-CSDR
constraints in~\cite{Aschieri:2003vy}, provided that appropriate $\o_{a}\hookrightarrow SU(2)$ embeddings are chosen.

However, there are important differences. First, the approach described here provides a
clear dynamical mechanism which chooses a unique vacuum. This depends crucially on the first term
in~\eq{V-general-2}, that removes the degeneracy of all possible embeddings of $SU(2)$, which have
vanishing field strength $F_{ab}$.  Moreover, it may provide an additional mechanism for
further symmetry breaking as discussed in subsection~\ref{sec:KK3}.  Another difference is that the
starting point in~\cite{Aschieri:2003vy} is a six-dimensional gauge theory with some given gauge
group, such as $U(1)$. Here the six-dimensional gauge group depends on the parameters of the model.

\section{Discussion}
\label{fuzzyextradim-PhD-Discussion}

I have presented a renormalizable four-dimensional $SU(\cN)$ gauge theory with a suitable multiplet
of scalars, which dynamically develops fuzzy extra dimensions that form a fuzzy sphere. The model
can then be interpreted as six-dimensional gauge theory, with gauge group and geometry depending on
the parameters in the original Lagrangian. I explicitly found the tower of massive Kaluza-Klein
modes, consistent with an interpretation as compactified higher-dimensional gauge theory, and
determine the effective compactified gauge theory. Depending on the parameters of the model the
low-energy gauge group can be $SU(n)$, or broken further e.g. to $SU(n_1) \times SU(n_2) \times
U(1)$, with mass scale determined by the extra dimension.

There are many remarkable aspects of this model.  First, it provides an extremely simple and
geometrical mechanism of dynamically generating extra dimensions. The model is based on a basic
lesson from noncommutative gauge theory, namely that noncommutative or fuzzy spaces can be obtained
as solutions of matrix models. The mechanism is quite generic, and does not require fine-tuning or
supersymmetry. This provides in particular a realisation of the basic ideas of compactification and
dimensional reduction within the framework of renormalizable quantum field theory. Moreover, I am
essentially considering a large $\cN$ gauge theory, which should allow to apply the analytical
techniques developed in this context.

One of the main features of the mechanism I presented here is that the effective properties of the
model including its geometry depend on the particular parameters of the Lagrangian, which are
subject to renormalisation. In particular, the  RG flow of these parameters depends on the specific
vacuum i.e. geometry, which in turn will depend on the energy scale. For example, it could be that
the model assumes a `type 3' vacuum as discussed in subsection \ref{sec:KK3} at low energies, which
might be quite close to the standard model. At higher energies, the parameter $\tilde b$ (which
determines the effective gauge group and which is expected to run quadratically under the RG flow)
will change, implying a very different vacuum with different gauge group etc. This suggests a rich
and complicated dynamical hierarchy of symmetry breaking, which remains to be elaborated.

In particular, I have shown that the low-energy gauge group is given by
$SU(n_1) \times SU(n_2)\times U(1)$ or $SU(n)$, while gauge groups which are products of more than
two simple components (apart from $U(1)$) do not seem to occur in this model. The values of $n_1$
and $n_2$ are determined dynamically. Moreover, the existence of a magnetic flux in the vacua with
non-simple gauge group is very interesting in the context of fermions, since internal fluxes
naturally lead to chiral massless fermions~\cite{Steinacker:2007ay}. However this not possible for a
minimal set of fermions and their spectrum have to be doubled.

There is also an intriguing analogy between the model under discussion and string theory, in the
sense that as long as $a=0$, there are a large number of possible vacua (given by all possible
partitions \eq{solution-general}) corresponding to compactifications, with no dynamical selection
mechanism to choose one from the other. Remarkably this analog of the `string vacuum problem' is
simply solved by adding a term to the action.

Finally I should point out some potential problems or shortcomings of the model. First, the
existence of the most interesting vacuum structure of type 3 [\eq{vacuum-mod4} or \eq{vacuum-mod5}]
has not been yet fully established. Furthermore, the presented results are valid only for a small
range of the parameter space of the theory. A complete analysis is expected to give a rich hierarchy
of symmetry breaking patterns.
Finally, the use of scalar Higgs fields $\phi_a$ without
supersymmetry may seem somewhat problematic due to the strong
renormalisation behaviour of scalar fields. This is in some sense consistent with the interpretation
as higher-dimensional gauge theory, which would be non-renormalizable in the classical case.
Moreover, a large value of the quadratically divergent term $\tilde b$ is quite desirable here as
explained in subsection \ref{sec:emergence}, and does not require particular fine-tuning.

\chapter{Noncommutative Spacetime and Gravity}\label{linPhD}

In the previous chapters, I followed a rather modest generalisation
keeping the continuum characteristic for the ordinary spacetime whereas promoting the extra
dimensional space to noncommutative `manifold'. It is evident that these ideas can be tested also in
ordinary spacetime. By following this approach I assume that in the energy regime where the
noncommutativity is expected to be valid, the ordinary spacetime cannot be described as a continuum.
Elementary cells of length scale $\mu^{-1}_{P}$ (with $\mu_{P}$ denoting the Planck mass) are
expected to form out; the ordinary spacetime then is just a limit of this noncommutative `phase'.

The purpose of the study of noncommutative spacetime  is three-fold. First to explore new
mathematical ideas and generalisations of the Einstein gravity. Secondly, it
has been claimed~\cite{Madore:2000aq} that the noncommutative generalisation of extra-dimensions can
help to avoid quantum divergences generated by operators living over them; the minimum length scale
of the noncommutative space elementary cells stand as an ultraviolet cutoff of the otherwise
divergent quantities. Of course this is an idea that could be tested in the ordinary spacetime also.
Last and probably most importantly, having assumed the extra dimensions in the Planck mass regime to
appear noncommutative characteristics would  be no profound reason for this behaviour not to be the
case in the ordinary dimensions too.

The search for consistent noncommutative deformations of  Einstein gravity has been a subject of
interest for considerable amount of time. An incomplete list of references
is~\cite{Chamseddine:1992yx,Madore:1993br,Chamseddine:1996zu,Chamseddine:2000si}. Particularly
noteworthy are the gravity theories built on fuzzy spaces. A simple model in two dimensions with
Euclidean signature has been developed~\cite{Madore:1991bw,Grosse:1992bm,Madore:1995ms}.
More recently, a noncommutative deformation of gravity has been investigated for
the case of two-dimensional fuzzy space~\cite{Buric:2005yi}, and some conclusions
concerning the emergence of gravity as a macroscopic phenomenon of noncommutativity were
made~\cite{Buric:2006di,Buric:2007hb}. In~\cite{Aschieri:2005yw,Aschieri:2005zs}  the
problem has been studied as deformation of the algebra of diffeomorphisms
whereas in~\cite{Steinacker:2007dq} a gravitational gravitational  was emerged from a noncommutative
gauge theory of a matrix model.

In the following sections, I provide a noncommutative  generalisation of Einstein gravity for
the approximation of `small' noncommutativity. For the extra assumption of linear
perturbations of both the Poisson structure which defines  the algebra and the tetrad of our
noncommutative space, a relation between the two notions is found. Linear perturbations of the
metric are found to receive corrections controlled by the noncommutative structure of the algebra.
Such a result suggests possible relation between gravity and the microscopic noncommutative
structure of space.

\section[Noncommutativity/Gravity correspondence. A Suggestion]{Gravity and its correspondence with
noncommutative algebras. A Suggestion}

It is known fact that to a noncommutative geometry one can associate in various ways a
gravitational field. This can be elegantly done~\cite{Connes:1988ym,Connes:1996gi} in the
imaginary-time formalism  and perhaps less so~\cite{Madore:1996bb} in the real-time formalism.
Here following~\cite{Buric:2006di,Buric:2007hb} I will show a gravitational field is intimately
associated with a lack of commutativity of the local coordinates of the space-time structure on
which the field is defined. The gravitational field can be described by a moving frame with local
components $e^\mu_\alpha$; the lack of commutativity by a commutator $J^{\mu\nu}$.  To a coordinate
$x^\mu$ one associates a conjugate momentum $p_\alpha$ and to this couple a commutator
\begin{equation}
[p_\alpha, x^\mu] = e^\mu_\alpha\,.                 \label{pxe}
\end{equation}

Let me introduce a set $J^{\mu\nu}$ of elements of an associative algebra $\cA$ (`noncommutative
space' or `fuzzy space') defined by commutation relations
\begin{equation}
[x^\mu, x^\nu] = i\kbar J^{\mu\nu}(x^\sigma)\,.      \label{xx}
\end{equation}
The constant $\kbar$ is a square of a real number which defines the length scale on which the
effects of noncommutativity become important.  The $J^{\mu\nu}$ are restricted by Jacobi identities;
I shall show below that there are two other requirements which also restrict them.

I suppose the differential calculus over $\cA$ to be defined by a frame, a set of one-forms
$\theta^\alpha$ which commute with the elements of the algebra. We assume the derivations dual to
these forms to be inner, given by momenta $p_\alpha$ as in ordinary quantum mechanics
\begin{equation}
e_\alpha = \adj (p_\alpha)\,.
\end{equation}
Recall that here the momenta $p_\alpha$ include a factor $(i\kbar)^{-1}$.  The momenta stand in
duality to the position operators by the relation~(\ref{pxe}).  However, now consistency relations
in the algebra restrict $\theta^\alpha$ and $J^{\mu\nu}$. Most important thereof is the Leibniz rule
which defines differential relations
\begin{equation}
i\kbar [p_\alpha, J^{\mu\nu}] = [x^{[\mu},[p_\alpha,x^{\nu]}]]
=  [x^{[\mu}, e^{\nu]}_\alpha]                      \label{Jg}
\end{equation}
between the $J^{\mu\nu}$ on the left and the frame components $e_\alpha^\mu$ on the right.

Now one can state the relation (\ref{pxe}) between noncommutativity and gravity more precisely. The
right-hand side of this identity defines the gravitational field. The left-hand side must obey
Jacobi identities. These identities yield relations between quantum mechanics in the given curved
space-time and the noncommutative structure of the algebra. The three aspects of reality then, the
curvature of space-time, quantum mechanics and the noncommutative structure are intimately
connected.

We resume the various possibilities in a diagram, starting with a classical
\nobreak{metric}~$\tld{g}_{\mu\nu}$.
\begin{equation}
\begin{array}{ccccccc}
& &\tld{g}_{\mu\nu} &\longrightarrow &\tld{\theta}^\alpha
&\longleftrightarrow &\tld{\Lambda}^\alpha_\mu
\\[8pt]
& & & &\downarrow & &
\\[8pt]
& &g_{\mu\nu} &\longleftarrow &\theta^\alpha
&\longleftrightarrow &\Omega(\cA)
\\[8pt]
& & & &\downarrow & &
\\[8pt]
& & & &J^{\mu\nu} &\longrightarrow &\cA
\end{array}
\end{equation}
The most important flow of information is from the classical metric $\tld{g}_{\mu\nu}$ to the
commutator $J^{\mu\nu}$, defined in three steps.  The first step is to associate to the metric a
moving frame $\tld{\theta}^\alpha$, which can be written in the form 
$\tld{\theta}^\alpha = \tld{\theta}^\alpha_\mu dx^\mu$.  The frame is then `quantised' according to
the ordinary rules of quantum mechanics; the dual derivations $\tld{e}_\alpha$ are replaced by inner
derivations $e_\alpha = \adj (p_\alpha)$ of a noncommutative algebra. The commutation
relations are defined by the $J^{\mu\nu}$, obtained from the $\theta^\alpha$ by solving a
differential equation.  If the space is flat and the frame is the canonical flat frame then the
right-hand side of~(\ref{Jg}) vanishes and it is possible to consistently choose $J^{\mu\nu}$ to be
constant or zero. The same is also possible if the $e_{\a}^{\mu}$ happens to
belong in the center of algebra $\cA$, $\cZ(\cA)$. Note that the map~(\ref{map}) is not single
valued since any constant $J^{\mu\nu}$ has flat space as inverse image. Conversely a non-trivial
noncommutative algebra defined by a non-constant Poisson structure
$J^{\mu\nu}=J^{\mu\nu}(x^{\sigma})$ generally makes the noncommutative space defined by $x^{\mu}$
curved.

Here, following~\cite{Buric:2006di}, I define the map
 \begin{equation}
J^{\mu\nu} \mapsto   \mbox{\Curv} (\theta^\alpha) \label{map}
\end{equation}
from the Poisson structure $J^{\mu\nu}$ to the curvature of a frame and discuss some interesting
conclusions concerning the model I suggest; these were further improved in~\cite{Buric:2007hb} .
The existence of~\eq{map} allows me to express the {Ricci tensor}\footnote{More correctly the
noncommutative corrections that the `would-be' Ricci tensor receives at its commutative limit.} in
terms of $J^{\mu\nu}$. The interest at the moment of this point is limited by the fact that I have
no `equations of motion' for $J^{\mu\nu}$.

However, since the constraints of our Jacobi system cannot be solved exactly I work in the
first order approximation of noncommutativity, controlled by $\kbar$, and in linear perturbation of
both flat space and some constant Poisson structure $J_{0}^{\mu\nu}$; I prove that the former
receive corrections from the latter suggesting a deeper relation between noncommutativity and
gravity. I denote the order parameters for the perturbation of the metric and the perturbation of
the Poisson Structure as $\e_{GF}$ and $\e$ respectively.

To be more specific let $\mu$ be a typical `large' source mass with `Schwarzschild radius'
$G_N\mu$. Then one has two length scales, determined by respectively the square $G_N\hbar$ of
the Planck length and by $\kbar$. The gravitational field is weak if the dimensionless parameter
$\epsilon_{GF} = G_N \hbar \mu^2$ is small; the space-time is almost commutative if the
dimensionless parameter $\epsilon = \kbar \mu^2$ is small. These two parameters are not necessarily
related but since I am interested to allow constant noncommutativity in the flat space limit (c.f.
subsection~\ref{gf}), I shall here assume that they are of the same order of magnitude
\begin{equation}
\epsilon_{GF} \simeq \epsilon\,.             \label{ekm}
\end{equation}

In section~\ref{WKB} I give an example of explicit calculation of the~\eq{map} for the case of
covariant WKB approximation which I mimic here as far as possible. An interesting related
conclusion concerns the mode decomposition of the image metric. This was investigated
in~\cite{Buric:2007hb}. It was found there that in the linear approximation there are three modes in all;
two dynamical modes of a spin-2 particle plus a scalar mode. They need however not all be
present: the graviton will be polarised by certain background noncommutative `lattice' structure.
This leads to the problem of the propagation of the modes in the `lattice'. Furthermore  an
energy-momentum for the Poisson structure defined and its eventual contribution of
this energy-momentum to the gravitational field equations was studied.

The motivation for considering noncommutative geometry as an `avatar' of gravity is the belief that
it sheds light on the role of the gravitational field as the universal regulator of ultra-violet
divergences.  Details on these ideas can be found elsewhere~\cite{Madore:2000aq} and a simple
explicit solution in~\cite{Buric:2005yi}. The model I present is in definite overlap
with an interesting recent interpretation~\cite{Steinacker:2007dq} of the map~(\ref{map}) as a
redefinition of the gravitational field in terms of noncommutative electromagnetism.

\newpage
\section{The Correspondence}

To fix the notation let me briefly recall some elements of the noncommutative frame formalism.

\subsection{Preliminary formalism}

Let me start with a noncommutative $*$-algebra $\cA$ generated by four hermitean elements $x^\mu$
which satisfy the commutation relations~(\ref{xx}).  Assume that over $\cA$ is a differential
calculus which is such that the module of one-forms is free and possesses a preferred frame
$\theta^\alpha$ which commutes,
\begin{equation}
[x^\mu, \theta^\alpha] = 0                             \label{mod}
\end{equation}
with the algebra. The metric on this preferred frame, since  $[f(x^{\m}),\th^{\a}]=0$ also, takes
the form
\begin{equation}
g=g_{\m\n}(dx^{\m}\otimes dx^{\n})
=g_{\m\n}\Big[(e_{\a}^{\m}\th^{\a})\otimes(e_{\b}^{\n}\th^{\b})\Big]
=\underbrace{\Big(g_{\m\n}e_{\a}^{\m}e_{\b}^{\n}\Big)}_{=g_{\a\b}}\Big(\th^{\a}\otimes\th^{\b}\Big)
\end{equation}
with the $g_{\a\b}$ to be rescalable in the flat Minkowski metric, $diag(-1,1,1,1)$. The space one
obtains in the commutative limit is therefore parallelizable with a global moving frame
$\tld{\theta}^\alpha$ defined to be the commutative limit of $\theta^\alpha$. Then the differential
is given by
\begin{equation}
dx^\mu = e_\alpha^\mu \theta^\alpha, \qquad e_\alpha^\mu = e_\alpha x^\mu\,.
\label{tetrad}
\end{equation}

The differential calculus is defined as the largest one consistent with the module structure of the
one-forms so constructed.  The algebra is defined by a product which is restricted by the matrix of
elements $J^{\mu\nu}$; the metric is defined by the matrix of elements $e_\alpha^\mu$.  Consistency
requirements, essentially determined by  Leibniz rules, impose relations between these two matrices
which in simple situations allow me to find a one-to-one correspondence between the structure of the
algebra and the metric.  The input of which I shall make the most use is the Leibniz
rule~(\ref{Jg}) which can also be written as relation between one-forms
\begin{equation}
i\kbar d J^{\mu\nu} = [dx^\mu, x^\nu] + [x^\mu, dx^\nu]\,.   \label{lr}
\end{equation}
or
\begin{equation}
i\kbar e_{\a} J^{\mu\nu}=[e_{\a}^{[\mu},x^{\nu]}]\,,
\label{lrn}
\end{equation}
using the definitions of \eq{tetrad}. One can see here a differential equation for $J^{\mu\nu}$ in
terms of $e^\mu_\alpha$. It is obvious that if the space is flat and the frame is the canonical
flat frame then the right-hand side of~\eq{lrn} vanishes and it is possible to consistently choose
$J^{\mu\nu}$ to be constant or zero; the same is also possible if the $e_{\a}^{\mu}$ happens to
belong in the center of $\cA$ algebra, $\cZ(\cA)$. On the other hand, the noncommutative
algebra \textit{must} be defined by a non-constant Poisson structure
$J^{\mu\nu}=J^{\mu\nu}(x^{\sigma})$, i.e. being \textit{non-trivial}, when $e_{\a}^{\mu}$ does not
commute with the elements of the algebra and in addition the $[e_{\a}^{\mu},x^{\nu}]$ matrix has
antisymmetric component under $\mu\lra\nu$ exchange. Relation \eq{lrn} cannot be solved
exactly but in important special cases reduces to a simple differential equation of one variable.

It is important to note that~\eq{lrn} or in its equivalent form~\eq{Jg} is essentially the Jacobi
identity concerning one momentum and two noncommutative `coordinates'. Keeping the associativity of
the noncommutative algebra three other Jacobi identities must be satisfied. The one
concerning the $x$'s
\begin{equation}
[x^{\mu},[x^{\nu},x^{\l}]]+[x^{\nu},[x^{\l},x^{\mu}]]+[x^{\l},[x^{\mu},x^{\nu}]]=0\,,\label{JIxxx}
\end{equation}
constrain additionally our noncommutative system. In subsection \ref{PhaseSpace} I show that the
Jacobi identities concerning two or three momenta are automatically satisfied at least for our
working approximations.

In addition, I must insure that the differential is well defined. A necessary condition is that
$d[x^\mu, \theta^\alpha] = 0$, from which it follows that the momenta $p_\alpha$ must satisfy
the quadratic relation~\cite{Madore:2000aq}
\begin{equation}
2p_{\a}p_{\b}P^{\a\b}{}_{\c\d}-p_{\a}F^{\a}{}_{\c\d}-K_{\a\d}=0\,
\end{equation}
with $P^{\a\b}{}_{\c\d}$, $F^{\a}{}_{\c\d}$, and $K_{\a\d}$ in the center of $\cA$, $\cZ(\cA)$. On
the other hand, from~(\ref{mod}) it follows that
\begin{equation}
d[x^{\mu}, \theta^{\alpha}] =
[dx^{\mu}, \theta^{\alpha}] + [x^\mu, d\theta^{\alpha}] =
e^\mu_\beta [\theta^{\beta}, \theta^{\alpha}] -
\tfrac 12 [x^\mu, C^\alpha{}_{\beta\gamma}] \theta^{\beta}\theta^{\gamma}\,,
\end{equation}
where I  have  introduced the Ricci rotation coefficients
\begin{equation}
d\theta^\alpha =-\frac 12 C^\alpha{}_{\beta\gamma} \theta^\beta \theta^\gamma\,.
\end{equation} 
Therefore I find that multiplication of one-forms must satisfy
\begin{equation}
[\theta^{\alpha}, \theta^{\beta}] =
\tfrac 12 \theta^\beta_\mu [x^\mu, C^\alpha{}_{\gamma\delta}]
\theta^{\gamma}\theta^{\delta}\,.                               \label{bffp}
\end{equation}
Using the consistency conditions I obtain that
\begin{equation}
\theta^{[\beta}_\mu [x^\mu, C^{\alpha]}{}_{\gamma\delta}] = 0\,,                 \label{xC}
\end{equation}
and also that the expression $ \theta^{(\alpha}_\mu [x^\mu, C^{\beta)}{}_{\gamma\delta}] $ must be
central. Note that these conditions are valid for our working approximations of the next
two subsections.

\subsection{The quasi-commutative approximation}         \label{g2a}

{To lowest order in $\e$ the partial derivatives are well defined} and the approximation,
which I shall refer to as the quasi-commutative,
\begin{equation}
[x^\lambda, f] = i\kbar J^{\lambda\sigma} \prt_\sigma f + \cO\left((i\kbar)^{2}\right), \qquad
e_{\alpha}(f)=[p_\alpha, f] =\prt_\alpha f + \cO(i\kbar).
\label{xCom-e(f)-approx}
\end{equation}
is valid.  The Leibniz rule~\eq{lrn} and the Jacobi identity~\eq{JIxxx} can be written in this
approximation as
\begin{align}
& e_\alpha J^{\mu\nu} = \partial_\sigma e^{[\mu}_\alpha J^{\sigma\nu ]}\,,         \label{L*}\\
&\vare_{\kappa\lambda\mu\nu} J^{\gamma\lambda}e_\gamma J^{\mu\nu}=0\,. \label{J*}
\end{align}
I shall refer to these equations including their integrability conditions as the Jacobi equations.

Written in frame components the Jacobi equations become
\begin{align}
&e_\gamma J^{\alpha\beta} - C^{[\alpha}{}_{\gamma\delta} J^{\beta]\delta} = 0\,,      \label{cl}\\
& \vare_{\alpha\beta\gamma\delta}J^{\gamma\eta} (e_\eta J^{\alpha\beta}
+ C^\alpha{}_{\eta\zeta} J^{\beta\zeta}) = 0\,.              \label{mb***}
\end{align}
I have used here the expression for the rotation coefficients, valid also in the quasi-commutative
approximation:
\begin{equation}
C^\alpha{}_{\beta\gamma} = \theta^\alpha_\mu e_{[\beta}e^\mu_{\gamma]}
=-e^\nu_\beta e^\mu_\gamma\partial_{[\nu}\theta^\alpha_{\mu]}\,.\label{rc}
\end{equation}
Taking the antisymmetric part of \eq{mb***} over the $\a$ and $\b$ indices and taking in account
the \eq{cl} one finds the relation
\begin{equation}
\epsilon_{\alpha\beta\gamma\delta}
J^{\alpha\zeta}J^{\beta\eta}C^{\gamma}{}_{\eta\zeta} = 0\,.   \label{b**}
\end{equation}
Now it is possible to solve the \eq{cl} for the rotation coefficients which are found to be
\begin{equation}
J^{\gamma\eta}e_\eta J^{\alpha\beta} = J^{\alpha\eta}J^{\beta\zeta}C^\gamma{}_{\eta\zeta}\,,
\end{equation}
or, provided $J^{-1}$ exists,
\begin{equation}
C^\alpha{}_{\beta\gamma}
 = J^{\alpha\eta} e_\eta J^{-1}_{\beta\gamma}\,.            \label{mbb}
\end{equation}
Note that from general considerations also follows that the rotation coefficients must
satisfy the gauge condition
\begin{equation}
e_\alpha C^{\alpha}{}_{\beta\gamma} =  0\,.                 \label{gauge-co}
\end{equation}

Equation \eq{mbb} means that in the quasi-classical approximation the linear connection and
therefore the curvature can be directly expressed in terms of the commutation relations.  This is
the content of the map \eq{map}. Indeed since
\begin{equation}
\omega_{\alpha\beta\gamma}
=\tfrac 12 (C_{\alpha\beta\gamma}
- C_{\beta\gamma\alpha} + C_{\gamma\alpha\beta})
\end{equation}
for the Ricci curvature tensor for example one obtains
\begin{align}
&2R_{\beta\zeta}
= J_{(\zeta\delta }e^\alpha e^\delta J^{-1}_{\beta)\alpha }
+ J^{\alpha\delta }e_{(\zeta}e_\delta J^{-1}_{\alpha\beta)}\nonumber\\
&\phantom{2R_{\alpha\zeta}}
- J_{(\zeta }{}^\eta e^\alpha J^{-1}_{\eta\gamma}
J^{\gamma\delta }e_\delta J^{-1}_{\beta)\alpha}
+ J^{\alpha\delta }e_\delta J^{-1}_{\eta\beta}
J^{\eta\gamma }e_\gamma J^{-1}_{\alpha\zeta}\nonumber\\
&\phantom{2R_{\alpha\zeta}}
+ J_{\eta\delta }e^\delta J^{-1}_{\beta\alpha}
J^{\eta\gamma }e_\gamma J^{-1}_{\zeta}{}^\alpha
+ J^{\alpha\eta }e_{(\zeta} J^{-1}_{\eta\gamma}
J^{\gamma\delta }e_\delta J^{-1}_{\beta)\alpha}\nonumber\\
&\phantom{ 2R_{\alpha\zeta}}-\tfrac 12
J_{\zeta\delta }e^\delta J^{-1 \alpha\eta}
J_{\beta\gamma }e^\gamma J^{-1}_{\alpha\eta}
+ J^{ \alpha\delta}e_\delta J^{-1}_{\alpha\eta}
J_{(\zeta\gamma }e^\gamma J^{-1}_{\beta)}{}^\eta\,,     \label{R2ndo}
\end{align}
where the ordering on the right-hand side, have been neglected as it gives the corrections of
second-order in $\e$, which in our current investigation are omitted. To understand better the
relation between the commutator and the curvature in the following section I shall consider a
linearisation about a fixed `ground state'.

\subsection{Weak field approximation}   \label{gf}

If I consider the $J^{\mu\nu}$ of the previous sections as the components of a classical field on a
curved manifold then in the limit when the manifold becomes flat the `equations of motion'
\begin{equation}
e_{\alpha}J^{\mu\nu}=0\,,
\end{equation}
become Lorentz invariant and obtain as the possible solutions either zero or constant
noncommutativity. However, I suppose that as $e^\lambda_\alpha \to e^\lambda_{0\alpha}$ I obtain
\begin{equation}
J^{\mu\nu} \to J_0^{\mu\nu}, \qquad \det J_0 \neq 0\,, \label{SLB}
\end{equation}
which breaks Lorentz Invariance in vacuum; $J_{0}^{\mu\nu}=0$ would be a stronger assumption.

Let me now consider fluctuations around a particular given solution to the problem I have set. I
suppose that I have a reference solution comprising a frame
 $e^\mu_{0\alpha} =\delta^\mu_{\alpha}$ and a commutation relation $J_0^{\mu\nu}$ which I perturb as
\begin{equation}
J^{\mu\nu} = J_0^{\mu\nu} + \epsilon\, I^{\mu\nu}\,, \qquad
e^\mu_\alpha = \delta^\mu_{\beta}
(\delta^\beta_\alpha + \epsilon_{GF}\,\Lambda^\beta_\alpha)\,.     \label{pert}
\end{equation}
Due to the assumption (\ref{SLB}) as $e^{\l}_{\a}$ reach the constant `curvature' frame limit it is
reasonable to assume $\epsilon\backsim\epsilon_{GF}$ with the gravitational field order parameter to
be $\epsilon_{GF}=G_{N}\hbar\mu^{2}$. Recall that $\mu$ is some typical large gravitational mass and
$G_{N}\hbar$ the square Planck length. In order the second and higher order approximation to be
negligible I furthermore assume $\epsilon_{GF}\ll 1$. Then in terms of the unknowns $I$ and
$\Lambda$ the Jacobi and Leibniz constraints~\eq{lrn} and~\eq{JIxxx} become
respectively
\begin{align}
&\vare_{\lambda\mu\nu\sigma} [x^\lambda, I^{\mu\nu}] = 0\,,    \label{*1a}\\
&i\kbar e_\alpha I^{\mu\nu} =
[\Lambda^\mu_\alpha, x^\nu] - [\Lambda^\nu_\alpha, x^\mu]\,.  \label{*2a}
\end{align}

In the quasi-commutative approximation [relations~\eq{xCom-e(f)-approx}], the two
constraint equations become
\begin{align}
&\vare_{\lambda\mu\nu\sigma}
J_{0}^{\lambda\sigma} \prt_\sigma I^{\mu\nu} = 0\,,                  \label{*1b}\\
&e_\alpha I^{\mu\nu} =
\prt_\sigma \Lambda^{[\mu}_\alpha J_0^{\sigma\nu]}\,.             \label{*2b}
\end{align}
These two equations are the origin of the particularities  of the construction I present here, they
and the fact that the `ground-state' value of $J^{\mu\nu}$ is an invertible matrix.

The constraint equations become particularly transparent if one first rewrite them in frame
components and then introduce the new unknowns
\begin{equation}
\hat{I}_{\alpha\beta} =
J^{-1}_{0\alpha\gamma} J^{-1}_{0\beta\delta} I^{\gamma\delta}\,, \qquad
\hat{\Lambda}_{\alpha\beta} = J^{-1}_{0\beta\gamma}\Lambda^\gamma_\alpha\,. \label{Hat}
\end{equation}
Let me also decompose $\hat{\Lambda}$ as the sum
\begin{equation}
\hat{\Lambda}_{\alpha\beta} =
\hat{\Lambda}^+_{\alpha\beta} + \hat{\Lambda}^-_{\alpha\beta}
\end{equation}
of a symmetric and antisymmetric term. The constraints become
\begin{gather}
\partial_{\a}\hat{I}_{\b\c}+\partial_{\b}\hat{I}_{\c\a}+\partial_{\c}\hat{I}_{\a\b}=0 \label{L1b} \\
e_{\a}\hat{I}_{\b\c}=
\partial_{[\b}\hat{\L}^{+}_{\c]\a}+\partial_{[\b}\hat{\L}^{-}_{\c]\a}\label{L2b}\,.
\end{gather}
I introduce
\begin{equation}
\hat{I} = \frac{1}{2} \hat{I}_{\alpha\beta}\theta^\alpha \theta^\beta\,, \qquad
\hat{\Lambda}^- = \frac{1}{2} \hat{\Lambda}^-_{\alpha\beta}\theta^\alpha \theta^\beta\,.
\end{equation}
However, due to the second of (\ref{xCom-e(f)-approx}), the (\ref{L1b}) constraint can be written in
a more compact form (`cocycle' condition)
\begin{equation}
\vare^{\a\b\c\d}e_{\a}\hat{I}_{\b\c}=0,\qquad \mbox{or}\qquad d\hat{I}=0\,, \label{L1b-new}
\end{equation}
which is correct up to the first order of my working approximation. Then by taking the
(\ref{L2b}) into account one finds
\begin{equation}
\vare^{\a\b\c\d}\partial_{\a}\hat{\L}^{-}_{\b\c}=0\,,
\end{equation}
since the symmetric part of $\hat{\L}$ has vanishing contribution in the relation. For the
approximation of almost flat metric this can be written as
\begin{equation}
\vare^{\a\b\c\d}e_{\a}\hat{\L}_{\b\c}^{-}=0,\qquad\mbox{or}\qquad d\hat{\L}^{-}=0\,, \label{L1b-L2b}
\end{equation}
which can be rewritten as
\begin{equation}
e_{[\b}\hat{\L}_{\c]\a}^{-}=-e_{\a}\hat{\L}_{\b\c}^{-}\,. \label{L1b-L2b-new}
\end{equation}
Substituting the last relation in the (\ref{L2b}) constraint one obtains
\begin{equation}
e_{\a}(\hat{I}+\hat{\L}^{-})_{\b\c}=\partial_{[\b}\hat{\L}_{\c]\a}^{+} \label{+3}
\end{equation}
as the final result.
This equation has the integrability conditions
\begin{equation}
e_\alpha e_{[\beta} \hat{\Lambda}^+_{\gamma]\delta} -
e_\delta e_{[\beta} \hat{\Lambda}^+_{\gamma]\alpha} = 0\,.            
\end{equation}
But the left-hand side is the linearised approximation to the curvature of a metric with components 
$g_{\mu\nu} +\epsilon\hat{\Lambda}^+_{\mu\nu}$. If it vanishes then the perturbation is a
derivative; for some one-form $A$
\begin{equation}
\hat{\Lambda}^+_{\beta\gamma} = \frac{1}{2} e_{(\beta} A_{\gamma)}\,.   \label{dL+}
\end{equation}
Equation~(\ref{+3}) becomes therefore
\begin{equation}
e_\alpha (\hat{I} + \hat{\Lambda}^- - dA)_{\beta\gamma} = 0\,.    \label{+}
\end{equation}
It follows then that for some two-form $c$ with constant components $c_{\beta\gamma}$
\begin{equation}
\hat{\Lambda}^- = - \hat{I} + dA + c\,.                         \label{LI}
\end{equation}
The remaining constraints are satisfied identically. The most important relation is
Equation~(\ref{LI}) which, in terms of the original `unhatted' quantities, becomes
\begin{equation}
\Lambda^\alpha_\beta =
-J^{-1}_{0\beta\gamma} I^{\alpha\gamma} +
J_0^{\alpha\gamma}(c_{\beta\gamma}+e_{\gamma} A_{\beta})\,.         \label{wh}
\end{equation}
This condition is much weaker than, but similar to Equation~(\ref{pJx}).

\subsection{The algebra to geometry map}                   \label{agm}

One can now be more precise about the map~(\ref{map}).  Let $\theta^\alpha$ be a frame which is a
small perturbation of a flat frame and let $J^{\alpha\beta}$ be the frame components of a small
perturbation of a constant `background' $J_0$. Then the map~(\ref{map}) is defined by
\begin{equation}
I^{\alpha\beta}\mapsto \Lambda^\alpha_\beta=
-J_0^{\alpha\gamma}( \hat{I}_{\gamma\beta}+ e_\beta A_\gamma)\,.
\label{why}
\end{equation}

The perturbation on the frame
$e^\mu_\alpha = \delta^\mu_{\beta}(\delta^{\beta}_{\alpha} + \e\Lambda^\beta_\alpha)$
engenders a perturbation of the metric. Indeed the {bilinearity of the metric implies that
the frame components of the metric $g^{\alpha\beta}$ are complex numbers.} For the choice of frame
I have made and for the working approximation of linear perturbation of flat space I have
\begin{equation}
g^{\alpha\beta} = \eta^{\alpha\beta} - \ep h^{\alpha\beta}\,.  \label{hh}
\end{equation}
or in coordinate indices
\begin{equation}
g^{\mu\nu} = g(dx^\mu \otimes dx^\nu) = e^\mu_\alpha e^\nu_\beta g^{\alpha\beta}\,.
\end{equation}
We write $g^{\mu\nu}$ as a sum
\begin{equation}
g^{\mu\nu} = g_+^{\mu\nu} + g_-^{\mu\nu}
\end{equation}
of symmetric and antisymmetric parts. In the previous section I proved that in lowest
order in the noncommutativity in general I have $h^{\alpha\beta} = -h^{\beta\alpha}$ so 
$g^{\mu\nu}_{+}$ and $g^{\mu\nu}_{-}$ can be decomposed as
\begin{equation}
g_+^{\mu\nu}
= \tfrac 12 \eta^{\alpha\beta} [e^{\mu}_\alpha, e^{\nu}_\beta]_+
- \tfrac 12 \ep h^{\alpha\beta} [e^{\mu}_\alpha, e^{\nu}_\beta]_{-}
\end{equation}
and
\begin{equation}
g_-^{\mu\nu}
= \tfrac 12 \eta^{\alpha\beta} [e^{\mu}_\alpha, e^{\nu}_\beta]_{-}
- \tfrac 12  \ep h^{\alpha\beta} [e^{\mu}_\alpha, e^{\nu}_\beta]_+\,.
\end{equation}
Then for the linear perturbations of the flat space consider one can set
\begin{equation}
g^{\mu\nu} = \eta^{\mu\nu} - \epsilon g_1^{\mu\nu}, \qquad
e^\mu_\alpha = \delta^\mu_\alpha + \epsilon \Lambda^\mu_\alpha
\end{equation}
and the leading order of perturbation in the metric is given as
\begin{equation}
g_1^{\mu\nu} =
 \eta^{\alpha\beta} \Lambda^{(\mu}_\alpha \delta^{\nu)}_\beta = \Lambda^{(\mu\nu)}\,.
\end{equation}
From~\eq{why} then
\begin{equation}
g_1{}_{\alpha\beta} = - J_{0(\alpha}{}^{\gamma} (\hat{I}_{\gamma\beta)} + e_{\beta)} A_\gamma)
\label{hI}
\end{equation}
The correction (\ref{hh}) will appear only in second order.

The frame itself is given by
\begin{equation}
\theta^\alpha = d ( x^\alpha -\e J_0^{\alpha\gamma}A_\gamma)
- \e J_0^{\alpha\gamma} \hat{I}_{\gamma\beta} dx^\beta\,.
\end{equation}
Therefore one finds  the following expressions
\begin{align}
& d\theta^\alpha
= - \e J_0^{\alpha\gamma}e_\delta \hat{I}_{\gamma\beta} dx^\delta dx^\beta
= \tfrac 12 \e J_0^{\alpha\delta}e_\delta \hat{I}_{\beta\gamma}dx^\gamma dx^\beta\,,\\
&C^\alpha{}_{\beta\gamma }=
\e J_0^{\alpha\delta} e_\delta\hat{I}_{\beta\gamma}\,,\\
&\omega_{\alpha\beta\gamma}
= \tfrac 12 \e (J_{0[\alpha}{}^\delta e_\delta\hat{I}_{\beta\gamma]}
+ J_{0\beta}{}^\delta e_\delta\hat{I}_{\alpha\gamma})\,.  \label{oJ}
\end{align}
The torsion obviously vanishes.

Then the linearised Riemann tensor, using~(\ref{oJ}) and the cocycle condition, is given by the
expression
\begin{equation}
R_{\alpha\beta\gamma\delta} =
\tfrac{1}{2}\e\,e^\eta \left(J_0{}_{\eta[\gamma} e_{\delta]}\hat{I}_{\alpha\beta}
+ J_0{}_{\eta[\alpha} e_{\beta]}\hat{I}_{\gamma\delta}\right)\,.
\label{vtj4}
\end{equation}
For the  Ricci curvature I find
\begin{equation}
R_{\beta\gamma } =
 - \tfrac{1}{2}\e\,e^\zeta \left( J_0{}_{\zeta(\beta}e^\alpha \hat{I}_{\gamma)\alpha}
+ J_0^{\alpha}{}_{\zeta} e_{(\beta} \hat{I}_{\gamma)\alpha} \right)\,.
\label{vtj5}
\end{equation}
One more contraction yields the expression
\begin{equation}
R = - 2\e\,J_0{}^{\zeta\alpha}e_\zeta e^\beta\hat{I}_{\alpha\beta}
\label{vtj6}
\end{equation}
for the Ricci scalar. Using again the cocycle condition permits me to write this in the form
\begin{equation}
R =  \e\,\Delta \chi\,,           \label{vtj7}
\end{equation}
where the scalar trace component is defined as
\begin{equation}
\chi = J_0^{\alpha\beta}\hat{I}_{\alpha\beta}\,.      \label{ch}
\end{equation}
The Ricci scalar is a divergence. Classically it vanishes when the field equations are satisfied.

\subsection{Phase space}\label{PhaseSpace}

It is obviously the case that in the commutative limit the four coordinate generators tend to the
space-time coordinates and the four momenta tend to the conjugate momenta. The eight generators
become the coordinates of phase space. For this to be consistent all Jacobi identities must be
satisfied, including those with two and three momenta.

I consider first the identities
\begin{equation}\label{JI-momenta}
[p_\alpha, [p_\beta, x^\mu] ] +
[p_\beta, [x^\mu, p_\alpha] ] +
[x^\mu, [p_\alpha, p_\beta] ] = 0\,.
\end{equation}
Using the duality of space-time coordinates and momenta $[p_{\a},x^{\mu}]=e^{\mu}_{\a}$ and the definition
$P_{\a\b}=[p_{\a},p_{\b}]$ the above equation is written in a more compact form
\begin{equation}
[x^{\mu},P_{\a\b}]=-\Big\{[p_{\a},e^{\mu}_{\b}]-[p_{\b},e^{\mu}_{\a}]\Big\}\,,
\end{equation}
which reminds the Leibniz rule (\ref{lr}) being in fact its conjugate equivalent. Indeed the (\ref{JI-momenta}) are satisfied by
\begin{equation}
i\kbar
P_{\a\b}=\underbrace{(-J_{0\a\b}^{-1}+\epsilon\hat{I}_{\a\b})}_{=-J_{\a\b}^{-1}}+\cO(\e^{2})\,.
\end{equation}
The remaining identities, involving only the momenta, are then satisfied by virtue of the fact that
the two-form $\hat{\Lambda}$ is closed. There is evidence to the fact that this relation is valid to
all orders in $\epsilon$.

We also find that
\begin{equation}
[p_\alpha - J^{-1}_{0\alpha\mu} x^\mu, x^\nu] =
\delta^\nu_\alpha - J^{-1}_{0\alpha\mu} (J_0^{\mu\nu} + \epsilon I^{\mu\nu})
+ \epsilon \Lambda^\nu_\alpha
= \epsilon (\Lambda^\nu_\alpha - J^{-1}_{0\alpha\mu}  I^{\mu\nu} ) = 0\,.
\end{equation}
For some set of constants $c_\alpha$ therefore, if the center of the algebra is trivial, I can write
\begin{equation}
i\kbar p_\alpha = J^{-1}_{0\alpha\mu} x^\mu + c_\alpha\,.           \label{pJx}
\end{equation}
The `Fourier transform' is linear.

Let $J_0^{\mu\alpha}$ be an invertible matrix of real numbers. For each such matrix there is an
obvious map from the algebra to the geometry given by
\begin{equation}
J^{\mu\nu}{} \mapsto
e^\nu_\alpha  =  J^{-1}_{0\alpha\mu} J^{\mu\nu}\,.              \label{JNe}
\end{equation}
For such frames one introduces momenta $p_\alpha$ and find that
\begin{equation}
[p_\alpha, x^\nu] = e^\nu_\alpha = J^{-1}_{0\alpha\mu} J^{\mu\nu} =
(i\kbar)^{-1} J^{-1}_{0\alpha\mu} [x^\mu, x^\nu]\,.
\end{equation}
That is
\begin{equation}
[i\kbar p_\alpha - J^{-1}_{0\alpha\mu} x^\mu, x^\nu] = 0\,.               \label{px}
\end{equation}
In conclusion then eq.~(\ref{pJx}) is satisfied. It is reasonable to interpret the results of the
previous section as the statement that this condition is stable under small perturbations of the
geometry or algebra.

\subsection{An Example}

Consider ($2-d$)-Minkowski space with coordinates $(t,x)$ which satisfy the commutation relations $[t,x] = ht$ and with a geometry
encoded in the frame $\theta^1= t^{-1} dx$, $\theta^0= t^{-1} dt$. These data describe~\cite{Madore:2000aq} a noncommutative
version of the Lobachevsky plane. The region around the line $t=1$ can be considered as a vacuum. For the approximations of the
previous section to be valid one must rescale $t$ so that in a singular limit the vacuum region
becomes the entire space. I can do this by setting
\begin{equation}
t = 1 + c t^\prime
\end{equation}
and consider the limit $c\to 0$. So that the geometry remain invariant I must scale the metric. I
do this by rescaling
$\theta^0$
\begin{equation}
\theta^0 \mapsto c^{-1}\theta^0\,.
\end{equation}
The commutation relations become then
\begin{equation}
[t^\prime,x] = c^{-1} h + h t^\prime
\end{equation}
and to leading order in $c$ the frame becomes
\begin{equation}
\theta^0 = (1-ct^\prime) dt^\prime\,, \qquad
\theta^1 = (1-ct^\prime) dx\,.
\end{equation}
From the definitions~(\ref{pert}) I find that
\begin{equation}
\begin{array}{ll}
J_0^{01} = c^{-1} h\,, &
\epsilon I^{01} = ht^\prime\,,\\[6pt]
J_{0,01} = - c h^{-1}\,, &
\epsilon \Lambda^\alpha_\beta = c t^\prime \delta^\alpha_\beta
\end{array}
\end{equation}
and therefore I obtain the map~(\ref{map}) as defined in the previous sections. This example is not
quite satisfactory since the cocycle conditions~(\ref{L1b-new}),~(\ref{L1b-L2b}) are vacuous in
dimension two.

\section{The WKB Approximation}\label{WKB}

Let me now suppose that the algebra $\cA$ is a tensor product
\begin{equation}
\cA = \cA_0 \otimes \cA_\omega
\end{equation}
of a `slowly-varying' factor $\cA_0$ in which all amplitudes lie and a `rapidly-varying' phase factor which is of
order-of-magnitude $\epsilon$ so that only functions linear in this factor can appear. The generic element $f$ of the algebra is
of the form then
\begin{equation}
f(x^\lambda, \phi) = f_{0}(x^\lambda) +\epsilon \bar{f}(x^\lambda) e^{i\omega\phi}
\end{equation}
where $f_{0}$ and $\bar{f}$ belong to $\cA_{0}$. Because of the condition on $\epsilon$ the
factor order does not matter and these elements form an algebra. I suppose that both $\Lambda$ and
$I$ belong to $\cA_\omega$:
\begin{equation}
\Lambda^\alpha_\beta
= {\bar{\Lambda}}^\alpha_\beta e^{i\omega\phi}, \qquad
I^{\alpha\beta} = \bar{I}^{\alpha\beta} e^{i\omega\phi}\,,
\end{equation}
where $\bar{\Lambda}^\alpha_\beta$ and $\bar{I}^{\alpha\beta}$ belong to
$\cA_0$. Therefore I have also
\begin{equation}
g_1^{\mu\nu} = \bar{g}^{\mu\nu}e^{i\omega\phi}\,.
\end{equation}
Introducing the normal $\xi_\alpha = e_\alpha \phi$ to the surfaces of constant phase and
$\h^{\a}=J_{0}^{\a\b}\xi_{\b}$ I have
\begin{align}
&e_{\alpha} I_{\beta\gamma}
= (i\omega\xi_\alpha \bar{I}_{\beta\gamma}
+ e_{\alpha}\bar{I}_{\beta\gamma}) e^{i\omega\phi}\,\\
&e_{\alpha} \Lambda_{\beta\gamma}
= (i\omega\xi_\alpha \bar{\Lambda}_{\beta\gamma}
+ e_{\alpha}\bar{\Lambda}_{\beta\gamma}) e^{i\omega\phi}\,.
\end{align}

In the WKB approximation the cocycle condition becomes
\begin{equation}
\xi_\alpha \hat{I}_{\beta\gamma}
+ \xi_\beta \hat{I}_{\gamma\alpha}
+ \xi_\gamma \hat{I}_{\alpha\beta} = 0\,.                    \label{cyc}
\end{equation}
I multiply this equation by $\xi^\alpha$ and obtain
\begin{equation}
\xi^2 \hat{I}_{\beta\gamma} 
+ \xi_{[\beta} \hat{I}_{\gamma]\alpha}\xi^\alpha = 0\,.      \label{cyc2}
\end{equation}
If $\xi^2 \neq 0$ then I conclude that
\begin{equation}
\hat{I}_{\beta\gamma} 
= - \xi^{-2}\xi_{[\beta} \hat{I}_{\gamma]\alpha}\xi^\alpha\,. 
\end{equation}
If on the other hand $\xi^2 = 0$ then I conclude that
\begin{equation}
\xi_{[\beta} \hat{I}_{\gamma]\alpha}\xi^\alpha = 0\,,
\end{equation}
which restrict the possible perturbation over the noncommutativite algebra.
In terms of the scalar $\chi$ I obtain the relation
\begin{equation}
\hat{I}_{\alpha\beta}\eta^\beta
= -\tfrac 12 \chi \xi_\alpha\,.                           \label{ec}
\end{equation}

Using the definition of $\eta$ I find in the WKB approximation to
first order
\begin{align}
&\omega_{\alpha\beta\gamma}
= \tfrac 12 i\omega\e \left(\eta_{[\alpha} \hat{I}_{\beta\gamma]}
+\eta_{\beta} \hat{I}_{\alpha\gamma}\right),\label{oJ1}\,\\
&R_{\alpha\beta\gamma\delta} =- \tfrac 12\e (i\omega)^2
\left(\eta_{[\gamma}\xi_{\delta]} \hat{I}_{\alpha\beta}
- \eta_{[\alpha}\xi_{\beta]} \h{I}_{\gamma\delta} \right)\,,\label{R}\\
&R_{\beta\gamma} =- \tfrac 12\e (i\omega)^2 
\left(\xi_{(\beta} \eta^\alpha 
- \xi^{\alpha} \eta_{(\beta}\right) \hat{I}_{\gamma)\alpha}\,,
\label{cts2b}\\
& R =  \e(i\omega)^2\chi\xi^2\,.              \label{cts}
\end{align}
In average, the linear-order expressions  vanish. One can calculate to second order if I average
over several wavelengths. I use the approximations
\begin{equation}
\vev{\hat{I}^{\alpha\beta}} = 0\,, \qquad
 \vev{\hat{I}^{\alpha\beta} \hat{I}^{\gamma\delta}}
= \frac 12 \hat{\bar{I}}^{\alpha\beta} \hat{\bar{I}}^{\gamma\delta}\,.
\end{equation}
Also as $e_\delta J^{-1}_{\beta\gamma}
= -J^{-1}_{\beta\eta}e_\delta J^{\eta\zeta}\, J^{-1}_{\zeta\gamma}$
I can write 
$e_\delta J^{-1}_{\beta\gamma} = \e e_\delta \hat{I}_{\beta\gamma}$\,.
Therefore I find  expanding~(\ref{R2ndo}) to second order
the expression
\begin{equation}
\vev{R_{\beta\gamma}}
= \tfrac 12\e^2(i\omega)^2
\Big( \bar{\chi} \xi^\alpha\eta_{(\gamma }\hat{\bar{I}}_{\beta )\alpha}
+ \tfrac 34  \bar{\chi}^2 \xi_\beta\xi_\gamma
+\eta^2\hat{\bar{I}}_{\eta\beta}\hat{\bar{I}}^\eta{}_\gamma
-\tfrac 12 \eta_\beta \eta_\gamma
\hat{\bar{I}}_{\alpha\eta}\hat{\bar{I}}^{\alpha\eta} \Big) \label{Ric}
\end{equation}
for the Ricci tensor and the expression
\begin{equation}
\vev{R} = \tfrac 18 \e^2 (i\omega)^2
(2 \eta^2 \hat{\bar{I}}_{\alpha\beta}\hat{\bar{I}}^{\alpha\beta}
+ 7\bar{\chi}^2\xi^2)
\end{equation}
for the Ricci scalar.

\bigskip
\noindent
{\large\textit{Note on WKB  cohomology}}

I briefly motivate here the notation used in this section.  I introduced the algebra of de~Rham
forms with a different differential inspired from the WKB approximation. The differential can be
introduced for all forms but I give the construction only for the case of two-forms. Let
$f_{\alpha\beta}$ be a two-form and define the differential $d_\xi$ of $f$ by the
formulae~(\ref{L1b-new}),(\ref{L1b-L2b}). The interesting point is that the rank of the cohomology
module $\HH^2$, an elementary form of Spencer cohomology, depends on the norm of $\xi$. Let $c$ be a
2-cocycle. Then
\begin{equation}
\xi_\alpha c_{\beta\gamma} +
\xi_\beta c_{\gamma\alpha} +\xi_\gamma c_{\alpha\beta} = 0.
\end{equation}
Multiplying this by $\xi^\alpha$ one obtains the condition~\eq{cyc2}. There are two possibilities.
If $\xi^2 \neq 0$ then it follows immediately that the 2-cocycle is exact. That is, $\HH^2 = 0$. If
on the other hand $\xi^2 = 0$ then there are cocycles which are not exact. One can think of these
as plane-wave solutions to Maxwell's equations. Main result of this section is the dependence of
the Riemann tensor uniquely on the cohomology $\HH$.

\section{Discusion}

I have derived, following~\cite{Buric:2006di,Buric:2007hb}, a relation between the structure of an
associated algebra as defined by the Poisson structure $J^{\mu\nu}$ of the commutation relations
between the generators $x^\mu$ on the one hand and  the metrics which the algebra can support, that
is, which are consistent with the structure of a  differential calculus over the algebra on the
other. I have expressed this relation as the map~(\ref{map}) from the $J$ which defines the algebra
to the frame. The essential ingredients in the definition of the map are the Leibniz rules and the
assumption~(\ref{mod}) on the structure of the differential calculus. Although there have
been found~\cite{Cerchiai:2000uz,Maceda:2003xr,Buric:2004rm} numerous particular examples, there is
not yet a systematic discussion of either the range or kernel of the map. I have here to a certain
extent alleviated this, but only in the context of perturbation theory around a vacuum and even
then, only in the case of a high-frequency wave. A somewhat similar relation has been
found~\cite{Madore:1997ec} in the case of radiative, asymptotically-flat space-times.

One starts with a consistent flat-space solution to the constraints of the algebra and
of the geometry, a solution with the unusual property that its momenta and position stand in a
relation of simple duality, a consequence of which is the fact that the Fourier-transformation is
local. Then she (he) perturbed both structures, the geometric and the algebraic, in a seemingly
arbitrary manner, but within the context of linear-perturbation theory and requiring that the
constraints remain valid. Finally she (he) completely solve the constraints of the perturbation and
exhibit a closed solution as I presented here. I have shown that the degrees-of-freedom or basic
modes of the resulting theory of gravity can be put in  correspondence with those of the
noncommutative structure. As an application of the formalism I have considered a high-frequency
perturbation of the metric an assumption which mimic WKB approximation and I have calculated the
noncommutative corrections that the Ricci tensor receives both in WKB approximation but also in the
more general case of quasi-commutative and weak field approximation.

In~\cite{Buric:2007hb} the model was investigated further. It turned out that the
perturbation of the Poisson structure contributes to the energy-momentum as an additional effective
source of the gravitational field. It was stressed there  that because of the identification of the
gravitational field with the Poisson structure the perturbation of the latter is in fact a
reinterpretation of a perturbation of the former and not an extra field. The difference with
classical gravity lie in the choice of field equations and in the WKB approximation  this amounts
only to a modification of the conserved quantity.

\chapter{Conclusions}
\label{Conclusions}

The emergence of the SM of particle physics from a more fundamental theory is still an open problem.
Theoretical constructions of describing the fundamental interactions of nature, such as string
theory, suggests the existence of extra dimensions, which can be either large or compactified, as I
have assumed here. Then, in order to obtain a reasonable low energy model, a suitable scheme of
dimensional reduction have to be applied. Among them the CSDR scheme was the first which, making use of higher
dimensions, incorporates in a unified manner the gauge and the ad-hoc Higgs sector of the spontaneously broken
gauge theories in four dimensions~\cite{Forgacs:1979zs,Kapetanakis:1992hf,Kubyshin:1989vd}.

It is worth noting that of particular interest for the construction of fully realistic theories in the
framework of CSDR are the following virtues that complemented the original suggestion: 
(i) The possibility to obtain chiral fermions in four dimensions resulting from vector-like reps of
the higher dimensional gauge theory~\cite{Chapline:1982wy,Kapetanakis:1992hf},
(ii) The possibility to deform the metric of certain non-symmetric coset spaces and thereby obtain
more than one scales~\cite{Kapetanakis:1992hf,Farakos:1986sm,Hanlon:1992mn},
(iii) The possibility that the popular softly broken supersymmetric four-dimensional chiral gauge
theories might have their origin in a higher dimensional supersymmetric theory, which is
dimensionally reduced over non-symmetric coset
spaces~\cite{Manousselis:2000aj,Manousselis:2001xb,Manousselis:2001re}.

Main objective of the present dissertation was the investigation to which extent applying both CSDR
and Wilson flux breaking mechanism one can obtain reasonable low energy models. Applying the
CSDR scheme over the six-dimensional compact coset spaces leads to anomaly-free $SO(10)$ and
$E_{6}$ GUTs. However, their gauge symmetry cannot be broken further towards the SM group structure
by an ordinary Higgs mechanism; their four-dimensional scalars belong in the fundamental
representation of the $SO(10)$ and $E_{6}$ gauge groups. As a way out it has been 
suggested~\cite{Zoupanos:1987wj} to additionally apply the Wilson flux breaking
mechanism~\cite{Zoupanos:1987wj,Hosotani:1983xw,Hosotani:1983vn,Witten:1985xc}.
Conclusion of the present research (chapter~\ref{Classification}) was that even though the
application of the method can lead to reasonable low energy models, none of them has the SM gauge
group structure. This point in the direction that either the application of the Hosotani breaking
mechanism is not appropriate enough when used with CSDR scheme, or some field content is missing
from the higher-dimensional theory. A possible way out could be the addition of some scalar fields
which have to transform in suitable representations of the higher dimensional gauge group.
Although that such an assumption is not very appealing for a gauge-Higgs unification scheme it could
be imposed by some other more fundamental theory and therefore cannot be omitted from further
investigation. Furthermore, relaxing the requirement of the higher dimensional gauge group, i.e.
being $E_{8}$ and/or the dimensionality of the initial theory, there are more possibilities of
reasonable low energy models. It is worth noting that these two assumptions are suggested by the
Heterotic string theory but other choices can be also considered~\cite{Jittoh:2008jc}. The full
study of the problem could be a subject of a future publication. Finally, one could repeat the
calculation for the complete problem of the dimensional reduction of an $\cN=1$, $E_{8}$
supergravity Chapline\,-\,Manton action~\cite{Chatzistavrakidis:2007pp,Chatzistavrakidis:2008ii}.

In chapter~\ref{FuzzyCSDR}, I discussed a generalisation of the CSDR scheme by the assumption of
providing the extra dimensional spaces with noncommutative characteristics (`fuzzy spaces'). Even
though the work presented there was not an original one of the author it was an opportunity to
discuss how the ordinary CSDR scheme and its generalised image as applied over Fuzzy spaces differ.
Interesting advantages of this generalised scheme of dimensional reduction was (i) the enhancement
of the initial gauge symmetry due to the noncommutative characteristics of the internal manifold and
(ii) the emergence of renormalizable theories in four dimensions.

In chapter~\ref{FuzzyExtraDim}, the connection of the Fuzzy-CSDR with the renormalizability of the
resulting four-dimensional models was investigated. I examined the inverse problem and assumed a
four-dimensional gauge theory with scalar fields and with the most general renormalizable potential.
I concluded that fuzzy extra dimensions can be dynamically generated as a vacuum solution of this
potential at least for some range of its parameter space. Noteworthy, the initial gauge symmetry was
broken by the same vacuum solution into the SM gauge group structure. In~\cite{Steinacker:2007ay}
the model was generalised to include fermions.

Finally, in chapter~\ref{linPhD} the ordinary space-time was also assumed to be a noncommutative
`manifold',  but for the energy scale for which this could be possible. As as small contribution on
the possible generalisations that the Einstein gravity can have under this assumption, I examined
how the curvature and the noncommutative structure of the algebra are related. The problem was
studied up to the linear approximation of both notions and resulted in circumstantial evidence that
the gravity could be a macroscopic phenomenon of a space-time noncommutative structure
which may be assumed in the Planck energy regime.

\end{mainmatter}
\begin{appendices}

\chapter{Tables of low-energy models}
\label{sec:CSDR-Classification-tables}


\section{Dimensional reduction over symmetric $6D$ coset spaces}\label{CSDR-SCosets-Results}

\setlength{\tabcolsep}{0.5pt}
\setlength{\arraycolsep}{0.5pt}
\begin{scriptsize}
\begin{longtable}[b]{|c!{\vrule width 1pt}c|c!{\vrule width1pt}c|c|c|}
\caption{{\bf Dimensional reduction over symmetric $6D$ coset spaces.}
\emph{Particle physics models leading to $SO(10)$ GUTs in four dimensions.}
\label{SO10ChannelSCosetsTable-CSDR}}\\
\hline
{\bf Case}
&
{\bf Embedding}
&$\begin{array}{c} 
\mbf{6D}~{\bf Coset~Space}
\end{array}$
& $\mbf{H}$ 
&$\begin{array}{c}
{\bf Surviving}\\
{\bf scalars}\\
{\bf under}~\mbf{H}
\end{array}$
&$\begin{array}{c}
{\bf Surviving}\\
{\bf fermions}\\
{\bf under}~\mbf{H}
\end{array}$
\\[10pt]\hhline{*{6}{=}}                                          
\endfirsthead
\hline
\multicolumn{6}{|l|}{\tiny continued from previous page}
\\\hline
{\bf Case}
&{\bf Embedding}
&$\begin{array}{c} 
\mbf{6D}~{\bf Coset~Space}
\end{array}$
& $\mbf{H}$ 
&$\begin{array}{c}
{\bf Surviving}\\
{\bf scalars}\\
{\bf under}~\mbf{H}
\end{array}$
&$\begin{array}{c} 
{\bf Surviving}\\
{\bf fermions}\\
{\bf under}~\mbf{H}
\end{array}$
\\[10pt]\hhline{*{6}{=}}
\endhead                                                           
\hline\multicolumn{6}{|r|}{\tiny\slshape{continued on next page}}
\\\hline
\endfoot                                                           
\hline
\endlastfoot
$\mbf{1a}$
& 
$\begin{array}{l@{}l@{}l@{}}                                                    
E_{8}&\supset& SO(6)\times SO(10)\\                                             
\mbf{248}&=&(\mbf{1}, \mbf{45})+(\mbf{6},\mbf{10})+(\mbf{15},\mbf{1})\\
              &+& (\mbf{4},\mbf{16})+(\overline{\mbf {4}},\overline{\mbf {16}})
\end{array}$
& 
$\frac{SO(7)}{SO(6)}$
&
 $SO(10)$
&
 $\mbf{10}$
&
$\begin{array}{l}
\mbf{16}_{L}\\
\mbf{16}_{L}'
\end{array}$
\\\hline 
$\mbf{2b}$   
&
 $\begin{array}{l}                                                        
 E_{8}\supset SO(6)\times SO(10)\\                                        
 SO(6)\backsim SU(4)\supset SU(3)\times U^{I}(1)\\
 \begin{array}{l@{}l@{}l@{}}
 E_{8}&\supset& (SU(3)\times U^{I}(1))\times SO(10)\\
 \mbf{248}&=&(\mbf{1},\mbf{1})_{(0)}+(\mbf{1}, \mbf{45})_{(0)}+(\mbf{8},\mbf{1})_{(0)}\\
                   &+&(\mbf{3},\mbf{10})_{(-2)}+(\overline{\mbf{3}},\mbf{10})_{(2)}\\  
                   &+&(\mbf{3},\mbf{1})_{(4)}+(\overline{\mbf{3}},\mbf{1})_{(-4)}\\
                   &+&(\mbf{1},\mbf{16})_{(-3)}+(\mbf{1},\overline{\mbf{16}})_{(3)}\\
                   &+&(\mbf{3},\mbf {16})_{(1)}+(\overline{\mbf {3}},\overline{\mbf {16}})_{(-1)}
 \end{array}
 \end{array}$
&
$\frac{SU(4)}{SU(3)\times U^{I}(1)}$
&
   $ SO(10)\,\Big(\times U^{I}(1)\Big)$
&
$\begin{array}{l}
\mbf{10}_{(-2)}\\
\mbf{10}_{(2)}
 \end{array}$
&
$\begin{array}{l}
 \mbf{16}_{L(3)}\\
 \mbf{16}_{L(-1)}\\
\mbf{16}_{L(3)}'\\
\mbf{16}_{L(-1)}'
 \end{array}$
\\\hline
$\mbf{3d},~\mbf{6d}$
&
$\begin{array}{l}
E_{8}\supset SO(6)\times SO(10)\\                             
 SO(6)\backsim SU(4)\supset SU'(3)\times U^{II}(1)\\          
 SU'(3)\supset SU^{a}(2)\times U^{I}(1)\\                     
Y^{I'}=\frac{b}{3}Y^{I}+\frac{b}{3}Y^{II}\\                   
Y^{II'}=\frac{2a}{3}Y^{I}-\frac{a}{3}Y^{II}\\[3pt]
\mbox{\textbf{or}}\\[3pt]
E_{8}\supset SO(6)\times SO(10)\\                                 
SO(6)\backsim SU(4)\\                                             
SU(4) \supset SU^{a}(2)\times SU^{b}(2)\times U^{II}(1)\\         
SU^{b}(2)\supset U^{I}(1)\\
Y^{I'}=-b\,Y^{I}\\                                               
Y^{II'}=a\,Y^{II}\\                                              
\begin{array}{l@{}l@{}l@{}}
 E_{8} &\supset& (SU^{a}(2)\times U^{I'}(1)\times U^{II'}(1))\\
          &\times& SO(10)\\
\mbf{248}&=&(\mbf{1}, \mbf{1})_{(0, 0)}+(\mbf{1}, \mbf{1})_{(0, 0)}\\             
 	      &+&(\mbf{3}, \mbf{1})_{(0, 0)}+(\mbf{1}, \mbf{45})_{(0, 0)}\\       
  	      &+&(\mbf{1}, \mbf{1})_{(-2 b, 0)}+(\mbf{1}, \mbf{1})_{(2 b, 0)}\\
              &+&(\mbf{2}, \mbf{1})_{(-b, 2 a)}+(\mbf{2}, \mbf{1})_{(b, -2 a)}\\
              &+&(\mbf{2}, \mbf{1})_{(-b, -2 a)}+(\mbf{2}, \mbf{1})_{(b, 2 a)}\\
              &+&(\mbf{1}, \mbf{10})_{(0, -2 a)}+(\mbf{1}, \mbf{10})_{(0, 2 a)}\\
              &+&(\mbf{2}, \mbf{10})_{(b, 0)}+(\mbf{2}, \mbf{10})_{(-b, 0)}\\
              &+&(\mbf{1}, \mbf{16})_{(b, -a)}+(\mbf{1}, \overline{\mbf{16}})_{(-b, a)}\\
              &+&(\mbf{1}, \mbf{16})_{(-b, -a)}+(\mbf{1}, \overline{\mbf{16}})_{(b, a)}\\
              &+&(\mbf{2}, \mbf{16})_{(0, a)}+(\mbf{2}, \overline{\mbf{16}})_{(0, -a)}
\end{array}
\end{array}$
&
$\begin{array}{l}
\left(\frac{SU(3)}{SU^{a}(2)\times U^{I'}(1)}\right)\\
\hfill\times\left(\frac{SU(2)}{U^{II'}(1)}\right)
\end{array}$
&
$\begin{array}{l@{}}
    SO(10)\\
    \Big(\times U^{I'}(1)\times U^{II'}(1)\Big)
\end{array}$
&
$\begin{array}{l}
\mbf{10}_{(0, -2 a)}\\
\mbf{10}_{(0, 2 a)}\\
\mbf{10}_{(b, 0)}\\
\mbf{10}_{(-b, 0)}
\end{array}$
&
$\begin{array}{l}
\mbf{16}_{L(b, -a)}\\
\mbf{16}_{L(-b, -a)}\\
\mbf{16}_{L(0, a)}\\
\mbf{16}_{L(b, -a)}'\\
\mbf{16}_{L(-b, -a)}'\\
\mbf{16}_{L(0, a)}'
\end{array}$
\\\hline
$\mbf{4f},~\mbf{7f}$ 
& 
$\begin{array}{l}                                                         
 E_{8}\supset SO(6)\times SO(10)\\                                        
 SO(6)\backsim SU(4)\supset SU'(3)\times U^{III}(1)\\                     
 SU'(3)\supset SU^{a}(2)\times U^{II}(1)\\
SU^{a}(2)\supset U^{I}(1)\\
Y^{I'}=aY^{I}+\frac{a}{3}Y^{II}+\frac{a}{3}Y^{III}\\
Y^{II'}=-bY^{I}+\frac{b}{3}Y^{II}+\frac{b}{3}Y^{III}\\
Y^{III'}=-\frac{2c}{3}Y^{I}+\frac{c}{3}Y^{III}\\[3pt]
\mbox{\textbf{or}}\\[3pt]
E_{8}\supset SO(6)\times SO(10)\\                                                  
 SO(6)\backsim SU(4)\\                                                             
SU(4)\supset SU^{a}(2)\times SU^{b}(2)\times U^{III}(1)\\
SU^{a}(2)\supset U^{II}(1)\\
SU^{b}(2)\supset U^{I}(1)\\
Y^{I'}=a\,Y^{I}-a\,Y^{II}\\                                                 
Y^{II'}=-b\,Y^{I}-b\,Y^{II}\\                                               
Y^{III'}=-c\,Y^{III}\\                                                      
 \begin{array}{l@{}l@{}l@{}}
 E_{8}&\supset& SO(10)\\
         &\times& U^{I'}(1)\times U^{II'}(1)\times U^{III'}(1)\\
\mbf{248}&=&\mbf{1}_{(0, 0, 0)}+\mbf{1}_{(0, 0, 0)}+\mbf{1}_{(0, 0, 0)}\\    
              &+&\mbf{45}_{(0, 0, 0)}\\                                      
              &+&\mbf{1}_{(-2 a, 2 b, 0)}+\mbf{1}_{(2 a, -2 b, 0)}\\         
              &+&\mbf{1}_{(-2 a, -2 b, 0)}+\mbf{1}_{(2 a, 2 b, 0)}\\
              &+&\mbf{1}_{(-2 a, 0, -2 c)}+\mbf{1}_{(2 a, 0, 2 c)}\\
              &+&\mbf{1}_{(0, -2 b, -2 c)}+\mbf{1}_{(0, 2 b, 2 c)}\\
              &+&\mbf{1}_{(-2 a, 0, 2 c)}+\mbf{1}_{(2 a, 0, -2 c)}\\
              &+&\mbf{1}_{(0, -2 b, 2 c)}+\mbf{1}_{(0, 2 b, -2 c)}\\
              &+&\mbf{10}_{(0, 0, 2 c)}+\mbf{10}_{(0, 0, -2 c)}\\
              &+&\mbf{10}_{(0, 2 b, 0)}+\mbf{10}_{(0, -2 b, 0)}\\
              &+&\mbf{10}_{(2 a, 0, 0)}+\mbf{10}_{(-2 a, 0, 0)}\\
              &+&\mbf{16}_{(a, b, c)}+\overline{\mbf{16}}_{(-a, -b, -c)}\\
              &+&\mbf{16}_{(-a, -b, c)}+ \overline{\mbf{16}}_{(a, b, -c)}\\
              &+&\mbf{16}_{(-a, b, -c)}+\overline{\mbf{16}}_{(a, -b, c)}\\
              &+&\mbf{16}_{(a, -b, -c)}+\overline{\mbf{16}}_{(-a, b, c)}
\end{array}
\end{array}$
&
$\begin{array}{l}
\left(\frac{SU(2)}{U^{I'}(1)}\right)\times\left(\frac{SU(2)}{U^{II'}(1)}\right)\\
\hfill\times\left(\frac{SU(2)}{U^{III'}(1)}\right)
\end{array}$
&
$\begin{array}{l@{}}
    SO(10)\,\Big(\times U^{I'}(1)\\
    \times U^{II'}(1)\times U^{III'}(1)\Big)
\end{array}$
&
$\begin{array}{l}
\mbf{10}_{(2 a, 0, 0)}\\
\mbf{10}_{(-2 a, 0, 0)}\\
\mbf{10}_{(0, 2 b, 0)}\\
\mbf{10}_{(0, -2 b, 0)}\\
\mbf{10}_{(0, 0, 2 c)}\\
\mbf{10}_{(0, 0, -2 c)}
 \end{array}$
&
$\begin{array}{l}
\mbf{16}_{L(a, b, c)}\\
\mbf{16}_{L(-a, -b, c)}\\
\mbf{16}_{L(-a, b, -c)}\\
\mbf{16}_{L(a, -b, -c)}\\
\mbf{16}_{L(a, b, c)}'\\
\mbf{16}_{L(-a, -b, c)}'\\
\mbf{16}_{L(-a, b, -c)}'\\
\mbf{16}_{L(a, -b, -c)}'
\end{array}$
\\\hline
$\mbf{5e} $                                                        
& $\begin{array}{l}                                                
 E_{8}\supset SO(6)\times SO(10)\\                                 
 SO(6)\backsim SU(4)\\
 SU(4)\supset SU^{a}(2)\times SU^b{}(2)\times U^{I}(1)\\
 \begin{array}{l@{}l@{}l@{}}
 E_{8}&\supset& (SU^{a}(2)\times SU^{b}(2)\times U^{I}(1))\\
         &\times& SO(10)\\
 \mbf{248}&=&(\mbf{1},\mbf{1},\mbf{1})_{(0)}+(\mbf{1},\mbf{1},\mbf{45})_{(0)}\\
                  &+&(\mbf{3},\mbf{1},\mbf{1})_{(0)}+(\mbf{1},\mbf{3},\mbf{1})_{(0)}\\
                  &+&(\mbf{2},\mbf{2},\mbf{1})_{(2)}+(\mbf{2},\mbf{2},\mbf{1})_{(-2)}\\
                  &+&(\mbf{1},\mbf{1},\mbf{10})_{(2)}+(\mbf{1},\mbf{1},\mbf{10})_{(-2)}\\
                  &+&(\mbf{2},\mbf{2},\mbf{10})_{(0)}\\                                          
                  &+&(\mbf{2},\mbf{1},\mbf{16})_{(1)}+(\mbf{2},\mbf{1},\overline{\mbf{16}})_{(-1)}\\
                  &+&(\mbf{1},\mbf{2},\mbf{16})_{(-1)}+(\mbf{1},\mbf{2},\overline{\mbf{16}})_{(1)}
 \end{array}
 \end{array}$
& $\frac{Sp(4)\times SU(2)}{SU^{a}(2)\times SU^{b}(2)\times U^{I}(1)}$
&
   $SO(10)\,\Big(\times U^{I}(1)\Big)$
&$\begin{array}{l}
\mbf{10}_{(0)}\\
\mbf{10}_{(2)}\\
\mbf{10}_{(-2)}
\end{array}$
& $\begin{array}{l}
 \mbf{16}_{L(1)}\\
 \mbf{16}_{L(-1)}\\
\mbf{16}_{L(1)}'\\
\mbf{16}_{L(-1)}'
 \end{array}$
\end{longtable}
\end{scriptsize}
 \newpage
\begin{landscape}
\setlength{\tabcolsep}{2pt} 
\setlength{\arraycolsep}{1.5pt} 
\begin{scriptsize}
\begin{longtable}{|c!{\vrule width 1pt}c|c|c|c!{\vrule width 1pt}c|c|}
\caption{{\bf Application of Hosotani breaking mechanism on particle physics models which are listed in
table~\ref{SO10ChannelSCosetsTable-CSDR}}.
\label{SO10ChannelSCosetsTable-CSDR-HOSOTANI}}\\
\hline
{\bf Case}
&
\begin{tabular}{c}
 {\bf Discrete}\\
 {\bf Symme-}\\
 {\bf tries}
\end{tabular}
&
$\mbf{K'}$
&
$\begin{array}{c}
{\bf Surviving}\\
{\bf scalars}\\
{\bf under}~\mbf{K'}
\end{array}$
&
$\begin{array}{c}
{\bf Surviving~fermions}\\
{\bf under}~\mbf{K'}
\end{array}$
&
$\mbf{K}$
&
$\begin{array}{c}
{\bf Surviving~fermions}\\
{\bf under}~\mbf{K}
\end{array}$
\\[10pt]\hhline{*{7}{=}}                               
\endfirsthead
\hline
\multicolumn{7}{|l|}{\tiny continued from previous page}
\\\hline
{\bf Case}
&
\begin{tabular}{c}
 {\bf Discrete}\\
 {\bf Symme-}\\
 {\bf tries}
\end{tabular}
&
$\mbf{K'}$
&
$\begin{array}{c}
{\bf Surviving}\\
{\bf scalars}\\
{\bf under}~\mbf{K'}
\end{array}$
&
$\begin{array}{c}
{\bf Surviving~fermions}\\
{\bf under}~\mbf{K'}
\end{array}$
&
$\mbf{K}$
&
$\begin{array}{c}
{\bf Surviving~fermions}\\
{\bf under}~\mbf{K}
\end{array}$
\\[10pt]\hhline{*{7}{=}}
\endhead                                                                        
\hline\multicolumn{7}{|r|}{\tiny\slshape{continued on next page}}
\\\hline
\endfoot                                                                        
\hline
\endlastfoot
$\mbf{1a}$
&
$\begin{array}{c}
\mathrm{W}\\
\scriptscriptstyle{(\bb{Z}_{2})}
\end{array}$                                                                      
&
 $\begin{array}{l@{}}
SU^{(i)}(2) \times SU^{(ii)}(2) \\
\times SU(4)
\end{array}$
&
$(\mbf{1},\mbf{1}, \mbf{6})$
&
 $\begin{array}{l}
 (\mbf{2},\mbf{1},\mbf{4})_{L}-(\mbf{2},\mbf{1}, \mbf{4})'_{L} \\
 (\mbf{1},\mbf{2},\mbf{4})_{L}+(\mbf{1},\mbf{2}, \mbf{4})'_{L}
 \end{array}$
&
$\begin{array}{c}
\mbox{Not}\\
\mbox{Interesting}
\end{array}$
&\\
\ \
&
$\begin{array}{c}
\mathrm{W\times Z}\\
\scriptscriptstyle{(\bb{Z}_{2}\times\bb{Z}_{2})}
\end{array}$
&
 $\mbox{''}$
&
$ (\mbf{2},\mbf{2},\mbf{1})$
&
 $+\lra -$
&
$SU^{diag}(2)\times SU(4)$
&
 $\begin{array}{l}
 (\mbf{2},\mbf{4})_{L}+(\mbf{2}, \mbf{4})'_{L} \\
 (\mbf{2},\mbf{4})_{L}-(\mbf{2}, \mbf{4})'_{L}
 \end{array}$
\\\hline 
$\mbf{2b}$   
 &
$\begin{array}{c}
\mathrm{W}\\
\scriptscriptstyle{(\one)}\\
\end{array}$                                                                     
 &
 $\mbox{$H$ Unbroken}$
 &
 &
 &
 & \\
\ \
&
 $\begin{array}{c}
\mathrm{W\times Z}\\
\scriptscriptstyle{(\bb{Z}_{4})}
   \end{array}$
&
$\begin{array}{c}
\mbox{Not}\\
\mbox{ Interesting}
\end{array}$
&
&
&
&
\\\hline
$\mbf{3d}$, $\mbf{6d}$
&
$\begin{array}{c}
\mathrm{W}\\
\scriptscriptstyle{(\bb{Z}_{2})}                                                          
 \end{array}$
&
$\begin{array}{l@{}}
SU^{(i)}(2)\times SU^{(ii)}(2)\\
\times SU(4)\\
 \Big(\times U^{I'}(1)\times U^{II'}(1)\Big)
\end{array}$
 &
$\begin{array}{l}
(\mbf{1},\mbf{1},\mbf{6})_{(b, 0)}\\
(\mbf{1},\mbf{1},\mbf{6})_{(-b, 0)}
\end{array}$
&
$\begin{array}{l}
(\mbf{2},\mbf{1},\mbf{4})_{L(b,-a)}-(\mbf{2},\mbf{1},\mbf{4})'_{L(b,-a)}\\
(\mbf{1},\mbf{2},\mbf{4})_{L(b, -a)}+(\mbf{1},\mbf{2},\mbf{4})'_{L(b, -a)}\\
(\mbf{2},\mbf{1},\mbf{4})_{L(-b,-a)}-(\mbf{2},\mbf{1},\mbf{4})'_{L(-b, -a)}\\
(\mbf{1},\mbf{2},\mbf{4})_{L(-b, -a)}+(\mbf{1},\mbf{2},\mbf{4})'_{L(-b, -a)}\\
(\mbf{2},\mbf{1},\mbf{4})_{L(0, a)}-(\mbf{2},\mbf{1},\mbf{4})'_{L(0, a)}\\
(\mbf{1},\mbf{2},\mbf{4})_{L(0, a)}+(\mbf{1},\mbf{2},\mbf{4})'_{L(0, a)}
\end{array}$
&
$\begin{array}{c}
\mbox{Not}\\
\mbox{Interesting}
\end{array}
$
&\\
\ \
& 
$\begin{array}{c}
\mathrm{W\times Z}\\
\scriptscriptstyle{(\bb{Z}_{2}\times\bb{Z}_{2})} 
\end{array}$
& 
$\mbox{''}$
&
$\begin{array}{l}
(\mbf{2},\mbf{2},\mbf{1})_{(b, 0)}\\
(\mbf{2},\mbf{2},\mbf{1})_{(-b, 0)}
\end{array}$
&
$+\lra-$
&
$\begin{array}{l@{}}
SU^{diag}(2)\times SU(4)\\
\Big(\times U^{I'}(1)\times U^{II'}(1)\Big)
\end{array}$
&
$\begin{array}{l}
(\mbf{2},\mbf{4})_{L(b,-a)}+(\mbf{2},\mbf{4})'_{L(b,-a)}\\
(\mbf{2},\mbf{4})_{L(b, -a)}-(\mbf{2},\mbf{4})'_{L(b, -a)}\\
(\mbf{2},\mbf{4})_{L(-b,-a)}+(\mbf{2},\mbf{4})'_{L(-b, -a)}\\
(\mbf{2},\mbf{4})_{L(-b, -a)}-(\mbf{2},\mbf{4})'_{L(-b, -a)}\\
(\mbf{2},\mbf{4})_{L(0, a)}+(\mbf{2},\mbf{4})'_{L(0, a)}\\
(\mbf{2},\mbf{4})_{L(0, a)}-(\mbf{2},\mbf{4})'_{L(0, a)}
\end{array}$
\\\hline
$\mbf{4f}$, $\mbf{7f}$ 
&
$\begin{array}{c}
\mathrm{W}\\
\scriptscriptstyle{(\bb{Z}_{2})}                                                                              
 \end{array}$
 &
$\begin{array}{l@{}}
 SU^{(i)}(2)\times SU^{(ii)}(2)\\
\times SU(4)\\
\Big(\times U^{I'}(1)\times U^{II'}(1) \\
\times U^{III'}(1)\Big)
\end{array}$
 &
$\begin{array}{l}
(\mbf{1},\mbf{1},\mbf{6})_{(0, 2 b, 0)}\\
(\mbf{1},\mbf{1},\mbf{6})_{(0, -2 b, 0)}\\
(\mbf{1},\mbf{1},\mbf{6})_{(0, 0, 2 c)}\\
(\mbf{1},\mbf{1},\mbf{6})_{(0, 0, -2 c)}
\end{array}$
&
$-$
&
$\begin{array}{c}
\mbox{Not}\\
\mbox{Interesting}
\end{array}$
&\\
\ \
&
&
&
&
&
&\\
\ \
&
$\begin{array}{c}
\mathrm{W}\\
\scriptscriptstyle{(\bb{Z}_{2})^{2}}                                                
 \end{array}$
&
$\mbox{"}$
&
$\begin{array}{l}
(\mbf{1},\mbf{1},\mbf{6})_{(0, 0, 2 c)}\\
(\mbf{1},\mbf{1},\mbf{6})_{(0, 0, -2 c)}
\end{array}$
&
$\mbox{"}$
&
$\mbox{"}$
&\\
&\ \
&
&
&
&
&\\
\ \
&
$\begin{array}{c}
\mathrm{W\times Z}\\
\scriptscriptstyle{(\bb{Z}_{2}\times \bb{Z}_{2})}                                         
 \end{array}$
&
$\mbox{"}$
&
$\begin{array}{l}
(\mbf{2},\mbf{2},\mbf{1})_{(0, 2 b, 0)}\\
(\mbf{2},\mbf{2},\mbf{1})_{(0, -2 b, 0)}\\
(\mbf{2},\mbf{2},\mbf{1})_{(0, 0, 2 c)}\\
(\mbf{2},\mbf{2},\mbf{1})_{(0, 0, -2 c)}
\end{array}$
&
$\mbox{"}$
&
$\mbox{"}$
&
\\\hline
$\mbf{5e} $                                                     
&
$\begin{array}{c}                                               
\mathrm{W} \\
\scriptscriptstyle{(\bb{Z}_{2})}
\end{array}$
&
 $\begin{array}{l@{}}
 SU^{(i)}(2)\times SU^{(ii)}(2)\\
 \times SU(4)\\
 \Big(\times U(1)\Big)
\end{array}$
 &
 $\begin{array}{l@{}}
(\mbf{1},\mbf{1},\mbf{6})_{(0)}
\end{array}$
 & 
$\begin{array}{l@{}}
({\mbf 2},{\mbf 1},{\mbf 4})_{L(1)}-({\mbf 2},{\mbf 1},{\mbf 4})'_{L(1)}\\
({\mbf 1},{\mbf 2},{\mbf 4})_{L(1)}+({\mbf 1},{\mbf 2},{\mbf 4})'_{L(1)}\\
({\mbf 2},{\mbf 1},{\mbf 4})_{L(-1)}-({\mbf 2},{\mbf 1},{\mbf 4})'_{L(-1)}\\
({\mbf 1},{\mbf 2},{\mbf 4})_{L(-1)}+({\mbf 1},{\mbf 2},{\mbf 4})'_{L(-1)}
\end{array}$
&
$\begin{array}{c}
\mbox{Not}\\
\mbox{Interesting}
 \end{array}$
&\\
\ \
&
$\begin{array}{c}
\mathrm{W} \\
\scriptscriptstyle{(\bb{Z}_{2})^{2}}
 \end{array}$
&
$\mbox{"}$
&
$\begin{array}{l@{}}
(\mbf{2},\mbf{2},\mbf{1})_{(0)}
\end{array}$
&
$\mbox{"}$
&
$\begin{array}{l@{}}
SU^{diag}(2)\times SU(4)\\
\Big(\times U(1)\Big)
\end{array}$
&
$\begin{array}{l@{}}
({\mbf 2},{\mbf 4})_{L(1)}-({\mbf 2},{\mbf 4})'_{L(1)}\\
({\mbf 2},{\mbf 4})_{L(1)}+({\mbf 2},{\mbf 4})'_{L(1)}\\
({\mbf 2},{\mbf 4})_{L(-1)}-({\mbf 2},{\mbf 4})'_{L(-1)}\\
({\mbf 2},{\mbf 4})_{L(-1)}+({\mbf 2},{\mbf 4})'_{L(-1)}
\end{array}$\\
\ \
&
&
&
&
&
&\\
\ \
&
$\begin{array}{c}
\mathrm{W\times Z} \\
\scriptscriptstyle{(\bb{Z}_{2}\times\bb{Z}_{2})}
 \end{array}$
&
$\mbox{"}$
&
$\mbox{"}$
&
$+\lra-$
&
$\mbox{"}$
&
$\begin{array}{l@{}}
({\mbf 2},{\mbf 4})_{L(1)}+({\mbf 2},{\mbf 4})'_{L(1)}\\
({\mbf 2},{\mbf 4})_{L(1)}-({\mbf 2},{\mbf 4})'_{L(1)}\\
({\mbf 2},{\mbf 4})_{L(-1)}+({\mbf 2},{\mbf 4})'_{L(-1)}\\
({\mbf 2},{\mbf 4})_{L(-1)}-({\mbf 2},{\mbf 4})'_{L(-1)}
\end{array}$
\end{longtable}
\end{scriptsize}
\end{landscape}

\newpage
\section{Dimensional reduction over non-symmetric $6D$ coset spaces}\label{CSDR-NSCosets-Results}

\setlength{\tabcolsep}{1pt}
\setlength{\arraycolsep}{1pt}
\begin{scriptsize}
\begin{longtable}[t]{ |c!{\vrule width 1pt}c|c !{\vrule width 1pt} c|c|c|}
\caption{{\bf Dimensional reduction over symmetric $6D$ coset spaces}.
\emph{Particle physics models leading to  $E_{6}$ GUTs in four dimensions.}
\label{E6ChannelTable-CSDR}}\\
\hline
{\bf Case}
&
{\bf Embedding}
&
$\begin{array}{c} 
\mbf{6D}\\
{\bf Coset~Space}
\end{array}$
& 
$\mbf{H}$ 
&
$\begin{array}{c}
{\bf Surviving}\\
{\bf scalars}\\
{\bf under}~\mbf{H}
\end{array}$
&
$\begin{array}{c} 
{\bf Surviving}\\
{\bf fermions}\\
{\bf under}~\mbf{H}
\end{array}$
\\[10pt]\hhline{*{6}{=}}                                      
\endfirsthead
\hline
\multicolumn{6}{|l|}{\tiny continued from previous page}
\\\hline
{\bf Case}
&
{\bf Embedding}
&
$\begin{array}{l} 
\mbf{6D}\\
{\bf Coset~Space}
\end{array}$
&
 $\mbf{H}$ 
&
$\begin{array}{c}
{\bf Surviving}\\
{\bf scalars}\\
{\bf under}~\mbf{H}
\end{array}$
&
$\begin{array}{c} 
{\bf Surviving}\\
{\bf fermions}\\
{\bf under}~\mbf{H}
\end{array}$
\\[10pt]\hhline{*{6}{=}}
\endhead                                                     
\hline\multicolumn{6}{|r|}{\tiny\slshape{continued on next page}}
\\\hline
\endfoot                                                     
\hline
\endlastfoot
$\mbf{2a'}$ 
&
 $\begin{array}{l}                                                    
E_{8}\supset SU'(3)\times E_{6}\\                                     
 \begin{array}{l@{}l@{}l@{}}
 E_{8}&\supset& SU'(3) \times E_{6}\\
\mbf{248}&=&(\mbf{8},\mbf{1})+(\mbf{1},\mbf{78})\\
              &+&(\mbf{3},\mbf{27})+(\overline{\mbf{3}},\overline{\mbf{27}})
\end{array}
\end{array}$
&
$\frac{G_{2}}{SU'(3)}$
&
$E_{6}$
&
$\begin{array}{l}
\mbf{27}\\
\overline{\mbf{27}}
 \end{array}$
&
$\begin{array}{l}
\mbf{78}\\
\mbf{27}_{L}\\
\mbf{27}_{L}'
\end{array}
$
\\\hline
$\mbf{3b'}$
&
$\begin{array}{l}                                                         
 E_{8}\supset SU'(3)\times E_{6}\\                                        
SU'(3) \supset SU^{a}(2)\times U^{I}(1)\\
Y^{I'}=-Y^{I}\\                                                           
 \begin{array}{l@{}l@{}l@{}}
 E_{8}&\supset& SU^{a}(2)\times U^{I'}(1) \times E_{6}\\
\mbf{248}&=&(\mbf{1},\mbf{1})_{(0)}+(\mbf{1},\mbf{78})_{(0)}\\     
              &+&(\mbf{3},\mbf{1})_{(0)}\\                         
              &+&(\mbf{2},\mbf{1})_{(-3)}+(\mbf{2},\mbf{1})_{(3)}\\
              &+&(\mbf{1},\mbf{27})_{(2)}+(\mbf{1},\overline{\mbf{27}})_{(-2)}\\
              &+&(\mbf{2},\mbf{27})_{(-1)}+(\mbf{2},\overline{\mbf{27}})_{(1)}
\end{array}
\end{array}$
&
$\left(\frac{Sp(4)}{SU^{a}(2)\times U^{I}(1)}\right)_{nonmax}$
&
$E_{6}\,\Big(\times U^{I'}(1)\Big)$
&
$\begin{array}{l}
\mbf{27}_{(2)}\\
\mbf{27}_{(-1)}\\
\overline{\mbf{27}}_{(-2)}\\
\overline{\mbf{27}}_{(1)}
\end{array}$
&
$\begin{array}{l}
\mbf{1}_{(0)}\\
\mbf{78}_{(0)}\\
\mbf{27}_{L(2)}\\
\mbf{27}_{L(-1)}\\
\mbf{27}_{L(2)}'\\
\mbf{27}_{L(-1)}'\\
\end{array}$
\\\hline       
$\mbf{4c'}$
&
$\begin{array}{l}                                                            
 E_{8}\supset SU'(3)\times E_{6}\\                                           
SU'(3) \supset SU^{a}(2)\times U^{II}(1)\\
SU^{a}(2)\supset U^{I}(1)\\
 \begin{array}{l@{}l@{}l@{}}
 E_{8}&\supset&  E_{6}\times U^{I}(1) \times U^{II}(1)\\
\mbf{248}&=&\mbf{1}_{(0,0)}+\mbf{1}_{(0,0)}\\
              &+&\mbf{78}_{(0)}\\
              &+&\mbf{1}_{(-2,0)}+\mbf{1}_{(2,0)}\\
              &+&\mbf{1}_{(-1,3)}+\mbf{1}_{(1,-3)}\\
              &+&\mbf{1}_{(1,3)}+\mbf{1}_{(-1,-3)}\\
              &+&\mbf{27}_{(0,-2)}+\overline{\mbf{27}}_{(0,2)}\\
              &+&\mbf{27}_{(-1,1)}+\overline{\mbf{27}}_{(1,-1)}\\
              &+&\mbf{27}_{(1,1)}+\overline{\mbf{27}}_{(-1,-1)}
\end{array}
\end{array}$
&
$\frac{SU(3)}{U^{I}(1)\times U^{II}(1)}$
&
$E_{6}\,\Big(\times U^{I}(1)\times U^{II}(1)\Big)$
&
$\begin{array}{l}                
\mbf{27}_{(0,-2)}\\              
\mbf{27}_{(-1,1)}\\              
\mbf{27}_{(1,1)}\\
\overline{\mbf{27}}_{(0,2)}\\
\overline{\mbf{27}}_{(1,-1)}\\
\overline{\mbf{27}}_{(-1,-1)}\\
\scriptstyle{(a=0,c=-2)}\\
\scriptstyle{(b=-1,d=1)}
\end{array}$
&
$\begin{array}{l}
\mbf{1}_{(0,0)}\\
\mbf{1}_{(0,0)}\\
\mbf{78}_{(0,0)}\\
\mbf{27}_{L(0,-2)}\\
\mbf{27}_{L(-1,1)}\\
\mbf{27}_{L(1,1)}\\
\mbf{27}_{L(0,-2)}'\\
\mbf{27}_{L(-1,1)}'\\
\mbf{27}_{L(1,1)}'
\end{array}$
\\\hline
\end{longtable}
\end{scriptsize}

\setlength{\tabcolsep}{2pt}
\setlength{\arraycolsep}{1.5pt}
\begin{scriptsize}
\begin{longtable}[b]{|c!{\vrule width 1pt}c|c|c|}
\caption{\textbf{Application of Hosotani breaking mechanism on particle physics models which are listed in
table~\ref{E6ChannelTable-CSDR}.}
\emph{The surviving fields are calculated for the embeddings $\bb{Z}_{2}\hookrightarrow E_{6}$ and
$(\bb{Z}_{2}\times\bb{Z}_{2})\hookrightarrow E_{6}$, discussed in
subsections~\ref{TopologicallyInducedGaugeGroupBreaking-E6-Z2} and~\ref{TopologicallyInducedGaugeGroupBreaking-E6-Z2Z2}}
\label{E6ChannelTable-CSDR-HOSOTANI}}\\
\hline
{\bf Case}
&
\begin{tabular}{c}
{\bf Discrete}\\
{\bf Symmetries}
\end{tabular}
&
$\mbf{K'}$
&
$\begin{array}{c}
{\bf Surviving~fermions}\\
{\bf under}~\mbf{K'}
\end{array}$
\\[10pt]\hhline{*{4}{=}}                                
\endfirsthead
\hline
\multicolumn{4}{|l|}{\tiny continued from previous page}
\\\hline
{\bf Case}
&
\begin{tabular}{c}
{\bf Discrete}\\
{\bf Symmetries}
\end{tabular}
&
$\mbf{K'}$
&
$\begin{array}{c}
{\bf Surviving~fermions}\\
{\bf under}~\mbf{K'}
\end{array}$
\\[10pt]\hhline{*{4}{=}}
\endhead                                                       
\hline\multicolumn{4}{|r|}{\tiny\slshape{continued on next page}}
\\\hline
\endfoot                                                       
\hline
\endlastfoot
$\mbf{2a'}$ 
&
$\begin{array}{c}
\mathrm{W}\\
\scriptstyle{(\bb{Z}_{2})}\\
\scriptstyle{[embedding~(\mbf{1})]}
\end{array}$                                                                                              
&
$SO(10)\times U(1)$
&
$\begin{array}{c}
\mbf{1}_{(0)}\\
\mbf{45}_{(0)}\\
\begin{array}{l}
\mbf{1}_{L(-4)}+\mbf{1}'_{L(-4)}\\
\mbf{10}_{L(-2)}+\mbf{10}'_{L(-2)}\\
\mbf{16}_{L(1)}-\mbf{16}'_{L(1)}
\end{array}
\end{array}$\\
\ \
&
&
&\\
\ \
&
$\begin{array}{c}
\mathrm{W}\\
\scriptstyle{(\bb{Z}_{2})}\\
\scriptstyle{[embeddings}\\
\scriptstyle{(\mbf{2}),~(\mbf{3})]}
\end{array}$                                                                                              
&
$SU(2)\times SU(6)$
&
$\begin{array}{c}
(\mbf{3},\mbf{1})\\
(\mbf{1},\mbf{35})\\
\begin{array}{l}
(\mbf{1},\mbf{15})_{L}+(\mbf{1},\mbf{15})'_{L}\\
(\mbf{2},\overline{\mbf{6}})_{L}-(\mbf{2},\overline{\mbf{6}})'_{L}
 \end{array}
 \end{array}$
\\\hline
$\mbf{3b'}$
&
$\begin{array}{c}
\mathrm{W}\\
\scriptstyle{(\bb{Z}_{2})}\\
\scriptstyle{[embedding~(\mbf{1})]}
\end{array}$                                                                                           
&
$SO(10)\times U(1)\,\Big(\times U^{I}(1)\Big)$
&
$\begin{array}{c}
\mbf{1}_{(0,0)}\\
\mbf{1}_{(0,0)}\\
\mbf{45}_{(0,0)}\\
\begin{array}{l}
\mbf{1}_{L(-4,2)}+\mbf{1}'_{L(-4,2)}\\
\mbf{10}_{L(-2,2)}+\mbf{10}'_{L(-2,2)}\\
\mbf{16}_{L(1,2)}-\mbf{16}'_{L(1,2)}\\
\mbf{1}_{L(-4,-1)}+\mbf{1}'_{L(-4,-1)}\\
\mbf{10}_{L(-2,-1)}+\mbf{10}'_{L(-2,-1)}\\
\mbf{16}_{L(1,-1)}-\mbf{16}'_{L(1,-1)}
\end{array}
\end{array}$\\
\ \
&
&
&\\
\ \
&
$\begin{array}{c}
\mathrm{W}\\
\scriptstyle{(\bb{Z}_{2})}\\
\scriptstyle{[embeddings}\\
\scriptstyle{(\mbf{2}),~(\mbf{3})]}
\end{array}$                                                                                              
&
$SU(2) \times SU(6)\,\Big(\times U^{I}(1)\Big)$
&
$\begin{array}{c}
(\mbf{1},\mbf{1})_{(0)}\\
(\mbf{3},\mbf{1})_{(0)}\\
(\mbf{1},\mbf{35})_{(0)}\\
\begin{array}{l}
(\mbf{1},\mbf{15})_{L(2)}+(\mbf{1},\mbf{15})'_{L(2)}\\
(\mbf{2},\overline{\mbf{6}})_{L(2)}-(\mbf{2},\overline{\mbf{6}})'_{L(2)}\\
(\mbf{1},\mbf{15})_{L(-1)}+(\mbf{1},\mbf{15})'_{L(-1)}\\
(\mbf{2},\overline{\mbf{6}})_{L(-1)}-(\mbf{2},\overline{\mbf{6}})'_{L(-1)}
 \end{array}
\end{array}$\\
\ \
&
&
&\\
\ \
& 
$\begin{array}{c}                                
\mathrm{W\times Z}\\                             
\scriptstyle{(\bb{Z}_{2}\times \bb{Z}_{2})}\\
\scriptstyle{[embedding~(\mbf{2'})]}\\
\end{array}$
&
$\begin{array}{c}
SU(2)\times SU(6)\,\Big(\times U^{I}(1)\Big)
\end{array}$
&
$\begin{array}{c}
(\mbf{1},\mbf{1})_{(0)}\\
(\mbf{3},\mbf{1})_{(0)}\\
(\mbf{1},\mbf{35})_{(0)}\\
\begin{array}{l}
(\mbf{1},\mbf{15})_{L(2)}-(\mbf{1},\mbf{15})'_{L(2)}\\
(\mbf{2},\overline{\mbf{6}})_{L(2)}+(\mbf{2},\overline{\mbf{6}})'_{L(2)}\\
(\mbf{1},\mbf{15})_{L(-1)}-(\mbf{1},\mbf{15})'_{L(-1)}\\
(\mbf{2},\overline{\mbf{6}})_{L(-1)}+(\mbf{2},\overline{\mbf{6}})'_{L(-1)}
 \end{array}
\end{array}$\\
\ \
&
&
&\\
\ \
&
$\begin{array}{c}                                                    
\mathrm{W\times Z}\\                                                 
\scriptstyle{(\bb{Z}_{2}\times \bb{Z}_{2})}\\
\scriptstyle{[embedding~(\mbf{3'})]}
\end{array}$
&
$\begin{array}{c}
SU^{(i)}(2)\times SU^{(ii)}(2)\times SU(4)\times U(1)
\,\Big(\times U^{I}(1)\Big)
\end{array}$
&
$\begin{array}{c}
(\mbf{1},\mbf{1},\mbf{1})_{(0,0)}\\
(\mbf{1},\mbf{1},\mbf{1})_{(0,0)}\\
(\mbf{3},\mbf{1},\mbf{1})_{(0,0)}\\
(\mbf{1},\mbf{3},\mbf{1})_{(0,0)}\\
(\mbf{1},\mbf{1},\mbf{15})_{(0,0)}\\
\begin{array}{l}
(\mbf{1},\mbf{1},\mbf{1})_{L(4,2)}+(\mbf{1},\mbf{1},\mbf{1})'_{L(4,2)}\\    %
(\mbf{2},\mbf{2},\mbf{1})_{L(2,2)}+(\mbf{2},\mbf{2},\mbf{1})'_{L(2,2)}\\
(\mbf{1},\mbf{1},\mbf{6})_{L(-2,2)}+(\mbf{1},\mbf{1},\mbf{6})'_{L(-2,2)}\\      %
(\mbf{2},\mbf{1},\mbf{4})_{L(-1,2)}-(\mbf{2},\mbf{1},\mbf{4})'_{L(-1,2)}\\
(\mbf{1},\mbf{2},\overline{\mbf{4}})_{L(1,2)}-(\mbf{1},\mbf{2},\overline{\mbf{4}})_{L(1,2)}\\      %
(\mbf{1},\mbf{1},\mbf{1})_{L(4,-1)}+(\mbf{1},\mbf{1},\mbf{1})'_{L(4,-1)}\\       %
(\mbf{2},\mbf{2},\mbf{1})_{L(2,-1)}+(\mbf{2},\mbf{2},\mbf{1})'_{L(2,-1)}\\
(\mbf{1},\mbf{1},\mbf{6})_{L(-2,-1)}+(\mbf{1},\mbf{1},\mbf{6})'_{L(-2,-1)}\\         %
(\mbf{2},\mbf{1},\mbf{4})_{L(-1,-1)}-(\mbf{2},\mbf{1},\mbf{4})'_{L(-1,-1)}\\
(\mbf{1},\mbf{2},\overline{\mbf{4}})_{L(1,-1)}-(\mbf{1},\mbf{2},\overline{\mbf{4}})_{L(1,-1)}\\         
\end{array}
\end{array}$
\\\hline  
$\mbf{4c'}$
& 
$\begin{array}{c}
\mathrm{W}\\
\scriptstyle{(\bb{Z}_{2})}\\
\scriptstyle{[embedding~(\mbf{1})]}
\end{array}$                                                                                              
&
$SO(10)\times U(1)\,\Big(\times U^{I}(1)\times U^{II}(1)\Big)$
&
$\begin{array}{c}
\mbf{1}_{(0,0,0)}\\
\mbf{1}_{(0,0,0)}\\
\mbf{1}_{(0,0,0)}\\
\mbf{45}_{(0,0,0)}\\
\mbox{and}\\
\begin{array}{l}
\mbf{1}_{L(-4,0,-2)}+\mbf{1}'_{L(-4,0,-2)}\\
\mbf{10}_{L(-2,0,-2)}+\mbf{10}'_{L(-2,0,-2)}\\
\mbf{16}_{L(1,0,-2)}-\mbf{16}'_{L(1,0,-2)}\\
\end{array}\\
\mbox{or}\\
\begin{array}{l}
\mbf{1}_{L(-4,-1,1)}+\mbf{1}'_{L(-4,-1,1)}\\
\mbf{10}_{L(-2,-1,1)}+\mbf{10}'_{L(-2,-1,1)}\\
\mbf{16}_{L(1,-1,1)}-\mbf{16}'_{L(1,-1,1)}\\
\end{array}\\
\mbox{or}\\
\begin{array}{l}
\mbf{1}_{L(-4,1,1)}+\mbf{1}'_{L(-4,1,1)}\\
\mbf{10}_{L(-2,1,1)}+\mbf{10}'_{L(-2,1,1)}\\
\mbf{16}_{L(1,1,1)}-\mbf{16}'_{L(1,1,1)}
\end{array}1
\end{array}$\\
\ \
&
&
&\\
\ \
&$\begin{array}{c}
\mathrm{W}\\
\scriptstyle{(\bb{Z}_{2})}\\
\scriptstyle{[embeddings}\\
\scriptstyle{(\mbf{2}),~(\mbf{3})]}
\end{array}$                                                                                              
&
$SU(2)\times SU(6)\,\Big(\times U^{I}(1)\times U^{II}(1)\Big)$
&
$\begin{array}{c}
\mbf{1}_{(0,0)}\\
\mbf{1}_{(0,0)}\\
(\mbf{3},\mbf{1})_{(0,0)}\\
(\mbf{1},\mbf{35})_{(0,0)}\\
\mbox{and}\\
\begin{array}{l}
(\mbf{1},\mbf{15})_{L(0,-2)}+(\mbf{1},\mbf{15})'_{L(0,-2)}\\
(\mbf{2},\overline{\mbf{6}})_{L(0,-2)}-(\mbf{2},\overline{\mbf{6}})'_{L(0,-2)}
\end{array}\\
\mbox{or}\\
\begin{array}{l}
(\mbf{1},\mbf{15})_{L(-1,1)}+(\mbf{1},\mbf{15})'_{L(-1,1)}\\
(\mbf{2},\overline{\mbf{6}})_{L(-1,1)}-(\mbf{2},\overline{\mbf{6}})'_{L(-1,1)}
\end{array}\\
\mbox{or}\\
\begin{array}{l}
(\mbf{1},\mbf{15})_{L(1,1)}+(\mbf{1},\mbf{15})'_{L(1,1)}\\
(\mbf{2},\overline{\mbf{6}})_{L(1,1)}-(\mbf{2},\overline{\mbf{6}})'_{L(1,1)}
\end{array}
\end{array}$
\\\hline
\end{longtable}
\end{scriptsize}

\chapter{Matrix Models}

\section{Gauge theory on the fuzzy sphere (multi-matrix model)}
\label{sec:fuzzygaugetheory}

Here I briefly review the construction of Yang-Mills gauge theory on $S^2_N$ as multi-matrix
model~\cite{Steinacker:2003sd,Presnajder:2003ak,Carow-Watamura:1998jn}. Consider the action 
\begin{eqnarray}
S &=& \frac{4\pi}{\cN}\, \Tr \Big(a^2 (\phi_a\phi_a + C_2(N))^2 
+  \frac 1{\tilde g^2}\, F_{ab}^\dagger F_{ab}\Big)
\label{YM-FS}
\end{eqnarray}
where $\phi_a = - \phi_a^{\dagger}$ is an antihermitean $\cN \times \cN$ matrix,
and define\footnote{This can indeed be seen as components of the two-form $F = dA + AA$.}
\begin{equation}
F_{{a}{b}} = [\phi_{{a}}, \phi_{{b}}] 
-  \varepsilon^{c}{}_{ab} \phi_{{c}}\,,
\end{equation}
as I did in section~\ref{FuzzySphere}. This action is invariant under the $U(\cN)$ `gauge' symmetry
acting as
$$
\phi_a \to U^{-1} \phi_a U\,
$$
A priori, I do not assume any underlying geometry, which arises dynamically. I claim that it
describes $U(n)$ Yang-Mills gauge theory on the fuzzy sphere $S^2_{N}$, assuming that $\cN = N n$.

To see this, note first that the action is positive definite, with global minimum $S=0$ for the
`vacuum' solution
\begin{equation}
\phi_a = X_a^{(N)}\otimes \one_n
\end{equation}
where $X_a \equiv X_a^{(N)}$ are the generators of the $N$- dimensional irrep. of $SU(2)$. This is a
first indication that the model `dynamically generates' its own geometry, which is the fuzzy sphere
$S^2_{N}$. In any case, it is natural to write a general field $\phi_a$ in the form
\begin{equation}
\phi_a = (X_a\otimes \one_{n}) + A_a,
\label{gaugefield-S2N}
\end{equation}
and to consider $A_a = \sum_\a A_{a}^{\a}(x) \,{\cal T}^{\a}$ as functions 
$A_{a}^{\a}(x) = -A_{a}^{\a}(x)^\dagger$ on the fuzzy sphere $S^2_{N}$, taking value in $u(n)$ with
generators $ T_\a$. The gauge transformation then takes the form
\begin{eqnarray}
{A}_a &\to&  U^{-1} {A}_a U +  U^{-1} [X_a, U]\nn\\
 &=&  U^{-1} {A}_a U -i  U^{-1} J_{a} U,
\end{eqnarray}
which is the transformation rule of a  $U(n)$ gauge field. The field strength becomes
\begin{eqnarray}
F_{ab} &=& [X_a,{ A}_b] -  [X_b,{A}_a] + [{A}_a,{A}_b] - \varepsilon^{c}{}_{ab}{A}_c \nn\\
  &=&  -i J_{a} A_b  +i J_{b} A_a + [{A}_a,{A}_b] - \varepsilon^{c}{}_{ab} {A}_c.
\end{eqnarray}
This look like the field strength of a nonabelian $U(n)$ gauge field, with the caveat that one has
three degrees of freedom rather than two. To solve this puzzle, consider again the action,
writing it in the form
\begin{eqnarray}
S &=& \frac{4\pi}{\cN}\, \Tr \Big(a^2 \varphi^2 + \frac 1{\tilde g^2}\, F_{ab}^\dagger F_{ab}\Big),
\label{YM-FS-2}
\end{eqnarray}
where I introduce the scalar field
\begin{equation}
\varphi := \phi_a \phi_a + C_2(N)= X_a {A}_{a} + {A}_{a} X_a + {A}_{a}{A}_{a}.
\end{equation}
Since only configurations where $\varphi$ and $F_{ab}$ are small will significantly contribute to
the action, it follows that
\begin{equation}
x_a {A}_a + {A}_a x_a = O(\frac{\varphi}N)
\label{A-constraint}
\end{equation}
is small. This means that ${A}_{a}$ is tangential in the (commutative) large $N$ limit, and two
tangential gauge degrees of freedom\footnote{To recover the familiar form of gauge theory, one needs
to rotate the components locally by $\frac{\pi} 2$ using the complex structure of $S^2$. A more
elegant way to establish the interpretation as Yang-Mills action can be given  using differential
forms on $S^2_{N}$.} survive. Equivalently, one can use the scalar field $\phi = N \varphi$, which
would acquire a mass of order $N$ and decouple from the theory.

I have thus established that the matrix model~\eq{YM-FS} is indeed a fuzzy version of pure $U(n)$
Yang-Mills theory on the sphere, in the sense that it reduces to the commutative model in the large
$N$ limit. Without the term $(\phi_a \phi_a + C_2(N))^2$, the scalar field corresponding to the
radial component of $A_a$ no longer decouples and leads to a different model.

The main message to be remembered is the fact that the matrix model~\eq{YM-FS} without any further
geometrical assumptions dynamically generates the space $S^2_{N}$, and the fluctuations
turn out to be gauge fields governed by a $U(n)$ Yang-Mills action. Furthermore, the vacuum has no
flat directions\footnote{The excitations turn out to be monopoles as
expected~\cite{Steinacker:2003sd}.}

\section{Stability of the `fuzzy sphere solution'}\label{sec:stability}

To establish stability of the vacua~\eq{vacuum-mod1},~\eq{vacuum-mod2} one should work out the
spectrum of excitations around this solution and check whether there are flat or unstable modes.
This is a formidable task in general, and here I only consider the simplest case of the irreducible
vacuum~\eq{vacuum-mod1} for the case $\tilde b = C_2(N)$ and $d=0$ here. Once I 
have established that all fluctuation modes have strictly positive eigenvalues, the same will hold
in a neighbourhood of this point in the moduli space of couplings $(a,b,d,\tilde g, g_6)$.

An intuitive way to see this is by noting that the potential $V(\phi_a)$ can be interpreted as
Yang-Mills theory on $S^2_{N}$ with gauge group $U(n)$. Since the sphere is compact, I expect
that all fluctuations around the vacuum  $\phi_a = X_a^{(N)} \otimes \one_n$ have positive energy.
I fix $n=1$ for simplicity. Thus I write
\begin{equation}
\phi_a =  X_a + A_a(x)
\end{equation}
where $A_a(x)$ is expanded into a suitable basis of harmonics of $S^2_{N}$, which I should find. 
It turns out that a convenient way of doing this is to consider the antihermitean $2N \times 2N$
matrix~\cite{Steinacker:2003sd}
\begin{equation}
\Phi = -\frac{i}{2} + \phi_a \sigma_a = \Phi_{0} \, + A
\end{equation}
which satisfies
\begin{equation}
\Phi^2=  \phi_a \phi_a - \frac{1}{4}+ \frac i2 \varepsilon_{abc} F_{bc} \sigma_a\,.
\label{Phi2-F}
\end{equation}
Thus $\Phi^2 = -\frac{N^2}{4}$ for $A=0$, and in general I have
\begin{equation}
\tilde S_{YM} := \Tr (\Phi^2 + \tilde b + \frac{1}{4})^2
 = \Tr \Big((\phi_a \phi_a  + \tilde b)^2 + F_{ab}^\dagger F_{ab} \Big).
\label{Stilde}
\end{equation}
The following maps turn out to be useful:
\begin{equation}
\cD(f) := i\{\Phi_0, f\}, \qquad \cJ(f) := [\Phi_0, f]
\end{equation}
for any matrix $f$. The maps $\cD$ and $\cJ$ satisfy
\begin{equation}
\cJ \cD = \cD \cJ =i[\Phi_0^2,.], \qquad \cD^2 - \cJ^2 = -2\{\Phi_0^2,.\},
\end{equation}
which for the vacuum under consideration become
\begin{equation}
\cJ \cD = \cD \cJ =0, \qquad \cD^2 - \cJ^2 = N^2 , \qquad \cJ^3 = -N^2\, \cJ.
\end{equation}
Note also that
\begin{equation}
\cJ^2(f) = [\phi_a,[\phi_a,f]]  =: -\Delta f
\end{equation}
is the Laplacian, with eigenvalues $\Delta f_l =  l(l+1) f_l$  (for the vacuum).

It turns out that the following is a natural basis of fluctuation modes:
\begin{eqnarray}
\delta \Phi^{(1)} &=& A^{(1)}_a \sigma_a = \cD(f) - f, \nn \\
\delta \Phi^{(2)} &=&   A^{(2)}_a \sigma_a = \cJ^2(f') -\cJ^2(f')_0 
   = \cJ^2(f') +\Delta f'  \nn \\
\delta \Phi^{(g)} &=&  A^{(g)}_a \sigma_a = \cJ(f'') 
\label{basis}
\end{eqnarray}
for antihermitean $N \times N$ matrices $f,f',f''$, which will be expanded into orthonormal modes $f
= \sum  f_{l,m}\, Y_{lm}$. Using orthogonality it is enough to consider these modes separately, i.e.
$f = f_l = - f_l^\dagger$ with $\Tr(f_l^\dagger f_l) =1$. One can show that these modes form a
complete set of fluctuations around $\Phi_0$ (for the vacuum). Here $A_{(g)}$ corresponds to gauge
transformations, which I will omit from now on.
Using
\begin{equation}
\Tr (f \cJ(g)) = - \Tr (\cJ(f) g),, \qquad \Tr (f \cD(g)) = \Tr f (\cD(f) g)
\end{equation}
I can now compute the inner product matrix $\Tr A^{(i)}A^{(j)}$:
\begin{eqnarray}
\Tr (A^{(1)}A^{(1)}) &=&  \Tr (((N^2-1) f - \Delta(f)) g), \nn\\
\Tr (A^{(1)}A^{(2)}) &=&   \Tr(\Delta(f) g), \nn\\
\Tr (A^{(2)}A^{(2)}) &=& \Tr((N^2\Delta(f) -\Delta^2 f))g).
\end{eqnarray}
It is convenient to introduce the matrix of normalizations for the modes $A^{(i)}$,
\begin{eqnarray}
G_{ij}\equiv \Tr((A^{(i)})^\dagger A^{(j)})= \left(\begin{array}{cc} (N^2-1)  - \Delta,& \Delta \\
\Delta, & N^2\Delta -\Delta^2
\end{array}\right)
\end{eqnarray}
which is positive definite except for the zero mode $l=0$ where $A^{(2)}$ is not defined.

We can now expand the action \eq{YM-FS} up to second order in these fluctuations. Since $F_{ab}=0$
and $(\phi_a \phi_a + \tilde b)=0$ for the vacuum, I have\footnote{Note that $\delta 
\Tr(\phi\cdot\phi)=0$ except for the zero mode $A^{(1)}_0$ with $l=0$ where
$\delta^{(1)} \Tr(\phi\cdot\phi) \neq 0$, as follows from \eq{phi2-fluct}. This mode corresponds 
to fluctuations of the radius, which will be discussed separately.}
\begin{equation}
\delta^2 S_{YM}  =
 \Tr \Big(-\frac 1{\tilde g^2}\, \delta F_{ab} \delta F_{ab} 
 + a^2 \delta(\phi_a \phi_a)\delta(\phi_b \phi_b)\Big).
\end{equation}
If $a^2\geq \frac 1{\tilde g^2}$, this can be written as
\begin{eqnarray}
\delta^2 S_{YM}  &=&
 \Tr \Big(\frac 1{\tilde g^2}\, (-\delta F_{ab} \delta F_{ab} 
 + a^2 \delta(\phi_a \phi_a)\delta(\phi_a \phi_a))
  + (a^2-\frac 1{\tilde g^2})\delta(\phi_a \phi_a)\delta(\phi_a
  \phi_a) \Big) \nn\\
&=&  \Tr\Big(\frac 1{\tilde g^2}\,\delta \Phi^2 \delta \Phi^2
+ (a^2-\frac 1{\tilde g^2})\delta(\phi_a \phi_a)\delta(\phi_a
  \phi_a)\Big) 
\end{eqnarray}
and similarly for 
$a^2 < \frac 1{\tilde g^2}$. It is therefore enough to show that
\begin{equation}
\delta^2 \tilde S_{YM}  = \Tr(\delta \Phi^2 \delta \Phi^2)
= \Tr(-\delta^{(i)} F_{ab} \delta^{(j)} F_{ab} 
+ \delta^{(i)} (\phi\cdot\phi) \delta^{(j)} (\phi\cdot\phi) ) 
\end{equation}
has a finite gap in the excitation spectrum. This spectrum can be computed efficiently as follows:
note first
\begin{eqnarray}
\delta^{(1)} \Phi^2 &=& -i\cD^2(f) +i \cD(f) 
                    = -i\cJ^2(f) +i \cD(f) -i N^2 f, \nn\\
\delta^{(2)} \Phi^2 &=& -i\cD(\Delta f), \nn\\
\delta^{(g)} \Phi^2 &=& -i\cD\cJ(f) = [\Phi_0^2,f] =0
\label{phi2-fluct}
\end{eqnarray}
for the vacuum. One then finds
\begin{eqnarray}
\Tr (\delta^{(1)} (\Phi^2) \delta^{(1)} (\Phi^2)) 
&=& - \Tr(f)((-(N^2+1)\Delta  + (N^2 -1)N^2) g), \nn\\
\Tr (\delta^{(1)} (\Phi^2) \delta^{(2)} (\Phi^2)) 
&=& -  \Tr(f)  (\Delta^2)( g) , \nn\\
\Tr (\delta^{(2)} (\Phi^2) \delta^{(2)} (\Phi^2)) 
&=& - \Tr (g) (-\Delta^3 + N^2\Delta^2)g).
\end{eqnarray}
Noting that the antihermitean modes satisfy $\Tr(f_l f_l) =-1$,
this gives
\begin{equation}
\delta^2 \tilde S_{YM}  
= \left(\begin{array}{cc} 
-(N^2+1)\Delta + N^4-N^2, &  \Delta^2 \\ 
  \Delta^2 , & -\Delta^3  + N^2\Delta^2
\end{array}\right) = G T
\label{Stilde-fluct}
\end{equation}
where the last equality defines $T$. The eigenvalues of $T$ are found to be
$N^2$ and $\Delta$. These eigenvalues coincide\footnote{To see this, assume that I use an 
orthonormal basis  $A^o_{(i)}$ instead of the basis \eq{basis}, i.e. $A=b_1 A^o_{(1)}+b_2
A^o_{(2)}$. Then I can write $G=g^T g$ and $b_i=g_{ij} a_j$. Thus \eq{Stilde-fluct} becomes $a^T\ 
G T\ a=b^T\ g\ T\ g^{-1} b$, and the eigenvalues of $ g\ T\ g^{-1}$ coincide with those of $T$,
which therefore gives the masses.} with the spectrum of the fluctuations of $\tilde S_{YM}$.
In particular, all modes with $l>0$ have positive mass. The $l=0$ mode 
\begin{equation}
A^{(1)}_0 = \cD(f_0) - f_0 = (2 i \Phi_0 -1) f_0
 = 2 i f_0\, \sigma_a \phi_a 
\end{equation}
requires special treatment, and corresponds precisely to the fluctuations of the normalization $\a$,
i.e. the radius of the sphere. We have shown explicitly in \eq{Higgs-potential} that this $\a =
\a(y)$ has a positive mass. Therefore I conclude that  all modes have positive mass, and 
there is no flat or unstable direction. This establishes the stability of this vacuum.

The more general case  $\tilde b = C_2(N) + \epsilon$ with $\a \neq 1$ could be analysed
with the same methods, which however will not be presented in this dissertation. For the reducible
vacuum~\eq{vacuum-mod2} or~\eq{vacuum-mod3} the analysis is more complicated, and will not be
carried out here.

\section{Stability of `type-1' and `type-2' vacua}
\label{stability-VAC1-VAC2}

To verify the stability of  `type-1' and `type-2' vacua solution I used the following
Mathematica code:

\newpage
\begin{framed}
\begin{scriptsize}
\begin{verbatim}

<< LinearAlgebra`MatrixManipulation`
<< Algebra`ReIm`

tensorProductIdentity[Amatrix_, n_] :=
  Module[{aux, result},
    aux = Amatrix;
    If[n > 1,
      Do[
        result = 
          BlockMatrix[{{aux,ZeroMatrix[Length[aux], Length[Amatrix]]}, 
                      {ZeroMatrix[Length[Amatrix], Length[aux]], Amatrix}}];
        aux = result,
        {i, 1, n - 1}
      ],
      If[n == 1,
        result = Amatrix,
        Print[
          "Unexpected value for (Dim of Identity tensor product term)=n.\\
           VACconstruction[Np>=2,n>=1] is expected."];
          Abort[]
      ]
    ];
    Return[result]
  ]

VACIconstruction[Np_,n_] := 
  Module[{J0, Jplus, Jminus, x, X, j, m, alpha, phi, vac}, 
    Clear[phi]; 
    Im[jt] ^= 0; 
    j = (Np - 1)/2; 
    Jminus = Table[
               KroneckerDelta[mp,mm + 1](((j + mp)!(j - mm)!)/((j + mm)!(j - mp)!))^(1/2), 
               {mp, -j, j}, {mm, -j, j}
             ];
    Jplus = Table[
              KroneckerDelta[mp,mm - 1](((j + mm)!(j - mp)!)/((j + mp)!(j - mm)!))^(1/2), 
              {mp, -j, j}, {mm, -j, j}
            ];
    J0 = Table[KroneckerDelta[mp, m] (m), {mp, -j, j}, {m, -j, j}]; 
    x[1] = (I/2)*(Jplus + Jminus); 
    x[2] = (1/2)*(Jplus - Jminus); 
    x[3] = (I) *J0; 
    m = 2*(j - jt); 
    alpha =1 - m/(2 * j); 
    Do[
      If[Np >= 2 , 
         X[k] = x[k], 
         Print["\<Unexpected value for (Dim of rep)=Np.\\
                  VACconstruction[Np>=2,n>=1] is expected.\>"]; 
         Abort[]
      ]; 
      phi[k] = Simplify[alpha*tensorProductIdentity[X[k], n]],
      {k, 1, 3}
    ]; 
   vac = {phi[1], phi[2], phi[3]}
 ]

\end{verbatim}

\newpage

\begin{verbatim}

A1reps[j_] := 
  Module[{J0, Jminus, Jplus, x, xlist}, 
     Jminus = Table[
                KroneckerDelta[mp, mm + 1] (((j + mp)!(j - mm)!)/((j + mm)!(j - mp)!))^(1/2), 
                {mp, -j, j}, {mm, -j, j}
               ];
     Jplus = Table[
               KroneckerDelta[mp, mm - 1](((j + mm)!(j - mp)!)/((j + mp)!(j - mm)!))^(1/2), 
               {mp, -j, j}, {mm, -j, j}
              ];
     J0 = Table[KroneckerDelta[mp, m] (m), {mp,-j, j}, {m, -j,j}]; 
     x[1] = (I/2)*(Jplus + Jminus); 
     x[2] = (1/2)* (Jplus - Jminus); 
     x[3] = (I)* J0; 
     xlist = {x[1], x[2], x[3]}
 ]

VACIIconstruction[Np1_, n1_, n2_] :=
   Module[{x1, x2, X1, X2, j1, m1, alpha1, Np2, j2, m2, alpha2, vac},
    Clear[phi];
    Im[jt] ^= 0;
    j1 = (Np1 - 1)/2;
    x1 = A1reps[j1];
    m1 = 2(j1 - jt);
    alpha1 = 1 - m1/(2*j1);
    Np2 = Np1 + 1;
    j2 = (Np2 - 1)/2;
    x2 = A1reps[j2];
    m2 = 2(j2 - jt);
    alpha2 = 1 - m2/(2*j2);
    Do[If[Np1 >= 2 ,
         X1[k] = x1[[k]],
         Print["Unexpected value for (Dim of rep)=Np. VACconstruction[Np>=2,n>=1] is expected."];
         Abort[]
        ];
      If[Np2 >= 2 ,
        X2[k] = x2[[k]],
        Print["Unexpected value for (Dim of rep)=Np. VACconstruction[Np>=2,n>=1] is expected."];
        Abort[]
      ];
      phi1[k] = Simplify[alpha1*tensorProductIdentity[X1[k], n1]];
      phi2[k] = Simplify[alpha2*tensorProductIdentity[X2[k], n2]];
      phi[k] = 
        BlockMatrix[{{phi1[k],ZeroMatrix[Length[ phi1[k] ], Length[ phi2[k] ] ]},
                    {ZeroMatrix[Length[phi2[k]], Length[phi1[k]]], phi2[k]}}],
      {k, 1, 3}
    ];
    vac = {phi[1], phi[2], phi[3]}
  ]


VACIperturbation[Np_, n_] :=
  Module[{perturbedMatrix, NBig, f},
    NBig = Np*n;
    Clear[perturbedMatrix];
    perturbedMatrix = ZeroMatrix[NBig];
    Do[
      f = Random[Real, {-1, +1}];
      perturbedMatrix = ReplacePart[perturbedMatrix, f, {{i, j}, {j, i}}],
      {i, 1, NBig}, {j, 1, NBig}
      ];
    Return[- I *perturbedMatrix]
   ]

\end{verbatim}

\newpage

\begin{verbatim}

VACIIperturbation[Np1_, n1_, n2_] :=
  Module[{perturbedMatrix, Np2, NBig, f},
    Np2 = Np1 + 1;
    NBig = Np1*n1 + Np2*n2;
    Clear[perturbedMatrix];
    perturbedMatrix = ZeroMatrix[NBig];
    Do[
      f = Random[Real, {-1, +1}];
      perturbedMatrix = ReplacePart[perturbedMatrix, f, {{i, j}, {j, i}}],
      {i, 1, NBig}, {j, 1, NBig}
      ];
    Return[- I *perturbedMatrix]
   ]

perturbedVACI[Np_, n_] :=
  Module[{},
    Clear[e];
    e={e1,e2,e3};
    Im[e1] ^= 0;
    Im[e2] ^= 0;
    Im[e3] ^= 0;
    Table[
      VACIconstruction[Np, n][[k]] + e[[k]] *VACIperturbation[Np, n],
      {k, 1, 3}
      ]
    ]

perturbedVACII[Np1_, n1_, n2_] :=
  Module[{},
    Clear[e];
    e={e1,e2,e3};
    Im[e1] ^= 0;
    Im[e2] ^= 0;
    Im[e3] ^= 0;
    Table[
      VACIIconstruction[Np1, n1, n2][[k]] + 
        e[[k]] *VACIIperturbation[Np1, n1, n2],
      {k, 1, 3}
      ]
    ]

perturbedFieldStrengthI[Np_, n_] :=
  Module[{perturbedphi},
    perturbedphi = perturbedVACI[Np, n];
    Table[
      (Dot[perturbedphi[[a]], perturbedphi[[b]] ] - 
            Dot[perturbedphi[[b]], perturbedphi[[a]]]) -
        Sum[Signature[{a, b, c}]perturbedphi[[c]], {c, 1, 3}],
      {a, 1, 3}, {b, 1, 3}
      ]
    ]

perturbedFieldStrengthII[Np1_, n1_, n2_] :=
  Module[{perturbedphi},
    perturbedphi = perturbedVACII[Np1, n1, n2];
    Table[
      (Dot[perturbedphi[[a]], perturbedphi[[b]] ] - 
            Dot[perturbedphi[[b]], perturbedphi[[a]]]) -
        Sum[Signature[{a, b, c}]perturbedphi[[c]], {c, 1, 3}],
      {a, 1, 3}, {b, 1, 3}
      ]
    ]

\end{verbatim}

\newpage

\begin{verbatim}

perturbedPotentialI[Np_, n_, InvgtSqr_] :=
  Module[{perturbedF},
    perturbedF = perturbedFieldStrengthI[Np, n];
    Collect[
      Simplify[
        Tr[
          Sum[-InvgtSqr*Dot[perturbedF[[a, b]], perturbedF[[a, b]] ] ,
            {a, 1, 3}, {b, 1, 3}]
          ]
        ],
      e]
    ]

perturbedPotentialII[Np1_, n1_, n2_, InvgtSqr_] :=
  Module[{perturbedF},
    perturbedF = perturbedFieldStrengthII[Np1, n1, n2];
    Collect[
      Simplify[
        Tr[
          Sum[-InvgtSqr*Dot[perturbedF[[a, b]], perturbedF[[a, b]] ] ,
            {a, 1, 3}, {b, 1, 3}]
          ]
        ],
      e]
    ]

potentialI[Np_, n_, InvgtSqr_, jmin_] :=
  Module[{perturbedPotential, perturbedPotentialRe, perturbedPotentialIm, 
    potentialVacAux},
    Clear[e1,e2,e3,jt];
    Im[e1] ^= 0;
    Im[e2] ^= 0;
    Im[e3] ^= 0;
    perturbedPotential = perturbedPotentialI[Np, n, InvgtSqr];
    perturbedPotentialIm = Im[perturbedPotential];
    perturbedPotentialRe = Expand[Re[perturbedPotential]];
    
    potential1[e1_,e2_,e3_,jt_] = perturbedPotentialRe;
    
    potentialVacAux = perturbedPotentialRe /. {e1 -> 0, e2 -> 0, e3 ->0};
    potentialVac1[jt_] := potentialVacAux; 

    Print["Perturbed Potential"];
    Print[perturbedPotential];
    Print[" ********************************************"];
    Print["Imaginary Part of Perturbed Potential"];
    Print[perturbedPotentialIm];
    Print[" ********************************************"];
    Print["Real Part of Perturbed Potential"];
    Print[perturbedPotentialRe];
    Print["*********************************************"];
    Print["N=", Np*n, "," , " ", "Np=", Np, ",", " ", "n=", n ];
    DeleteFile[{"potentialI","potentialVACI"}];
    Save["potentialI",potential1];
    Save["potentialVACI",potentialVac1]
  ]  

\end{verbatim}

\newpage

\begin{verbatim}

PotentialII[Np1_, n1_, n2_, InvgtSqr_, jmin_] :=
  Module[{perturbedPotential, perturbedPotentialRe, perturbedPotentialIm, 
    potentialVacAux},
    Clear[e1,e2,e3,jt];
    Im[e1] ^= 0;
    Im[e2] ^= 0;
    Im[e3] ^= 0;    
    perturbedPotential = perturbedPotentialII[Np1, n1, n2, InvgtSqr];
    perturbedPotentialIm = Im[perturbedPotential];
    perturbedPotentialRe = Expand[Re[perturbedPotential]];
    
    potential2[e1_,e2_,e3_,jt_] = perturbedPotentialRe;
    
    potentialVacAux = perturbedPotentialRe /. {e1 -> 0, e2 -> 0, e3 ->0};
    potentialVac2[jt_] := potentialVacAux;

    Print["Perturbed Potential"];  
    Print[perturbedPotential];
    Print["*********************************************"];
    Print["Imaginary Part of Perturbed Potential"];
    Print[perturbedPotentialIm];
    Print["*********************************************"];
    Print["Real Part of Perturbed Potential"];
    Print[perturbedPotentialRe];
    Print["*********************************************"];
    Print["N=", Np1*n1 + (Np1 + 1)*n2, "," , " ", "N1=", Np1, ",", " ", "n1=",
           n1, ",", " ", "N2=", Np1 + 1, ",", " ", "n2=", n2 ];
    DeleteFile[{"potentialII","potentialVACII"}];
    Save["potentialII",potential2];
    Save["potentialVACII",potentialVac2]
 ]

\end{verbatim}

\end{scriptsize}
\end{framed}

I run the code on the HET-cluster of the Physics Department of N.T.U.A. with the choices:

\noindent
\medskip
{\large\textbf{type-1}}\\
$N=39$ and $n =3$\\
being the $\cN=Nn=117$ partition of $\cN\times\cN$ matrices.

\noindent
\medskip
{\large\textbf{{type-2}}}\\
$N_{1}=23$, $n_{1}=3$ and $N_{2}=N_{1}+1$, $n_{2}=2$\\
being the $\cN=n_{1}N_{1}+n_{2}N_{2}=117$ partition of $\cN\times\cN$ matrices.

These are within the approximation of large matrices assumed in
chapter~\ref{FuzzyExtraDim} up to second decimal digit precision.
We have also choose  $\frac{1}{\tld{g}^{2}}=0.9$ for the coupling constant of the $F_{ab}F^{ab}$
term of the matrix model potential presented in chapter~\ref{FuzzyExtraDim}. Finally the $\tld{j}$
has been chosen to be in the $N_{1}<\tld{N}<N_{2}=N_{1}+1$ interval.

\newpage
The following results prove our arguments of the stability of `type-1' and `type-2' solutions


\bigskip
\noindent
{\large\textbf{type-1}}
\begin{small}
\begin{verbatim}

potential1[e1$_, e2$_, e3$_, jt$_] = 8310.196242138307*e1^2 + 0.*e1^4 + 
     8171.922774302844*e2^2 + 675171.794769941*e1^2*e2^2 + 0.*e2^4 + 
     0.*e1*e2*e3 + 8245.841526584754*e3^2 + 642124.806455661*e1^2*e3^2 + 
     648257.0741220384*e2^2*e3^2 + 0.*e3^4 + 8.730939070642787*e1*jt + 
     0.*e1^3*jt + 0.*e1^2*e2*jt + 5.302009621272134*e1*e2^2*jt + 
     7.694136324951049*e3*jt - 419.89704482079856*e1*e3*jt - 
     210.06603550339628*e1^2*e3*jt + 0.*e2*e3*jt - 4000.7618414361873*e2^2*e3*
      jt - 0.4532475695330902*e1*e3^2*jt + 0.*e2*e3^2*jt + 0.*e3^3*jt + 
     221.68421052631604*jt^2 - 1.3785693269434898*e1*jt^2 + 
     11703.294608077817*e1^2*jt^2 + 0.*e1*e2*jt^2 + 
     11446.737752030098*e2^2*jt^2 - 1.2148636302556433*e3*jt^2 - 
     140.10470181632246*e1*e3*jt^2 + 0.*e2*e3*jt^2 + 
     11299.83308575747*e3^2*jt^2 - 23.335180055401562*jt^3 + 
     0.0483708535769628*e1*jt^3 + 0.04262679404404693*e3*jt^3 + 
     0.6140836856684655*jt^4
 
Attributes[e1$] = {Temporary}
 
Attributes[e2$] = {Temporary}
 
Attributes[e3$] = {Temporary}
 
Attributes[jt$] = {Temporary}
 
Im[e1] ^= 0
 
Im[e2] ^= 0
 
Im[e3] ^= 0
 
Im[jt] ^= 0

\end{verbatim}

\end{small}

\begin{small}
\begin{verbatim}
Local Minima

{4686.55,

{e1 -> 2.44977x10^{-6}  , e2 -> 1.14775x10^{-19}   , e3 -> 2.23931x10^{-6}  }}

\end{verbatim}
\end{small}

\newpage

\bigskip
\noindent
{\large\textbf{type-2}}
\begin{small}
\begin{verbatim}

potential2[e1$_, e2$_, e3$_, jt$_] = 8239.417547028384*e1^2 + 0.*e1^4 + 
     8203.264441042475*e2^2 + 630723.2670988459*e1^2*e2^2 + 0.*e2^4 + 
     0.*e1*e2*e3 + 8127.235004958384*e3^2 + 644909.1812379633*e1^2*e3^2 + 
     605788.2695466514*e2^2*e3^2 + 0.*e3^4 + 13.585420318861358*e1*jt + 
     0.*e1^3*jt + 0.*e1^2*e2*jt + 978.7956924200289*e1*e2^2*jt + 
     11.315311245671497*e3*jt + 549.6356339064939*e1*e3*jt - 
     1629.9495936550218*e1^2*e3*jt + 0.*e2*e3*jt + 265.54923446855196*e2^2*e3*
      jt + 1156.5670778145934*e1*e3^2*jt + 0.*e2*e3^2*jt + 0.*e3^3*jt + 
     229.4039525691708*jt^2 - 3.8127881735926827*e1*jt^2 + 
     12036.411465663265*e1^2*jt^2 + 0.*e1*e2*jt^2 + 
     11971.194007631077*e2^2*jt^2 - 3.0508573620735935*e3*jt^2 - 
     19.438795293342935*e1*e3*jt^2 + 0.*e2*e3*jt^2 + 
     11587.559130487403*e3^2*jt^2 - 40.96741395741206*jt^3 + 
     0.23732001554284943*e1*jt^3 + 0.18286354102580768*e3*jt^3 + 
     1.8298771102902616*jt^4
 
Attributes[e1$] = {Temporary}
 
Attributes[e2$] = {Temporary}
 
Attributes[e3$] = {Temporary}
 
Attributes[jt$] = {Temporary}
 
Im[e1] ^= 0
 
Im[e2] ^= 0
 
Im[e3] ^= 0
 
Im[jt] ^= 0

\end{verbatim}

\end{small}

\begin{small}
\begin{verbatim}
Local Minima

{13.6354, 
{e1 -> -2.40332x10^{-6}  , e2 -> -1.60605x10^{-20}   , e3 -> -3.19194x10^{-7}  }}

\end{verbatim}
\end{small}

\newpage

\begin{figure}
\begin{center}
\ifpdf
\includegraphics[viewport= 0 0 300 120]{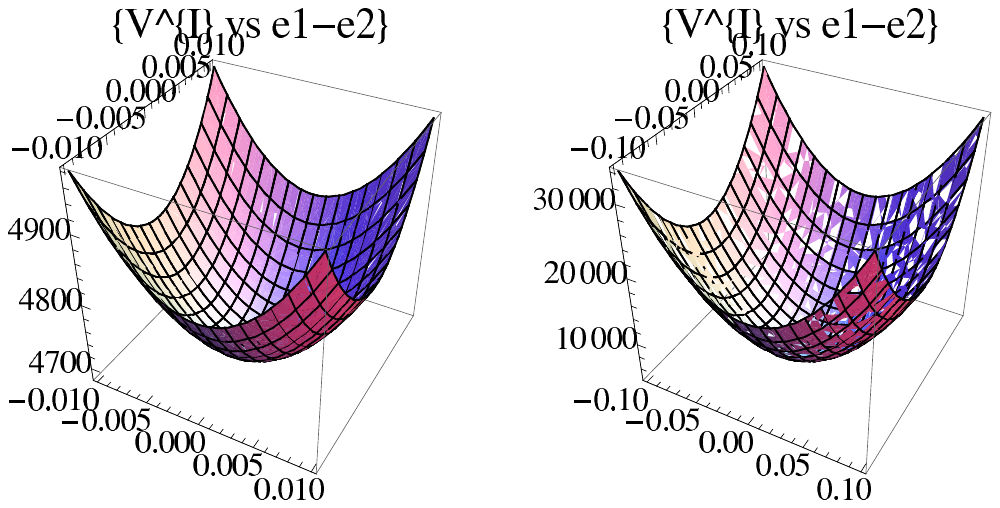}
\else
\includegraphics[width=\hugefigwidth,scale=0.90]{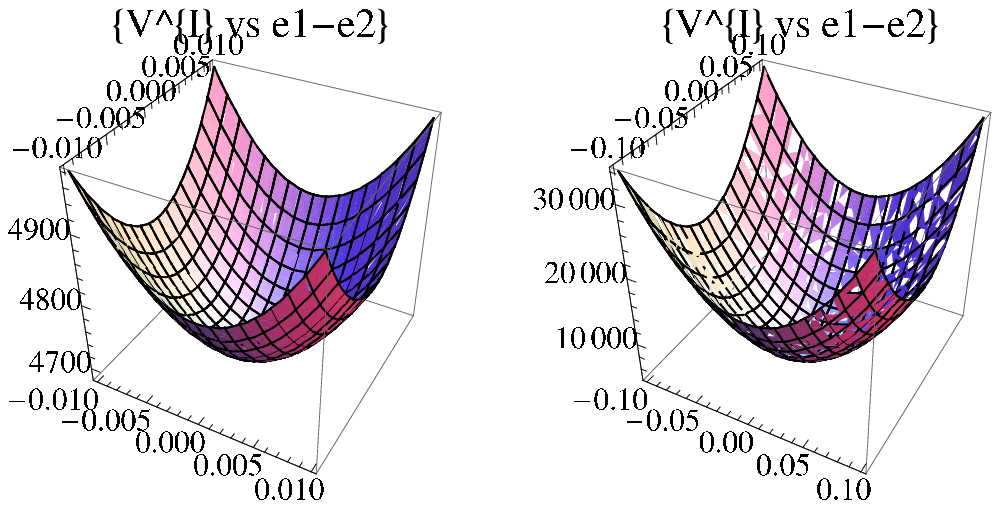}
\fi
\end{center}
\caption{\emph{`type-1'~Potential}}
\end{figure}
\begin{figure}
\begin{center}
\ifpdf
\includegraphics[viewport=0 0 300 120]{plotPotentialI-e1e2-Helvetica.pdf}
\else
\includegraphics[width=\hugefigwidth,scale=0.90]{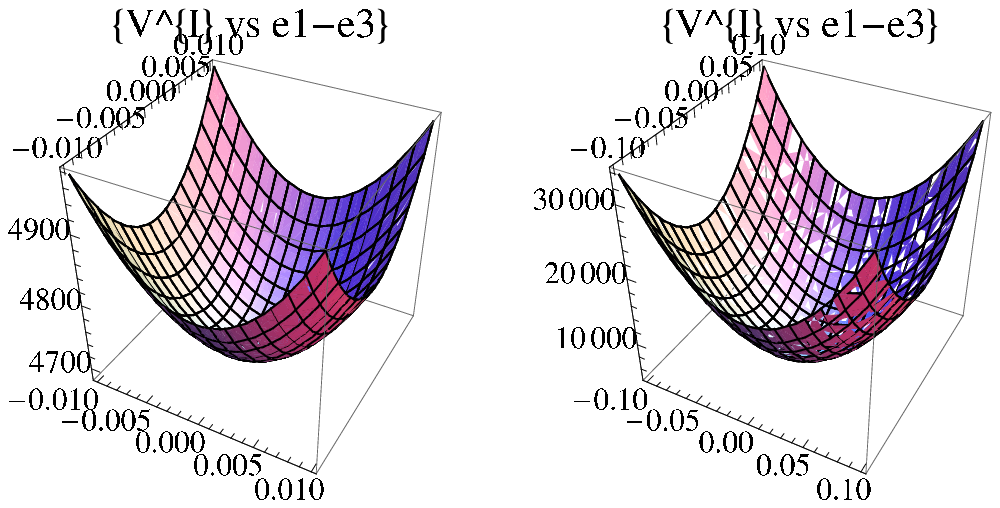}
\fi
\end{center}
\caption{\emph{`type-1'~Potential}}
\end{figure}
\begin{figure}
\begin{center}
\ifpdf
\includegraphics[viewport=0 0 300 120]{plotPotentialI-e1e2-Helvetica.pdf}
\else
\includegraphics[width=\hugefigwidth,scale=0.90]{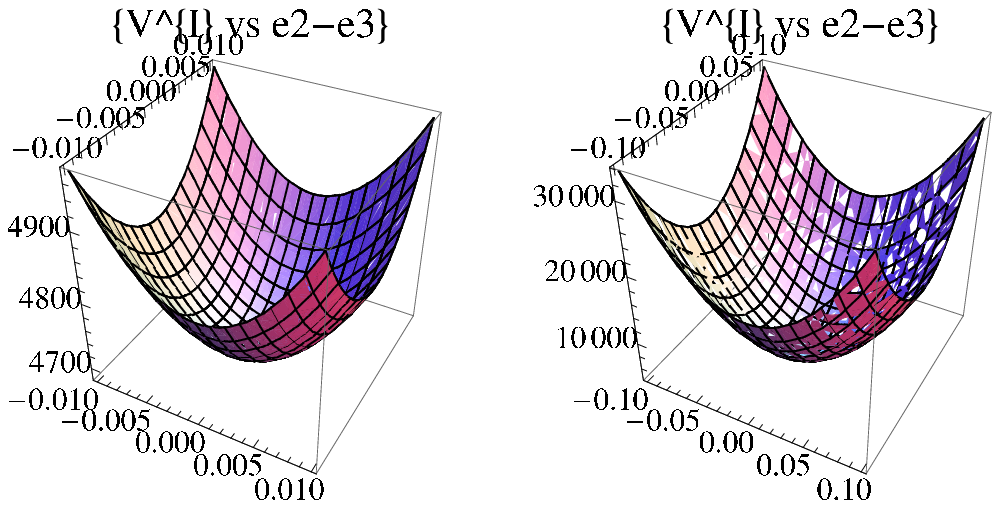}
\fi
\end{center}
\caption{\emph{`type-1'~Potential}}
\end{figure}


\begin{figure}
\begin{center}
\ifpdf
\includegraphics[viewport=0 0 300 120]{plotPotentialI-e1e2-Helvetica.pdf}
\else
\includegraphics[width=\hugefigwidth,scale=0.90]{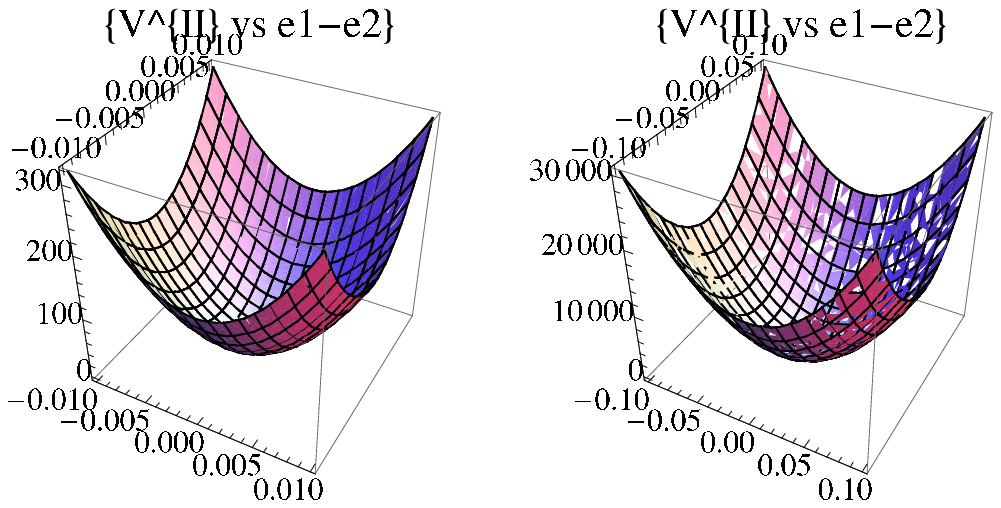}
\fi
\end{center}
\caption{\emph{`type-2'~Potential}}
\end{figure}
\begin{figure}
\begin{center}
\ifpdf
\includegraphics[viewport=0 0 300 120]{plotPotentialI-e1e2-Helvetica.pdf}
\else
\includegraphics[width=\hugefigwidth,scale=0.90]{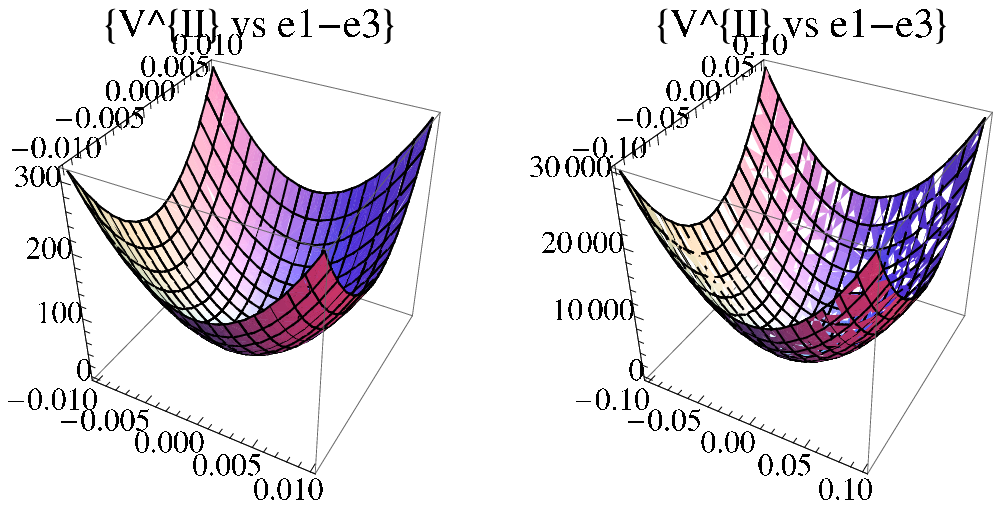}
\fi
\end{center}
\caption{\emph{`type-2'~Potential}}
\end{figure}
\begin{figure}
\begin{center}
\ifpdf
\includegraphics[viewport=0 0 300 120]{plotPotentialI-e1e2-Helvetica.pdf}
\else
\includegraphics[width=\hugefigwidth,scale=0.90]{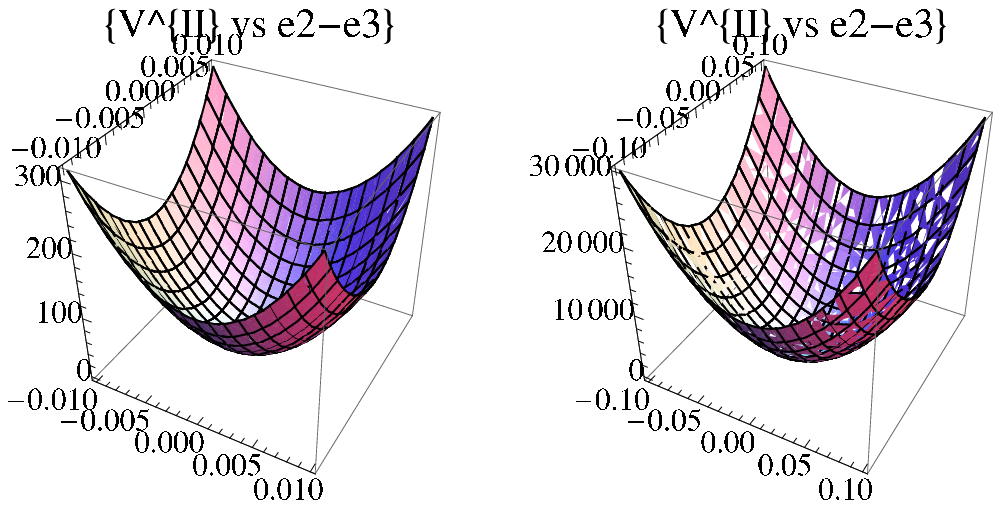}
\fi
\end{center}
\caption{\emph{`type-2'~Potential}}
\end{figure}

\end{appendices}
\begin{backmatter}

\begin{thebibliography}{10%
0}

\bibitem{Weinberg:1967tq}
S.~Weinberg, ``{A Model of Leptons},''
\href{http://dx.doi.org/10.1103/PhysRevLett.19.1264}{{\em Phys. Rev. Lett.}
  {\bf 19} (1967)  1264--1266}.

\bibitem{Glashow:1970gm}
S.~L. Glashow, J.~Iliopoulos, and L.~Maiani, ``{Weak Interactions with
  Lepton-Hadron Symmetry},''
\href{http://dx.doi.org/10.1103/PhysRevD.2.1285}{{\em Phys. Rev.} {\bf D2}
  (1970)  1285--1292}.

\bibitem{Georgi:1974sy}
H.~Georgi and S.~L. Glashow, ``{Unity of All Elementary Particle Forces},''
\href{http://dx.doi.org/10.1103/PhysRevLett.32.438}{{\em Phys. Rev. Lett.} {\bf
  32} (1974)  438--441}.

\bibitem{Fritzsch:1974nn}
H.~Fritzsch and P.~Minkowski, ``{Unified Interactions of Leptons and
  Hadrons},''
\href{http://dx.doi.org/10.1016/0003-4916(75)90211-0}{{\em Ann. Phys.} {\bf 93}
  (1975)  193--266}.

\bibitem{Mohapatra:1985xm}
P.~K. Mohapatra, R.~N. Mohapatra, and P.~B. Pal, ``{Implications of E6 Grand
  Unification},''
\href{http://dx.doi.org/10.1103/PhysRevD.33.2010}{{\em Phys. Rev.} {\bf D33}
  (1986)  2010}.

\bibitem{Weinberg:1978ym}
S.~Weinberg, ``{Gauge Hierarchies},''
\href{http://dx.doi.org/10.1016/0370-2693(79)90248-X}{{\em Phys. Lett.} {\bf
  B82} (1979)  387}.

\bibitem{Kaluza:1921tu}
T.~Kaluza, ``{On the Problem of Unity in Physics},''
{\em Sitzungsber. Preuss. Akad. Wiss. Berlin (Math. Phys.)} {\bf 1921} (1921)
  966--972.

\bibitem{Klein:1926tv}
O.~Klein, ``Quantum theory and five-dimensional theory of relativity,''
{\em Z. Phys.} {\bf 37} (1926)  895--906.

\bibitem{Bailin:1987jd}
D.~Bailin and A.~Love, ``{Kaluza-Klein theories},''
{\em Rept. Prog. Phys.} {\bf 50} (1987)  1087--1170.

\bibitem{Salam:1981xd}
A.~Salam and J.~A. Strathdee, ``{On Kaluza-Klein Theory},''
{\em Annals Phys.} {\bf 141} (1982)  316--352.

\bibitem{Witten:1983ux}
E.~Witten, ``{Fermion Quantum Numbers in Kaluza-Klein Theory},''. Lecture given
  at Shelter Island II Conf., Shelter Island, N.Y., 1-2 Jun 1983.

\bibitem{Horvath:1977st}
Z.~Horvath, L.~Palla, E.~Cremmer, and J.~Scherk, ``{Grand Unified Schemes and
  Spontaneous Compactification},''
\href{http://dx.doi.org/10.1016/0550-3213(77)90351-0}{{\em Nucl. Phys.} {\bf
  B127} (1977)  57}.

\bibitem{Chapline:1982wy}
G.~Chapline and R.~Slansky, ``{Dimensional reduction and flavor chirality},''
{\em Nucl. Phys.} {\bf B209} (1982)  461.

\bibitem{Duff:1986hr}
M.~J. Duff, B.~E.~W. Nilsson, and C.~N. Pope, ``{Kaluza-Klein Supergravity},''
{\em Phys. Rept.} {\bf 130} (1986)  1--142.

\bibitem{Green:1987sp}
M.~B. Green, J.~H. Schwarz, and E.~Witten, {\em {S}uperstring theory. vol. 1:
  {I}ntroduction}.
\newblock {C}ambridge {M}onographs {O}n {M}athematical {P}hysics. {C}ambridge,
  {U}k: {U}niv. {P}r., 1987.

\bibitem{Green:1987mn}
M.~B. Green, J.~H. Schwarz, and E.~Witten, {\em {S}uperstring theory. vol. 2:
  {L}oop amplitudes, anomalies and phenomenology}.
\newblock {C}ambridge {M}onographs {O}n {M}athematical {P}hysics. {C}ambridge,
  {U}k: {U}niv. {P}r., 1987.

\bibitem{Forgacs:1979zs}
P.~Forgacs and N.~S. Manton, ``{Space-time symmetries in gauge theories},''
{\em Commun. Math. Phys.} {\bf 72} (1980)  15.

\bibitem{Kapetanakis:1992hf}
D.~Kapetanakis and G.~Zoupanos, ``{Coset Space Dimensional Reduction of Gauge
  Theories},''
{\em Phys. Rept.} {\bf 219} (1992)  1--76.

\bibitem{Kubyshin:1989vd}
Y.~A. Kubyshin, I.~P. Volobuev, J.~M. Mourao, and G.~Rudolph, {\em
  {D}imensional {R}eduction of {G}auge {T}heories, {S}pontaneous
  {C}ompactification and {M}odel {B}uilding}.
\newblock Leipzig Univ. - KMU-NTZ-89-07 (89,REC.SEP.) 80p, 1989.

\bibitem{Zoupanos:1987wj}
G.~Zoupanos, ``{Wilson flux breaking and Coset Space Dimensional Reduction},''
{\em Phys. Lett.} {\bf B201} (1988)  301.

\bibitem{Hosotani:1983xw}
Y.~Hosotani, ``{Dynamical mass generation by compact extra dimensions},''
{\em Phys. Lett.} {\bf B126} (1983)  309.

\bibitem{Hosotani:1983vn}
Y.~Hosotani, ``{Dynamical gauge symmetry breaking as the Casimir effect},''
{\em Phys. Lett.} {\bf B129} (1983)  193.

\bibitem{Witten:1985xc}
E.~Witten, ``{Symmetry breaking patterns in superstring models},''
{\em Nucl. Phys.} {\bf B258} (1985)  75.

\bibitem{Kapetanakis:1989gd}
D.~Kapetanakis and G.~Zoupanos, ``{Discrete Symmetries and Coset Space
  Dimensional Reduction},''
{\em Phys. Lett.} {\bf B232} (1989)  104.

\bibitem{Castellani:1999fz}
L.~Castellani, ``{On G/H geometry and its use in M-theory compactifications},''
  {\em Annals Phys.} {\bf 287} (2001)  1--13,
\href{http://arxiv.org/abs/hep-th/9912277}{{\tt hep-th/9912277}}.

\bibitem{Chatzistavrakidis:2007by}
A.~Chatzistavrakidis, P.~Manousselis, N.~Prezas, and G.~Zoupanos, ``{On the
  consistency of Coset Space Dimensional Reduction},'' {\em Phys. Lett.} {\bf
  B656} (2007)  152--157,
\href{http://arxiv.org/abs/0708.3222}{{\tt arXiv:0708.3222 [hep-th]}}.

\bibitem{Manton:1981es}
N.~S. Manton, ``{Fermions and Parity Violation in dimensional reduction
  schemes},''
{\em Nucl. Phys.} {\bf B193} (1981)  502.

\bibitem{Manousselis:2000aj}
P.~Manousselis and G.~Zoupanos, ``{Supersymmetry breaking by dimensional
  reduction over coset spaces},'' {\em Phys. Lett.} {\bf B504} (2001)
  122--130,
\href{http://arxiv.org/abs/hep-ph/0010141}{{\tt hep-ph/0010141}}.

\bibitem{Manousselis:2001xb}
P.~Manousselis and G.~Zoupanos, ``{Soft supersymmetry breaking due to
  dimensional reduction over non-symmetric coset spaces},'' {\em Phys. Lett.}
  {\bf B518} (2001)  171--180,
\href{http://arxiv.org/abs/hep-ph/0106033}{{\tt hep-ph/0106033}}.

\bibitem{Manousselis:2001re}
P.~Manousselis and G.~Zoupanos, ``{Dimensional reduction over coset spaces and
  supersymmetry breaking},'' {\em JHEP} {\bf 03} (2002)  002,
\href{http://arxiv.org/abs/hep-ph/0111125}{{\tt hep-ph/0111125}}.

\bibitem{Lust:1989tj}
D.~Lust and S.~Theisen, ``{Lectures on String Theory},''
{\em Lect. Notes Phys.} {\bf 346} (1989)  1--346.

\bibitem{Taylor:1988vt}
T.~R. Taylor and G.~Veneziano, ``{Strings and D=4},''
{\em Phys. Lett.} {\bf B212} (1988)  147.

\bibitem{Dienes:1998vg}
K.~R. Dienes, E.~Dudas, and T.~Gherghetta, ``{Grand Unification at intermediate
  mass scales through Extra Dimensions},'' {\em Nucl. Phys.} {\bf B537} (1999)
  47--108,
\href{http://arxiv.org/abs/hep-ph/9806292}{{\tt hep-ph/9806292}}.

\bibitem{Kobayashi:1998ye}
T.~Kobayashi, J.~Kubo, M.~Mondragon, and G.~Zoupanos, ``{Running of soft
  parameters in extra space-time dimensions},'' {\em Nucl. Phys.} {\bf B550}
  (1999)  99--122,
\href{http://arxiv.org/abs/hep-ph/9812221}{{\tt hep-ph/9812221}}.

\bibitem{Kubo:1999ua}
J.~Kubo, H.~Terao, and G.~Zoupanos, ``{Kaluza-Klein thresholds and
  regularization (in)dependence},'' {\em Nucl. Phys.} {\bf B574} (2000)
  495--524,
\href{http://arxiv.org/abs/hep-ph/9910277}{{\tt hep-ph/9910277}}.

\bibitem{Lust:1986ix}
D.~Lust, ``{Compactification of ten-dimensional superstring theories over Ricci
  flat coset spaces},''
{\em Nucl. Phys.} {\bf B276} (1986)  220.

\bibitem{Castellani:1986rg}
L.~Castellani and D.~Lust, ``{Superstring compactification on homogeneous coset
  spaces with torsion},''
{\em Nucl. Phys.} {\bf B296} (1988)  143.

\bibitem{Gavrilik:1999xr}
A.~M. Gavrilik, ``{Coset-space string compactification leading to 14
  subcritical dimensions},'' {\em Heavy Ion Phys.} {\bf 11} (2000)  35--41,
\href{http://arxiv.org/abs/hep-th/9911120}{{\tt hep-th/9911120}}.

\bibitem{MuellerHoissen:1987cq}
F.~Mueller-Hoissen and R.~Stuckl, ``{Coset spaces and ten-dimensional Unified
  Theories},''
{\em Class. Quant. Grav.} {\bf 5} (1988)  27.

\bibitem{Batakis:1989gb}
N.~A. Batakis, K.~Farakos, D.~Kapetanakis, G.~Koutsoumbas, and G.~Zoupanos,
  ``{Compactification over coset spaces with torsion and vanishing cosmological
  constant},''
{\em Phys. Lett.} {\bf B220} (1989)  513.

\bibitem{Wetterich:1982ed}
C.~Wetterich, ``{Dimensional reduction of Weyl, Majorana and Majorana-Weyl
  spinors},''
{\em Nucl. Phys.} {\bf B222} (1983)  20.

\bibitem{Palla:1983re}
L.~Palla, ``{On dimensional reduction of gauge theories: symmetric fields and
  harmonic expansion},''
{\em Z. Phys.} {\bf C24} (1984)  195.

\bibitem{Pilch:1984xx}
K.~Pilch and A.~N. Schellekens, ``{Formulae for the eigenvalues of the
  Laplacian on tensor harmonics on symmetric coset spaces},''
{\em J. Math. Phys.} {\bf 25} (1984)  3455.

\bibitem{Forgacs:1985vp}
P.~Forgacs, Z.~Horvath, and L.~Palla, ``{Spontaneous compactification to
  nonsymmetric spaces},''
{\em Z. Phys.} {\bf C30} (1986)  261.

\bibitem{Barnes:1986ea}
K.~J. Barnes, P.~Forgacs, M.~Surridge, and G.~Zoupanos, ``{On fermion masses in
  a dimensional reduction scheme},''
{\em Z. Phys.} {\bf C33} (1987)  427.

\bibitem{Bott:1965}
R.~Bott, {\em {Differential and Combinatorial Topology}}.
\newblock {P}rinceton {U}niv. {P}ress, 1965.

\bibitem{Witten:1984dg}
E.~Witten, ``{Some properties of O(32) superstrings},''
{\em Phys. Lett.} {\bf B149} (1984)  351--356.

\bibitem{Pilch:1985qf}
K.~Pilch and A.~N. Schellekens, ``{Fermion spectra from superstrings},''
{\em Nucl. Phys.} {\bf B259} (1985)  637.

\bibitem{Green:1984bx}
M.~B. Green, J.~H. Schwarz, and P.~C. West, ``{Anomaly free chiral theories in
  six-dimensions},''
{\em Nucl. Phys.} {\bf B254} (1985)  327--348.

\bibitem{Chapline:1980mr}
G.~Chapline and N.~S. Manton, ``{The geometrical significance of certain Higgs
  Potentials: An approach to Grand Unification},''
{\em Nucl. Phys.} {\bf B184} (1981)  391.

\bibitem{Bais:1985yd}
F.~A. Bais, K.~J. Barnes, P.~Forgacs, and G.~Zoupanos, ``{Dimensional reduction
  of gauge theories yielding Unified Models spontaneously broken to
  ${SU}(3)\times{U}(1)$},''
{\em Nucl. Phys.} {\bf B263} (1986)  557.

\bibitem{Farakos:1986sm}
K.~Farakos, G.~Koutsoumbas, M.~Surridge, and G.~Zoupanos, ``{Dimensional
  reduction and the Higgs Potential},''
{\em Nucl. Phys.} {\bf B291} (1987)  128.

\bibitem{Farakos:1986cj}
K.~Farakos, G.~Koutsoumbas, M.~Surridge, and G.~Zoupanos, ``{Geometrical
  Hierarchy and Spontaneous Symmetry Breaking},''
{\em Phys. Lett.} {\bf B191} (1987)  135.

\bibitem{Kapetanakis:1990tk}
D.~Kapetanakis and G.~Zoupanos, ``{Fermion masses from dimensional
  reduction},''
{\em Phys. Lett.} {\bf B249} (1990)  73--82.

\bibitem{Kozimirov:1989kn}
N.~G. Kozimirov, V.~A. Kuzmin, and I.~I. Tkachev, ``{Dimensional reduction on
  not simply connected homogeneous spaces},''
{\em Sov. J. Nucl. Phys.} {\bf 49} (1989)  164.

\bibitem{Kozimirov:1989xp}
N.~G. Kozimirov, V.~A. Kuzmin, and I.~I. Tkachev, ``{Dimensional reduction on
  multiply connected homogeneous spaces},''
{\em Phys. Rev.} {\bf D43} (1991)  1949--1955.

\bibitem{Slansky:1981yr}
R.~Slansky, ``{Group Theory for Unified Model Building},''
{\em Phys. Rept.} {\bf 79} (1981)  1--128.

\bibitem{Chatzistavrakidis:2007pp}
A.~Chatzistavrakidis, P.~Manousselis, N.~Prezas, and G.~Zoupanos, ``{Coset
  Space Dimensional Reduction of Einstein--Yang--Mills theory},'' {\em Fortsch.
  Phys.} {\bf 56} (2008)  389--399,
\href{http://arxiv.org/abs/0712.2717}{{\tt arXiv:0712.2717 [hep-th]}}.

\bibitem{Coquereaux:1984ca}
R.~Coquereaux and A.~Jadczyk, ``{Symmetries of Einstein Yang-Mills fields and
  dimensional reduction},''
\href{http://dx.doi.org/10.1007/BF01211045}{{\em Commun. Math. Phys.} {\bf 98}
  (1985)  79}.

\bibitem{Chaichian:1986mf}
M.~Chaichian, A.~P. Demichev, N.~F. Nelipa, and A.~Y. Rodionov, ``{Geometrical
  method of spontaneous symmetry breaking in terms of fiber bundles and the
  standard model of electroweak interactions},''
\href{http://dx.doi.org/10.1016/0550-3213(87)90005-8}{{\em Nucl. Phys.} {\bf
  B279} (1987)  452}.

\bibitem{Dvali:2001qr}
G.~R. Dvali, S.~Randjbar-Daemi, and R.~Tabbash, ``{The origin of spontaneous
  symmetry breaking in theories with large extra dimensions},''
  \href{http://dx.doi.org/10.1103/PhysRevD.65.064021}{{\em Phys. Rev.} {\bf
  D65} (2002)  064021},
\href{http://arxiv.org/abs/hep-ph/0102307}{{\tt arXiv:hep-ph/0102307}}.

\bibitem{Manousselis:2004xd}
P.~Manousselis and G.~Zoupanos, ``{Dimensional reduction of ten-dimensional
  supersymmetric gauge theories in the N = 1, D = 4 superfield formalism},''
  \href{http://dx.doi.org/10.1088/1126-6708/2004/11/025}{{\em JHEP} {\bf 11}
  (2004)  025},
\href{http://arxiv.org/abs/hep-ph/0406207}{{\tt arXiv:hep-ph/0406207}}.

\bibitem{Behrndt:2004km}
K.~Behrndt and M.~Cvetic, ``{General N = 1 supersymmetric flux vacua of
  (massive) type IIA string theory},''
  \href{http://dx.doi.org/10.1103/PhysRevLett.95.021601}{{\em Phys. Rev. Lett.}
  {\bf 95} (2005)  021601},
\href{http://arxiv.org/abs/hep-th/0403049}{{\tt arXiv:hep-th/0403049}}.

\bibitem{Koerber:2008rx}
P.~Koerber, D.~Lust, and D.~Tsimpis, ``{Type IIA AdS4 compactifications on
  cosets, interpolations and domain walls},''
  \href{http://dx.doi.org/10.1088/1126-6708/2008/07/017}{{\em JHEP} {\bf 07}
  (2008)  017},
\href{http://arxiv.org/abs/0804.0614}{{\tt arXiv:0804.0614 [hep-th]}}.

\bibitem{House:2005yc}
T.~House and E.~Palti, ``{Effective action of (massive) IIA on manifolds with
  SU(3) structure},'' \href{http://dx.doi.org/10.1103/PhysRevD.72.026004}{{\em
  Phys. Rev.} {\bf D72} (2005)  026004},
\href{http://arxiv.org/abs/hep-th/0505177}{{\tt arXiv:hep-th/0505177}}.

\bibitem{Govindarajan:1986kb}
T.~R. Govindarajan, A.~S. Joshipura, S.~D. Rindani, and U.~Sarkar,
  ``{Supersymmetric compactification of the Heterotic String on Coset
  Spaces},''
\href{http://dx.doi.org/10.1103/PhysRevLett.57.2489}{{\em Phys. Rev. Lett.}
  {\bf 57} (1986)  2489}.

\bibitem{Govindarajan:1986iz}
T.~R. Govindarajan, A.~S. Joshipura, S.~D. Rindani, and U.~Sarkar, ``{Coset
  spaces as alternatives to Calabi-Yau spaces in the presence of gaugino
  condensation},''
{\em Int. J. Mod. Phys.} {\bf A2} (1987)  797.

\bibitem{Micu:2004tz}
A.~Micu, ``{Heterotic compactifications and nearly-Kaehler manifolds},''
  \href{http://dx.doi.org/10.1103/PhysRevD.70.126002}{{\em Phys. Rev.} {\bf
  D70} (2004)  126002},
\href{http://arxiv.org/abs/hep-th/0409008}{{\tt arXiv:hep-th/0409008}}.

\bibitem{Frey:2005zz}
A.~R. Frey and M.~Lippert, ``{AdS strings with torsion: Non-complex heterotic
  compactifications},''
  \href{http://dx.doi.org/10.1103/PhysRevD.72.126001}{{\em Phys. Rev.} {\bf
  D72} (2005)  126001},
\href{http://arxiv.org/abs/hep-th/0507202}{{\tt arXiv:hep-th/0507202}}.

\bibitem{Manousselis:2005xa}
P.~Manousselis, N.~Prezas, and G.~Zoupanos, ``{Supersymmetric compactifications
  of heterotic strings with fluxes and condensates},''
  \href{http://dx.doi.org/10.1016/j.nuclphysb.2006.01.008}{{\em Nucl. Phys.}
  {\bf B739} (2006)  85--105},
\href{http://arxiv.org/abs/hep-th/0511122}{{\tt arXiv:hep-th/0511122}}.

\bibitem{Douzas:2008va}
G.~Douzas, T.~Grammatikopoulos, and G.~Zoupanos, ``{Coset Space Dimensional
  Reduction and Wilson Flux Breaking of Ten-Dimensional N=1, E(8) Gauge
  Theory},'' \href{http://arxiv.org/abs/0808.3236}{{\tt arXiv:0808.3236
  [hep-th]}}.
(a shorter version will be published in Eur. Phys. J. C).

\bibitem{Douzas:2007zz}
G.~Douzas, T.~Grammatikopoulos, J.~Madore, and G.~Zoupanos, ``{Coset space
  dimensional reduction and classification of semi-realistic particle physics
  models},''
\href{http://dx.doi.org/10.1002/prop.200710515}{{\em Fortsch. Phys.} {\bf 56}
  (2008)  424--429}.

\bibitem{Koca:1984dr}
M.~Koca, ``{Dimensional reduction of exceptional E(6), E(8) gauge groups and
  flavor chirality},''
\href{http://dx.doi.org/10.1016/0370-2693(84)90271-5}{{\em Phys. Lett.} {\bf
  B141} (1984)  400--402}.

\bibitem{Jittoh:2008jc}
T.~Jittoh, M.~Koike, T.~Nomura, J.~Sato, and T.~Shimomura, ``{Model building by
  coset space dimensional reduction scheme using ten-dimensional coset
  spaces},''
\href{http://arxiv.org/abs/0803.0641}{{\tt arXiv:0803.0641 [hep-ph]}}.

\bibitem{Kim:2006hv}
J.~E. Kim and B.~Kyae, ``{String MSSM through flipped SU(5) from Z(12)
  orbifold},''
\href{http://arxiv.org/abs/hep-th/0608085}{{\tt arXiv:hep-th/0608085}}.

\bibitem{Kim:2006hw}
J.~E. Kim and B.~Kyae, ``{Flipped SU(5) from Z(12-I) orbifold with Wilson
  line},'' \href{http://dx.doi.org/10.1016/j.nuclphysb.2007.02.008}{{\em Nucl.
  Phys.} {\bf B770} (2007)  47--82},
\href{http://arxiv.org/abs/hep-th/0608086}{{\tt arXiv:hep-th/0608086}}.

\bibitem{Kim:2007mt}
J.~E. Kim, J.-H. Kim, and B.~Kyae, ``{Superstring standard model from Z(12-I)
  orbifold compactification with and without exotics, and effective R-
  parity},'' {\em JHEP} {\bf 06} (2007)  034,
\href{http://arxiv.org/abs/hep-ph/0702278}{{\tt arXiv:hep-ph/0702278}}.

\bibitem{Kim:2007dx}
J.~E. Kim, ``Z(12-i) {O}rbifold {C}ompactification toward {SUSY} {S}tandard
  model,'' \href{http://dx.doi.org/10.1142/S0217751X07038876}{{\em Int. J. Mod.
  Phys.} {\bf A22} (2008)  5609--5621},
\href{http://arxiv.org/abs/0706.3498}{{\tt arXiv:0706.3498 [hep-ph]}}.

\bibitem{Forste:2005gc}
S.~Forste, H.~P. Nilles, and A.~Wingerter, ``{The Higgs mechanism in heterotic
  orbifolds},'' \href{http://dx.doi.org/10.1103/PhysRevD.73.066011}{{\em Phys.
  Rev.} {\bf D73} (2006)  066011},
\href{http://arxiv.org/abs/hep-th/0512270}{{\tt arXiv:hep-th/0512270}}.

\bibitem{Nilles:2006np}
H.~P. Nilles, S.~Ramos-Sanchez, P.~K.~S. Vaudrevange, and A.~Wingerter,
  ``{Exploring the SO(32) heterotic string},'' {\em JHEP} {\bf 04} (2006)  050,
\href{http://arxiv.org/abs/hep-th/0603086}{{\tt arXiv:hep-th/0603086}}.

\bibitem{Lebedev:2006kn}
O.~Lebedev {\em et al.}, ``{A mini-landscape of exact MSSM spectra in heterotic
  orbifolds},'' \href{http://dx.doi.org/10.1016/j.physletb.2006.12.012}{{\em
  Phys. Lett.} {\bf B645} (2007)  88--94},
\href{http://arxiv.org/abs/hep-th/0611095}{{\tt arXiv:hep-th/0611095}}.

\bibitem{Lebedev:2006tr}
O.~Lebedev {\em et al.}, ``{Low Energy Supersymmetry from the Heterotic
  Landscape},'' \href{http://dx.doi.org/10.1103/PhysRevLett.98.181602}{{\em
  Phys. Rev. Lett.} {\bf 98} (2007)  181602},
\href{http://arxiv.org/abs/hep-th/0611203}{{\tt arXiv:hep-th/0611203}}.

\bibitem{Lebedev:2007hv}
O.~Lebedev {\em et al.}, ``{The Heterotic Road to the MSSM with R parity},''
  \href{http://dx.doi.org/10.1103/PhysRevD.77.046013}{{\em Phys. Rev.} {\bf
  D77} (2008)  046013},
\href{http://arxiv.org/abs/0708.2691}{{\tt arXiv:0708.2691 [hep-th]}}.

\bibitem{Blumenhagen:2005pm}
R.~Blumenhagen, G.~Honecker, and T.~Weigand, ``{Supersymmetric (non-)abelian
  bundles in the type I and SO(32) heterotic string},'' {\em JHEP} {\bf 08}
  (2005)  009,
\href{http://arxiv.org/abs/hep-th/0507041}{{\tt arXiv:hep-th/0507041}}.

\bibitem{Blumenhagen:2005zg}
R.~Blumenhagen, G.~Honecker, and T.~Weigand, ``{Non-abelian brane worlds: The
  heterotic string story},'' {\em JHEP} {\bf 10} (2005)  086,
\href{http://arxiv.org/abs/hep-th/0510049}{{\tt arXiv:hep-th/0510049}}.

\bibitem{Blumenhagen:2006ux}
R.~Blumenhagen, S.~Moster, and T.~Weigand, ``{Heterotic GUT and standard model
  vacua from simply connected Calabi-Yau manifolds},''
  \href{http://dx.doi.org/10.1016/j.nuclphysb.2006.06.005}{{\em Nucl. Phys.}
  {\bf B751} (2006)  186--221},
\href{http://arxiv.org/abs/hep-th/0603015}{{\tt arXiv:hep-th/0603015}}.

\bibitem{Blumenhagen:2006wj}
R.~Blumenhagen, S.~Moster, R.~Reinbacher, and T.~Weigand, ``{Massless spectra
  of three generation U(N) heterotic string vacua},'' {\em JHEP} {\bf 05}
  (2007)  041,
\href{http://arxiv.org/abs/hep-th/0612039}{{\tt arXiv:hep-th/0612039}}.

\bibitem{Madore:1989ma}
J.~Madore, ``{On a modification of Kaluza-Klein Theory},''
{\em Phys. Rev.} {\bf D41} (1990)  3709.

\bibitem{Carlson:2001bk}
C.~E. Carlson and C.~D. Carone, ``Discerning noncommutative extra dimensions,''
  {\em Phys. Rev.} {\bf D65} (2002)  075007,
\href{http://arxiv.org/abs/hep-ph/0112143}{{\tt hep-ph/0112143}}.

\bibitem{Jurco:2001my}
B.~Jurco, P.~Schupp, and J.~Wess, ``Nonabelian noncommutative gauge theory via
  noncommutative extra dimensions,'' {\em Nucl. Phys.} {\bf B604} (2001)
  148--180,
\href{http://arxiv.org/abs/hep-th/0102129}{{\tt hep-th/0102129}}.

\bibitem{Szabo:2006wx}
R.~J. Szabo, ``Symmetry, gravity and noncommutativity,'' {\em Class. Quant.
  Grav.} {\bf 23} (2006)  R199--R242,
\href{http://arxiv.org/abs/hep-th/0606233}{{\tt hep-th/0606233}}.

\bibitem{Madore:2000aq}
J.~Madore, ``An introduction to noncommutative differential geometry and
  physical applications,''
{\em Lond. Math. Soc. Lect. Note Ser.} {\bf 257} (2000)  1--371.

\bibitem{Aschieri:2003vy}
P.~Aschieri, J.~Madore, P.~Manousselis, and G.~Zoupanos, ``Dimensional
  reduction over fuzzy coset spaces,'' {\em JHEP} {\bf 04} (2004)  034,
\href{http://arxiv.org/abs/hep-th/0310072}{{\tt hep-th/0310072}}.

\bibitem{Aschieri:2006uw}
P.~Aschieri, T.~Grammatikopoulos, H.~Steinacker, and G.~Zoupanos, ``Dynamical
  generation of fuzzy extra dimensions, dimensional reduction and symmetry
  breaking,'' {\em JHEP} {\bf 09} (2006)  026,
\href{http://arxiv.org/abs/hep-th/0606021}{{\tt hep-th/0606021}}.

\bibitem{Connes:1994yd}
A.~Connes, {\em Noncommutative geometry}.
\newblock Academic Press, 1994.

\bibitem{Connes:2000ti}
A.~Connes, ``Noncommutative geometry: Year 2000,''
\href{http://arxiv.org/abs/math.qa/0011193}{{\tt math.qa/0011193}}.

\bibitem{Castellani:2000xt}
L.~Castellani, ``Noncommutative geometry and physics: A review of selected
  recent results,'' {\em Class. Quant. Grav.} {\bf 17} (2000)  3377--3402,
\href{http://arxiv.org/abs/hep-th/0005210}{{\tt hep-th/0005210}}.

\bibitem{Ydri:2001pv}
B.~Ydri, ``Fuzzy physics,''
\href{http://arxiv.org/abs/hep-th/0110006}{{\tt hep-th/0110006}}.

\bibitem{DuboisViolette:1988ir}
M.~Dubois-Violette, R.~Kerner, and J.~Madore, ``{Noncommutative Differential
  Geometry of Matrix Algebras},''
{\em J. Math. Phys.} {\bf 31} (1990)  316.

\bibitem{Rothe:2005nw}
H.~J. Rothe, ``{Lattice gauge theories: An Introduction},''
{\em World Sci. Lect. Notes Phys.} {\bf 74} (2005)  1--605.

\bibitem{Madore:1996cb}
J.~Madore and J.~Mourad, ``{Noncommutative Kaluza-Klein Theory},''
\href{http://arxiv.org/abs/hep-th/9601169}{{\tt hep-th/9601169}}.

\bibitem{Coquereaux:1988ne}
R.~Coquereaux and A.~Jadczyk, ``{Riemannian geometry, fiber bundles,
  Kaluza-Klein theories and all that},''
{\em World Sci. Lect. Notes Phys.} {\bf 16} (1988)  1--345.

\bibitem{Jordan:1933vh}
P.~Jordan, J.~von Neumann, and E.~P. Wigner, ``{On an Algebraic generalization
  of the quantum mechanical formalism},''
{\em Annals Math.} {\bf 35} (1934)  29--64.

\bibitem{DuboisViolette:1990xq}
M.~Dubois-Violette, ``Noncommutative differential geometry, quantum mechanics
  and gauge theory,''. Lecture given at 19th Int. Conf. on Differential
  Geometric Methods in Theoretical Physics, Rapallo, Italy, Jun 19-24, 1990.

\bibitem{Dimakis:1992fj}
A.~Dimakis and F.~Mueller-Hoissen, ``Quantum mechanics as noncommutative
  symplectic geometry,''
{\em J. Phys.} {\bf A25} (1992)  5625--5648.

\bibitem{Deruelle:1986ia}
N.~Deruelle and J.~Madore, ``{The Friedmann Universe as an attractor of a
  Kaluza-Klein Cosmology},''
{\em Mod. Phys. Lett.} {\bf A1} (1986)  237--253.

\bibitem{Madore:1991bw}
J.~Madore, ``The fuzzy sphere,''
{\em Class. Quant. Grav.} {\bf 9} (1992)  69--88.

\bibitem{Ramgoolam:2001zx}
S.~Ramgoolam, ``On spherical harmonics for fuzzy spheres in diverse
  dimensions,'' {\em Nucl. Phys.} {\bf B610} (2001)  461--488,
\href{http://arxiv.org/abs/hep-th/0105006}{{\tt hep-th/0105006}}.

\bibitem{Ramgoolam:2002wb}
S.~Ramgoolam, ``Higher dimensional geometries related to fuzzy odd- dimensional
  spheres,'' {\em JHEP} {\bf 10} (2002)  064,
\href{http://arxiv.org/abs/hep-th/0207111}{{\tt hep-th/0207111}}.

\bibitem{Hasebe:2003gx}
K.~Hasebe and Y.~Kimura, ``{Dimensional hierarchy in quantum Hall effects on
  fuzzy spheres},'' {\em Phys. Lett.} {\bf B602} (2004)  255--260,
\href{http://arxiv.org/abs/hep-th/0310274}{{\tt hep-th/0310274}}.

\bibitem{Aschieri:2004vh}
P.~Aschieri, J.~Madore, P.~Manousselis, and G.~Zoupanos, ``Unified theories
  from fuzzy extra dimensions,'' {\em Fortsch. Phys.} {\bf 52} (2004)
  718--723,
\href{http://arxiv.org/abs/hep-th/0401200}{{\tt hep-th/0401200}}.

\bibitem{Aschieri:2005wm}
P.~Aschieri, J.~Madore, P.~Manousselis, and G.~Zoupanos, ``Renormalizable
  theories from fuzzy higher dimensions,''
\href{http://arxiv.org/abs/hep-th/0503039}{{\tt hep-th/0503039}}.

\bibitem{Karabali:2006eg}
D.~Karabali and V.~P. Nair, ``{Quantum Hall effect in higher dimensions, matrix
  models and fuzzy geometry},'' {\em J. Phys.} {\bf A39} (2006)  12735--12764,
\href{http://arxiv.org/abs/hep-th/0606161}{{\tt hep-th/0606161}}.

\bibitem{Aschieri:2007fb}
P.~Aschieri, H.~Steinacker, J.~Madore, P.~Manousselis, and G.~Zoupanos,
  ``{Fuzzy Extra Dimensions: Dimensional Reduction, Dynamical Generation and
  Renormalizability},''
\href{http://arxiv.org/abs/arXiv:0704.2880 [hep-th]}{{\tt arXiv:0704.2880
  [hep-th]}}.

\bibitem{Iso:2001mg}
S.~Iso, Y.~Kimura, K.~Tanaka, and K.~Wakatsuki, ``Noncommutative gauge theory
  on fuzzy sphere from matrix model,'' {\em Nucl. Phys.} {\bf B604} (2001)
  121--147,
\href{http://arxiv.org/abs/hep-th/0101102}{{\tt hep-th/0101102}}.

\bibitem{Valtancoli:2002rx}
P.~Valtancoli, ``Stability of the fuzzy sphere solution from matrix model,''
  {\em Int. J. Mod. Phys.} {\bf A18} (2003)  967,
\href{http://arxiv.org/abs/hep-th/0206075}{{\tt hep-th/0206075}}.

\bibitem{Myers:1999ps}
R.~C. Myers, ``{Dielectric-branes},'' {\em JHEP} {\bf 12} (1999)  022,
\href{http://arxiv.org/abs/hep-th/9910053}{{\tt hep-th/9910053}}.

\bibitem{Balachandran:2006gf}
A.~P. Balachandran and B.~A. Qureshi, ``Noncommutative geometry: Fuzzy spaces,
  the groenwald-moyal plane,''
\href{http://arxiv.org/abs/hep-th/0606115}{{\tt hep-th/0606115}}.

\bibitem{Balachandran:2001dd}
A.~P. Balachandran, B.~P. Dolan, J.-H. Lee, X.~Martin, and D.~O'Connor, ``Fuzzy
  complex projective spaces and their star-products,'' {\em J. Geom. Phys.}
  {\bf 43} (2002)  184--204,
\href{http://arxiv.org/abs/hep-th/0107099}{{\tt hep-th/0107099}}.

\bibitem{Grosse:1995jt}
H.~Grosse, C.~Klimcik, and P.~Presnajder, ``Topologically nontrivial field
  configurations in noncommutative geometry,'' {\em Commun. Math. Phys.} {\bf
  178} (1996)  507--526,
\href{http://arxiv.org/abs/hep-th/9510083}{{\tt hep-th/9510083}}.

\bibitem{Carow-Watamura:1996wg}
U.~Carow-Watamura and S.~Watamura, ``{Chirality and Dirac operator on
  noncommutative sphere},'' {\em Commun. Math. Phys.} {\bf 183} (1997)
  365--382,
\href{http://arxiv.org/abs/hep-th/9605003}{{\tt hep-th/9605003}}.

\bibitem{Dolan:2002af}
B.~P. Dolan and C.~Nash, ``{Chiral fermions and Spin(C) structures on matrix
  approximations to manifolds},'' {\em JHEP} {\bf 07} (2002)  057,
\href{http://arxiv.org/abs/hep-th/0207007}{{\tt hep-th/0207007}}.

\bibitem{Trivedi:2000mq}
S.~P. Trivedi and S.~Vaidya, ``Fuzzy cosets and their gravity duals,'' {\em
  JHEP} {\bf 09} (2000)  041,
\href{http://arxiv.org/abs/hep-th/0007011}{{\tt hep-th/0007011}}.

\bibitem{Grosse:1999ci}
H.~Grosse and A.~Strohmaier, ``{Noncommutative geometry and the regularization
  problem of 4D quantum field theory},'' {\em Lett. Math. Phys.} {\bf 48}
  (1999)  163--179,
\href{http://arxiv.org/abs/hep-th/9902138}{{\tt hep-th/9902138}}.

\bibitem{Madore:2000en}
J.~Madore, S.~Schraml, P.~Schupp, and J.~Wess, ``Gauge theory on noncommutative
  spaces,'' {\em Eur. Phys. J.} {\bf C16} (2000)  161--167,
\href{http://arxiv.org/abs/hep-th/0001203}{{\tt hep-th/0001203}}.

\bibitem{Connes:1990qp}
A.~Connes and J.~Lott, ``{Particle Models and Noncommutative Geometry (Expanded
  version)},''
{\em Nucl. Phys. Proc. Suppl.} {\bf 18B} (1991)  29--47.

\bibitem{Martin:1996wh}
C.~P. Martin, J.~M. Gracia-Bondia, and J.~C. Varilly, ``The standard model as a
  noncommutative geometry: The low- energy regime,'' {\em Phys. Rept.} {\bf
  294} (1998)  363--406,
\href{http://arxiv.org/abs/hep-th/9605001}{{\tt hep-th/9605001}}.

\bibitem{DuboisViolette:1989at}
M.~Dubois-Violette, J.~Madore, and R.~Kerner, ``{Gauge Bosons in a
  Noncommutative Geometry},''
{\em Phys. Lett.} {\bf B217} (1989)  485--488.

\bibitem{DuboisViolette:1988ps}
M.~Dubois-Violette, J.~Madore, and R.~Kerner, ``{Classical Bosons in a
  Noncommutative Geometry},''
{\em Class. Quant. Grav.} {\bf 6} (1989)  1709.

\bibitem{DuboisViolette:1989vq}
M.~Dubois-Violette, R.~Kerner, and J.~Madore, ``{Noncommutative Differential
  Geometry and New Models of Gauge Theory},''
{\em J. Math. Phys.} {\bf 31} (1990)  323.

\bibitem{Seiberg:1999vs}
N.~Seiberg and E.~Witten, ``String theory and noncommutative geometry,'' {\em
  JHEP} {\bf 09} (1999)  032,
\href{http://arxiv.org/abs/hep-th/9908142}{{\tt hep-th/9908142}}.

\bibitem{Chaichian:2001py}
M.~Chaichian, P.~Presnajder, M.~M. Sheikh-Jabbari, and A.~Tureanu,
  ``{Noncommutative standard model: Model building},'' {\em Eur. Phys. J.} {\bf
  C29} (2003)  413--432,
\href{http://arxiv.org/abs/hep-th/0107055}{{\tt hep-th/0107055}}.

\bibitem{Calmet:2001na}
X.~Calmet, B.~Jurco, P.~Schupp, J.~Wess, and M.~Wohlgenannt, ``The standard
  model on non-commutative space-time,'' {\em Eur. Phys. J.} {\bf C23} (2002)
  363--376,
\href{http://arxiv.org/abs/hep-ph/0111115}{{\tt hep-ph/0111115}}.

\bibitem{Aschieri:2002mc}
P.~Aschieri, B.~Jurco, P.~Schupp, and J.~Wess, ``{Non-commutative GUTs,
  standard model and C, P, T},'' {\em Nucl. Phys.} {\bf B651} (2003)  45--70,
\href{http://arxiv.org/abs/hep-th/0205214}{{\tt hep-th/0205214}}.

\bibitem{Brandt:2003fx}
F.~Brandt, C.~P. Martin, and F.~R. Ruiz, ``{Anomaly freedom in Seiberg-Witten
  noncommutative gauge theories},'' {\em JHEP} {\bf 07} (2003)  068,
\href{http://arxiv.org/abs/hep-th/0307292}{{\tt hep-th/0307292}}.

\bibitem{Carlson:2001sw}
C.~E. Carlson, C.~D. Carone, and R.~F. Lebed, ``{Bounding noncommutative
  QCD},'' {\em Phys. Lett.} {\bf B518} (2001)  201--206,
\href{http://arxiv.org/abs/hep-ph/0107291}{{\tt hep-ph/0107291}}.

\bibitem{Behr:2002wx}
W.~Behr {\em et al.}, ``{The Z --> gamma gamma, g g decays in the
  noncommutative standard model},'' {\em Eur. Phys. J.} {\bf C29} (2003)
  441--446,
\href{http://arxiv.org/abs/hep-ph/0202121}{{\tt hep-ph/0202121}}.

\bibitem{Hinchliffe:2002km}
I.~Hinchliffe, N.~Kersting, and Y.~L. Ma, ``Review of the phenomenology of
  noncommutative geometry,'' {\em Int. J. Mod. Phys.} {\bf A19} (2004)
  179--204,
\href{http://arxiv.org/abs/hep-ph/0205040}{{\tt hep-ph/0205040}}.

\bibitem{Schupp:2002up}
P.~Schupp, J.~Trampetic, J.~Wess, and G.~Raffelt, ``The photon neutrino
  interaction in non-commutative gauge field theory and astrophysical bounds,''
  {\em Eur. Phys. J.} {\bf C36} (2004)  405--410,
\href{http://arxiv.org/abs/hep-ph/0212292}{{\tt hep-ph/0212292}}.

\bibitem{Barrett:2006zh}
J.~W. Barrett and R.~A. Dawe~Martins, ``Non-commutative geometry and the
  standard model vacuum,'' {\em J. Math. Phys.} {\bf 47} (2006)  052305,
\href{http://arxiv.org/abs/hep-th/0601192}{{\tt hep-th/0601192}}.

\bibitem{Barrett:2006qq}
J.~W. Barrett, ``A lorentzian version of the non-commutative geometry of the
  standard model of particle physics,'' {\em J. Math. Phys.} {\bf 48} (2007)
  012303,
\href{http://arxiv.org/abs/hep-th/0608221}{{\tt hep-th/0608221}}.

\bibitem{Connes:2006qv}
A.~Connes, ``Noncommutative geometry and the standard model with neutrino
  mixing,'' {\em JHEP} {\bf 11} (2006)  081,
\href{http://arxiv.org/abs/hep-th/0608226}{{\tt hep-th/0608226}}.

\bibitem{Martins:2006rp}
R.~A.~D. Martins, ``Noncommutative geometry, topology and the standard model
  vacuum,'' {\em J. Math. Phys.} {\bf 47} (2006)  113507,
\href{http://arxiv.org/abs/hep-th/0609140}{{\tt hep-th/0609140}}.

\bibitem{Lizzi:2006te}
F.~Lizzi, ``Internal space for the noncommutative geometry standard model and
  string,'' {\em Int. J. Mod. Phys.} {\bf A22} (2007)  1317--1334,
\href{http://arxiv.org/abs/hep-th/0610023}{{\tt hep-th/0610023}}.

\bibitem{Dubois-Violette:1989at}
M.~Dubois-Violette, J.~Madore, and R.~Kerner, ``{Gauge Bosons in a
  Noncommutative Geometry},''
{\em Phys. Lett.} {\bf B217} (1989)  485--488.

\bibitem{Dubois-Violette:1988ps}
M.~Dubois-Violette, J.~Madore, and R.~Kerner, ``{Classical Bosons in a
  Noncommutative Geometry},''
{\em Class. Quant. Grav.} {\bf 6} (1989)  1709.

\bibitem{Dubois-Violette:1989vq}
M.~Dubois-Violette, R.~Kerner, and J.~Madore, ``{Noncommutative Differential
  Geometry and New Models of Gauge Theory},''
{\em J. Math. Phys.} {\bf 31} (1990)  323.

\bibitem{Madore:1992dk}
J.~Madore, ``On a quark - lepton duality,''
{\em Phys. Lett.} {\bf B305} (1993)  84--89.

\bibitem{Madore:1992ej}
J.~Madore, ``On a noncommutative extension of electrodynamics,''
\href{http://arxiv.org/abs/hep-ph/9209226}{{\tt hep-ph/9209226}}.

\bibitem{Steinacker:2007ay}
H.~Steinacker and G.~Zoupanos, ``Fermions on spontaneously generated spherical
  extra dimensions,'' {\em JHEP} {\bf 09} (2007)  017,
\href{http://arxiv.org/abs/arXiv:0706.0398 [hep-th]}{{\tt arXiv:0706.0398
  [hep-th]}}.

\bibitem{Steinacker:2003sd}
H.~Steinacker, ``Quantized gauge theory on the fuzzy sphere as random matrix
  model,'' {\em Nucl. Phys.} {\bf B679} (2004)  66--98,
\href{http://arxiv.org/abs/hep-th/0307075}{{\tt hep-th/0307075}}.

\bibitem{Steinacker:2004yu}
H.~Steinacker, ``Gauge theory on the fuzzy sphere and random matrices,'' {\em
  Springer Proc. Phys.} {\bf 98} (2005)  307--311,
\href{http://arxiv.org/abs/hep-th/0409235}{{\tt hep-th/0409235}}.

\bibitem{Andrews:2005cv}
R.~P. Andrews and N.~Dorey, ``Spherical deconstruction,'' {\em Phys. Lett.}
  {\bf B631} (2005)  74--82,
\href{http://arxiv.org/abs/hep-th/0505107}{{\tt hep-th/0505107}}.

\bibitem{Andrews:2006aw}
R.~P. Andrews and N.~Dorey, ``{Deconstruction of the Maldacena-Nunez
  compactification},'' {\em Nucl. Phys.} {\bf B751} (2006)  304--341,
\href{http://arxiv.org/abs/hep-th/0601098}{{\tt hep-th/0601098}}.

\bibitem{Azuma:2004ie}
T.~Azuma, K.~Nagao, and J.~Nishimura, ``Perturbative dynamics of fuzzy spheres
  at large n,'' {\em JHEP} {\bf 06} (2005)  081,
\href{http://arxiv.org/abs/hep-th/0410263}{{\tt hep-th/0410263}}.

\bibitem{Azuma:2005bj}
T.~Azuma, S.~Bal, and J.~Nishimura, ``{Dynamical generation of gauge groups in
  the massive Yang- Mills-Chern-Simons matrix model},'' {\em Phys. Rev.} {\bf
  D72} (2005)  066005,
\href{http://arxiv.org/abs/hep-th/0504217}{{\tt hep-th/0504217}}.

\bibitem{Azuma:2004zq}
T.~Azuma, S.~Bal, K.~Nagao, and J.~Nishimura, ``{Nonperturbative studies of
  fuzzy spheres in a matrix model with the Chern-Simons term},'' {\em JHEP}
  {\bf 05} (2004)  005,
\href{http://arxiv.org/abs/hep-th/0401038}{{\tt hep-th/0401038}}.

\bibitem{Aoki:2004sd}
H.~Aoki, S.~Iso, T.~Maeda, and K.~Nagao, ``Dynamical generation of a nontrivial
  index on the fuzzy 2- sphere,'' {\em Phys. Rev.} {\bf D71} (2005)  045017,
\href{http://arxiv.org/abs/hep-th/0412052}{{\tt hep-th/0412052}}.

\bibitem{Aoki:2006zi}
H.~Aoki, J.~Nishimura, and Y.~Susaki, ``{Suppression of topologically
  nontrivial sectors in gauge theory on 2D non-commutative geometry},''
\href{http://arxiv.org/abs/hep-th/0604093}{{\tt hep-th/0604093}}.

\bibitem{Abel:2005rh}
S.~A. Abel, J.~Jaeckel, V.~V. Khoze, and A.~Ringwald, ``Noncommutativity, extra
  dimensions, and power law running in the infrared,'' {\em JHEP} {\bf 01}
  (2006)  105,
\href{http://arxiv.org/abs/hep-ph/0511197}{{\tt hep-ph/0511197}}.

\bibitem{Lim:2006bx}
C.~S. Lim, N.~Maru, and K.~Hasegawa, ``{Six dimensional gauge-Higgs unification
  with an extra space S**2 and the hierarchy problem},''
\href{http://arxiv.org/abs/hep-th/0605180}{{\tt hep-th/0605180}}.

\bibitem{Antoniadis:2002ns}
I.~Antoniadis, K.~Benakli, and M.~Quiros, ``Supersymmetry and electroweak
  breaking by extra dimensions,''
{\em Acta Phys. Polon.} {\bf B33} (2002)  2477--2488.

\bibitem{Scrucca:2003ra}
C.~A. Scrucca, M.~Serone, and L.~Silvestrini, ``Electroweak symmetry breaking
  and fermion masses from extra dimensions,'' {\em Nucl. Phys.} {\bf B669}
  (2003)  128--158,
\href{http://arxiv.org/abs/hep-ph/0304220}{{\tt hep-ph/0304220}}.

\bibitem{Steinacker:2007dq}
H.~Steinacker, ``{Emergent Gravity from Noncommutative Gauge Theory},''
\href{http://arxiv.org/abs/arXiv:0708.2426 [hep-th]}{{\tt arXiv:0708.2426
  [hep-th]}}.

\bibitem{Carow-Watamura:1998jn}
U.~Carow-Watamura and S.~Watamura, ``Noncommutative geometry and gauge theory
  on fuzzy sphere,'' {\em Commun. Math. Phys.} {\bf 212} (2000)  395--413,
\href{http://arxiv.org/abs/hep-th/9801195}{{\tt hep-th/9801195}}.

\bibitem{Presnajder:2003ak}
P.~Presnajder, ``Gauge fields on the fuzzy sphere,''
{\em Mod. Phys. Lett.} {\bf A18} (2003)  2431--2438.

\bibitem{Karabali:2001te}
D.~Karabali, V.~P. Nair, and A.~P. Polychronakos, ``{Spectrum of Schroedinger
  field in a noncommutative magnetic monopole},'' {\em Nucl. Phys.} {\bf B627}
  (2002)  565--579,
\href{http://arxiv.org/abs/hep-th/0111249}{{\tt hep-th/0111249}}.

\bibitem{Balachandran:1999hx}
A.~P. Balachandran and S.~Vaidya, ``Instantons and chiral anomaly in fuzzy
  physics,'' {\em Int. J. Mod. Phys.} {\bf A16} (2001)  17--40,
\href{http://arxiv.org/abs/hep-th/9910129}{{\tt hep-th/9910129}}.

\bibitem{Chamseddine:1992yx}
A.~H. Chamseddine, G.~Felder, and J.~Frohlich, ``Gravity in noncommutative
  geometry,'' {\em Commun. Math. Phys.} {\bf 155} (1993)  205--218,
\href{http://arxiv.org/abs/hep-th/9209044}{{\tt hep-th/9209044}}.

\bibitem{Madore:1993br}
J.~Madore and J.~Mourad, ``{A Noncommutative extension of gravity},'' {\em Int.
  J. Mod. Phys.} {\bf D3} (1994)  221--224,
\href{http://arxiv.org/abs/gr-qc/9307030}{{\tt gr-qc/9307030}}.

\bibitem{Chamseddine:1996zu}
A.~H. Chamseddine and A.~Connes, ``The spectral action principle,'' {\em
  Commun. Math. Phys.} {\bf 186} (1997)  731--750,
\href{http://arxiv.org/abs/hep-th/9606001}{{\tt hep-th/9606001}}.

\bibitem{Chamseddine:2000si}
A.~H. Chamseddine, ``{Deforming Einstein's gravity},'' {\em Phys. Lett.} {\bf
  B504} (2001)  33--37,
\href{http://arxiv.org/abs/hep-th/0009153}{{\tt hep-th/0009153}}.

\bibitem{Grosse:1992bm}
H.~Grosse and J.~Madore, ``{A Noncommutative version of the Schwinger model},''
{\em Phys. Lett.} {\bf B283} (1992)  218--222.

\bibitem{Madore:1995ms}
J.~Madore, ``Linear connections on fuzzy manifolds,'' {\em Class. Quant. Grav.}
  {\bf 13} (1996)  2109--2120,
\href{http://arxiv.org/abs/hep-th/9506183}{{\tt hep-th/9506183}}.

\bibitem{Buric:2005yi}
M.~Buric and J.~Madore, ``A dynamical 2-dimensional fuzzy space,'' {\em Phys.
  Lett.} {\bf B622} (2005)  183--191,
\href{http://arxiv.org/abs/hep-th/0507064}{{\tt hep-th/0507064}}.

\bibitem{Buric:2006di}
M.~Buric, T.~Grammatikopoulos, J.~Madore, and G.~Zoupanos, ``Gravity and the
  structure of noncommutative algebras,'' {\em JHEP} {\bf 04} (2006)  054,
\href{http://arxiv.org/abs/hep-th/0603044}{{\tt hep-th/0603044}}.

\bibitem{Buric:2007hb}
M.~Buric, J.~Madore, and G.~Zoupanos, ``{The Energy-momentum of a Poisson
  structure},''
\href{http://arxiv.org/abs/arXiv:0709.3159 [hep-th]}{{\tt arXiv:0709.3159
  [hep-th]}}.

\bibitem{Aschieri:2005yw}
P.~Aschieri {\em et al.}, ``A gravity theory on noncommutative spaces,'' {\em
  Class. Quant. Grav.} {\bf 22} (2005)  3511--3532,
\href{http://arxiv.org/abs/hep-th/0504183}{{\tt hep-th/0504183}}.

\bibitem{Aschieri:2005zs}
P.~Aschieri, M.~Dimitrijevic, F.~Meyer, and J.~Wess, ``Noncommutative geometry
  and gravity,'' {\em Class. Quant. Grav.} {\bf 23} (2006)  1883--1912,
\href{http://arxiv.org/abs/hep-th/0510059}{{\tt hep-th/0510059}}.

\bibitem{Connes:1988ym}
A.~Connes, ``{The Action Functional in Noncommutative Geometry},''
{\em Commun. Math. Phys.} {\bf 117} (1988)  673--683.

\bibitem{Connes:1996gi}
A.~Connes, ``Gravity coupled with matter and the foundation of non- commutative
  geometry,'' {\em Commun. Math. Phys.} {\bf 182} (1996)  155--176,
\href{http://arxiv.org/abs/hep-th/9603053}{{\tt hep-th/9603053}}.

\bibitem{Madore:1996bb}
J.~Madore and J.~Mourad, ``Quantum space-time and classical gravity,'' {\em J.
  Math. Phys.} {\bf 39} (1998)  423--442,
\href{http://arxiv.org/abs/gr-qc/9607060}{{\tt gr-qc/9607060}}.

\bibitem{Cerchiai:2000uz}
B.~L. Cerchiai, G.~Fiore, and J.~Madore, ``{Geometrical techniques for the
  N-dimensional Quantum Euclidean Spaces},''
\href{http://arxiv.org/abs/math.qa/0002215}{{\tt math.qa/0002215}}.

\bibitem{Maceda:2003xr}
M.~Maceda, J.~Madore, P.~Manousselis, and G.~Zoupanos, ``Can noncommutativity
  resolve the big-bang singularity?,'' {\em Eur. Phys. J.} {\bf C36} (2004)
  529--534,
\href{http://arxiv.org/abs/hep-th/0306136}{{\tt hep-th/0306136}}.

\bibitem{Buric:2004rm}
M.~Buric and J.~Madore, ``Noncommutative 2-dimensional models of gravity,''
\href{http://arxiv.org/abs/hep-th/0406232}{{\tt hep-th/0406232}}.

\bibitem{Madore:1997ec}
J.~Madore, ``{On Poisson structure and curvature},'' {\em Rept. Math. Phys.}
  {\bf 43} (1999)  231--238,
\href{http://arxiv.org/abs/gr-qc/9705083}{{\tt gr-qc/9705083}}.

\bibitem{Hanlon:1992mn}
B.~E. Hanlon and G.~C. Joshi, ``{A Three generation Unified Model from Coset
  Space Dimensional Reduction},''
{\em Phys. Lett.} {\bf B298} (1993)  312--317.

\bibitem{Chatzistavrakidis:2008ii}
A.~Chatzistavrakidis, P.~Manousselis, and G.~Zoupanos, ``{Reducing the
  Heterotic Supergravity on nearly-Kahler coset spaces},''
\href{http://arxiv.org/abs/0811.2182}{{\tt arXiv:0811.2182 [hep-th]}}.

\end{thebibliography}
\providecommand{\href}[2]{#2}\begingroup\raggedright\endgroup

\listoffigures
\listoftables

\end{backmatter}
\end{document}